\documentclass[a4paper]{JHEP3}
\usepackage[centertags]{amsmath}
\usepackage{amssymb,amstext}
\usepackage{graphicx}
\usepackage{epsfig}
\usepackage[stable]{footmisc}
\usepackage{appendix}

\def\be{\begin{equation}}
\def\ee{\end{equation}}
\def\bea{\begin{eqnarray}}
\def\eea{\end{eqnarray}}
\def\ie{\begin{equation}\begin{aligned}}
\def\fe{\end{aligned}\end{equation}}

\newcommand{\m}{\mu}

\newcommand{\A}{{\alpha}}
\newcommand{\B}{{\beta}}
\newcommand{\C}{{\gamma}}

\newcommand{\da}{{\dot\alpha}}
\newcommand{\db}{{\dot\beta}}
\newcommand{\dc}{{\dot\gamma}}
\newcommand{\dd}{{\dot\delta}}

\makeatletter\@addtoreset{equation}{section}\makeatother

\newcommand{\vev}[1]{{\left< {#1} \right>}}

\newcommand{\tr}{{\rm tr\,}}
\newcommand{\cA}{{\mathcal A}}

\newcommand{\cN}{{\mathcal N}}
\newcommand{\cO}{{\mathcal O}}

\newcommand{\half}{\frac{1}{2}}
\newcommand{\Tr}{\text{Tr}}

\def\ie{\begin{equation}\begin{aligned}}
\def\fe{\end{aligned}\end{equation}}

\title{ABJ Triality: from Higher Spin Fields to Strings}

\author{Chi-Ming Chang$^a$, Shiraz Minwalla$^b$, Tarun Sharma$^b$, Xi Yin$^a$ \\
$^a$Center for the Fundamental Laws of Nature, 
Jefferson Physical Laboratory, Harvard University,
Cambridge, MA 02138, USA
\\
$^b$Dept. of Theoretical Physics, Tata Institute of Fundamental Research, Homi Bhabha Rd,
Mumbai 400005, India. \\
Email:\ \ {\bf cmchang@physics.harvard.edu, minwalla@theory.tifr.res.in, 
tarun@theory.tifr.res.in, xiyin@fas.harvard.edu}}

\abstract{ We demonstrate that a supersymmetric and parity violating version 
of Vasiliev's higher spin gauge theory in AdS$_4$ admits boundary conditions 
that preserve ${\cal N}=0,1,2,3,4$ or $6$ supersymmetries. In particular, 
we argue that the Vasiliev theory with $U(M)$ Chan-Paton and 
${\cal N}=6$ boundary condition is holographically dual to the 2+1 
dimensional $U(N)_k\times U(M)_{-k}$ ABJ theory in the limit of large 
$N,k$ and finite $M$. In this system all  bulk higher 
spin fields transform in the adjoint of the $U(M)$ gauge group, whose 
bulk t'Hooft coupling is $\frac{M}{N}$.  
Analysis of boundary conditions in Vasiliev theory 
allows us to determine exact relations between the parity breaking 
phase of Vasiliev theory and the coefficients of two and three point 
functions in Chern-Simons vector models at large $N$.  Our picture 
suggests that the supersymmetric Vasiliev theory can be obtained as a 
limit of type IIA string theory in AdS$_4\times \mathbb{CP}^3$, and that the 
non-Abelian Vasiliev theory at strong bulk 't Hooft coupling smoothly turn 
into a string field theory. The fundamental string is 
a singlet bound state of Vasiliev's higher spin particles held together 
by $U(M)$ gauge interactions. This is illustrated  by the thermal 
partition function of free ABJ theory on a two sphere at large $M$ and $N$ 
even in the analytically tractable free limit.
In this system the traces or strings of the low temperature phase  break  up 
into their Vasiliev particulate constituents at a $U(M)$ deconfinement phase 
transition of order unity. At a higher temperature
of order $T=\sqrt{\frac{N}{M}}$ Vasiliev's higher spin fields 
themselves break up into more elementary constituents at a $U(N)$ 
deconfinement temperature, in a process described in the bulk as 
black hole nucleation.}

\preprint{TIFR/TH/12-29}

\begin{document}

%

\section{Introduction and Summary}

It has long been speculated that the tensionless limit of string theory is a 
theory of higher spin gauge fields. One of the most important explicit and 
nontrivial construction of interacting higher spin gauge theory is Vasiliev's 
system in $AdS_4$. It was conjectured by Klebanov and Polyakov \cite{Klebanov:2002ja}, 
and by Sezgin and Sundell \cite{Sezgin:2002rt, Sezgin:2003pt}, that the parity 
invariant A-type and B-type Vasiliev theories 
are dual to 2+1 dimensional bosonic and fermionic $O(N)$ or $U(N)$ vector 
models in the singlet sector. Substantial evidence for these conjectures 
has been provided by comparison of three-point functions 
\cite{Giombi:2009wh, Giombi:2010vg}, and analysis of higher spin 
symmetries \cite{Giombi:2011rz, Giombi:2011ya, Maldacena:2011jn, 
Maldacena:2012sf}. 

It was noted in \cite{Aharony:2011jz, Giombi:2011kc} that, at large $N$, 
the free 
$O(N)$ and $U(N)$ 
theories described above each have a family of one parameter conformal 
deformations, corresponding to turning on a finite Chern-Simons level 
for the $O(N)$ or $U(N)$ gauge group. It was conjectured  
in \cite{Giombi:2011kc} that the bulk duals of the resultant 
Chern-Simons vector models is given by a one parameter
family of parity violating Vasiliev theories. In the bulk description 
parity is broken by a nontrivial phase in function 
$f$ in Vasiliev's theory that controls bulk interactions. This conjecture
appeared to pass some nontrivial checks \cite{Giombi:2011kc} but 
also faced some puzzling challenges \cite{Giombi:2011kc}. In this paper 
we will find significant additional evidence in support of the proposal 
of \cite{Giombi:2011kc} from 
the study of the bulk duals of supersymmetric vector Chern-Simons theories. 

The duality between Vasiliev theory and 3d Chern-Simons boundary 
field theories does not rely on supersymmetry, and, indeed, most 
studies of this duality have been carried out in the non-supersymmetric 
context. However it is possible to construct supersymmetric 
analogues of the Type A and type B bosonic Vasiliev theories 
\cite{Vasiliev:1992av, Vasiliev:1995dn, Vasiliev:1999ba, Sezgin:2003pt, 
Engquist:2002vr, Leigh:2003gk}.
With appropriate boundary conditions, these supersymmetric Vasiliev 
theories preserve all higher spin symmetries and are conjectured to be 
dual to free boundary supersymmetric gauge theories. In the spirit 
of \cite{Giombi:2011kc} it is natural to attempt to construct bulk duals 
of the one parameter set of interacting supersymmetric Chern-Simons vector 
theories obtained by turning on a finite level $k$ for the Chern-Simons terms 
(recall that Chern Simons coupled gauge fields are free only in the limit 
$k \to \infty$). Interacting supersymmetric Chern-Simons theories differ 
from their free 
counterparts in three ways. First, as emphasized above, their Chern-Simons 
level is taken to be finite.  According to the conjecture of  
\cite{Giombi:2011kc} this 
is accounted for by turning on the appropriate phase in Vasiliev's 
equations. Second the Lagrangian includes potential terms of the schematic 
form $\phi^6$ and Yukawa terms of the schematic form $\phi^2 \psi^2$, 
where $\phi$ and $\psi$ are fundamental and antifundamental scalars and 
fermions in the field theory. These terms may be regarded as double 
and triple trace deformations of the field theory; as is well known, the
effect of such terms on the dual bulk theory may be accounted for 
by an appropriate modification of boundary conditions \cite{Witten:2001ua}. Lastly, 
supersymmetric field theories with ${\cal N}=4$ and ${\cal N}=6$ supersymmetry
necessarily have two gauge groups with matter in the bifundamental. 
Such theories may be obtained by from theories with a single Chern-Simons 
coupled gauge group at level $k$ and fundamental matter by gauging a global symmetry
with Chern-Simons level $-k$. In the dual bulk theory 
this gauging is implemented by a modification of the boundary conditions 
of the bulk vector gauge field \cite{Witten:2003ya}.

These elements together suggest that it should be possible to find 
one parameter families of Vasiliev theories that preserve some 
supersymmetry upon turning on the parity violating bulk phase, 
if and only if one also modifies the boundary conditions of all 
bulk scalars, fermions and sometimes gauge fields in a coordinated way. 
In this paper we find that this is indeed the case. We are able to 
formulate one parameter families of parity violating Vasiliev 
theory (enhanced with Chan-Paton factors, see below)  that 
preserve ${\cal N}=0,1,2,3,4$ or $6$ supersymmetries depending 
on boundary conditions. In every case we 
identify conjectured dual Chern-Simons vector models dual to our bulk 
constructions.\footnote{A similar 
analysis of the breaking of higher spin symmetry by boundary conditions 
allows us to demonstrate that all deformations of type A or type B 
Vasiliev theories break all higher spin symmetries other than the 
conformal symmetry. We are also able to use this analysis to determine the 
functional form of the double trace part of higher spin currents
that contain a scalar field.} 

The identification of parity violating Vasiliev theory with prescribed 
boundary conditions as the dual of Chern-Simons vector models pass a number 
of highly nontrivial checks. By considering of boundary conditions alone, 
we will be able to determine the exact relation between the parity 
breaking phase $\theta_0$ of Vasiliev theory, and two and three point 
function coefficients of Chern-Simons vector models at large $N$. These 
imply non-perturbative relations among purely field theoretic quantities 
that are previously unknown (and presumably possible to prove by generalizing the 
computation of correlators in Chern-Simons-scalar vector model of 
\cite{OferUpcoming} 
using Schwinger-Dyson \footnote{See \cite{Giombi:2011kc} for these equations 
in the Chern-Simons fermion model.} equations to the supersymmetric theories). 
The results 
also agree with the relation between $\theta_0$ and Chern-Simons 't Hooft 
coupling $\lambda = N/k$ determined in \cite{Giombi:2011kc} by explicit 
perturbative computations at one-loop and two-loop order.

From a physical viewpoint, the most interesting Vasiliev theory presented
in this paper is the ${\cal N}=6$ theory. It was already 
suggested in \cite{Giombi:2011kc} that 
a supersymmetric version of the parity breaking Vasiliev theory 
in $AdS_4$ should be dual to the vector model limit of the ${\cal N}=6$ 
ABJ theory, that is, a $U(N)_k\times U(M)_{-k}$ Chern-Simons-matter 
theory in the limit of large $N,k$ but finite $M$. Since the ABJ 
theory is also dual to type IIA string theory in $AdS_4\times 
\mathbb{CP}^3$ with flat $B$-field, it was speculated that the 
Vasiliev theory must therefore be a limit of this string theory. 
The concrete supersymmetric ${\cal N}=6$ Vasiliev system presented 
in this paper allows us to turn the suggestion of \cite{Giombi:2011kc}
into a precise conjecture for a duality between three distinct 
theories that are autonomously well defined atleast in particular limits.  

The ${\cal N} =6$ Vasiliev theory, conjectured below to be dual to 
$U(N) \times U(M)$ $ABJ$ 
theory has many elements absent in more familiar bosonic Vasiliev systems. 
First theory is `supersymmetric' in the bulk. This means that 
all fields of the theory are functions of fermionic variables $\psi_i$
$(i= 1 \ldots 6)$  which obey Clifford algebra commutation relations 
$\{ \psi_i, \psi_j \}= 2 \delta_{ij}$ (all bulk fields are also functions 
of the physical spacetime variables $x_\mu$ ($\mu=1 \ldots 4$) as well 
as Vasiliev's twistor variables $y_\alpha$, $z_\alpha$, ${\bar y}_\da$, 
$\bar{z}_\db$, as in bosonic Vasiliev theory). Next the star product used 
in the bulk equations is the usual Vasiliev star product times 
matrix multiplication in an auxiliary $M\times M$ space. The physical 
effect of this maneuver is to endow the bulk theory with a $U(M)$ 
gauge symmetry under which all bulk fields transform in the adjoint.
Finally, for the reasons described above, interactions of the theory 
are also modified by a bulk phase, and bulk scalars, fermions and gauge 
fields obey nontrivial boundary conditions that depend on this phase. 

The triality between $U(N) \times U(M)$ ABJ theory, type IIA string theory on 
$AdS_4\times \mathbb{CP}^3$, and supersymmetric parity breaking Vasiliev 
theory may qualitatively be understood in the following manner. The 
propagating degrees of freedom of ABJ theory consist of bifundamental 
fields that we  denote by $A_i$ and antibifundamental fields that we will 
call $B_i$. A basis for the gauge singlet operators of the theory 
is given by the traces ${\rm Tr} (A_1 B_1 A_2 B_2 \ldots A_m B_m)$. As is well 
known from the study of ABJ duality, these single trace operators 
are dual to single string states. The basic `partons' (the $A$ and $B$ 
fields) out of which this trace is composed are held together 
in this string state by the `glue' of $U(N)$ and $U(M)$ gauge interactions. 

Let us now study the limit $M \ll N$. In this limit the glue that joins 
$B$ type fields to $A$ type fields (provided by the gauge group $U(M)$) 
is significantly weaker than the glue that joins $A$ fields to $B$ fields 
(this glue is supplied by $U(N)$ interactions). In this limit the 
trace effectively breaks up into $m$ weakly interacting particles $A_1 B_1$, 
$A_2 B_2$ ... $A_m B_m$. These particles, which transform in the adjoint of 
$U(M)$, are the dual to the $U(M)$ adjoint fields of the dual ${\cal N}=6$ 
Vasiliev theory. Indeed the spectrum of operators of field theory 
operators of the form $A B$ precisely matches the spectrum of bulk 
fields of the dual Vasiliev system.

If our picture is correct, the fields of Vasiliev's theory must bind
together to make up fundamental IIA strings as $M/N$ is increased. 
We now describe a qualitative way in which this might happen. 
The bulk Vasiliev theory has gauge coupling $g\sim 1/\sqrt{N}$, 
 It follows that the 
{\it bulk} 't Hooft coupling is $\lambda_{bulk} = g^2M\sim M/N$. In the 
limit $M/N\ll 1$, the bulk Vasiliev theory is effectively weakly coupled. 
As $M/N$ increases, a class of multi-particle states of higher spin 
fields acquire large binding energies due to interactions, and are 
mapped to the single closed string states in type IIA string theory.
Roughly speaking, the fundamental string of string theory is simply 
the flux tube string of the non abelian bulk Vasiliev theory. 
 
Note that although we claim a family of supersymmetric Vasiliev theory 
with Chan-Paton factors and certain prescribed boundary conditions is 
equivalent to string theory on $AdS_4$, we are {\it not} suggesting that 
Vasiliev's equations are the same as the corresponding limit of closed 
string field equations. Not all single closed string states are 
mapped to single higher spin particles; infact the only closed strings 
that are mapped to Vasiliev's particles are those dual to the operators 
of the form ${\rm Tr} AB$. Closed string field theory is the weakly interacting 
theory of the `glueball' bound states of the Vasiliev fields; it is not
a weakly interacting description of Vasiliev's fields themselves.

We have asserted above that the glue between $B$ and $A$ 
partons is significantly weaker than the glue between $A$ and $B$ 
partons in the limit $M \ll N$. This claim may be made quantitatively 
precise in a calculation in the ${\it free}$ ABJ theory with 
$\frac{M}{N}$ taken to be an arbitrary parameter. The computation in 
question is the partition function of free ABJ field theories on a sphere 
in the t'Hooft large $N$ and $M$ limit. We use the fact that
the path integral that computes this partition function, even in the 
limit $k \to \infty$, is not completely free \cite{Aharony:2003sx}. This 
$k=\infty$ path integral 
includes the effects of strong interactions between matter and the 
Polyakov line of $U(N)$ and $U(M)$ gauge fields. This computation of 
the partition function is a straightforward application of the techniques 
described in \cite{Aharony:2003sx}, but yields an interesting result 
(see Section \ref{abjfpf}, and see
\cite{Shenker:2011zf, Giombi:2011kc} for related earlier work in the 
context of models with fundamental matter). We find that the 
theory undergoes {\it two} phase transitions 
as a function of temperature. At low temperature the theory is in a
confined phase. This phase may be thought of as a gas 
of traces of the form ${\rm Tr}(A_1 B_1 A_2 B_2 \ldots A_m B_m)$, or, 
roughly, closed strings.  Upon raising the 
temperature the field theory undergoes a first order phase transition 
at a temperature of order unity. Above the phase transition temperature, 
group $U(M)$ deconfines while the group $U(N)$ continues to completely 
confine\footnote{Throughout this paper we assume without loss 
of generality that $M \leq N$.}   (we make this statement precise below.)  The intermediate temperature 
phase has an effective description in terms of the partition function of 
a $U(M)$ gauge theory whose effective matter degrees of freedom are 
simply the set of adjoint `mesons' of the form $ A_i B_j$. These 
adjoint degrees of freedom are deconfined. In other words the traces of the 
low temperature phase 
(dual to fundamental strings of ABJ theory) split up into a free gas of 
smaller - but not yet indivisible units, i.e. the fields of Vasiliev's 
theory. Upon further raising the temperature, the theory 
undergoes yet another phase transition, this time of 
third order. This transition occurs at a temperature of order 
$\sqrt{\frac{N}{M}}$ and is associated with the complete `deconfinement' 
of the gauge group $U(N)$. At temperatures much higher than the second 
phase transition temperature, the system may be thought of as a plasma 
of the bifundamental and anti-bifundamental letters $A_i$ and $B_j$. 
In other words the basic units, ${\rm Tr}(A_i B_j)$, of the intermediate 
temperature phase, split up into their basic building blocks in the high
temperature phase. This extreme high temperature phase is presumably 
dual to a black hole in the bulk theory. \footnote{In the very high temperature 
limit, this phase has recently been studied in closely related 
supersymmetric Chern Simons theories even away from the free limit 
\cite{Jain:2012cs} (generalizing earlier computations in nonsupersymmetric theories 
in \cite{Giombi:2011kc}.}In the special case $M=N$
the intermediate phase never exists; the system directly transits from 
the string to the black hole phase. The fact that the $U(M)$ deconfinement
temperature is much smaller than the $U(N)$ deconfinement temperature 
demonstrates that the glue between $B$ and $A$ type partons is much 
weaker than than between $A$ and $B$ type partons. Our computations 
also strongly suggests that the string dual to ABJ theory has a new 
finite temperature phase - one composed of a gas of Vasiliev's particles - 
even at finite values of $\lambda$. 

Let us note a curious aspect of the conjectured duality between Vasiliev's
theory and ABJ theory. The gauge groups $U(N)$ and $U(M)$ appear on an even footing 
in the ABJ field theory. In the bulk Vasiliev description, however, 
the two gauge groups play a very different role. 
The gauge group $U(M)$ is manifest as a gauge symmetry in the bulk. However 
$U(N)$ symmetry is not manifest in the bulk (just as the $U(N)$ symmetry is not manifest 
in the bulk dual of ${\cal N}=4$ Yang Mills); the dynamics of this gauge 
group that leads to the emergence 
of the background spacetime for Vasiliev theory. The deconfinement transition for $U(M)$ is simply 
a deconfinement transition of the adjoint bulk degrees of freedom, while the deconfinement transition 
for $U(N)$ is associated with the very different process of 
`black hole formation'. If our proposal for the dual description is correct, 
the gauged Vasiliev theory must enjoy an $N \leftrightarrow M$ symmetry, 
which, from the bulk viewpoint is a sort of level -- rank duality. Of course
even a precise statement for the claim of such a level rank duality
only makes sense if Vasiliev theory is well defined `quantum mechanically'
(i.e. away from small $\frac{M}{N}$) at least in the large $N$ limit.

We have noted above that Vasiliev's theory should not be identified with 
closed string field theory. There may, however, be a sense in which 
it might be thought of as an open string field theory. 
We use the fact that there is an alternative 
way to engineer Chern-Simons vector models using string theory 
\cite{Gaiotto:2009tk}, that is by adding $N_f$ D6-branes wrapped 
on $AdS_4\times\mathbb{RP}^3$ inside the $AdS_4\times\mathbb{CP}^3$, 
which preserves ${\cal N}=3$ supersymmetry and amounts to adding 
fundamental hypermultiplets of the $U(N)_k$ Chern-Simons gauge group. 
In the ``minimal radius" limit where we send $M$ to zero, with flat $B$-field 
flux ${1\over 2\pi\alpha
}\int_{\mathbb{CP}^1}B = {N\over k}+{1\over 2}$, the geometry is entirely 
supported by the $N_f$ D6-branes \cite{Aharony:2009fc}.\footnote{We thank Daniel Jafferis for 
making this important suggestion and O. Aharony for related discussions.} 
This type IIA open+closed string theory 
is dual to ${\cal N}=3$ Chern-Simons vector model with $N_f$ hypermultiplet 
flavors. The duality suggests that the open+closed string field theory of 
the D6-branes reduces to precisely a supersymmetric Vasiliev theory in the 
minimal radius limit. Note that unlike the ABJ triality, here the open 
string fields on the D6-branes and the nonabelian higher spin gauge fields 
in Vasiliev's system both carry $U(N_f)$ Chan-Paton factors, and we expect 
one-to-one correspondence between single open string states and single higher 
spin particle states.

\section{Vasiliev's higher spin gauge theory in $AdS_4$ and its 
supersymmetric extension}\label{bulk}

The Vasiliev systems that we that we study in this paper 
are defined by a set of bulk equations of motion together with boundary 
conditions on the bulk fields. In this section we review the structure 
of the bulk equations. We turn to the consideration of boundary 
conditions in the next section. 

In this section we first present a detailed review of 
bulk equations of the `standard' Vasiliev theory. We then describe 
nonabelian and supersymmetric extensions of these equations.
Throughout this paper we work with the so-called 
non-minimal version of Vasiliev's equations, which describe the interactions 
of a field of {\it each}  non-negative integer spin $s$ in $AdS_4$. 
Under the AdS/CFT correspondence 
non-minimal Vasiliev equations are conjectured to be dual to gauged  $U(N)$ 
Chern-Simons-matter boundary theories.\footnote{
The non minimal equations admit a consistent truncation to the so-called 
minimal version of Vasiliev's equations; this truncation projects out the 
gauge fields for odd spins and are conjectured to supply the 
dual to $SO(N)$ Chern-Simons boundary theories.} 

There are exactly two `standard' non-minimal Vasiliev theories that preserve 
parity symmetry. These are the type A/B theories, which are conjectured to be 
dual to bosonic/fermionic $SU(N)$ vector models, restricted to the $SU(N)$-singlet sector. Parity invariant Vasiliev theories are particular examples of 
a larger class of generically parity violating Vasiliev theories. These 
theories appear to be labeled by a real even function of one 
real variable. In subsection \ref{bos} we present a review of 
these theories.  It was conjectured in 
\cite{Giombi:2011kc} that a class of these parity violating theories 
are dual to $SU(N)$ Chern-Simons vector models. 

In subsection \ref{nonabelian} we then present a straightforward nonabelian 
extension of Vasiliev's system, by introducing $U(M)$ Chan-Paton factors into 
Vasiliev's star product. The result of this extension is to promote the 
bulk gauge field to a $U(M)$ gauge field; all other bulk fields transform 
in the adjoint of $U(M)$. The local gauge transformation parameter 
of Vasiliev's theory is also promoted to a local $M \times M$ matrix field 
that transforms in the adjoint of $U(M)$. The nature of the 
boundary CFT dual to the non abelian Vasiliev theory depends on boundary 
conditions. With `standard'  magnetic type boundary 
conditions for all gauge fields (that set prescribed values for the field 
strengths restricted to the boundary) the dual boundary CFT  
is obtained simply by coupling $M$ copies of (otherwise non interacting) 
matter multiplets to the same boundary Chern-Simons gauge field.  The 
boundary theory has a `flavour' $U(M)$ global symmetry that 
acts on the $M$ identical matter multiplets.

In subsection \ref{fermion} we then introduce the so called $n$-extended 
supersymmetric Vasiliev theory (generalizing the special cases 
studied earlier in \cite{Vasiliev:1992av, Vasiliev:1995dn, Vasiliev:1999ba, Sezgin:2003pt, Engquist:2002vr}). 
The main idea is to 
enhance Vasiliev's fields to functions of $n$ fermionic fields 
$\psi_i$ ($i = 1 \ldots n$; we assume $n$ to be even) which obey a Clifford 
algebra\footnote{We emphasize that $n$ should not be confused with the number of globally conserved 
supercharges $4 {\cal N}$ (equivalently $4 {\cal N}$ is the number of 
supercharges in the superconformal algebra of the dual three-dimensional CFT). 
$n$ characterizes only the local structure of Vasiliev's equations of motion. 
${\cal N}$ on the other hand depends on the choice of boundary condition 
for bulk fields of spin $0$, $1/2$ and $1$. As we will see is 
${\cal N} \leq 6$  for parity violating Vasiliev theories, as expected from 
the dual CFT${}_3$ ($n$, or course, can be arbitrarily large
). }. This extension promotes the usual Vasiliev's fields to 
$2^{\frac{n}{2}} \times 2^{\frac{n}{2}}$ 
dimensional matrices (or operators) that act on the $2^{\frac{n}{2}}$ 
dimensional representation of the Clifford algebra. The local 
Vasiliev gauge transformations are also promoted to functions 
of $\psi_i$, and so $2^{\frac{n}{2}} \times 2^{\frac{n}{2}}$ matrices or 
operators\footnote{The bulk equations of motion the $n$ extended supersymmetric Vasiliev 
theory is identical 
to the $n=2$ theory extended by $U(2^{\frac{n}{2} -1})$ Chan Paton 
factors. However,  the language of $n$ extended supersymmetric Vasiliev theory
is more convenient when the boundary conditions of the problem 
break part of this $U(2^{\frac{n}{2} -1})$ symmetry, as will be the case 
later in this paper. }. Half of the resultant fields (and 
gauge transformations) are fermionic; the other half are bosonic.

\subsection{The standard parity violating bosonic Vasiliev theory}\label{bos}

In this section we present the `standard' non minimal Vasiliev equations, 
allowing, however, for parity violation. 

\subsubsection{Coordinates}

In Euclidean space the
fields of Vasiliev's theory are functions of a collection of 
bosonic variables 
$(x,Y,Z)=(x^\mu, y^\A,\bar y^\da, z^\A, \bar z^\da)$. 
$x^\mu$  ($\mu = 1 \ldots 4$) are an arbitrary set of coordinates on the 
four dimensional spacetime manifold. $y^\A$ and $z^\A$ are spinors under 
$SU(2)_L$ while ${\bar y}^\da$ and ${\bar z}^\da$ are spinors under 
a separate $SU(2)_R$.  As we will see below, Vasiliev's equations enjoy 
invariance under {\it local} (in spacetime) $SO(4)=SU(2)_L \times SU(2)_R$ 
rotations of $y^\A$, 
$z^\A$, ${\bar y}^\da$ and ${\bar z}^\da$. This local $SO(4)$ rotational 
invariance, which, as we will see below is closely related to the 
tangent space symmetry of the first order formulation of general relativity, 
is only a small part of the much larger gauge symmetry of Vasiliev's theory.

\subsubsection{Star Product}

Vasiliev's equations are formulated in terms of a star product. 
This is just the usual local product in coordinate space; whereas 
in auxiliary space it is given by 
\ie \label{defstar}
& f(Y,Z)*g(Y,Z)
\\
& = f(Y,Z) \exp\left[ \epsilon^{\A\B} \left( \overleftarrow\partial_{y^\A} + \overleftarrow\partial_{z^\A} \right) \left( \overrightarrow\partial_{y^\B} - \overrightarrow\partial_{z^\B} \right) + \epsilon^{\da\db} \left( \overleftarrow\partial_{y^\da} + \overleftarrow\partial_{z^\da} \right) \left( \overrightarrow\partial_{y^\db} - \overrightarrow\partial_{z^\db} \right)  \right] g(Y,Z)
\\
&= \int d^2u d^2v d^2\bar u d^2\bar v e^{u^\A v_\A + \bar u^\da \bar v_\da} f(y+u,\bar y + \bar u,z+u,\bar z+\bar u)
g(y+v,\bar y+\bar v,z-v,\bar z-\bar v).
\fe
In the last line, the integral representation of the star product is defined by the contour for $(u^\A, v^\A)$ along $e^{\pi i/4}\mathbb{R}$ in the complex plane, and $(\bar u^\da, \bar v^\da)$ along the contour $e^{-\pi i/4}\mathbb{R}$. 
It is obvious from the first line of \eqref{defstar} that 
$1*f=f*1=f$; this fact may be used to set the normalization of the integration measure 
in the second line. The star product is associative but non commutative; 
in fact it may be shown to be isomorphic to the usual Moyal star product 
under an appropriate change of variables. In Appendix \ref{star} we describe 
our conventions for lowering spinor indices and present
some simple identities involving the star product. 

Below we will make extensive use of the so called Kleinian operators 
$K$ and $\overline K$ defined as
\ie
K = e^{z^\A y_\A}, ~~~~\overline K = e^{\bar z^\da \bar y_\da}.
\fe
They have the property (see Appendix \ref{star} for a proof)
\ie \label{kprop}
&K*K=\overline K*\overline K=1,~~~~~\\
&K*f(y,z,\bar y,\bar z)*K = f(-y,-z,\bar y,\bar z),~~~\overline K*f(y,z,\bar y,\bar z)*\overline K = f(y,z,-\bar y,-\bar z).
\fe

\subsubsection{Master fields}

Vasiliev's master fields consists of an $x$-space 1-form 
$$W=W_\mu dx^\mu,$$ 
a $Z$-space 1-form 
$$S=S_\A dz^\A + S_\da d\bar z^\da,$$ 
and a scalar $B$, all of which depend on spacetime as well as the internal twistor coordinates which we denote collectively as $(x,Y,Z)=(x^\mu, y^\A,\bar y^\da, z^\A, \bar z^\da)$. It is sometimes convenient to write $W$ and $S$ together as a 1-form on $(x,Z)$-space
$${\cal A}=W+S=W_\mu dx^\mu + S_\A dz^\A + S_\da d\bar z^\da.$$ 
${\cal A}$ will be regarded as a gauge connection with respect to the 
$*$-algebra. 

We also define
\ie
{\hat S}&= S - \half z_\alpha dz^\alpha - \half {\bar z}_{\da} d{\bar z}^\da,\\
\hat{\cal A}&=W + {\hat S} 
= {\cal A} - \half z_\A dz^\A - \half \bar z_\da d\bar z^\da
=W_\mu dx^\mu 
+(-\half z_\alpha +S_\alpha) dz^\alpha +(-\half{\bar z}_{{\dot \alpha}}+S_{{\dot \alpha}})
d{\bar z}^{{\dot \alpha}}.
\fe

Let $d_x$ be the exterior derivative with respect to the spacetime coordinates 
$x^\mu$ and denote by $d_Z$ the exterior derivative with respect to the twistor
variables $(z^\alpha,\bar z^\da)$. We will write $d=d_x+d_Z$.
We will also find it useful to define the field strength
\begin{equation}\label{fs}\begin{split}
{\cal F}&=d_x \hat{\cal A}+ \hat{\cal A} \star \hat{\cal A}\\
        &=\left( d_x W + W*W \right) + 
           \left(d_x \hat S+\{W, \hat S \}_*\right) +
          \left(\hat S * \hat S \right).
\end{split}
\end{equation}
Note also that 
\begin{equation}
{\hat S} *{\hat S}= d_zS + S*S + \frac{1}{4} \left( \epsilon_{\alpha \beta} dz^\alpha dz^\beta 
     + \epsilon_{\da\db} d{\bar z}^\da d{\bar z}^\db \right).
\end{equation}

\subsubsection{Gauge Transformations}

Vasiliev's master fields transform under a large set of gauge symmetries. 
We will see later that the $AdS_4$ vacuum solution partially Higgs or 
breaks this gauge symmetry group down to a subgroup of large 
gauge transformations - either the higher spin symmetry group or the 
conformal group depending on boundary conditions. 

Infinitesimal gauge transformations
are generated by an arbitrary real function $\epsilon(x,Y,Z)$. 
By definition under gauge transformations
\begin{equation}\begin{split}\label{gt}
\delta \hat{\cal A}&= d_x \epsilon +\hat{\cal A} * \epsilon-
\epsilon * \hat{\cal A},\\
\delta B & = - \epsilon* B + B*\pi(\epsilon).\\
\end{split}
\end{equation}
In other words the 1-form master field 
transforms as a gauge connection under the star algebra while 
$B$ transforms as a `twisted' adjoint field.  The operation $\pi$ 
that appears in \eqref{gt} is defined as follows 
$$
\pi \left(y, z, dz, {\overline y}, {\overline z}, d{\overline z} \right)
=\left(-y, -z, -dz, {\overline y}, {\overline z}, d{\overline z} \right)
$$
Since $\epsilon$ does not involve differentials in $(z,\bar z)$, 
the action of $\pi$ on $\epsilon$ is equivalent to conjugation by $K$, 
namely $\pi(\epsilon) = K* \epsilon* K$. ($\pi$ 
acting on a 1-form in $(z_\A, \bar z_\da)$ acts by  
conjugation by $K$ together with flipping the sign of $dz$).

It follows from \eqref{gt} that the field strength ${\cal F}$ 
( and so each of the three brackets on the RHS of 
the second line of \eqref{fs}) transform in the adjoint. The same 
is true of $B*K$.
\begin{equation}\begin{split}
\delta {\cal F}&= [{\cal F}, \epsilon]_*,\\
\delta (B\star K) & = - \epsilon* (B * K) + (B*K) * \epsilon, \\
\end{split}
\end{equation}
When expanded in components the first line of \eqref{gt} implies that
\begin{equation}\begin{split}\label{gtc}
\delta W_\mu&=\partial_\mu \epsilon + W_\mu * \epsilon-
\epsilon * W_\mu, \\
\delta {\hat S}_\alpha&={\hat S}_\alpha * \epsilon-
\epsilon * {\hat S}_\alpha. \\
\end{split}
\end{equation}
In terms of unhatted variables, 
\begin{equation} \label{uhat} \begin{split}
\delta {\cal A}&= d \epsilon +  {\cal A}* \epsilon-
\epsilon * {\cal A}, \\
\delta S_\alpha&=\frac{\partial \epsilon}{\partial z^\alpha} 
+ S_\alpha * \epsilon-
\epsilon * S_\alpha. \\
\end{split}
\end{equation}

\subsubsection{Truncation}

The following truncation is imposed on the master fields and gauge 
transformation parameter $\epsilon$. Define 
$$R=K\overline K.$$ 
We require
\ie\label{Rtrunc}
{}[R, W]_* = \{R, S\}_*=[R, B]_*=[R, \epsilon]_*=0.
\fe
More explicitly, this is the statement that $W_\mu$, $B$ and $\epsilon$ 
are even functions of $(Y,Z)$ whereas $S_\A, S_\da$ are odd in $(Y,Z)$,
\begin{equation}\label{proj} \begin{split}
& W_\mu(x,y,\bar y,z,\bar z) = W_\mu(x,-y,-\bar y,-z,-\bar z),\\
& S_\A(x,y,\bar y,z,\bar z) = -S_\A(x,-y,-\bar y,-z,-\bar z),\\
& S_\da(x,y,\bar y,z,\bar z) = -S_\da(x,-y,-\bar y,-z,-\bar z),\\
& B(x,y,\bar y,z,\bar z) = B(x,-y,-\bar y,-z,-\bar z).\\
& \epsilon (x,y,\bar y,z,\bar z) = \epsilon(x,-y,-\bar y,-z,-\bar z).
\end{split}
\end{equation}
A physical reason for the imposition of this truncation is the spin 
statistics theorem. As the physical fields of Vasiliev's theory are
all commuting, they must also transform in the vector (rather than spinor)
conjugacy class of the $SO(4)$ tangent group; the projection \eqref{proj} 
ensures that this is the case. One might expect from this remark that 
the consistency of Vasiliev's equations requires this truncation; we will 
see explicitly below that this is the case.

\subsubsection{Reality Conditions}

It turns out that Vasiliev's master fields admit two consistent projections 
that may be used to reduce their number of degrees of freedom. These two 
projections are a generalized reality projection (somewhat analogous to the 
Majorana
condition for spinors) and a so called `minimal' truncation (very loosely 
analogous to a chirality truncation for spinors). These two truncations 
are defined in terms of two natural operations defined on the master 
field; complex conjugation and an operation defined by the symbol 
$\iota$. In this subsection we first define these two operations, and then 
use them to define the generalized reality projection. We will also briefly 
mention the minimal projection, even though we will not use the later in 
this paper.

Vasiliev's fields master fields admit a straightforward 
complex conjugation operation, ${\cal A}\to {\cal A}^*$, 
defined by complex conjugating each of the 
component fields of Vasiliev theory and also the spinor variables $Y,Z$\footnote{As complex conjugation
of $SO(3,1)$ interchanges left and right moving spinors, our definition of 
complex conjugation (the analytic continuation of the Lorentzian 
notion) must also have this property.}
\ie
(y^\A)^*=\bar y_{\dot \A},~~~~(\bar y_{\dot\A})^*= y^{ \A},~~~~(z^\A)^*=\bar z_{\dot \A},~~~~(\bar z_{\dot\A})^*= z^{ \A}.
\fe
It is easily verified that 
\begin{equation}\label{cco}
\left( M*N \right)^*=M^* * N^*
\end{equation}
where $M$ is an arbitrary $p$ form and $N$ and arbitrary $q$ form. 
In other words complex conjugation commutes with the star and wedge product, 
without reversing the order of either of these products. Note also that 
the complex conjugation operation squares to the identity.

We now turn to the definition of the operation $\iota$;  this operation 
is defined by  
\ie \label{iota}
\iota: ~(y,\bar y,z,\bar z,dz,d\bar z) \to (iy, i\bar y,-iz,-i\bar z,-idz,-id\bar z),
\fe
The signs in \eqref{iota} are chosen\footnote{Changing the RHS of 
\eqref{iota} by an overall sign makes no difference to fields that 
obey \eqref{proj}} to ensure 
\begin{equation}\label{ipr}
\iota(f *g)=\iota(g)* \iota(f)
\end{equation}
(see \eqref{iotaprop} for a proof).
In other words $\iota$ reverses the order of the $*$ product. 
Note however that $\iota$ by definition does not affect the order of wedge 
products of forms. As a consequence $\iota$ picks up an extra minus sign 
when acting on the product of two oneforms
$$ \iota(C*D)=- \iota(D)*\iota(C)$$
(see \eqref{iotaform} for a proof; 
the same equation is true if $C$ is a $p$ form and $D$ a $q$ form 
provided $p$ and $q$ are both odd; if atleast one of $p$ and $q$ is 
even we have no minus sign).

We now define the generalized reality projection that we will require 
Vasiliev's master fields to obey throughout this paper (this projection 
defines the non-minimal Vasiliev theory which we study through this paper). 
The projection is defined by the conditions 
\begin{equation} \label{RCD} 
\iota(W)^*=-W,~~~\iota(S)^* = -S,~~~\iota(B)^* = \overline K*B* \overline K =  K*B*K
\end{equation}
The equality of the two different expressions supplied for 
$\iota(B)^*$ in \eqref{RCD} follows upon using the fact $B$ commutes with 
$R=K \overline K$ (see \eqref{Rtrunc}). 

It is easily verified that \eqref{RCD} implies that  
\begin{equation} \label{et}
\iota \left( {\cal F} \right)^*=-{\cal F}
\end{equation}
(see \eqref{fsra} for an expansion in components) and that 
\begin{equation}\label{bsn}
\iota(B*K)^* =B*{\overline K}, ~~~\iota(B*{\overline K})^*=B*{K} .
\end{equation}

\eqref{RCD}
may be thought of as a combination of two separate projections. 
The first is the `standard' reality projection (see \eqref{nreal}).
The second is the `minimal truncation'\eqref{mintrunc}. As discussed in 
Appendix \ref{proj}, it is consistent to simultaneously impose invariance 
of Vasiliev's master field under both these projections. This operation 
defines the minimal Vasiliev theory (dual to $SO(N)$ Chern-Simons field 
theories). We will not study the minimal theory in this paper.

\subsubsection{Equations of motion}

Vasiliev's gauge invariant equations of motion take the form
\ie\label{veqns}
& {\cal F}= d_x \hat {\cal A} + \hat {\cal A}*\hat {\cal A} = f_*(B*K) dz^2 + \overline f_*(B*\overline K) d\bar z^2,
\\
&d_x B + \hat{\cal A}*B-B*\pi(\hat{\cal A})=0 .
\fe
where $f(X)$ is a holomorphic function of $X$, $\overline f$ its complex conjugate, 
and $f_*(X)$ the corresponding $*$-function of $X$. Namely, $f_*(X)$ is defined by 
replacing all products of $X$ in the Taylor series of $f(X)$ by the corresponding star products. 

Note that both sides of the first of \eqref{veqns} are gauge adjoints, 
while the 
second line of that equation transforms in the twisted adjoint.
In  Appendix \ref{veq} we have demonstrated that
 the second equation of \eqref{veqns} may be derived 
from the first  (assuming that $f(X)$ is a non-degenerate function) 
using the Bianchi identity
\begin{equation} \label{bianchi}
d_x{\cal F}+[A,{\cal F}]_*=0
\end{equation} 
In Appendix \ref{veq} we have also expanded Vasiliev's equations in components
to clarify their physical content. As elaborated in \eqref{veqnsb}
and \eqref{veqnsc},  it follows from \eqref{veqns} that the field 
strength $d W+ W*W$ is flat and that the adjoint fields $B*K$, $S_\alpha$ and 
$S_{\da}$ are covariantly constant. In addition, various components of
these adjoint fields commute or anticommute with each other under the 
star product  (see Appendix \ref{commut} for a listing). The fields 
${\hat S}_\A$ and ${\hat S}_\B$, however, fail to commute with each 
other; their commutation relations are given by 
\begin{equation}\begin{split}\label{St}
[\hat{S}_\alpha , \hat{S}_\beta]_* &= \epsilon_{\alpha \beta} f_*(B*K)\\
[\hat{S}_\da , \hat{S}_\db]_* &= \epsilon_{\da \db} {\bar f}_*(B*\bar{K})\\
\end{split}
\end{equation}

Using various formulae presented in the Appendix (see e.g. \eqref{fsr})
 it is easily verified that the Vasiliev equations, (expanded in the 
Appendix as \eqref{veqnsb} and \eqref{veqnsc}) map to themselves under the 
reality projection \eqref{RCD}. The same is true of the minimal truncation 
projection.

\subsubsection{Equivalences from field redefinitions}

Vasiliev's equations are characterized by a single complex holomorphic 
function $f$. In this subsection we address the following question: to what 
extent to different functions $f$ label different theories? 

Any field redefinition that preserves the gauge and Lorentz 
transformation properties of all fields, but changes the form of $f$ clearly 
demonstrates an equivalence of the theories with the corresponding choices 
of $f$. 
The most general field redefinitions consistent with gauge and Lorentz
transformations and the form of Vasiliev's equations are 
\begin{equation} \label{frd}\begin{split}
& B\to g_*(B*K)*K\\
& \widehat S_{z} \equiv (-\half z_\A+S_\A)dz^\A \to \widehat S_{z}*h_*(B*K),\\
& \widehat S_{\bar{z}}\equiv (-\half \bar z_\da + \bar S_\da)*d\bar z^\da \to 
  \widehat S_{\bar{z}}*{\tilde h}_*(-B *\overline K).
\end{split}
\end{equation}
Several comments are in order. First note that the field redefinitions above obviously
preserve form structure and gauge transformations properties. 
In particular these redefinitions preserve the fact that 
$B*K$, $S_z$ and $S_{\overline z}$ transform in the adjoint representation of the gauge group. 
Second the field redefinitions above are purely holomorphic (e.g. 
$g_*$ is a function only of $B*K$ but not of $B*{\overline K}$). It is 
not difficult to convince oneself that this is necessary in order to 
preserve the holomorphic form of Vasiliev's equations. Finally we have 
chosen to multiply the redefined functions $S_z$ and $S_{\overline z}$ 
with functions from the right rather than the left. There is no lack 
of generality in this, however, as 
\ie \label{scomut}
& \widehat S_{z}*h_*(B*K) = h_*(-B*K)* \widehat S_{z},~~~~ \widehat S_{z}* \overline h_*(B*\overline K) = \overline h_*(B* \overline K)* \widehat S_{z},
\\
& \widehat S_{\overline{z}}*h_*(B*K) = h_*(B*K)* \widehat S_{\overline{z}},~~~~ \widehat S_{\overline{z}}* \overline h_*(B*\overline K) = \overline h_*(-B* \overline K)* \widehat S_{\overline{z}},
\fe
(\eqref{scomut} follows immediately from \eqref{commut} derived in the 
Appendix). 
Finally, we have inserted a minus sign into the argument of the function 
${\tilde h}$ for future convenience.

The reality conditions \eqref{RCD} impose constraints on the functions 
$g$, $h$ and $\tilde h$. It is not difficult to verify that 
$g$ is forced to be an {\sl odd real} function $g(X)$. $g(X)$ is 
forced to be odd because the complex conjugation operation turns 
$K$ into $\overline K$. When $g$ is odd, however, the truncation 
\eqref{Rtrunc} may be used to turn $\overline K$ back into $K$. 
For instance, with 
$g_*(X)=g_1 X + g_3 X*X*X+\cdots$, the field redefinition is
\ie
B\to g_1 B + g_3 B*K*B*K*B + \cdots
\fe
The RHS is still real because $K*B*K=\overline K*B*\overline K$ 
(it would not be real if $g(X)$ were not odd).

In order to examine the constraints of \eqref{RCD} on the functions 
$h$ and ${\tilde h}$ note that 
\begin{equation}
\begin{split}
\iota( S_z * h(B*K) + S_{\overline{z}}* {\tilde h}_*(-B*\overline K))^* 
& = {\overline h}(B*{\overline K})*(-S_{\overline z})+ {\overline {\tilde h}}
(-B*K)*(-S_z) \\
& = - \left( S_{\overline z}*{\overline h}(-B*{\overline K})+ S_z* 
{\overline {\tilde h
}}(B*K) \right) \\
\end{split}
\label{iinv}
\end{equation}
 (where in the last step we have used \eqref{scomut}). 
 It follows that the redefined function ${\widehat S}$ obeys the reality 
condition \eqref{RCD} if and only if 
$${\tilde h}= {\overline h}$$
where ${\overline h}$ is the complex conjugate of the function $h$. 

The effect of the field redefinition of $B$ is simply to permit a 
redefinition of the argument of the function $f$ in Vasiliev's equations 
by an arbitrary odd real function. The effect of the field redefinition 
of ${\widehat  S}$ may be deduced as follows. The $dx^\mu\wedge dx^\nu$ 
component of Vasiliev's - the assertion 
that $W$ is a flat connection (see \eqref{veqnsb}) - is clearly 
preserved by this field redefinition. The $dx\wedge dZ$ components 
of the equation asserts that
${\widehat S}_z$ and ${\widehat S}_{\overline z}$ are covariantly constant. 
As $B*K$ and $B*{\overline K}$ are also covariantly constant 
(see \eqref{veqnsc}) the redefinition \eqref{frd} clearly preserves 
this equation as well. However the $dZ^2$ components of the equations become
\ie
& \widehat S_{z}*h_*(B*K)* \widehat S_{z}*h_*(B*K) = f_*(B*K)dz^2 ,
\\
&\left\{\widehat S*h_*(B*K), \widehat S_{\overline{z}}*\overline h_*(-B*\overline K)\right\}_*=0,
\\
& \widehat S_{\overline{z}}*\overline h_*(-B*\overline K)* \widehat S_{\overline{z}}*\overline h_*(-B*\overline K) = \overline f_*(B*\overline K)d\bar z^2.
\fe
Using \eqref{scomut} and the fact that $B*K$ commutes with $B*\overline{K}$ 
(this is obvious as $K$ and ${\overline K}$ commute), these equations may be 
recast as
\ie \label{hmp}
& h_*(-B*K)* \left( \widehat S_{z}* \widehat S_{z} \right)*h_*(B*K) = f_*(B*K)dz^2 ,
\\
&h_*(-B*K)* \left( \left\{\widehat S, \widehat S_{\overline{z}}\right\}_* 
\right)* \overline h_*(-B*\overline K)=0,
\\ & \overline h_*(B*\overline K)*  \left( \widehat S_{\overline{z}}* 
\widehat S_{\overline{z}} \right) *\overline h_*(-B*\overline K)
=\overline f_*(B*\overline K)d\bar z^2.
\fe
\eqref{hmp} is precisely the $dZ^2$ component of the 
Vasiliev equation (the third equation in  \eqref{veqnsb} ) with 
the replacement
\ie
f_*(X) \to h_*(-X)^{-1}*f_*(X)* h_*(X)^{-1},
\fe
or simply $f(X)\to h(X)^{-1} h(-X)^{-1} f(X)$.

So we see that the theory is really defined by $f(X)$ up to a change of 
variable $X\to g(X)$ for some odd real function $g(X)$ and multiplication 
by an invertible holomorphic even function. 
Provided that the function $f(X)$ admits a power series expansion about $X=0$ 
and that $f(0)\neq 0$,\footnote{This condition can probably be weakend, but 
cannot be completely removed. For example if $f(X)$ is an odd function, it 
is easy to convince oneself that it cannot be cast into the form 
\eqref{fxsol}. In this paper we will be interested in the Vasiliev duals 
to field theories. In the free limit, the dual Vasiliev theories to the 
field theory in question are given by $f(X)$ of the form \eqref{fxsol} 
with $\theta=0$. It follows that, atleast in a power series in the field 
theory coupling, the Vasiliev duals to the corresponding field theories 
are defined by an $f(X)$ that can be put in the form \eqref{fxsol}. }
in Appendix \ref{redef} we demonstrate that we can  can use these field 
redefinitions 
to put $f(X)$ in the form
\ie\label{fxsol}
f(X) = {1\over 4}+X \exp(i\theta(X))
\fe
where $\theta(X)=\theta_0 + \theta_2 X^2 + \cdots$ is an arbitrary real 
even function. 

Ignoring the special cases for which $f(X)$ cannot be cast into the form 
\eqref{fxsol}, the function $\theta(X)$ determines the general parity-violating Vasiliev theory.

\subsubsection{The AdS solution}

While Vasiliev's system is formulated in terms of a set of background 
independent equations, the perturbation theory is defined by expanding 
around the $AdS_4$ vacuum. In order to study this solution  it is useful 
to establish some conventions. Let $e_0^a$ and $w_0^{ab}$ ($a, b=1 \ldots 4$)
denote the usual vielbein and spin connection one-forms on any space 
(the index $a$ transforms under the vector representation of the tangent 
space $SO(4)$). We define the corresponding bispinor objects  
\begin{equation}\label{trans}
e_{\alpha {\dot \beta}}= \frac{1}{4} e^a \sigma^a_{\alpha {\dot \beta}},~~~~
w_{\alpha \beta}=\frac{1}{16} w^{ab} \sigma^{ab}_{\alpha \beta},~~~~
w_{\da \db}=-\frac{1}{16} w^{ab} {\bar \sigma}^{ab}_{\da \db}.
\end{equation}
(see Appendix \ref{spconv} for definitions of the $\sigma$ matrices that 
appear in this equation.) Let $e_0$ and $\omega_0$ be 
the vielbein and spin connection of Euclidean $AdS_4$ with unit radius.
It may be shown that (see Appendix \ref{solverf} for some details)
\ie \label{adss}
&{\cal A} = W_0(x|Y) \equiv e_0(x|Y) + \omega_0(x|Y) \\
&~~= (e_0)_{\A\db} y^\A\bar y^\db + (\omega_0)_{\A\B}y^\A y^\B + (\omega_0)_{\da\db} 
\bar y^\da \bar y^\db,~~~B=0.
\fe
solves Vasiliev's equations. We refer to this solution as the $AdS_4$ 
vacuum (as we will see below this preserves the 
$SO(2,4)$ invariance of $AdS$ space).

In the sequel we will find it convenient to work with a specific 
choice of coordinates and a specific choice of the vielbein field. 
For the metric on AdS space we work in Poincar\'e coordinates; the metric 
written in Euclidean signature takes the form 
\ie \label{metric}
ds^2 = {d\vec x^2 + dz^2\over z^2},
\fe
We also define the vielbein oneform fields 
\begin{equation}\label{em}
e_0^i= -\frac{dx^i}{z}, ~~~e_0^4= -\frac{dz}{z}
\end{equation} 
($a$ runs over the index $i=1 \ldots 3$ and $a=4$). 
The corresponding spin connection one form fields are given by 
\begin{equation}\label{sc}
w_0^{ab}= \frac{dx^i}{4z}\left[ \Tr(\sigma^{iz}\sigma^{ab}) + 
          \Tr(\bar{\sigma}^{iz}\bar{\sigma}^{ab}) \right]
\end{equation} 
Using \eqref{trans} we have explicitly
\ie\label{ewsol}
&\omega_0(x|Y) = -{1\over  8} {dx^i\over z}\left( y\sigma^{iz}y 
                 + \bar{y} \bar{\sigma}^{iz}\bar y \right),\\
& e_0(x|Y) = -{1\over 4} {dx_\mu\over z} y\sigma^\mu \bar y.
\fe
Here our convention for contracting spinor indices is 
$y\sigma^\mu \bar y = y^\A (\sigma^\mu)_\A{}^\db \bar y_\db$, etc 
(see Appendix \ref{spconv}).

\subsubsection{Linearization around AdS}

The linearization of Vasiliev's equations around the $AdS$ solution of the 
previous subsection, yields Fronsdal's equations for the fields 
of all spins $s = 1,2, \cdots, \infty$ together with the free minimally coupled
equation for an $m^2=-2$  scalar field. The demonstration of this fact is rather 
involved; we will not review it here but instead refer the reader to 
\cite{Vasiliev:1999ba, Bekaert:2005vh} for details. In this subsubsection we content ourselves with 
reviewing a few structural features of linearized solutions that will be 
of use to us in the sequel.

In the linearization of Vasiliev's equations around $AdS$, 
it turns out that the  the physical degrees of freedom 
are contained entirely in the master fields restricted to 
$Z\equiv (z_\A,\bar z_\da)=0$. The spin-$s$ degrees of freedom are contained in 
\ie
&\Omega^{(s-1+m,s-1-m)} = W_\mu(x,Y,Z=0)|_{y^{s-1+m}\bar y^{s-1-m}},
\\
&C^{(2s+n,n)}=B(x,Y,Z=0)|_{y^{2s+n}\bar y^n},
\\
& C^{(n,2s+n)}=B(x,Y,Z=0)|_{y^n\bar y^{2s+n}},
\fe
for $-(s-1)\leq m\leq (s-1)$ and $n\geq 0$.
In particular, $W(x,Y,Z=0)|_{y^{s-1}\bar y^{s-1}} = \Omega_{\A\db|\A_1\cdots \A_{s-1}\db_1\cdots\db_{s-1}} y^{\A_1}\cdots y^{A_{s-1}} \bar y^{\db_1}\cdots \bar y^{\db_{s-1}}dx^{\A\db}$ contains the rank-$s$ symmetric (double-)traceless (metric-like) tensor gauge field\footnote{In order to formulate Fronsdal type equations with higher spin gauge symmetry of the form $\delta\varphi_{\mu_1\cdots\mu_s} = \nabla_{(\mu_1}\epsilon_{\mu_2\cdots\mu_s)}+\cdots$, the spin-$s$ gauge field is taken to be a rank-$s$ symmetric double-traceless tensor field $\varphi_{\mu_1\cdots\mu_s}$. The trace part can be gauged away, however, leaving a symmetric rank-$s$ traceless tensor.}, and $B|_{y^{2s}}, B|_{\bar y^{2s}}$ contain the self-dual and anti-self-dual parts of the higher spin generalization of the Weyl curvature tensor (and involve up to $s$ spacetime derivatives on the symmetric tensor field). 
While the components of $W_\mu$ and $B$ listed above are sufficient to 
recover all information
about the spin $s$ fields, they are not the only components of the Vasiliev
field that are turned on in the linearized solution. 
The linearized Vasiliev equations relate the components
\ie \label{lefr}
&\cdots\leftarrow C^{(1,2s+1)}\leftarrow C^{(0,2s)}\leftarrow\Omega^{(0,2s-2)}\cdots\leftarrow\Omega^{(s-2,s)} \leftarrow\Omega^{(s-1,s-1)}\rightarrow
\\
&~~~~~~~~~\hookrightarrow \Omega^{(s,s-2)}\rightarrow\cdots \Omega^{(2s-2,0)}\rightarrow 
C^{(2s,0)}\rightarrow C^{(2s+1,1)}\rightarrow\cdots
\fe
Starting from $\Omega^{(s-1, s-1)}$, the arrows (to the left as well as to the 
right) are generated by the action of derivatives. This may schematically 
be understood as follows. $\Omega^{(s-1, s-1)}$ has 
$s-1$ symmetrized $\alpha$ type and $s-1$ symmetrized $\da$ type indices. 
Acting with the derivative $\partial_{\gamma \db}$, symmetrizing $\gamma$ 
with all the $\alpha$ type indices but contracting $\db$ with one of the
$\da$ type indices yields an object with $s$ $\alpha$ type indices but 
only $s-2$ $\da$ type indices, taking us along the right arrow from 
$\Omega^{(s-1, s-1)}$ in \eqref{lefr}. A similar operation, interchanging 
the role of dotted and undotted indices takes us along to the left. 

The equations for the metric-like fields $\varphi_{\mu_1\cdots\mu_s}$ of the 
standard form $(\Box - m^2)\varphi_{\mu_1\cdots\mu_s} + \cdots = ({\rm nonlinear~terms})$ 
can be extracted from Vasiliev's equation by solving the auxiliary fields 
in terms of the metric-like fields order by order.

\subsubsection{Parity}

We wish to study Vasiliev's equations in an expansion around $AdS$ space 
(with asymptotically $AdS$ boundary conditions, as we will detail in the 
next section). Consider the action of a parity operation. In the coordinates 
of \eqref{metric} this operation acts as $x^i \rightarrow -x^i$  (for 
$i=1 \ldots 3$). In order to fix the action of parity 
on the spinors $y^\alpha$, ${\bar y}^\da$ and $z^\alpha$ and ${\bar z}^\da$ 
we adopt the choice of vielbein \eqref{em}. With this choice  the vielbein's
are oriented along the coordinate axes and the parity operator on spinors 
takes the standard flat space form 
$\Gamma_5 \Gamma_1 \Gamma_2 \Gamma_3=\Gamma_4$. Using the explicit form 
for $\Gamma_4$ listed in \eqref{gd}, it follows that under parity 
\ie \label{parity}
&{\bf P}(W(\vec x,z,d\vec x,dz|y_\A, z_\A, \bar y_\da, \bar z_\da)) = W(-\vec x,z,-d\vec x,dz|i(\sigma_z \bar y)_\A,i(\sigma_z \bar z)_\A,i(\sigma_z y)_\da,i(\sigma_z z)_\da),
\\
&{\bf P}(S(\vec x,z|y_\A, z_\A, \bar y_\da, \bar z_\da)) = S(-\vec x,z|i(\sigma_z \bar y)_\A,i(\sigma_z \bar z)_\A,i(\sigma_z y)_\da,i(\sigma_z z)_\da),
\\
&{\bf P}(B(\vec x,z|y_\A, z_\A, \bar y_\da, \bar z_\da)) = \pm B(-\vec x,z|i(\sigma_z \bar y)_\A,i(\sigma_z \bar z)_\A,i(\sigma_z y)_\da,i(\sigma_z z)_\da)\
\fe 
(while the parity transformation of the oneform fields $W$ and $S$ are fixed 
by the transformations of $dx^\mu$ and $dZ$, the scalar $B$ can be either 
parity odd or parity even). 
With the choice of conventions adapted in Appendix \ref{spconv}, 
$i\sigma_z=-I$. 
Consequently parity symmetry acts on $(Y,Z)$ by exchanging $y_\A\leftrightarrow -\bar y_\da$, $z_\A\leftrightarrow -\bar z_\da$, and so exchanges the two terms $f_*(B*K)dz^2$ and $\overline f_*(B*\overline K)d\bar z^2$ in the equation of 
motion. 

When are Vasiliev's equations invariant under parity transformations?
As we have seen above, $B$ may be either parity even or odd. Thus we 
need either $f(X)= \overline f(X)$ or $f(X)= \overline f(-X)$. 
Combined with (\ref{fxsol}), we have
\ie
f_A(X)={1\over 4}+X, ~~~({\rm ~A~type )}~~{\rm or}~~
f_B(X)={1\over 4}+iX ~~~({\rm~ B~ type)}
\fe
They define the A-type and B-type Vasiliev theories, respectively.

Without imposing parity symmetry, however, the interactions of Vasiliev's system is governed by 
the function $f(X)$, or the phase $\theta(X)$. If $\theta(X)$ is not $0$ or $\pi/2$, 
parity symmetry is violated. Parity symmetry is formally restored, however if 
we assign nontrivial parity transformation on $\theta(X)$ (i.e. on the coupling 
parameters $\theta_{2n}$) as well; there are two ways of doing this, with 
the scalar master field $B$ being parity even or odd:
\ie
& P_A:~~B\to B,~~~\theta(X) \to -\theta(X),~~~{\rm or}\\
& P_B: ~~B\to -B,~~~\theta(X)\to \pi - \theta(X).
\fe
This will be useful in constraining the dependence of correlation functions 
on the coupling parameters $\theta_{2n}$.

\subsubsection{The duals of free theories}\label{dft}

The bulk scalar of Vasiliev's theory turns out to have an effective mass
$m^2=-2$ in units of the AdS radius. Near the boundary 
$z=0$ in the coordinates of \eqref{metric} the equation of motion 
the bulk scalar field $S$ to take the form  
\begin{equation} \label{sfo}
S\simeq a z + bz^2
\end{equation}
while the bulk vector field takes the form 
\begin{equation}\label{vfo}
A_\mu \simeq  a_\mu +j_\mu z
\end{equation}
In order to completely specify Vasiliev's dynamical system we need 
to specify boundary conditions for the bulk scalar and vector fields 
(the unique consistent boundary condition of fields of higher spin 
is that they decay near the boundary like $z^{s+1})$.) We postpone a 
systematic study of boundary conditions to the next section. In this
subsubsection we specify the boundary conditions that define, respectively, 
the Vasiliev dual to the theory of free bosons and free fermions. 

The type A bosonic Vasiliev theory with $b=0$ (for the unique bulk scalar) 
and $a_\mu=0$ (for the unique bulk vector field) is conjectured 
to be dual to the theory of a single fundamental $U(N)$ boson 
coupled to $U(N)$ Chern-Simons theory at infinite level $k$. The 
primary single trace operators of this theory have quantum numbers  
$$ \sum_{s=0}^\infty (s+1, s)$$
(the first label above refers to the scaling dimension of the operator, 
while the second label its spin), exactly matching the linearized 
spectrum of type A Vasiliev theory.
 In subsection \ref{breaking} below we demonstrate 
that these are the only boundary conditions for the type A theory 
that preserve higher spin symmetry, the necessary and sufficient 
condition for these equations to be dual to the theory of free scalars 
\cite{Maldacena:2011jn}.

The spectrum of primaries of a theory of free fermions subject to a 
$U(N)$ singlet condition is given by 
$$ (2,0) + \sum_{s=1}^\infty (s+1, s)$$
This is exactly the spectrum of the type B  Vasiliev theory 
with boundary conditions $a=a_\mu=0$. It is not difficult 
to convince oneself that these are the unique boundary conditions 
for the type B theory that preserve conformal invariance; in subsection
\ref{breaking} below that they also preserve the full 
the higher spin symmetry algebra, 
demonstrating that this Vasiliev system is dual to a theory of free 
fermions.

\subsection{Nonabelian generalization} \label{nonabelian}

Vasiliev's system in $AdS_4$ admits an obvious generalization to non-abelian 
higher spin fields, through the introduction of Chan-Paton factors, much like 
in open string field theory. We simply replace the master fields $W, S, B$ by 
$M\times M$ matrix valued fields, and replace the $*$-algebra in the gauge 
transformations and equations of motion by its tensor product with the algebra 
of $M\times M$ complex matrices. In making this generalization we 
modify neither the truncation \eqref{Rtrunc} nor the reality condition 
\eqref{RCD} (except that the complex conjugation in \eqref{RCD} is now defined with Hermitian conjugation on the $M\times M$ matrices). We will refer to this system as Vasiliev's 
theory with $U(M)$ Chan-Paton factors. 

One consequence of this replacement is that the $U(1)$ gauge field in the bulk 
turns into a $U(M)$ gauge field, and all other bulk fields are $M \times 
M$ matrices that transform in the 
adjoint of this gauge group. 

It is natural to conjecture that the non-minimal bosonic Vasiliev theory 
with $U(M)$ Chan-Paton factors is then dual to $SU(N)$ vector model with $M$ 
flavors. Take the example of A-type theory in $AdS_4$ with $\Delta=1$ 
boundary condition. The dual CFT is that of $NM$ free massless complex 
scalars $\phi_{ia}$, $i=1,\cdots,N$, $a=1,\cdots M$, restricted to the 
$SU(N)$-singlet sector. The conserved higher spin currents are single 
trace operators in the adjoint of the $U(M)$ global flavor symmetry. 
The dual bulk theory has a coupling constant $g\sim 1/\sqrt{N}$. The 
{\it bulk} 't Hooft coupling is then
\ie
\lambda = g^2M \sim {M\over N}.
\fe
We thus expect the bulk theory to be weakly coupled when $M/N\ll 1$. The 
latter will be referred to as the ``vector model limit" of quiver type 
theories. 

At the classical level the non abelian generalization of Vasiliev's theory
has $M^2$ different massless spin $s$ fields, and in particular $M^2$ 
different massless gravitons. This might appear to suggest that the dual 
field theory has $M^2$ exactly conserved stress tensors, in contradiction 
with general field theory lore for interacting field theories. 
In fact this is not the case. In 
Appendix \ref{bdry} we argue that $\frac{1}{N}$ effects lift the scaling 
dimension of all but one of the $M^2$ apparent stress tensors for every 
choice of boundary conditions except the one that is dual to a theory 
of $M^2$ decoupled free scalar or fermionic boundary fields.

\subsection{Supersymmetric extension}\label{fermion}

To construct Vasiliev's system with extended supersymmetry, we introduce 
Grassmannian auxiliary variables $\psi_i$, $i=1,\cdots,n$, that obey 
Clifford algebra $\{\psi_i, \psi_j\} = 2\delta_{ij}$, and commute with all 
the twistor variables $(Y,Z)$. By definition, the $\psi_i$'s do not 
participate in the $*$-algebra.  The master fields $W, S, B$, as 
well as the gauge transformation parameter $\epsilon$,  are now all 
functions of $\psi_i$'s as well as of 
$(x^\mu, y_\A, \bar y_\da, z_\A, \bar z_\da)$. 

The operators $\psi_i$ may be thought of as $\Gamma$ matrices that 
act on an auxiliary $2^{\frac{n}{2}}$ dimensional `spinor' space (we 
assume from now on that $n$ is even). Note that an arbitrary 
$2^{\frac{n}{2}} \times 2^{\frac{n}{2}}$ dimensional matrix can be 
written as a linear sum of products of $\Gamma$ matrices.\footnote{This fact gives a map from the space of $2^{\frac{n}{2}} \times 2^{\frac{n}{2}}$ dimensional matrices to constant forms on an $n$ dimensional 
space, where $\psi_i$ is regarded as a basis one-form. Every  
 $2^{\frac{n}{2}} \times 2^{\frac{n}{2}}$ dimensional matrix can be uniquely 
decomposed into the sum of a zero form $a_0 I$, a one form $a^i \psi_i$, 
a two form $a^{ij} \psi_i \psi_j$ $\ldots$ an $n$ form $a_n \psi_1 \psi_2 ...
\psi_n$. The number of basis forms is $(1+1)^n=2^n$, precisely matching the 
number of independent matrix elements.}
Consequently at this stage the 
extension of Vasiliev's system to allow for all fields to be functions of 
$\psi_i$ is simply identical to the non abelian extension of the previous 
subsection, for the special case $M=2^{\frac{n}{2}}$. The construction of this 
subsection differs from that of the previous one in the truncation we 
apply on fields. The condition \eqref{Rtrunc} continues to take the form 
\ie\label{newtrunc}
{}[R,W]_*=\{R,S\}_*=[R,B]_*=[R,\epsilon]_*=0.
\fe
but with $R$ now defined as 
\begin{equation}\label{Rdef}
R\equiv K \overline K\Gamma
\end{equation}
and where 
\begin{equation} \label{gdef} 
\Gamma\equiv i^{n(n-1)\over 2} \psi_1\psi_2\cdots\psi_n
\end{equation} 
(note that $\Gamma^2=1$ and that it is still true that $R*R=1$).

While the modified truncation \eqref{newtrunc} looks formally similar 
to \eqref{Rtrunc}, it has one very important difference. As with 
\eqref{Rtrunc} it ensures that those operators that commute with $\Gamma$
(i.e. are even functions of $\psi_i$) are also even functions of 
the spinor variables $Y, Z$. However odd functions of $\psi_i$, which 
anticommute with $\Gamma$, are now forced to be odd functions of 
$Y, Z$. Such functions transform in spinorial representations of 
the internal tangent space $SO(4)$. Consequently, the new projection
introduces bulk spinorial fields into Vasiliev's theory, and simultaneously
ensures that such fields are always anticommuting, in agreement with 
the spin statistics theorem.

The reality projection we impose on fields is almost unchanged compared 
to \eqref{RCD}. We demand 
\ie \label{imdef}
\iota(W)^*=-W,~~~\iota(S)^* = -S,~~~\iota(B)^* = \overline K*B* \overline K\Gamma = \Gamma K*B*K.
\fe
The operation $\iota$ and the complex conjugation on the master fields, $\cA\to\cA^*$, are defined in the section 2.1, in combination with $\iota:~\psi_i\to\psi_i$ but reverses the order of the product of $\psi_i$'s, and $\psi_i$'s are real under complex conjugation. We require $\iota$ to reverse 
the order of $\psi_i$ in order to ensure that 
$$\iota(\Gamma)^*=\Gamma^{-1}=\Gamma.$$
(the reversal in the order of $\psi_i$ compensates for the sign picked up 
by the factor of  $i^{n(n-1)\over 2}$ under complex conjugation in \eqref{gdef}). 
The only other modification in \eqref{imdef} compared to \eqref{RCD} 
is in the factor on $\Gamma$ in the action on $B$; this additional factor 
is necessary in order for the two terms on the RHS of $\iota(B)^*$ 
to be the same, after using the truncation equation \eqref{newtrunc}, 
given that $R$ in this section has an additional factor of $\Gamma$ as 
compared to the bosonic theory. 

Vasiliev's equations take the form
\ie\label{veqnss}
& {\cal F}= d_x \hat {\cal A} + \hat {\cal A}*\hat {\cal A} = f_*(B*K) dz^2 + \overline f_*(B*\overline K \Gamma) d\bar z^2,
\\
&d_x B + \hat{\cal A}*B-B*\pi(\hat{\cal A})=0 .
\fe
Compared to the bosonic theory, the only change in the first 
Vasiliev equation is the factor of $\Gamma$ in the 
argument of ${\overline f}$; this factor is needed in order to preserve 
the reality of Vasiliev equations under the operation \eqref{imdef}, as
it follows from \eqref{imdef} that 
$$\iota (B*K)^*={\overline K} *{\overline K} *B*{\overline K} \Gamma =
B*{\overline K} \Gamma.$$ 
The second Vasiliev equation is unchanged in form from the bosonic theory; 
however the operator $\pi$ is now taken to mean 
conjugation by $\Gamma \overline K$ together with $d\bar z\to -d\bar z$, or equivalently, by the truncation condition (\ref{newtrunc}) on the fields, 
conjugation by $K$ together with $dz\to -dz$. Note in particular that 
\ie
\pi(S) &= K*S_{\overline z}*K + \Gamma \overline K*S_z*\Gamma \overline K
\\
&= S_{\da}(x|-y,\bar y,-z,\bar z,\psi) d\bar z^\da + S_\A(x|y,-\bar y,z,-\bar z,-\psi)dz^\A
\\
&= S(x|y,-\bar y,z,-\bar z,-\psi,dz,-d\bar z).
\fe
As in the case of the bosonic theory, $f(X)$ can generically be 
cast into the form  $f(X)=\frac{1}{4}+X \exp(i\theta(X))$ 
by a field redefinition.

The expansion into components of the first of \eqref{veqnss} is given by 
\eqref{veqnsb}, with the last line of that equation replaced by 
\ie \label{llfe}
&{\hat S}*{\hat S} = f(B*K) dz^2 + {\bar f}(B*{\overline K} \Gamma)  \bar z^2,
\fe
The expansion in components of the second line of \eqref{veqnss} is given 
by \eqref{veqnsc} with no modifications. 

As in the case of the bosonic theory, the second equation in 
\eqref{veqnss} follows from the first using the Bianchi identity for the field 
strength. The details of the derivation differ in only minor ways from the 
bosonic derivation presented in Appendix \ref{veq}.\footnote{\eqref{fb} holds unchanged, \eqref{sb} holds with 
$\overline K \rightarrow \overline K \Gamma$ these two equations are equivalent
by \eqref{newtrunc}. Equation \eqref{sah} holds unchanged. \eqref{saht} 
applies with ${\overline K} \rightarrow \overline K \Gamma$. \eqref{sab}
holds unchanged.}

Parity acts as
\ie \label{susypar}
&{\bf P}(W(\vec x,z,d\vec x,dz|y_\A, z_\A, \bar y_\da, \bar z_\da)) = W(-\vec x,z,-d\vec x,dz|i(\sigma_z \bar y)_\A,i(\sigma_z \bar z)_\A,i(\sigma_z y)_\da,i(\sigma_z z)_\da),
\\
&{\bf P}(S(\vec x,z|y_\A, z_\A, \bar y_\da, \bar z_\da)) = S(-\vec x,z|i(\sigma_z \bar y)_\A,i(\sigma_z \bar z)_\A,i(\sigma_z y)_\da,i(\sigma_z z)_\da),
\\
&{\bf P}(B(\vec x,z|y_\A, z_\A, \bar y_\da, \bar z_\da)) = B(-\vec x,z|i(\sigma_z \bar y)_\A,i(\sigma_z \bar z)_\A,i(\sigma_z y)_\da,i(\sigma_z z)_\da)\,\Gamma.
\fe 
The factor of $\Gamma$ in the last of \eqref{susypar} is needed in order 
that the theory with $f(X)=\frac{1}{4} + X$ is parity invariant.

\subsection{The free dual of the parity preserving susy theory}\label{fdt}

In this subsection we consider the dual description of the 
parity preserving Vasiliev theory with 
appropriate boundary conditions. The equations we study have 
 $f(X)= \frac{1}{4}+ X$. 
Let us examine the bulk scalar fields which are given by the bottom 
component of the $B$ master field, namely $\Phi(x,\psi)=B(x|Y=Z=0,\psi)$, which obeys the truncation condition $\Gamma\Phi\Gamma=\Phi$, i.e. $\Phi$ is even in the $\psi_i$'s. There are $2^{n-1}$ real scalars, half of which are parity even, the other half parity odd. We impose boundary conditions to ensure 
that  $\Delta=1$ for the parity even scalars and $\Delta=2$ for the parity 
odd scalars (see the next sections for details).  In other words 
the fall off near the boundary is given by \eqref{sfo}, with  $b=0$ for 
parity even scalars, $a=0$ for all parity odd scalars. The 
boundary fall off for all gauge fields is given by \eqref{vfo} 
with $a_\mu=0$. 

The bulk theory has also has $m=0$ spin half bulk fermions, 
whose boundary conditions we now specify. Recall 
(see e.g. \cite{Iqbal:2009fd}) that the AdS/CFT dictionary for 
such fermions identifies the `source' with the coefficient of 
the $z^\frac{3}{2}$ fall off of the parity even part 
of the bulk fermionic field (the same information is also present in the 
$z^\frac{5}{2}$ fall off of the parity odd part of the fermion field), 
while the `operator vev' is identified with the coefficient of
the $z^\frac{3}{2}$ of the parity odd part of the bulk fermion field 
(the same information is also present in the 
$z^\frac{5}{2}$ fall off of the parity even part of the fermion field). 
We impose the standard boundary conditions that set all sources to zero, 
i.e. we demand that the leading ${\cal O}(z^\frac{3}{2})$ fall off 
of the fermionic field is entirely parity odd. We believe these 
boundary conditions preserve the fermionic higher spin symmetry 
(see section \ref{bdfcs} for a partial verification) and so yield the theory 
dual to a free field theory.

The field content of this dual field theory is 
as follows; we have $2^{{n\over2}-1}$ complex scalars in the fundamental representation and the same number of fundamental fermions (so that the singlets 
constructed out of bilinears of scalars or fermions match with the bulk scalars). We organize the fields in the boundary theory in the form
$$
\phi_{i A},~~~\psi_{i\dot B \A},
$$
where $i$ is the $SU(N)$ index, $A,\dot B$ are chiral and anti-chiral spinor indices of an $SO(n)$ global symmetry, and $\A$ denotes the spacetime spinor index of $\psi_{i\dot B}$. The $2^{n-2}+2^{n-2}$ $SU(N)$ singlet scalar operators, of dimension $\Delta=1$ and $\Delta=2$, are
\ie\label{scalarops}
\bar\phi^{iA}\phi_{iB},~~~ \bar\psi^{i\dot A}\psi_{i\dot B}.
\fe
They are dual to the bulk fields (projected to the parity even and parity odd components, respectively)
\ie
\Phi_+ = \Phi {1+\Gamma\over 2},~~~\Phi_- = -i\Phi {1-\Gamma\over 2}.
\fe
The free CFT has $U(2^{{n\over 2}-1})\times U(2^{{n\over 2}-1})$ bosonic flavor symmetry that act on the scalars and fermions separately, as well as $2^{n-2}$ 
complex fermionic symmetry currents
\ie
(J_{\A\mu})^{\dot B}{}_A = \bar \psi_{\A}^{i\dot B}\overleftrightarrow\partial_\mu \phi_{i A} +\cdots.
\fe

The Vasiliev bulk dual of the $U(2^{{n\over 2}-1})\times U(2^{{n\over 2}-1})$
global symmetry is given by Vasiliev gauge transformations with 
$\epsilon$ independent of $x$, $Y$ or $Z$, but an arbitrary real even 
function of $\psi_i$ (i.e. an arbitrary even Hermitian operator built out 
of $\psi_i$). Operators of this nature may be subdivided into 
parity even and parity odd Hermitian operators which mutually commute. 
The $2^{n-2}$ parity even operators of this nature generate one factor of
$U(2^{{n\over 2}-1})$ while the complementary parity even operators generate 
the second factor. The two central $U(1)$ elements are generated by 
$I+\Gamma$ and $I-\Gamma$ respectively; these operators clearly commute 
with all even functions of $\psi_i$, and so commute 
with all other generators, establishing their central nature.\footnote{ 
As an 
example let us consider the case $n=4$ that is of particular interest 
to us below. The parity even $U(2)=U(1) \times SU(2)$ is generated by 
$$(1+\Gamma), ~~~(1+\Gamma)\psi_4 \psi_i$$
while the parity odd $U(2)=U(1) \times SU(2)$ is generated by 
$$(1-\Gamma), ~~~(1-\Gamma)\psi_4 \psi_i$$
(where $i= 1 \ldots 3$).} It is easily verified that parity even 
Vasiliev scalars transform are neutral under the parity odd 
$U(2^{{n\over 2}-1})$ but transform in the adjoint of the parity even 
$U(2^{{n\over 2}-1})$ (the reverse statement is also true). 
On the other hand the parity even/odd spin half fields of Vasiliev theory 
transform in the (fundamental, antifundamental) and (fundamental, 
antifundamental), all in agreement with field theory expectations. 

With the boundary conditions described in this section, the bulk theory may  
be equivalently written as the $n'=2$ (i.e. minimally) extended 
supersymmetric Vasiliev theory with $U(2^{{n\over 2}-1})$ Chan-Paton factors 
and boundary conditions that preserved this symmetry. Our main interest 
in the bulk dual of the free theory, however, is as the starting point 
for the construction of the bulk dual of interacting theories. 
This will necessitate the introduction of parity violating phases into 
the theory and simultaneously modifying boundary conditions. The boundary 
conditions we will introduce break the $U(2^{{n\over 2}-1})$ global symmetry 
down to a smaller subgroup. In every case of interest the subgroup in 
question will turn out to be a subgroup of $U(2^{{n\over 2}-1})$
that is also a subgroup of the $SO(n)$\footnote{As we will see in the sequel, we will find it possible to 
choose boundary conditions to preserve up to ${\cal N}=6$ supersymmetries 
together with a flavour symmetry group which is a subgroup of 
$U(2^{{n\over 2}-1})\times U(2^{{n\over 2}-1})$.}
that rotates the $\psi_i$'s (here $\psi_i$ are the 
fermionic fields that enter Vasiliev's construction, not the fermions of 
the dual boundary theory). As the preserved symmetry algebras have a 
natural action on $\psi_i$, the language of extended supersymmetry 
will prove considerably more useful for us in subsequent sections than the 
language of the non abelian extension of the $n=2$ theory, which we will 
never adopt in the rest of this paper.  


\section{Higher Spin symmetry breaking by $AdS_4$ boundary conditions} 

In this technical section, we will demonstrate that 
higher spin bulk symmetries are broken by nontrivial values of the phase 
function $\theta$ and by generic boundary conditions. 

In this section we study mainly the bosonic Vasiliev theory. We 
demonstrate that  higher spin symmetry is broken by generic boundary 
conditions and generic values of the Vasiliev phase. Higher spin 
symmetry is preserved {\it only} for the type A and type B Vasiliev theories
with boundary conditions described in subsection \ref{dft}.
 We will see this explicitly by showing that, in every other case, 
the {\it nonlinear} (higher) 
spin-$s$ gauge transformation on the bulk scalar field, at the presence of 
a spin-$s'$ boundary source, violates the boundary condition for the scalar 
field itself for every other choice of phase or boundary condition. 
We also use 
this bulk analysis together with a Ward identity 
to compute the coefficient $c_{s s' 0}$ in the schematic equation 
$$\partial^\mu J^{(s)}_{\mu}=c_{s s' 0} J^{s'} O + \cdots $$
where the RHS includes the contributions of descendants of 
$J^{s'}$ and descendants of $O$.
The violation of the scalar boundary condition is directly related to a 
double trace term in the anomalous ``conservation" law of the boundary 
spin-$s$ current, via a Ward identity.

This section does not directly relate to the study of 
the bulk duals of supersymmetric Chern 
Simons theories. Apart from the basic formalism for the 
study of symmetries in Vasiliev theory (see subsection \ref{pa} below) 
the only result of this subsection that we will use later in the paper 
 are the identifications \eqref{ida} and \eqref{idb} presented below. 
The reader who is willing to take these results on faith, and 
who is uninterested in the bulk mechanism of higher spin symmetry 
breaking, could skip directly from subsection \eqref{pa} to the 
next section.

\subsection{Symmetries that preserve the AdS Solution} \label{pa}

The asymptotic symmetry group of Vasiliev theory in $AdS_4$ is generated 
by gauge parameters $\epsilon(x|Y, Z, \psi_i)$ that leave the $AdS_4$ vacuum 
solution \eqref{adss} invariant. $S=0$ in the solution 
\eqref{adss} is preserved if and only if the gauge transformation parameter 
is independent of $Z$, i.e it takes the form $\epsilon(x|Y, \psi_i)$. As $B$ 
transforms homogeneously under gauge transformations, 
$B=0$ (in \eqref{adss}) is preserved under arbitrary gauge transformations. 
The nontrivial conditions on $\epsilon(x|Y, \psi_i)$ arise from requiring 
that $W=W_0$ is preserved. For this to be the case $\epsilon(x|Y, \psi_i)$ is
required to obey the equation 
\ie\label{epseqn}
D_0 \epsilon(x|Y, \psi_i) \equiv d_x\epsilon(x|Y, \psi_i) + 
[W_0,\epsilon(x|Y, \psi_i)]_*=0.
\fe

As the gauge field $W_0$ in the $AdS_4$ vacuum obeys the equation 
$d_x W_0+W_0*W_0=0$, $W_0$ is a flat connection and so  may may be 
written in the ``pure gauge" form.
\ie
W_0 = L^{-1}*dL,
\fe
where $L^{-1}$ is the $*$-inverse of $L(x|Y)$. We may formally move to the gauge 
in which $W_0=0$;\footnote{Note that the formal gauge transformation by 
$L$ is not a true gauge symmetry of the theory, as it violates the AdS boundary 
condition. We regard it as merely a solution generating technique.} $W=0$ 
is preserved if and only if $\epsilon$ is independent of $x$. Transforming
back to the original gauge we conclude that the most general solution to 
(\ref{epseqn}) is given by $\epsilon(x|Y)$ of the form
\ie \label{ngf}
\epsilon(x|Y, \psi_i) = L^{-1}(x|Y)*\epsilon_0(Y, \psi_i)*L(x|Y).
\fe
where $\epsilon_0(Y)$ is independent of $x$ and is restricted, 
by the truncation condition, 
to be an even function of $y, \psi_i$.\footnote{This is obvious in the 
gauge in which $W$ vanishes. In the gauge \eqref{ngf} it follows from   
the truncation condition 
$[\epsilon, R]_*=0$, and that the fact that $[L(x|Y),R]_*=0$, 
we see that $\epsilon_0(Y)$.} 

The gauge function $L(x|Y)$ is not uniquely defined; it may be obtained 
by integrating the flat connection $W_0$ along a path from a base point $x_0$ 
to $x$. We would then have $L(x_0|Y)=1$ and 
$\epsilon_0(Y) = \epsilon(x_0|Y)$. See \cite{Bolotin:1999fa, Giombi:2009wh} 
for explicit formulae for $L(x|Y)$ in Poincar\'e coordinates. We have 
used the explicit form of $L(x|Y)$ to obtain an explicit form for 
$\epsilon(x|Y)$. We now describe our final result, which may easily 
independently be verified to obey \eqref{epseqn}

Let us define $y_\pm \equiv y\pm \sigma^z\bar y$. The $*$-contraction between 
$y_\pm$ and $y_\pm$ is zero, and is nonzero only between $y_\pm$ and $y_\mp$. 
Namely, we have
\ie
& (y_\pm)_\A* (y_\pm)_\B = (y_\pm)_\A (y_\pm)_\B,
\\
&(y_\pm)_\A* (y_\mp)_\B = (y_\pm)_\A (y_\mp)_\B + 2\epsilon_{\A\B}.
\fe
In Poincar\'e coordinates, $W_0$ may be written in terms of $y_\pm$ as
\ie
W_0 = -{dx^i\over 8z} y_+\sigma^{iz}y_+ + {dz\over 8z} y_+ y_-.
\fe
A generating function for solutions to (\ref{epseqn}) is given by
\ie\label{gaugpre}
\epsilon(x|Y) &= \exp\left[{z^{-{1\over 2}} \Lambda_+(\vec x) y_+ + z^{1\over 2}\Lambda_- y_- }\right]
\\
&= \exp\left[ \Lambda(x) y+ \overline{\Lambda}(x)\bar y \right],
\fe
where $\Lambda_+(\vec x), \Lambda(\vec x)$, and $\overline\Lambda(\vec x)$ are 
given in terms of constant spinors $\Lambda_0$ and $\Lambda_-$ by
\ie
&\Lambda_+(\vec x) = \Lambda_0 + \vec x\cdot \vec \sigma \sigma^z \Lambda_-,
\\
&\Lambda(x) = z^{-{1\over 2}} \Lambda_+(\vec x)  + z^{1\over 2}\Lambda_-,
\\
&\overline\Lambda(x) =- z^{-{1\over 2}}\sigma^z \Lambda_+(\vec x)  + z^{1\over 2}\sigma^z\Lambda_-.
\fe
$\epsilon(x|Y)$ as defined in \eqref{gaugpre} may directly be verified to 
obey the linear equation \eqref{epseqn}. \eqref{gaugpre} is a generating 
function for solutions to that equation in the usual: 
upon expanding $\epsilon(x|Y)$ in a power series in the 
arbitrary constant spinors $\Lambda_0$ and $\Lambda_-$ 
the coefficients of different powers in this Taylor expansion 
independently obey \eqref{epseqn} (this follows immediately from 
the linearity of \eqref{epseqn}).

Notice that the various Taylor coefficients in \eqref{gaugpre} contains 
precisely all generating parameters for the universal enveloping algebra of 
$so(3,2)$ (in the bosonic case) or its appropriate supersymmetric extension 
(in the susy case).
 
Let us first describe the bosonic case. Recall that, on the boundary, 
 the conserved currents of the higher spin algebra may be obtained by 
dotting a spin $s$ conserved current with $s-1$ conformal killing vectors. 
Let us define the `spin $s$ charges' as the charges obtained out of the 
spin $s$ conserved current by this dotting process. The spin-$s$ global 
symmetry generating parameter, $\epsilon^{(s)}(x|Y)$, 
is then obtained from the terms in (\ref{gaugpre}) of homogeneous degree 
$2s-2$ in $(y,\bar y)$ (or equivalently in $\Lambda_0$ and $\Lambda_-$). 

As a special case consider the `spin two' charges, i.e. the charges 
whose conserved currents correspond to the stress tensor dotted with 
a single conformal killing vector, i.e the conformal generators. 
These generators are quadratic in $(y,\bar y)$. These generators may be 
organized under the action of the boundary $SU(2)$ (i.e. the diagonal 
action of $SU(2)_L$ and $SU(2)_R$) as $3+3+3 +1$, corresponding to 3d 
angular momentum generators, momenta, boosts and dilations, in perfect 
correspondence with generators of the three dimensional conformal group 
$so(3,2)$.\footnote{It may be checked that The Poincar\'e generators are 
obtained by simply setting $\Lambda_-$ to zero.} Indeed the set of 
quadratic Hamiltonians in $Y$, with product
defined by the star algebra, provides an oscillator construction of 
$so(3,2)$.

Let us now turn to the supersymmetric theory. The generators of the 
full $n$ extended superconformal algebra are given by terms that 
are quadratic in $(y, \psi_i)$. Terms quadratic in $y$ are conformal 
generators. Terms quadratic in $\psi_i$ but independent of $y$ 
are $SO(n)$ $R$ symmetry generators. Terms linear in both $y$ and $\psi_i$ 
(we denote these by $\epsilon^{({3\over 2})}(x|Y)$) are supersymmetry and 
superconformal generators. More precisely the terms involving $\Lambda_0$ are 
Poincar\'e supersymmetry parameters, where the terms involving 
$\Lambda_-$ are special supersymmetry generators (in radial quantization 
with respect to the origin $\vec x=0$).

In the sequel we will will make use of the following easily verified 
algebraic 
property of the generating function $\epsilon(x|Y)$ (\ref{gaugpre}) under $*$ product,
\ie\label{epsid}
& \epsilon(x|Y) * f(y,\bar y) = \epsilon(x|Y) f(y+\Lambda, \bar y+\overline\Lambda),
\\
& f(y,\bar y) * \epsilon(x|Y) = \epsilon(x|Y) f(y-\Lambda, \bar y-\overline\Lambda).
\fe

\subsection{Breaking of higher spin symmetries by boundary conditions}
\label{breaking}

Any given Vasiliev theory is defined by its equations of motion together 
with boundary conditions for all fields. Given any particular boundary 
conditions one may ask the following question: which of the large gauge 
transformation described in the previous subsection 
preserve these boundary conditions? In other words 
which if any of the gauge transformations have the property that they 
return a normalizable state  (i.e. a solution of Vasiliev's
theory that obeys the prescribed boundary conditions)
when acting on an arbitrary normalizable state? Such gauge transformations 
are genuine global symmetries of the system.

In this paper we will study the exact action of the large 
gauge transformations of the previous section on an arbitrary 
{\it linearized} solution of Vasiliev's equations. The most general such 
solution may be obtained by superposition of the linearized responses 
to arbitrary boundary sources. Because of the linearity of the problem, 
it is adequate to study these sources one at a time. 
Consequently we focus on the linearized solution created by a spin $s$ 
source  at $x=0$ on the boundary of $AdS_4$. Such a source creates a response
of the $B$ field everywhere in $AdS_4$, and in particular in the 
neighborhood of the boundary at the point $x$. We study the 
higher spin gauge transformations  $\epsilon^{(s')}(x|Y)$ (for arbitrary 
$s'$) on the $B$ master field at this point.  The response to this 
gauge variation contains fields of various  spins $s''$. As we will see 
below the response for $s''>1$ always respects the standard boundary 
conditions for spin $s''$ fields. However the same is not true of 
the response of the fields of low spins, namely $s''=0, {1\over 2}$, or $1$. 
As we have seen in the previous section, 
for these fields it is possible to choose different boundary conditions, 
some of which turn out to be violated by the symmetry variation $\delta B$.

In the rest of this section we restrict our attention to 
the bosonic Vasiliev theory. The variation $\delta B$ under an asymptotic 
symmetry generated by $\epsilon(x|Y)$ in \eqref{gaugpre} is given by \eqref{gt}.

Let $B^{(s)}(x|Y)$ be the spin-$s$ component of the linearized $B(x|Y)$ 
sourced by a current $J^{(s)}$ on the boundary, i.e. the boundary to bulk 
propagator for the spin-$s$ component of the $B$ master field 
with the source inserted at ${\vec x}=0$. $B^{(s)}(x|Y)$ only 
contains terms of order $y^{2s+n}\bar y^n$ and 
$y^n \bar y^{2s+n}$, $n\geq 0$; as we have explained above, the 
coefficients of these terms are spacetime derivatives of the basic 
spin $s$ field.  We will work in 
Poincar\'e coordinates \eqref{metric}, with the spin-$s$ source located at 
$\vec x=0$. Without loss of generality, it suffices to consider the 
polarization tensor for $J^{(s)}$, a three-dimensional symmetric traceless 
rank-$s$ tensor, of the form $\varepsilon_{\A_1\cdots\A_{2s}} = 
\lambda_{\A_1}\cdots\lambda_{\A_{2s}}$, for an arbitrary polarization spinor 
$\lambda$. The corresponding boundary-to-bulk propagator is computed 
in \cite{Giombi:2009wh}. Here we generalize it slightly to the parity 
violating theory, by including the interaction phase $e^{i\theta_0}$, as
\ie
B^{(s)}(x|Y) 
&= {z^{s+1}\over (\vec x^2+z^2)^{2s+1}} e^{-y\Sigma\bar y} \left[ e^{i\theta_0} (\lambda {\bf x}\sigma^z y)^{2s} + e^{-i\theta_0} (\lambda \sigma^z{\bf x}\sigma^z \bar y)^{2s} \right],
\fe
where $\Sigma$ and ${\bf x}$ are defined as
\ie
\Sigma \equiv \sigma^z - {2z\over \vec x^2+z^2}{\bf x},~~~~~{\bf x}\equiv x^\mu\sigma_\mu = \vec x\cdot\vec \sigma + z\sigma^z.
\fe
\footnote{In the special case $s=0$ the terms in the square bracket reduce 
simply to $2 \cos \theta_0$. This observation is presumably related 
to the fact, discussed by Maldacena and Zibhoedov \cite{Maldacena:2012sf}, 
that the scalar and 
spin $s$ currents in the higher spin multiplets have different 
natural normalizations. In the sequel we will, indeed, identify the factor of 
$\cos \theta_0$ with the ratio of these normalizations.}

Note that this formula is valid for spin $s>1$, for the standard ``magnetic" boundary condition in the $s=1$ case and for $\Delta=1$ boundary condition in the $s=0$ case.
 The variation of $B$ under the asymptotic symmetry generated by $\epsilon(x|Y)$ is given by
\ie\nonumber
&\delta B = -\epsilon*B^{(s)}+B^{(s)}*\pi(\epsilon)
\\
&= - \epsilon(x|y,\bar y) B(x|y+\Lambda,\bar y+\overline\Lambda) + \epsilon(x|y,-\bar y) B(x|y-\Lambda, \bar y+\overline\Lambda),
\fe
where we made use of the properties (\ref{epsid}). Using the explicit expression of the boundary-to-bulk propagator, this is
\ie
\delta B &= - {z^{s+1}\over (\vec x^2+z^2)^{2s+1}} \left\{ e^{\Lambda y + \overline\Lambda\bar y} e^{-(y+\Lambda)\Sigma(\bar y+\overline\Lambda)} \left[ e^{i\theta_0} (\lambda {\bf x}\sigma^z (y+\Lambda))^{2s} 
+e^{-i\theta_0} (\lambda \sigma^z{\bf x}\sigma^z (\bar y+\overline\Lambda))^{2s} \right] \right.
\\
&~~~~ \left. - e^{\Lambda y - \overline\Lambda\bar y}  e^{-(y-\Lambda)\Sigma(\bar y+\overline\Lambda)} \left[ e^{i\theta_0} (\lambda {\bf x}\sigma^z (y-\Lambda))^{2s} 
+e^{-i\theta_0} (\lambda \sigma^z{\bf x}\sigma^z (\bar y+\overline\Lambda))^{2s} \right] \right\}
\\
&= - {z^{s+1}\over (\vec x^2+z^2)^{2s+1}} e^{-y\Sigma \bar y  + z^{-{1\over 2}}\Lambda_+ (1 - \sigma_z \Sigma) y + z^{{1\over 2}}\Lambda_- (1 + \sigma_z \Sigma) y}
\\
&~\times
\left\{ e^{  (z^{-{1\over 2}}\Lambda_+ + z^{1\over 2}\Lambda_-)\Sigma\sigma^z (z^{-{1\over 2}}\Lambda_+ - z^{1\over 2}\Lambda_-)+ z^{-{1\over 2}}\Lambda_+(\sigma^z-\Sigma)\bar y - z^{1\over 2}\Lambda_- (\sigma^z+\Sigma) \bar y } 
\right.\\
&~~~~\times \left.\left[ e^{i\theta_0} (\lambda {\bf x}\sigma^z (y+z^{-{1\over 2}}\Lambda_+ + z^{1\over 2}\Lambda_-))^{2s} 
+e^{-i\theta_0} (\lambda\sigma^z{\bf x}\sigma^z (\bar y-\sigma^z (z^{-{1\over 2}}\Lambda_+ - z^{1\over 2}\Lambda_-)))^{2s} \right] \right.
\\
&~ \left. - e^{ - (z^{-{1\over 2}}\Lambda_+ + z^{1\over 2}\Lambda_-)\Sigma\sigma^z (z^{-{1\over 2}}\Lambda_+ - z^{1\over 2}\Lambda_-)- z^{-{1\over 2}}\Lambda_+(\sigma^z-\Sigma)\bar y + z^{1\over 2}\Lambda_- (\sigma^z+\Sigma) \bar y }
\right.\\
&~~~~\times\left. \left[ e^{i\theta_0} (\lambda {\bf x}\sigma^z (y-z^{-{1\over 2}}\Lambda_+ - z^{1\over 2}\Lambda_-))^{2s} 
+e^{-i\theta_0} (\lambda \sigma^z{\bf x}\sigma^z (\bar y-\sigma^z (z^{-{1\over 2}}\Lambda_+ - z^{1\over 2}\Lambda_-)))^{2s} \right] \right\}.
\fe
Note that although the source is a spin-$s$ current, there are nonzero variation of fields of various spins in $\delta B$. The self-dual part of the higher spin Weyl tensor, in particular, is obtained by restricting $B(x|Y)$ to $\bar y=0$. The variation of the self-dual part of the Weyl tensors of various spins are given by
\ie\label{deltabs}
&\left.\delta B\right|_{\overline y=0} =  - {z^{s+1}\over (\vec x^2+z^2)^{2s+1}} e^{ z^{-{1\over 2}}\Lambda_+ (1 - \sigma_z \Sigma) y + z^{{1\over 2}}\Lambda_- (1 + \sigma_z \Sigma) y}
\\
&\times
\left\{ e^{  (z^{-{1\over 2}}\Lambda_+ + z^{1\over 2}\Lambda_-)\Sigma\sigma^z (z^{-{1\over 2}}\Lambda_+ - z^{1\over 2}\Lambda_-) } 
\left[ e^{i\theta_0} (\lambda {\bf x}\sigma^z (y+z^{-{1\over 2}}\Lambda_+ + z^{1\over 2}\Lambda_-))^{2s} 
+e^{-i\theta_0} (\lambda\sigma^z  {\bf x} (z^{-{1\over 2}}\Lambda_+ - z^{1\over 2}\Lambda_-))^{2s} \right] \right.
\\
& \left. - e^{ - (z^{-{1\over 2}}\Lambda_+ + z^{1\over 2}\Lambda_-)\Sigma\sigma^z (z^{-{1\over 2}}\Lambda_+ - z^{1\over 2}\Lambda_-) }
\left[ e^{i\theta_0} (\lambda {\bf x}\sigma^z (y-z^{-{1\over 2}}\Lambda_+ - z^{1\over 2}\Lambda_-))^{2s} 
+e^{-i\theta_0} (\lambda \sigma^z{\bf x} (z^{-{1\over 2}}\Lambda_+ - z^{1\over 2}\Lambda_-))^{2s} \right] \right\}.
\fe

Now let us examine the behavior of $\delta B$ near the boundary of $AdS_4$. In the $z\to 0$ limit, the leading order terms in $z$ are given by
\ie\label{bdryb}
&\left.\delta B\right|_{\overline y=0} ~\longrightarrow~  - {z\over |x|^{4s+2}} e^{ 2z^{1\over 2}\left({1\over |x|^2} \Lambda_+\sigma^z {\bf x} + \Lambda_-  \right) y}
\\
&~~~~\times
\left\{ e^{  {2\over x^2 }\Lambda_+ \sigma^z {\bf x} \Lambda_+  - 2\Lambda_+\Lambda_- } 
\left[ e^{i\theta_0} (\lambda {\bf x}\sigma^z (z^{{1\over 2}}y+\Lambda_+ ))^{2s} 
+e^{-i\theta_0} (\lambda \sigma^z{\bf x} \Lambda_+ )^{2s} \right] \right.
\\
& \left.~~~~~ - e^{ -{2\over x^2 }\Lambda_+ \sigma^z {\bf x} \Lambda_+  + 2\Lambda_+\Lambda_- }
\left[ e^{i\theta_0} (\lambda {\bf x}\sigma^z (z^{{1\over 2}}y - \Lambda_+ ))^{2s} 
+e^{-i\theta_0}  (\lambda \sigma^z {\bf x}\Lambda_+ )^{2s} \right] \right\}
\fe
The variation of the spin-$s''$ Weyl tensor, $\delta B^{(s'')}$, is extracted from terms of order $y^{2s''}$ in the above formula, which falls off like $z^{s''+1}$ as $z\to 0$. This is consistent with the boundary condition for fields of spin $s''>1$, independently of the phase $\theta_0$. As promised above, 
the spin $s''>1$ component of the response to an arbitrary gauge variation
{\it automatically} obeys the prescribed boundary conditions for such 
field and so appears to yield no restrictions on allowed boundary 
conditions for the theory.

\subsubsection{Anomalous higher spin symmetry variation of the scalar}

The main difference between the scalar field and fields of arbitrary 
spin is that the prescribed boundary conditions for scalars involve both 
the leading as well as the subleading fall off of the scalar field. 
So while the leading fall off of the scalar field will never be faster 
than $z^{1}$ (in agreement with the general analysis above upon setting 
$s''=0$), this is not sufficient to ensure that the scalar field variation
obeys its boundary conditions. 

Let us examine the variation of the scalar field due to a higher spin gauge transformation, at the presence of a spin-$s$ source at $\vec x=0$ on the boundary. The spin $s''=0$ component of the symmetry variation $\delta B$ is given by (\ref{deltabs}) with $(y,\bar y)$ set to zero,
\ie\label{bzero}
\delta B^{(0)} &=  -2 {z\over (\vec x^2+z^2)^{2s+1}} 
\sinh\left[  (z^{-{1\over 2}}\Lambda_+ + z^{1\over 2}\Lambda_-)\Sigma\sigma^z (z^{-{1\over 2}}\Lambda_+ - z^{1\over 2}\Lambda_-) \right]
\\
&~~~\times
\left[ e^{i\theta_0} (\lambda {\bf x}\sigma^z (\Lambda_+ + z\Lambda_-))^{2s} 
+e^{-i\theta_0} (\lambda \sigma^z  {\bf x}(\Lambda_+ - z\Lambda_-))^{2s} \right] 
\\
&=  {4 \over (\vec x^2+z^2)^{2s+1}} 
\sinh\left[ 2{\vec x^2-z^2\over \vec x^2+z^2}(\Lambda_+\Lambda_-) +2{\Lambda_+\vec x\cdot\vec\sigma\sigma^z\Lambda_+ - z^2\Lambda_-\vec x\cdot\vec\sigma\sigma^z\Lambda_-\over \vec x^2+z^2} \right]
\\
&~~~\times
\left[ \cos\theta_0 (\lambda {\vec x\cdot\vec\sigma}\sigma^z \Lambda_+ )^{2s} z +  i\sin\theta_0 \cdot 2s (\lambda (\Lambda_++{\vec x\cdot\vec\sigma} \sigma^z \Lambda_-))(\lambda {\vec x\cdot\vec\sigma}\sigma^z \Lambda_+)^{2s-1} z^2 + {\cal O}(z^3)\right] .
\fe
When expanded in a power series in $\Lambda$, the RHS of \eqref{bzero} 
has the schematic form 
$${\cal O}(\Lambda^{2s+2} ) \times {\rm ~ (Taylor ~expansion ~ in ~} 
\Lambda^4 )$$
Recall that the spin-$s'$ symmetry variation (see the previous subsection for 
a definition) is extracted from terms of order $2s'-2$ in $\Lambda_\pm$. 
It follows that we find a scalar response to spin $s'$ gauge transformations
only for $s'=s+2, s+4, ...$. 
When this is the case (i.e. when $s'-s$ is positive and even)
\ie
\delta_{(s')} B^{(0)} &= {4\over (\vec x^2)^{2s+1}} {2^{s'-s-1}\over (s'-s-1)!} \left(\Lambda_+\Lambda_- + {1\over \vec x^2}\Lambda_+\vec x\cdot\vec\sigma \sigma^z \Lambda_+ \right)^{s'-s-1}
\\
&~~~\times \left[ \cos\theta_0 (\lambda {\vec x\cdot\vec\sigma}\sigma^z \Lambda_+ )^{2s} z +  
i\sin\theta_0 \cdot 2s (\lambda (\Lambda_++{\vec x\cdot\vec\sigma} \sigma^z \Lambda_-))(\lambda {\vec x\cdot\vec\sigma}\sigma^z \Lambda_+)^{2s-1} z^2 + {\cal O}(z^3)\right] .
\fe
Recall that $\Lambda_+ = \Lambda_0 + \vec x\cdot\vec \sigma\sigma^z\Lambda_-$, 
and $\Lambda_0$, $\Lambda_-$ are arbitrary constant spinors. For generic 
parity violating phase $\theta_0$, and $s'>s>0$ with even $s'-s$, terms of order 
$z$ and $z^2$ are both nonzero, and so both $\Delta=1$ and $\Delta=2$ boundary 
conditions would be violated, leading to the breaking of spin-$s'$ 
symmetry. 

Note that the condition $s'>s>0$ and that $s'-s$ is even means that the broken 
symmetry has spin $s'>2$. In particular the $s'=2$ conformal symmetries 
are never broken.\footnote{Note that the extrapolation of this formula to the $s=0$ 
case assumes $\Delta=1$ boundary to bulk propagator, and the variation 
$\delta_{(s')}B^{(0)}$ is always consistent with the $\Delta=1$ boundary condition.}

The exceptional cases are when either $\cos\theta_0=0$ or $\sin\theta_0=0$. 
These are precisely the interaction phase of the parity invariant theories. 
In the A-type theory, $\theta_0=0$, we see that $\delta B^{(0,0)}\sim z+{\cal O}(z^3)$, 
and so $\Delta=1$ boundary condition is preserved while $\Delta=2$ boundary 
condition would be violated. This is as expected: the A-type theory with 
$\Delta=1$ boundary condition is dual to the free $U(N)$ or $O(N)$ theory 
which has exact higher spin symmetry, whereas the A-type theory with $\Delta=2$ 
boundary condition is dual to the critical theory, where the higher spin symmetry 
is broken at order $1/N$. For the B-type theory, $\theta_0=\pi/2$, we see that 
$\delta B^{(0,0)}\sim z^2+{\cal O}(z^3)$, and so the $\Delta=2$ boundary condition 
is preserved, while $\Delta=1$ boundary condition is violated. This is in agreement
with the former case being dual to free fermions, and the latter dual to critical 
Gross-Neveu model where the higher spin symmetry is broken.

In summary, the {\it only} conditions under which {\it any} higher 
spin symmetries are preserved are the type A theory with $\Delta=1$
or the type B theory with $\Delta=2$. These are precisely the theories
conjectured to be dual to the free boson and free fermion theory 
respectively, in agreement with the results of \cite{Maldacena:2011jn}.

\subsubsection{Ward identity and current non-conservation relation}

To quantify the breaking of higher spin symmetry, we now derive a sort 
of Ward identity that relates the anomalous spin-$s$ symmetry variation 
of the bulk fields, as seen above, to the non-conservation relation of the 
three-dimensional spin-$s'$ current that generates the corresponding 
global symmetry of the boundary CFT. 

Let us first word the argument in boundary field theory language. Let us 
consider the field theory quantity 
$$\langle J^{s}(0) \cdots  \rangle $$
where $\ldots$ denote arbitrary current insertions away from the point 
$x^\mu$, and $\langle~ \rangle$ denotes averaging with the measure 
of the field theory path integral. On the path integral we now perform 
the change of variables corresponding to a spin $s'$ `symmetry'. Let 
$J^{(s')}_\mu$ denote the corresponding current. When $J^{(s')}_\mu$ is conserved
this change of variables leaves the path integral unchanged in the 
neighborhood of $x$ (it acts on the insertions, but we ignore those 
as they are well separated from $x$). When the current is not conserved, 
however, it changes the action by $\epsilon \,\partial^\mu J^{(s')}_\mu(y)$. 
Let us suppose that 
\begin{equation}\label{nc} 
\partial^\mu J^{(s')}_\mu(y) = \frac{1}{2} 
\sum_{s_1,s_2} J^{(s_1)} {\cal D}^{(s')}_{s_1s_2} J^{(s_2)} + \cdots,
\end{equation}
where ${\cal D}^s_{s_1s_2}$ is a differential operator,
It follows that, in the large $N$ limit, the change in the path integral induced by this 
change of variables is given by 
$$ \int d^3y \,\langle J^{(s_1)}(y) \cdots   \rangle\,  {\cal D}^{(s')}_{s_1 s} 
\langle J^{s}(0) J^{(s)}(y) \rangle $$
(where we have used the fact that the insertion of canonically normalized 
double trace operator contributes in the large $N$ limit only under 
conditions of maximal factorization). 
In other words the symmetry transformation amounts to an effective 
operator insertion of $J^{(s_1)}$. Specializing to the case $s_1=0$ 
we conclude that, in the presence of a spin $s$ source $J^{(s)}$, a 
spin $s'$ symmetry transformation should turn on 
a non normalizable mode for the scalar field given by  
\begin{equation}\label{tpcc}
{\cal D}^{(s')}_{0 s} 
\langle J^{s}(0) J^{(s)}(y) \rangle .
\end{equation}

Before proceeding with our analysis, we pause to restate our derivation 
of \eqref{tpc} in bulk rather than field theory language.
Denote collectively by $\Phi$ all bulk fields, and by 
$\varphi_{\mu\cdots}^{(s)}$ a particular bulk field of some spin $s$. 
Consider the 
spin-$s'$ symmetry generated by gauge parameter 
$\epsilon(x)$, under which $\varphi_{\mu\cdots}\to\varphi_{\mu\cdots}+
\delta_\epsilon\varphi_{\mu\cdots} $. Let $\phi(\vec x)$ be the renormalized 
boundary value of $\varphi(\vec x,z)$, namely 
$\varphi(\vec x,z) \to z^\Delta \phi(\vec x)$ as $z\to 0$. Let us consider 
the expectation value of $\phi(\vec x)$ at the presence of some boundary 
source $j^{\mu\cdots}$ (of some other spin $s$) located away from $\vec x$. 
The path integral is invariant under an infinitesimal field redefinition 
$\Phi\to \Phi+\delta_\epsilon\Phi$, where $\delta_\epsilon$ takes the form of 
the asymptotic symmetry variation in the bulk, but vanishes for $z$ less than 
a small cutoff near the boundary, so as to preserve the prescribed boundary 
condition, $\Phi(\vec x',z) \to z^{3-\Delta} j(\vec x') + {\cal O}(z^\Delta)$. 
From this we can write
\ie\label{impt}
0 &= \left.\int D\Phi\right|_{\Phi(\vec x',z) \to z^{3-\Delta} j(\vec x') + {\cal O}(z^\Delta)} \delta_\epsilon \left[ \varphi^{(s_1)}(\vec x,z) \exp\left( -S[\Phi] \right) \right]
\\
&= \left\langle \delta_\epsilon \varphi^{(s_1)}(\vec x,z) \right\rangle_j - \left\langle \varphi^{(s_1)}(\vec x,z) \,\delta_\epsilon S \right\rangle_j .
\fe
The spin-$s$ source $j$ is subject to the transversality condition 
$\partial_{i_1} j_{(s)}^{i_1\cdots i_{s}}=0$. Now $\delta_\epsilon S$ should 
reduce to a boundary term,
\ie\label{deltsdbl}
\delta_\epsilon S = \int_{\partial AdS} dy\, \epsilon \,\partial^\mu J^{(s')}_\mu(y) = {1\over 2} \int_{\partial AdS} \epsilon \sum_{s_1,s_2}\phi^{(s_1)} 
{\cal D}^{s'}_{s_1s_2} \phi^{(s_2)} + \cdots,
\fe
where ${\cal D}^s_{s_1s_2}$ is a differential operator, and $J_\mu$ is the boundary current associated with the global symmetry generating parameter $\epsilon$ which is now a constant along the cutoff surface, which is then taken to $z\to 0$. On the RHS of (\ref{deltsdbl}), we omitted possible higher order terms in the fields. From (\ref{impt}) we then obtain the relation
\ie
\left\langle \delta_\epsilon\varphi^{(s_1)}(\vec x,z) \right\rangle_j &= \left\langle \varphi^{(s_1)}(\vec x,z) \int_{\partial AdS} d\vec x'\epsilon\,\phi^{(s_1)}(\vec x') {\cal D}^s_{s_1s_2}\phi^{(s_2)}(\vec x') \right\rangle_j + ({\rm higher~order})
\\
&= \epsilon\int_{\partial AdS} d\vec x'  \left\langle\varphi^{(s_1)}(\vec x,z)\phi^{(s_1)}(\vec x')\right\rangle {\cal D}^s_{s_1s_2}\left\langle \phi^{(s_2)}(\vec x')\right\rangle_j + ({\rm higher~order}).
\fe
Now specialize to the case $s_1=0$, i.e. $\varphi^{(s_1)}$ is the scalar field $\varphi$ subject to the boundary condition such that the dual operator has dimension $\Delta$. The anomalous symmetry variation shows up in terms of order $z^{3-\Delta}$ in $\delta_\epsilon \varphi(\vec x,z)$. After integrating out $\vec x'$ using the two-point function of $\varphi$ and taking the limit $z\to 0$, we obtain the relation
\ie\label{wardid}
\left.\left\langle \delta_\epsilon\varphi(\vec x,z) \right\rangle_j\right|_{z^{3-\Delta}} &= \epsilon\, {\cal D}^s_{0s_2}\left\langle \phi^{(s_2)}(\vec x) \right\rangle_j + ({\rm higher~order}),
\fe
Keep in mind that $j$ is the spin-$s_2$ transverse boundary source, and $\epsilon$ is the spin-$s$ global symmetry generating parameter. The differential operator ${\cal D}^{s'}_{s_1s_2}$ appears in the spin-$s'$ current non-conservation relation of the form
\ie
\partial^\mu J^{(s)}_{\mu\cdots} = {1\over 2}\sum_{s_1,s_2}J^{(s_1)}_{\cdots} {\cal D}^{s}_{s_1s_2}J^{(s_2)}_{\cdots} +({\rm total~derivative}) + ({\rm triple~trace}).
\fe 
In particular, the double trace term on the RHS that involves a scalar operator takes the form
\ie
J^{(0)}(\vec x) \,{\cal D}^s_{0s_2} J^{(s_2)}(\vec x) +({\rm total~derivative}).
\fe
Knowing the LHS of (\ref{wardid}) from the gauge variation of Vasiliev's bulk fields, and using that fact that $\left\langle \phi^{(s_2)}(\vec x) \right\rangle_j$ is given by the boundary two-point function of the spin-$s_2$ current, we can then derive ${\cal D}^s_{0s_2}$ using this Ward identity. 
In other words we have rederived \eqref{tpcc}.

\eqref{tpcc} applies to arbitrary sources $J^s$ and also to arbitrary 
spin $s'$ symmetry transformations. Let us assume that our sources 
is of the form specified in the previous subsection; all spinor 
indices on the source are dotted so with a constant spinor  $\lambda$
which is chosen so that  
$$\lambda {\vec \sigma} \sigma_z \lambda = {\vec \epsilon'}.$$
In other words our source is uniformly polarized in the $\epsilon$ direction.
Let us also choose the spin $s'$ variation to be generated by the current 
$J^\mu_{a_1 \ldots a_{2s'-2}} \Lambda_0^{a_1} \ldots \Lambda_0^{a_{2s'-2}}$
with
$$\Lambda_0{\vec \sigma} \sigma_z \Lambda_0 = {\vec \epsilon}$$
where ${\vec \epsilon}$ is a constant vector. In other words we have 
chosen to specialize attention to those symmetries generated by 
the spin $s'$ current contracted with $s'-1$ translations in the 
direction $\epsilon$ rather than with a generic conformal killing vector.
If we compare with the asymptotic symmetry variation the bulk scalar 
derived earlier we must set $\Lambda_-$ to zero and $\Lambda_+=\Lambda_0$. 
It follows from the previous subsection that 
\ie\label{bzz}
\delta B^{(0)} &=  {4 \over (\vec x^2)^{2s_2+1}} {1\over (s-s_2-1)!}\left( {2\over \vec x^2} \Lambda_0 \vec x\cdot\vec\sigma\sigma^z \Lambda_0 \right)^{s-s_2-1}
\\
&~~~\times
\left[ \cos\theta_0 (\lambda {\vec x\cdot\vec\sigma}\sigma^z \Lambda_0)^{2s_2} z +  i\sin\theta_0 \cdot 2s_2 (\lambda \Lambda_0)(\lambda {\vec x\cdot\vec\sigma}\sigma^z \Lambda_0)^{2s_2-1} z^2 + {\cal O}(z^3)\right] .
\fe
In the $\Delta=1$ case, the anomalous variation comes from the order $z^2$ term in (\ref{bzz}), giving
\ie\label{dss}
{\cal D}^s_{0s_2}\left \langle \phi^{(s_2)}(\vec x)\right\rangle_j = \sin\theta_0  C_{ss_2} {(\varepsilon\cdot x)^{s-s_2} (2 x \cdot\varepsilon x\cdot \varepsilon' - x^2 \varepsilon\cdot \varepsilon')^{s_2-1} \epsilon^{\mu\nu\rho}\varepsilon'_\mu \varepsilon_\nu x_\rho \over (\vec x^2)^{s+s_2+1}},
\fe
Here $C_{ss_2}$ is a numerical constant that depends only on $s$ and 
$s_2$.

\eqref{dss} gives a formula for the appropriate term in \eqref{nc} 
when the operators that appear in this equation have two point 
functions 
\begin{equation} \label{tpc} \begin{split}
\langle O(0) O(x)\rangle &=\frac{\alpha_0}{x^2}, \\
\langle J^{s}(0) J^s(x)\rangle &=\frac{\alpha_s x_-^{2s}}{x^{4s+2}}. \\
\end{split}
\end{equation}
Note in particular that these two point functions are independent of 
the phase $\theta$. Let us now compare this relation to the 
results of Maldacena and Zhiboedov \cite{Maldacena:2012sf}. Those authors determined the 
non-conservation relation of currents of spin $s$, which in the lightcone 
direction to take the form 
\ie\label{djsimp}
& \partial_\mu J^{(s)\mu}{}_{-\cdots-} = {\widetilde\lambda_b\over \sqrt{1+\widetilde\lambda_b^2}} \sum_{s'}\, a_{ss'}\, \epsilon_{-\mu\nu} J^{(0)} \partial_-^{s-s'-1} \partial^\mu J^{(s')\nu}{}_{-\cdots -}+\cdots,
\fe
where $\cdots$ stands for double trace terms involving two currents of nonzero spins, 
total derivatives, and triple trace terms. Note that the first term we exhibited on 
the RHS of (\ref{djsimp}) is not a primary by itself, but when combined with the total 
derivatives term in $\cdots$ becomes a double trace primary operator in the 
large $N$ limit. We have used the notation $\widetilde\lambda_b$ of 
\cite{Maldacena:2012sf} in the case of quasi-boson theory, but normalized 
the two-point function of $J^{(0)}$ to be independent of 
$\widetilde\lambda_b$. 

Indeed with $({\cal D}^s_{0s'} J^{(s')})_{-\cdots-}\sim \epsilon_{-\mu\nu}
\partial_-^{s-s'-1}\partial^\mu J^{(s')\nu}{}_{-\cdots -}$, and the identification
\ie \label{ida}
\widetilde\lambda_b = \tan\theta_0,
\fe
the structure of the divergence of the current agrees with (\ref{dss}) obtained 
from the gauge transformation of bulk fields.


Similarly,  in the $\Delta=2$ case, the anomalous variation comes from the order $z$ term in (\ref{bzz}). We have
\ie\label{dssii}
{\cal D}^s_{0s_2}\left \langle \phi^{(s_2)}(\vec x)\right\rangle_j = \cos\theta_0  \widetilde C_{ss_2}{(\varepsilon\cdot x)^{s-s'} (2 x \cdot\varepsilon x\cdot \varepsilon' - x^2 \varepsilon\cdot \varepsilon')^{s'} \over (\vec x^2)^{s+s'+1}}.
\fe
This should be compared to the current non-conservation relation in the quasi-fermion theory, of the form
\ie\label{djsimp}
& \partial_\mu J^{(s)\mu}{}_{-\cdots-} = {\widetilde\lambda_f\over \sqrt{1+\widetilde\lambda_f^2}} \sum_{s'}\, \widetilde a_{ss'} J^{(0)} \partial_-^{s-s'-1} J^{(s')}{}_{-\cdots -}+({\rm total~ derivative})+\cdots,
\fe
Once again, this agrees with the structure of (\ref{dssii}), with $({\cal D}^s_{0s'} J^{(s')})_{-\cdots-}\sim \partial_-^{s-s'-1} J^{(s')}{}_{-\cdots -}$, and the identification
\ie \label{idb}
\widetilde\lambda_f = \cot\theta_0.
\fe


Following the argument of \cite{Maldacena:2012sf}, the double trace terms involving a scalar operator in the current non-conservation relation we derived from gauge transformation in Vasiliev theory allows us to determine the violation of current conservation in the three-point function, $\left\langle (\partial\cdot J^{(s)})\, J^{(s')} J^{(0)}\right\rangle$, and hence fix the normalization of the parity odd term in the $s-s'-0$ three-point function.

Here we encounter a puzzle, however. By the Ward identity argument, we should also see an anomalous variation under global higher spin symmetry of a field $\varphi^{(s_1)}$ of spin $s_1>1$. This is not the case for our $\delta_\epsilon B^{(s_1)}$ as computed in (\ref{deltabs}). Presumably the resolution to this puzzle lies in the gauge ambiguity in extracting the correlators from the boundary expectation value of Vasiliev's master fields, which has not been properly understood thus far. This gauge ambiguity may also explain why one seems to find vanishing parity odd contribution to the three point function by naively applying the gauge function method of \cite{Giombi:2010vg}.\footnote{We thank S. Giombi for discussions on this.}

\subsubsection{Anomalous higher spin symmetry variation of spin-1 gauge fields}

Since one can choose a family of mixed electric-magnetic boundary conditions on the spin-1 gauge field in $AdS_4$, such a boundary condition will generically be violated by the nonlinear asymptotic higher spin symmetry transformation as well.

Let us consider the self-dual part of the spin-1 field strength, whose variation is given in terms of $\delta_\epsilon B^{(2,0)}(\vec x,z|y)$, i.e. the terms in $\delta_\epsilon B$ of order $y^2$ and independent of $\bar y$. According to (\ref{bdryb}), the leading order terms in $z$, namely order $z^2$ terms, of $\delta_\epsilon B^{(2,0)}(\vec x,z)$ in the presence of a spin-$s$ boundary source at $\vec x=0$ is given by
\ie\label{btwozero}
&\delta_\epsilon B^{(2,0)}(\vec x,z|y) \,\longrightarrow\,  - {z^2\over |x|^{4s+2}} \left[ 2\left({1\over |x|^2} \Lambda_+\sigma^z {\bf x} + \Lambda_-  \right) y\right]^2 \sinh\left[ {2\over x^2 }\Lambda_+ \sigma^z {\bf x} \Lambda_+  - 2\Lambda_+\Lambda_-  \right]
\\
&~~~~~~~~~~~~~~~~~~\times \left[e^{i\theta_0} (\lambda {\bf x}\sigma^z \Lambda_+ )^{2s}+e^{-i\theta_0}(\lambda \sigma^z{\bf x} \Lambda_+ )^{2s} \right]
\\
&~~~~~ - e^{i\theta_0} {4s z^2\over |x|^{4s+2}} \cdot \left[2 \left({1\over |x|^2} \Lambda_+\sigma^z {\bf x} + \Lambda_-  \right) y\right] \cosh\left[ {2\over x^2 }\Lambda_+ \sigma^z {\bf x} \Lambda_+  - 2\Lambda_+\Lambda_-  \right]  (\lambda{\bf x}\sigma^z y) (\lambda{\bf x}\sigma^z\Lambda_+)^{2s-1}
\\
&~~~~~ - e^{i\theta_0} {2s(2s-1) z^2\over |x|^{4s+2}} \sinh\left[ {2\over x^2 }\Lambda_+ \sigma^z {\bf x} \Lambda_+  - 2\Lambda_+\Lambda_-  \right]  (\lambda{\bf x}\sigma^z y)^2 (\lambda{\bf x}\sigma^z\Lambda_+)^{2s-2}.
\\
\fe
The anti-self-dual components, $\delta_\epsilon B^{(0,2)}(\vec x,z|\bar y)$, is related by complex conjugation.
Note that by the linearized Vasiliev equations with parity violating phase $\theta_0$, $B^{(2,0)}$ and $B^{(0,2)}$ are related to the ordinary field strength $F_{\mu\nu}$ of the vector gauge field by
\ie
& B^{(2,0)}(x|y) = e^{i\theta_0} z^2 F^+_{\mu\nu}(x) (\sigma^{\mu\nu})_{\A\B} y^\A y^\B,
\\
& B^{(0,2)}(x|\bar y) = e^{-i\theta_0} z^2 F^-_{\mu\nu}(x) (\sigma^{\mu\nu})_{\da\db} \bar y^\da \bar y^\db.
\fe
The factor $z^2$ here comes from the $z$-dependence of the vielbein in $e^\mu_{\A\dc}e^\nu_{\B\dd}\epsilon^{\dc\dd}$. The two point functions of the 
operators dual to the gauge field in the equation above are given by 
\begin{equation} \label{tpcg}
\left\langle J^\mu(0) J^\nu(x) \right\rangle = {1\over \pi^2 g^2} {\delta^{\mu\nu} - {2x^\mu x^\nu\over x^2}\over x^4},
\end{equation}
where $g$ is the bulk gauge coupling constant.
The mixed boundary condition 
$$ F_{ij}= i\zeta \epsilon_{ijk} F_{zi}~~~~{\rm at}~z=0$$ is equivalent to\footnote{In order to see this let us, for instance, take the special case 
$i=1$. The relation becomes $e^{i \rho} (F_{z1}-F_{23})
=e^{-i\rho} (F_{z1}+F_{23})$, so that 
$F_{23}=\frac{e^{2 i \rho}-1}{e^{2 i \rho}+1} F_{z1}.$}
\ie
 \left. e^{-i\rho} F^+_{zi}\right|_{z=0} = \left.e^{i\rho} F_{zi}^-\right|_{z=0},~~~~{\rm where}~~ e^{2i\rho} \equiv {1+i{\zeta}\over 1- i{\zeta}}.
\fe 
We see that precisely when $\theta_0=0$ or $\pi/2$, the standard magnetic boundary condition, i.e. $\rho=0$ ($k=\infty$), is consistent with higher spin gauge symmetry. For generic $\theta_0$, however, there is {\it no} choice of $\rho$ for the boundary condition to be consistent with the higher spin symmetry variation on $\delta_\epsilon B^{(2,0)}$ and $\delta_\epsilon B^{(0,2)}$. Therefore, we see again that the parity violating phase breaks all higher spin symmetries. From this one can also derive the double trace term involving a spin-1 current in the divergence of the spin-$s$ current of the boundary theory, using the method of the previous subsection.



\section{Partial breaking of supersymmetry by boundary conditions}\label{pssbbc}

In this very important section we now turn to 
supersymmetric Vasiliev theory. We investigate the 
action of asymptotic supersymmetry transformations on bulk fields of spin 0, 
$1/2$, and 1. As in the case of higher spin symmetries, we find that no 
supersymmetry transformation preserves generic boundary conditions. In other 
words generic boundary conditions on fields violate all supersymmetries. 
However we identify special classes of boundary conditions that 
that preserve ${\cal N}=1,2,3,4$ and $6$ supersymmetries\footnote{Theories 
with ${\cal N}=5$ supersymmetry involve $SO$ and $Sp$ gauge groups on the 
boundary. Such theories presumably have bulk duals in terms of the `minimal'
Vasiliev theory, which we, however, never study in this paper. We thank O. Aharony and
S. Yokoyama for related discussions.}
in the next section.
We go on present conjectures for CFT duals for these theories.

We emphasize that the boundary conditions presented in this section preserve 
supersymmetry when acting on {\it linearized} solutions of 
Vasiliev's theory. The study of arbitrary linearize solutions is insufficient
to completely determine the boundary conditions that preserve supersymmetry 
as we now explain. 

Consider a linearized solution of a bulk scalar dual 
to an operator of dimension unity. The solution to such a scalar field 
decays at small $z$ like ${\cal O}(z)$, and the boundary condition
on this scalar asserts the vanishing of the ${\cal O}(z^2)$ term. However 
terms quadratic in  ${\cal O}(z)$ are of ${\cal O}(z^2)$ at leading order, 
and so could potentially violate the boundary condition. It follows 
that the linearized boundary conditions studied presented in this 
section are not exact, but will be corrected at nonlinear order. Indeed 
we know one source of such corrections; the boundary condition deformations
dual to the triple trace deformations of the dual boundary Chern Simons 
theory. We ignore all such nonlinear deformations in this section 
(see the next section for some remarks).

\subsection{Structure of Boundary Conditions}

Consider the $n$-extended supersymmetric Vasiliev theory with parity violating 
phase $\theta_0$. We already know that all higher spin symmetries are broken by any 
choice of boundary condition on fields of low spins, as expected for any interacting 
CFT. We also expect that any parity non-invariant CFT to have at most ${\cal N}=6$ 
supersymmetry, and the question is whether the breaking of supersymmetries to 
${\cal N}\leq 6$ in the $n$-extended Vasiliev theory can be seen from the violating 
of boundary conditions by supersymmetry variations. The answer will turn out to be 
yes. In fact, we will be able to identify boundary conditions that preserve 
${\cal N}=0,1,2,3,4$ and $6$ supersymmetries, in precise agreement with the various 
${\cal N}$-extended supersymmetric Chern-Simons vector models that differ from one 
another by double and triple trace deformations.

To begin we shall describe a set of boundary condition assignments on all 
bulk 
fields of spin $0$, ${1\over 2}$, and $1$, that will turn out to preserve various 
number of supersymmetries and global flavor symmetries. The supersymmetry transformation 
of the bulk fields of spin $0$, ${1\over 2}$, and $1$ are derived explicitly in terms 
of the master field $B(x|Y)$ in Appendix \ref{susyappen}. For convenience we will speak of the 
$n$-extended parity violating supersymmetric Vasiliev theory with no extra Chan-Paton 
factors, though our discussion can be straightforwardly generalized to include $U(M)$ 
Chan-Paton factors.  The bulk theory together with the prescribed boundary conditions 
are then conjectured to be holographically dual to supersymmetric Chern-Simons vector 
models with various number of supersymmetries and superpotentials. 

\subsubsection{Scalars}

Vasiliev's theory contains $2^{n-2}$ parity even scalar fields and an equal 
number of parity odd scalar fields. We expect the most general allowed 
boundary condition for these fields to take the form \eqref{bcs} (with 
$d_{abc}$ set to zero, as we restrict attention to linear analysis in 
this section). If we view the collection of scalar fields as a linear vector 
space of dimension $2^{n-1}$ then \eqref{bcs} asserts that the $z$ 
component of scalars lies in a particular half dimensional subspace of this 
vector space, while the $z^2$ component of the scalars lies in a 
complementary half dimensional subspace (obtained from the first space 
by switching the role of parity even and parity odd scalars). Now the 
Vasiliev master field $B$ packs all  $2^{n-1}$ scalars into a single 
even function of $\psi_i$. In order to specify the boundary conditions 
on scalars, we must specify the $2^{n-2}$ dimensional subspace (of the 
$2^{n-1}$ dimensional space of even functions of $\psi^i$) that multiply 
$z$ in the small $z$ expansion of these fields. We must also choose 
out a half dimensional subspace of functions that multiply $z^2$ 
(as motivated above, this subspace will always turn out to be complementary 
to the first). 

How do we specify the subspaces of interest? The technique we adopt is the 
following. We choose any convenient reference subspace $S$ that has the 
property that $S+ \Gamma S$ is the full space. Let $\gamma$ be an arbitrary
hermitian operator (built out of the $\psi_i$ fields) that acts on the 
subspace $S$ - i.e. $\Gamma$ is the exponential of a  
linear combination of projectors for the basis states of $S$. An arbitrary 
real half dimensional subspace in the space of functions is 
given by $e^{i \gamma} S + \Gamma e^{-i \gamma} S$. The complementary subspace
(obtained by flipping parity even and parity odd functions) is given by 
 $e^{i \gamma} S - \Gamma e^{-i \gamma} S$. 
In other words the most general boundary conditions for the scalar part 
of $B$ takes the form   
\ie \label{sbc}
B^{(0)}(\vec x,z) = (e^{i\C}+ \Gamma e^{-i\C})\tilde f_1(\psi) z+(e^{i\C}- 
\Gamma e^{-i\C})\tilde f_2(\psi) z^2 + {\cal O}(z^3)
\fe
where $f_1(\psi)$  and $f_2(\psi)$ represent any function - not necessarily 
the same - that lie within the reference real half dimensional subspace 
on the space of functions of $\psi$, and $\gamma$ is an operator, to be 
specified, that acts on this subspace. It is not difficult to verify that 
\eqref{sbc} is consistent with the reality of $B$. 
\eqref{sbc} may also be rewritten as 
\ie \label{sbcn} \begin{split}
B^{(0)}(\vec x,z)& = z \left( (1+ \Gamma) \cos \C\, \tilde f_1 + (1-\Gamma) 
i \sin \C \, \tilde f_1 \right) \\
& + z^2 \left( (1- \Gamma) \cos \C \, \tilde f_2 + (1+\Gamma) 
i \sin \C \, \tilde f_2 \right) +{\cal O}(z^3),
\end{split}
\fe
a form that makes the connection with \eqref{bcs} more explicit.

In the special case $\gamma=0$, ${\tilde f_1}$ 
and ${\tilde f_2}$ can be arbitrary (i.e. the reference half dimensional 
space can be chosen arbitrarily) and  \eqref{sbc}
simply asserts that parity odd scalars have dimension 1 while parity 
even scalars have dimension 2. 

\subsubsection{Spin half fermions}

Boundary conditions for spin half fermions are specified more simply than 
for their scalar counterparts. The most general boundary condition
relates the parity even part of any given fermion (the `source') to the 
parity odd piece of all other fermions (`the vev'). The most general 
real boundary condition of this form is that the spin-${1\over 2}$ 
part of $B$ take the form 
\ie
B^{({1\over 2})}(\vec x,z|Y)\big|_{\cO(y,\bar y)}= z^{3\over 2}\left[e^{i\A}(\chi y) -\Gamma  e^{-i\A}(\bar \chi \bar y)\right] + {\cal O}(z^{5\over 2}),~~~~~\chi = \sigma^z\bar\chi.
\fe
where $\chi$ is an arbitrary spinor and $\alpha$ is an arbitrary hermitian 
operator (i.e. function of $\psi_i$). Reality of $B^{(\half)}$ imposes 
$(\chi^\alpha)^*=-i \bar\chi_{\dot\alpha}$.

In the limit $\alpha=0$ these boundary conditions simply assert that 
the $z^{\frac{3}{2}}$ fall off of the fermion is entirely parity odd. Recall
that according to the standard AdS/CFT rules,  the parity even component of 
the fermion field may be identified with the expectation value of the boundary operator, while the parity odd part is an operator deformation. When 
$\alpha$ (which in general is a linear operator that acts on $\chi, \bar\chi$, which are functions of $\psi$) is nonzero, 
the boundary conditions assert a linear relation between parity even and 
parity odd pieces, of the sort dual to a fermion-fermion double trace 
operator. 

\subsubsection{Gauge Fields}

The electric-magnetic mixed boundary condition on the spin-1 field is
\ie
B^{(1)}(\vec x,z|Y)\big|_{\cO(y^2,\bar y^2)} = z^2\left[e^{i\beta}(yFy)+ \Gamma e^{-i\beta}(\bar y \overline F\bar y)\right] + {\cal O}(z^3) ,
~~~~
F=-\sigma^z \overline F\sigma^z.
\fe
Here $\beta$ is equal to $\theta_0$ 
for the magnetic boundary condition, corresponding 
to ungauged flavor group in the boundary CFT (recall that $e^{i \theta} F$ 
is identified with the bulk  Maxwell field strength; see above). Once 
again $\beta$ is, in general, an operator that acts on $F, \overline F$. Reality 
of $B^{(1)}$ gives  $(F^\alpha_\beta)^*=\bar F_{\dot\alpha}^{\dot\beta}$

We will see that the ${\cal N}=4$ and 
${\cal N}=6$ boundary conditions requires taking $\beta$ to be a nontrivial linear 
operator that acts on $F, \overline{F}$, which amounts to gauging a flavor group with 
a finite Chern-Simons level.

Now to characterize the boundary condition, we simply need to give the linear 
operators $\A$, $\C$, $\B$ which act on $\tilde f_{1,2}(\psi)$, $\chi(\psi)$, 
$F(\psi)$, and a set of linear conditions on $\tilde f_{1,2}(\psi)$.

We now proceed to enumerate boundary conditions that preserve different 
degrees of supersymmetry. In each case we also conjecture a field theory 
dual for the resultant Vasiliev theory. For future use we present the 
Lagrangians of the corresponding field theories in Appendix \ref{lags}.

\subsection{The ${\cal N}=2$ theory with two $\Box$ chiral multiplets}
\label{net}

Let us start with $n=4$ extended supersymmetric Vasiliev theory. The master fields depend 
on the auxiliary Grassmannian variables $\psi_1,\psi_2,\psi_3,\psi_4$. With $\theta(X)=0$, 
$\A=0$ and $\C=0$ in the fermion and scalar boundary conditions, respectively, the dual 
CFT is the free theory of 2 chiral multiplets (in ${\cal N}=2$ language) in the fundamental 
representation of $SU(N)$, with a total number of 16 supersymmetries. Now we will 
turn on nonzero $\theta_0$, and describe a set of boundary conditions that preserve 
${\cal N}=2$ supersymmetry (4 supercharges) and $SU(2)$ flavor symmetry. The 
boundary condition for the spin-1 field is the standard magnetic one. The boundary 
condition for spin-${1\over 2}$ and spin-0 fields are given by (\ref{fbcc}), 
(\ref{chre}), (\ref{scbc}), with
\ie \label{ntbc}
\A= \C= \theta_0,~~~ [\psi_1, \tilde f_1] = [\psi_1, \tilde f_2] = 0~~\text{or}~~P_{1,\psi_2\psi_3,\psi_2\psi_4,\psi_3\psi_4}\tilde f_{1,2}=\tilde f_{1,2}.
\fe
where $P_{\psi_i,\cdots}$ stands for the projection onto the subspace spanned by the monomials $\psi_i,\cdots$;
$\tilde f_{1,2}$ are subject to the constraint that they commute with $\psi_1$, or equivalently, $\tilde f_{1,2}$ are spanned by $1,\psi_2\psi_3, \psi_2\psi_4, \psi_3\psi_4$. The 2 supersymmetry parameters are given by $\Lambda_+=\Lambda_0$, $\Lambda_-=0$, with
\ie
\Lambda_0 = \eta \psi_1~~{\rm and}~~\eta\psi_1\Gamma,
\fe
where $\Gamma=\psi_1\psi_2\psi_3\psi_4$. 
$\eta$ is a constant Grassmannian spinor parameter that anti-commutes with 
all $\psi_i$'s. 

Clearly, with $\A=\theta_0$, (\ref{VspV}) obeys the fermion boundary condition 
(\ref{fbcc}), (\ref{chre}), and (\ref{FspF}) obeys the magnetic boundary condition 
on the spin-1 fields (\ref{FbarFone}), (\ref{FbarF}). (\ref{fermiontoboson}) with 
$\A=\C$ obeys (\ref{scbc}) with $\tilde f_{1,2}$ of the form $ \{\psi_1,\lambda\}$, 
or $ \{\psi_1 \Gamma, \lambda\}$, both of which commute with $\psi_1$. Finally, 
in the RHS of (\ref{VgbV}), all commutators of $\tilde f_{1,2}$ vanish, leaving 
the terms with anti-commutators only, which satisfy (\ref{fbc}), (\ref{chre}) 
with $\C=\A$. Clearly, an $SU(2)\simeq SO(3)$ flavor symmetry rotating 
$\psi_2,\psi_3,\psi_4$ is preserved by this ${\cal N}=2$ boundary condition.

It is natural to propose that the $n=4$ extended parity violating Vasiliev 
theory with this boundary condition is dual to ${\cal N}=2$ Chern-Simons 
vector model with 2 fundamental chiral multiplets. There is no gauge 
invariant superpotential in this case, while there is an $SU(2)$ flavor 
symmetry\footnote{Note that 
the field theory is left invariant under a larger set of 
$U(2)$ transformations, which rotates the chiral multiplets into each 
other. However the diagonal 
$U(1)$ in $U(2)$ acts in the same way on all fundamental fields, and so 
is part of the $U(N)$ gauge symmetry. There is nonetheless a bulk 
 gauge field - with $\psi$ content I -formally 
corresponding to this $U(1)$ factor.} 
rotating the two chiral multiplets, which is identified with the $SO(3)$
symmetry of rotations in $\psi_1$, $\psi_2$ and $\psi_3$ preserved by 
the boundary conditions listed above. 

Let us elaborate on, for instance, the scalar boundary conditions. 
There are a total of eight scalars in the problem
(the number of even functions of $\psi_i$). A basis for parity even 
scalars is given by $(1+\Gamma)$ and $(1+\Gamma) \psi_1 \psi_i$ where 
$i  =1 \ldots 3$. A basis for parity odd scalars is given by 
$(1-\Gamma)$ and $(1-\Gamma) \psi_1 \psi_i$. In each case the 
scalars transform in the $1+3$ of $SU(2)$. Recall 
that the fundamental fields of the field theory (scalars as well as 
fermions)  transform in the $\half$ 
of the flavour symmetry $SU(2)$; it follows that bilinears in these 
fields also transform in the $1+3$ of $SU(2)$, establishing a 
natural map between bulk fields and field theory operators. 

The boundary conditions \eqref{ntbc} assert that the coefficient of the 
${\cal O}(z^2)$ term of the parity even scalars/vectors is equal 
to $\tan \theta_0$ times the coefficient of the ${\cal O}(z^2)$ of the 
corresponding parity odd scalars/vectors. Similarly the coefficient of the 
${\cal O}(z)$ term of the parity odd scalars/vectors  is equal 
to $\tan \theta_0$ times the coefficient of the ${\cal O}(z)$ of the 
corresponding parity even scalars/vectors. This is exactly the kind 
of boundary condition generated by a double trace deformation that couples
the dual dimension one and dimension two operators, with equal couplings 
in the scalar and vector (of $SU(2)$) channels. We will elaborate on this
in much more detail in the next section.

\subsection{A family of ${\cal N}=1$ theories with two $\Box$ chiral multiplets}

If we keep only the supersymmetry generator given by
\ie
\Lambda_0 = \eta \psi_1,
\fe
then a one-parameter family of boundary conditions that preserve ${\cal N}=1$ supersymmetry is given by
\ie \label{nobc}
\A = \theta_0 P^S_1 + \C P^A_1,~~~~\B=\theta_0,~~~ [\psi_1, \tilde f_1] = [\psi_1, \tilde f_2] = 0,
\fe
where $P^S_1$ and $P^A_1$ are the projection operators that projects an odd function of $\psi_i$'s onto the subspaces spanned by
\ie
\psi_1\Gamma, \psi_2,\psi_3,\psi_4 ~~~~({\rm all~anti-commute~with~}\psi_1)
\fe
and
\ie
\psi_1, \psi_2\Gamma, \psi_3\Gamma, \psi_4\Gamma ~~~~({\rm all~commute~with~}\psi_1)
\fe
respectively.
$\C$ is now an arbitrary phase (independent of $\psi_i$).

This family of boundary conditions is dual to ${\cal N}=1$ deformations of the ${\cal N}=2$ theory with two chiral flavors, by turning on an ${\cal N}=1$ (non-holomorphic) superpotential that preserves the $SU(2)$ flavor symmetry (corresponding to the bulk symmetry that rotates $\psi_2,\psi_3,\psi_4$).

The same theory can also be rewritten as the $n=2$ extended supersymmetric Vasiliev theory with 
$M=2$ matrix extension. The spin-1, fermion, and scalar boundary conditions are given by
\ie\label{N1n2bc}
\A=\theta_0P_{\psi_2}+\gamma P_{\psi_1},~~~~\B=\theta_0,~~~~[\psi_1, \tilde f_1] = [\psi_1, \tilde f_2] = 0.
\fe

It is natural to wonder about the relationship between the parameter $\gamma$
above and the field theory parameter $\omega$ (see \eqref{pN=1}). General 
considerations leave this relationship undetermined; however we will 
present a conjecture for this relationship in the next section.

\subsection{The ${\cal N}=2$ theory with a $\Box$ chiral multiplet and a $\overline\Box$ chiral multiplet}

Now let us describe a boundary condition that preserve the two supersymmetries generated by
\ie\label{twosusys}
\Lambda_-=0, ~~~\Lambda_0 = \eta \psi_1~~~{\rm and}~~~\eta \psi_2.
\fe
It is given by
\ie
\beta= \theta_0, ~~~\A = \theta_0 (1-P_{\psi_3\Gamma,\psi_4\Gamma}),~~~\C= \theta_0 P_{1,\psi_3\psi_4}.
\fe
where $P_{\psi_i,\cdots}$ stands for the projection onto the subspace spanned by the 
monomials $\psi_i,\cdots$, as before;
$\tilde f_{1,2}$ are now subject to the constraint that they commute with {\it either} 
$\psi_1$ {\it or} $\psi_2$, i.e. $\tilde f_{1,2}$ are spanned by $1,\psi_3\psi_4$, 
$\psi_{1}\psi_{3}$, $\psi_1\psi_4$, $\psi_2\psi_3$, $\psi_2\psi_4$. 
Note that when acting on the latter four monomials, $\C$ vanishes, and $\tilde f_1$ 
and $\tilde f_2$ may be replaced by ${1+\Gamma\over 2}\tilde f_1$ and 
${1-\Gamma\over 2}\tilde f_2$. Therefore, only half of the components of 
$\tilde f_{1,2}$ are independent, as required. One can straightforwardly 
verified that this set of boundary conditions preserve the two supersymmetries 
(\ref{twosusys}). Clearly, the $U(1)$ flavor symmetry that rotates $\psi_3,\psi_4$ 
is still preserved, but there is no $SU(2)$ flavor symmetry. We also have 
the $U(1)$ R symmetry corresponding to rotations of $\psi_1, \psi_2$. 

The $n=4$ Vasiliev theory with this boundary is then naturally proposed to be dual 
to ${\cal N}=2$ Chern-Simons vector model with a fundamental and an anti-fundamental 
chiral flavor, with $U(1)\times U(1)$ flavor symmetry \footnote{
One of these two $U(1)$ factors is actually part of the gauge group and 
so acts trivially on all gauge invariant operators.}
 (corresponding to the components of the bulk vector gauge field proportional to 1 
 and $\psi_3\psi_4$) besides the $U(1)$ R-symmetry, which means that the ${\cal N}=2$ 
 superpotential vanishes, since a nonzero superpotential would break the 
 $U(1)\times U(1)$ flavor symmetry to a single $U(1)$.

\subsection{A family of $\cN=2$ theories with a $\Box$ chiral multiplet and a $\overline\Box$ chiral multiplet}

The boundary condition in the above section is a special point inside a one-parameter 
family of boundary conditions which preserved the same set of supersymmetries. It is given by
\ie
&\beta=\theta_0, ~~~\A = \theta_0 (1-P_{\psi_3\Gamma,\psi_4\Gamma}) +\tilde\A (P_{\psi_3\Gamma}
- P_{\psi_4\Gamma}),\\
&\C = \theta_0P_{1,\psi_3\psi_4}+\tilde\A P_{\psi_2\psi_4,\psi_1\psi_4}
,\\
&P_{1,\psi_1\psi_4,\psi_2\psi_4,\psi_3\psi_4} \tilde f_{1,2} = \tilde f_{1,2}.
\fe
This one-parameter family of deformations is naturally identified with the superpotential 
deformation of the $\cN=2$ Chern-Simons vector model with a fundamental and an anti-fundamental 
chiral flavor. This superpotential is marginal at infinite $N$; at finite $N$ there are two 
inequivalent conformally invariant fixed points \cite{Gaiotto:2007qi}. The $\tilde\A=0$ point 
is the boundary condition on the above section, describing the $\cN=2$ theory with no 
superpotential, whereas $\tilde\A=\pm\theta_0$ give the ${\cal N}=3$ point, as will be 
discussed in the next subsection.

\subsection{The ${\cal N}=3$ theory}\label{N=3bc}

The ${\cal N}=3$ boundary condition that preserve supersymmetry generated by the parameters
\ie
\Lambda_-=0, ~~~\Lambda_0 = \eta \psi_1,~ \eta\psi_2,~{\rm and}~\eta \psi_3,
\fe
is given by
\ie\label{N3bc}
\beta=\theta_0, ~~~\A = \theta_0 (1-P_{\psi_1\psi_2\psi_3}) - \theta_0 P_{\psi_1\psi_2\psi_3},
~~~ \C = \theta_0,~~~ P_{1,\psi_1\psi_4,\psi_2\psi_4,\psi_3\psi_4} \tilde f_{1,2} = \tilde f_{1,2}.
\fe
This boundary condition is dual to the ${\cal N}=3$ Chern-Simons vector 
model with a single fundamental hypermultiplet, which may be obtained from the ${\cal N}=2$ 
theory with a fundamental and an anti-fundamental chiral multiplet by a turning on a 
superpotential. The $SO(3)$ symmetry of rotations 
in $\psi_1$, $\psi_2$ and $\psi_3$ maps to the $SO(3)$ R-symmetry of the 
model. Notice that unlike the case studied in section 
\ref{net}, $\alpha \neq \gamma$ reflecting 
the fact that the $SO(3)$ R symmetry, unlike a flavor symmetry, 
 acts differently on bosons and fermions.

\subsection{The ${\cal N}=4$ theory}\label{N=4bc}
The ${\cal N}=4$ boundary condition that preserve supersymmetry generated by the parameters
\ie
\Lambda_-=0, ~~~\Lambda_0 = \eta \psi_i,~~~i=1,2,3,4,
\fe
is given by
\ie\label{N4bc}
\B=\theta_0(1-P_{\Gamma}),~~~\A = \theta_0 P_{\psi_i},
~~~ \C = \theta_0P_1.
\fe
$\tilde f_{1,2}$ are subject to the constraint
\ie
P_\Gamma \tilde f_{1,2}=0.
\fe
Note also that the components of $\tilde f_{1,2}$ proportional to $\psi_i\psi_j$ are subject 
to the projection ${1\pm \Gamma\over 2}$ also, as follows automatically from (\ref{sbc}), 
(\ref{sbcn}). The boundary conditions above 
are invariant under the $SO(4)$ R symmetry of rotations in $\psi_1$, $\psi_2$,
$\psi_3$ and $\psi_4$.

This boundary condition is dual to the ${\cal N}=4$ Chern-Simons quiver theory with gauge 
group $U(N)_k\times U(1)_{-k}$ and a single bi-fundamental hypermultiplet. The latter can 
be obtained from the ${\cal N}=3$ $U(N)_k$ Chern-Simons vector model with one hypermultiplet 
flavor by gauging the $U(1)$ flavor current multiplet with another ${\cal N}=3$ Chern-Simons 
gauge field at level $-k$ \cite{Gaiotto:2008sd}.

\subsection{An one parameter family of $\cN=3$ theories}
There is an one parameter family of boundary conditions that preserves the 
same supersymmetry as in subsection \ref{N=3bc},
\ie\label{N=3bcP}
&\beta=\theta_0(1-P_\Gamma)+\tilde\beta P_\Gamma, ~~~\A = \theta_0 P_{\psi_i}+
\tilde\beta (P_{\psi_1\Gamma,\psi_2\Gamma,\psi_3\Gamma}- P_{\psi_4\Gamma}),\\
&\C = \theta_0P_1+\tilde\beta P_{\psi_1\psi_4,\psi_2\psi_4,\psi_3\psi_4}
,\\
&P_{1,\psi_1\psi_4,\psi_2\psi_4,\psi_3\psi_4} \tilde f_{1,2} = \tilde f_{1,2}.
\fe
The boundary condition in subsection \ref{N=3bc} is at $\tilde\beta=\theta_0$. At $\tilde\beta=0$, 
the (\ref{N=3bcP}) coincides with (\ref{N4bc}), and the $\cN=3$ supersymmetry is enhanced to $\cN=4$.

\subsection{The ${\cal N}=6$ theory}\label{N=6bc}

To construct the bulk dual of the ${\cal N}=6$ ABJ vector model \cite{abjm, Aharony:2008gk}, 
we need to double the number of matter fields in the boundary field theory, and correspondingly 
quadruple the number of bulk fields. This is achieved with the $n=6$ extended supersymmetric 
Vasiliev theory, which in the parity even case (dual to free CFT) can have up to 64 
supersymmetries. We are interested in the parity violating theory, with nonzero interaction 
phase $\theta_0$, with a set of boundary conditions that preserve ${\cal N}=6$ supersymmetries, 
generated by the parameters
\ie
\Lambda_0 = \eta \psi_i,~~~i=1,2,\cdots,6.
\fe
Similarly to the ${\cal N}=4$ theory with one hypermultiplet, here we need to take the 
boundary condition on the bulk spin-1 field to be
\ie\label{N6bc}
\B=\theta_0(1-P_{\Gamma}) - \theta_0 P_\Gamma. 
\fe
The spin-${1\over 2}$ and spin-0 boundary conditions are given by
\ie\label{N6bcf}
\A = \theta_0 (1-P_{\psi_i \Gamma}) - \theta_0 P_{\psi_i \Gamma},~~~~ \C = \theta_0 P_{1,\psi_i\psi_j},
\fe
where $P_{\psi_i\Gamma}$ for instance stands for the projection onto the subspace spanned 
by {\it all} $\psi_i\Gamma$'s, $i=1,2,\cdots,6$. $\tilde f_{1,2}$ are subject to the constraint
\ie
P_{\Gamma,\psi_i\psi_j\Gamma} \tilde f_{1,2}=0,
\fe
which projects out half of the components of $\tilde f_{1,2}$. 
Note that these boundary conditions enjoy invariance under the $SO(6)$ 
R symmetry rotations of the $\psi_i$ coordinates.

By comparing the difference between $\beta$ and $\theta_0$ with the Chern-Simons level 
of what would be the flavor group of the ${\cal N}=3$ Chern-Simons vector model with 
two hypermultiplets, we will be able to identify 
$\theta_0$ in terms of $k$ below. 

\subsection{Another one parameter family of $\cN=3$ theories}
There is another one parameter family of boundary conditions that preserves the 
same supersymmetry as in subsection \ref{N=3bc},
\ie\label{N=3bcP2}
&\beta=\theta_0(1-P_\Gamma)+\tilde\beta P_\Gamma, 
\\
&\A = \theta_0( P_{\psi_i,\psi_a}+P_{\psi_i\psi_j\psi_a,\psi_i\psi_a\psi_b,\psi_4\psi_5\psi_6}-
P_{\psi_a\Gamma})+\tilde\beta (P_{\psi_i\Gamma}- P_{\psi_1\psi_2\psi_3}),\\
&\C = \theta_0P_{1,\psi_i\psi_a,\psi_a,\psi_b}-\tilde\beta P_{\psi_i\psi_j}
,\\
&P_{1,\psi_i\psi_j,\psi_i\psi_a,\psi_a\psi_b} \tilde f_{1,2} = \tilde f_{1,2},
\fe
where $i,j=1,2,3$ and $a,b=4,5,6$. At $\tilde\beta=-\theta_0$, the (\ref{N=3bcP2}) coincides with 
the boundary condition in \ref{N=6bc}, and the $\cN=3$ supersymmetry is enhanced to $\cN=6$.

\section{Deconstructing the supersymmetric boundary conditions}

\subsection{The goal of this section}

As we have explained early in this paper, the  Vasiliev dual to free 
boundary superconformal Chern Simons theories is well known. In the 
previous section we have also conjectured phase and boundary condition 
deformations of this Vasiliev theory that describe the bulk duals of 
several fixed lines of superconformal Chern Simons theories with 
known Lagrangians. These interacting superconformal Chern Simons theories 
differ from their free counterparts in three important respects.
\begin{itemize}
\item{1.} The level $k$ of the $U(N)$ Chern-Simons theory is taken 
to infinity holding $\frac{N}{k}=\lambda$ fixed. The free theory is 
recovered on taking $\lambda \to 0$. 
\item{2.} The Lagrangian of the theory includes marginal triple trace 
interactions of the schematic form $(\phi^2)^3$ and double trace 
deformations of the form $(\phi^2)
(\psi^2)$ and $(\phi \psi)^2 $ (the brackets indicate the 
structure of color index contractions). 
\item{3.} In some examples including the ${\cal N}=6$ ABJ theory  
we will also gauge a subgroup of the global symmetry group of the 
theory with the aid of a new Chern-Simons gauge field.
\end{itemize}

In this section we carefully compare the supersymmetric boundary 
conditions, determined in the previous section, with the Lagrangian 
of the conjectured field theory duals of these systems. This analysis 
allows us to understand the separate contributions of each of the three 
factors listed above to the boundary conditions of the previous section. 
It also yields some information about the relationship between the bulk 
deformation parameters and field theoretic quantities.

The analysis presented in this section was partly motivated by the 
following quantitative goal. In the previous section we have presented 
two one parameter sets of ${\cal N}=3$ Vasiliev boundary conditions 
\eqref{N=3bcP} and \eqref{N=3bcP2} at any given fixed value of the 
Vasiliev phase $\theta_0$. The first of these fixed lines 
interpolates to an ${\cal N}=4$ theory while the second 
which interpolates to a ${\cal N}=6$ theory. For each line of boundary
conditions we have also conjectured a one parameter set of dual boundary 
field theories. In order to complete the statement of the duality 
between these systems we need to be propose an identification of the 
parameter that labels boundary conditions with the parameter that 
labels the dual field theories. The analysis of this section was 
undertaken partly in order to establish this map. We have been only 
partly successful in this respect. While we propose a tentative 
identification of parameters below, there is an unresolved puzzle 
in the analysis that leads to this identification; as a consequence 
we are not confident of this identification. We leave the resolution 
of this puzzle to future work. 

We begin  this lengthy section with a review of well known effects of 
items (2) and (3)
listed above on the bulk dual systems. With 
these preliminaries out of the way we then turn to the main 
topic of this section, namely the deconstruction of the supersymmetric
boundary conditions determined in the previous subsection.

\subsection{Marginal multitrace deformations from gravity}
\label{mtrace}

As we have reviewed in the previous section, the supersymmetric 
Vasiliev theory contains fields of every half integer spin, including 
scalars with $m^2=-2$, spin half fields with $m=0$, and massless vectors.  
It is well known that the only consistent boundary conditions
for the fields with spin $s > 1$ is that they decay near $z=0$ like 
$z^{s+1}$.\footnote{In other words the coefficient of the leading 
fall off is required to vanish.} On the other hand consistency permits 
more interesting boundary conditions for fields of spin zero,
spin half and  spin one. In this section we will review the subset of 
these boundary conditions that preserve conformal invariance, together 
with their dual boundary interpretations. The discussion in this 
subsection is an application of well known material (see for example the references \cite{Witten:2001ua,Gubser:2002vv,Mueck:2002gm,Sever:2002eg,
Sever:2002fk,Witten:2003ya} - we most closely follow the 
approach of the paper \cite{Mueck:2002gm}). 

\subsubsection{scalars}

The Vasiliev theories we study contain a set of scalar fields propagating 
in $AdS_4$, 
all of which have have $m^2=-2$ in $AdS$ units. In the free theory 
the boundary conditions for some of these scalars, $S_a$, are chosen so 
that the corresponding operator has dimension 1 (these are the so 
called alternate boundary conditions) while the boundary 
conditions for the remaining scalars, $F_\alpha$, are chosen so that its 
dual operator has dimension $2$ (these are the so called regular 
boundary conditions). See Appendix \ref{sbc} for a detailed 
discussion of these boundary conditions and their dual bulk 
interpretation. 

Let us suppose that the Lagrangian for these scalars at quadratic order 
takes the form\footnote{Vasiliev's theory is currently formulated in terms of equations of 
motion rather than an action. As a consequence, 
the values of the coupling constants
$g_a$ and $g_\alpha$, for the scalars that naturally appear in 
Vasiliev's equations, are undetermined by a linear analysis. 
The study of interactions would permit the determination 
of the relative values of coupling constants, but we do not perform
such a study in this paper. }
\begin{equation} \label{actsc} 
\sum_a \frac{1}{g_a^2} \int \sqrt{g}\left(  \partial_\mu {\bar S}_a 
\partial^\mu 
S_a -2 {\bar S_a} 
S_a  \right)+ 
\sum_\A 
\frac{1}{g_\A^2} \int \sqrt{g}\left(  \partial_\mu {\bar F}_\A \partial^\mu 
F_\A -2 {\bar F_\A} F_\A  \right).\\
\end{equation}
The redefinition 
$$S_a = g_a s_a, ~~~ F_\alpha = g_\alpha f_\alpha$$
sets all couplings to unity as in the discussion in Appendix 
\ref{sbc}.

As explained in detail in Appendix \ref{tsf} the action 
and boundary conditions of bulk scalars do not completely characterize
the boundary dynamics of the system. For instance in a theory with a single 
regular quantized scalar and one 
alternately quantized scalar there exist a one parameter set of 
inequivalent boundary actions, each of 
which lead to identical boundary conditions for (appropriately 
redefined) bulk fields. However there is a distinguished `simplest'  
set of boundary counterterms corresponding to any particular boundary 
condition (this is the undeformed or $\theta_0=0$ system described in 
Appendix \ref{tsf}). This simple counterterm has the following 
distinguishing property; it yields vanishing two point functions between 
any operator of dimension one and any other operator of dimension two. 
Every other choice of counterterms yields correlators between these 
operators that vanish at separated points but are have nonvanishing 
contact term contributions. 

In this section we assume that the counterterm action corresponding 
to the scalar boundary conditions above takes the simple ($\theta_0=0$)
form referred to above. We will then deduce the effect of a double 
and triple trace deformation on the boundary conditions of bulk fields.

The two point functions of the operators 
dual to $s_a$ and $f_\alpha$\footnote{i.e. the two point functions for the 
operators for which  coefficient of the $z^2$ fall off of the field $s_a$ 
is a source, and the operator for which the coefficient of the $z$ fall off 
of the field $f_\alpha$ is the source} are given by\footnote{The general formula for the nontrivial
prefactor is $\frac{\Gamma (\Delta +1)(2 \Delta -d)}{\pi^{\frac{d}{2}}\
\Gamma(\Delta-d/2) 
\Delta}$.} \cite{Freedman:1998tz}\footnote{ The Fourier transforms 
$$G(k) = \int d^3x e^{i k.x} G(x)$$
(appropriately regulated) evaluate to $\frac{1}{|k|}$ for the dimension one 
operator (alternate quantization), and to $-|k|$ for the dimension two 
operator (regular quantization). Note that these quantities are 
the negative inverses of each other, in agreement with the general 
analysis of Appendix \ref{sbc}. } 
\begin{equation}\label{tpf} \begin{split}
&\frac{1}{2 \pi^2}  
\frac{1}{x^2}~~~({\rm operators~dual~to~}s_a), \\
&\frac{1}{2 \pi^2}  
\frac{2}{x^4}~~~({\rm operators~dual~to~}f_\alpha).
\end{split}
\end{equation} 
Later in this paper we will be interested in determining the Vasiliev 
dual to large $N$ theories deformed by double and triple trace 
scalar operators. The field theory deformations we study are marginal 
in the large $N$ limit and take the form 
\begin{equation}\label{margdef}
\int d^3 x \left( \frac{\pi^2}{2 k^2} c_{abc} \sigma^a \sigma^b \sigma^c + 
\frac{2\pi}{k} d_{a \alpha} \sigma^a 
\phi^\A \right)
\end{equation}
where $\sigma^a$ is proportional to the operator dual to $s_a$ and 
$\phi^\A$ is proportional to the operator dual to $f_\A$ 
(the factors in \eqref{margdef} have been inserted for future 
convenience). We will assume 
that it is known from field theoretic analysis that 
\begin{equation} \label{norm}
\begin{split}
\langle \sigma^a(x) \sigma^b(0) \rangle &= \delta^{ab}\frac{2 N h^a_+}{(4\pi)^2 x^2}, \\
\langle \phi^\alpha(x) \phi^\beta(0) \rangle &= \delta^{\alpha \beta}
\frac{4N h^\alpha_-}{(4\pi)^2 x^4}, \\
\end{split}
\end{equation}
(the factors on the RHS have been inserted for later convenience; $h_+^a$ 
and $h_-^\alpha$ are numbers). It follows from a comparison of \eqref{norm} 
and \eqref{tpf} that the operator dual to $s^a$ is 
$\frac{2}{\sqrt{N h^a_+}} \sigma^a$ while the operator dual to $f^\A$ 
is $\frac{2}{\sqrt{N h^\alpha_-}} \phi^\A$

Let us suppose that at small $z$,\footnote{This expansion is in conformity with \eqref{falloff} because 
$\zeta=\frac{1}{2}$ for the $m^2=-2$ scalars of Vasiliev theory.}
\begin{equation}\label{smz}
s_a=s^{(1)}_a z+ s^{(2)}_a z^2 + {\cal O}(z^3), ~~~
f_\alpha= f^{(1)}_\alpha z+ f^{(2)}_\alpha z^2 + {\cal O}(z^3).
\end{equation}
It follows from the analysis of \ref{sbc} that the marginal deformation 
\eqref{margdef} induces the boundary conditions
\begin{equation} \label{bcs} \begin{split}
s^{(2)}_a&= \frac{\pi N \sqrt{h_+^a h_-^\alpha}}{2k} d_{a \alpha} f^{(2)}_\alpha + 
3 \frac{ \pi^2 N^{\frac{3}{2}} \sqrt{h_+^a h_+^b h_+^c} }{16 k^2} 
c_{abc} s^{(1)}_b s^{(1)}_c,\\
f^{(1)}_\alpha&= -\frac{\pi N \sqrt{h_+^a h_-^\alpha}}{2k} d_{a \alpha} s^{(1)}_a.
\end{split}
\end{equation}
If we denote the boundary expansion of the original bulk fields by  
\begin{equation}\label{smz}
S_a=S^{(1)}_a z+ S^{(2)}_a z^2 + {\cal O}(z^3), ~~~
F_\alpha= F^{(1)}_\alpha z+ F^{(2)}_\alpha z^2 + {\cal O}(z^3),
\end{equation}
then 
\begin{equation} \label{bcsf} \begin{split}
\frac{S^{(2)}_a}{g_a}&= 
\frac{\pi N \sqrt{h_+^a h_-^\alpha}}{2k} d_{a \alpha} 
\frac{ F^{(2)}_\alpha}{g_\alpha} + 
3 \frac{ \pi^2 N^{\frac{3}{2}} \sqrt{h_+^a h_+^b h_+^c} }{16 k^2} 
c_{abc} \frac{S^{(1)}_b}{g_b}  \frac{S^{(1)}_c}{g_c},\\
\frac{F^{(1)}_\alpha}{g_\alpha}&= 
-\frac{\pi N \sqrt{h_+^a h_-^\alpha}}{2k} d_{a \alpha} \frac{S^{(1)}_a}{g_a}.
\end{split}
\end{equation}

In summary the boundary conditions \eqref{bcsf} are the bulk dual of the 
field theory deformation \eqref{margdef}. 

In the rest of this subsection we ignore triple trace deformations 
and focus our attention entirely on the double trace deformations. 
As explained in Appendix \ref{sbc}, in this case the modified 
boundary condition in \eqref{smz} can be undone by a rotation in the space 
of scalar fields. This is most easily seen in the special case that we 
have a single $S$ type scalar and a single $F$ type scalar so that 
both the $a$ and $\alpha$ indices run over a single value and 
can be ignored. Let us define the rotated fields 
\begin{equation}\label{rotated}
\frac{S'}{g_a}=\cos \theta \frac{S}{g_a} + \sin \theta 
\frac{F}{g_\alpha}, ~~~\frac{F'}{g_\alpha}=\cos \theta 
\frac{F}{g_\alpha}-\sin \theta \frac{S}{g_a}
\end{equation}
with
\begin{equation}\label{ttf}
\tan \theta=\frac{\pi N \sqrt{h_+^a h_-^\alpha}}{2k} d_{a \alpha}.
\end{equation}
Notice that the field redefinition \eqref{rotated} leaves the bulk action 
invariant. Moreover, it follows from \eqref{bcsf} that 
$$(S')^{(2)}=(F')^{(1)}=0.$$
In other words the rotated fields $S'$ and $F'$ obey the same bulk 
equations and same boundary conditions in the presence of 
the double trace deformation as the unrotated fields $S$ and $F$ 
obey in their absence.

At first sight this observation leads to the following paradox. 
A double trace deformation by the parameter $d$ may be thought of 
as the result of compounding two double trace deformations of magnitude 
$d_1$ and $d_2$ respectively, such that $d_1+d_2=d$. As the system
after the deformation by $d_1$ is apparently self similar to the 
system in its absence, it would appear to follow that the rotation 
that results from the deformation with $d_1 +d_2$ is simply the sum 
of the rotations corresponding to $d_1$ and $d_2$ respectively;
in other words that the rotation angle $\theta$ is linear in 
$d$. This conclusion is in manifest contradiction with \eqref{ttf}. 

The resolution of this contradiction lies in the fact that the systems 
with and without the the double trace deformations are not, infact, 
isomorphic. The reason for this is that the boundary counterterm action 
does not take the simple $\theta=0$ form in terms of rotated fields 
in the system with the double trace deformation (see Appendix \ref{sbc}).
In the theory with double trace deformations there is, in particular, 
 a nonzero contact term in the two point functions of the two operators 
with distinct scaling dimensions; this contact term is absent in the 
original system.

\subsubsection{Spin half fermions}\label{fermbc}

The Vasiliev theories we study include a collection of 
real fermions $\psi^a_1$ and $\psi^a_2$ propagating in $AdS_4$ 
space. It is sometimes useful to work with the 
complex fermions $\psi^a=\frac{\psi^a_1+i \psi^a_2}{\sqrt{2}}$ and 
${\bar \psi}^a=\frac{\psi_a^1-i {\bar \psi}^a_2}{\sqrt{2}}$. 
Let us suppose that the bulk action takes the form 
\begin{equation}\label{bl}
\sum_{a}\frac{1}{g_a^2} 
\int {\bar \psi}^a D_\mu \Gamma^\mu \psi_a. 
\end{equation} 
Using the rules described for instance in 
\cite{Iqbal:2009fd}, the two point function for the operator dual to 
$\psi^a$ is easily computed and we find the answer 
\begin{equation}\label{ftpf}
\frac{1}{g_a^2}\frac{\vec x\cdot \vec \sigma}{\pi^2 x^4}.
\end{equation}
The same result also applies to the two point functions of the operators 
dual to $\psi_1^a$ and $\psi_2^a$ independently. 

In analogy with the bosonic case described in the previous subsection, 
the formula \eqref{ftpf} presumably applies only with the simplest 
choice of boundary counterterms 
\cite{Henningson:1998cd,Mueck:1998iz,Henneaux:1998ch,Laia:2011zn} - 
the analogue of $\theta_0=0$ in Appendix 
\ref{tsf}- consistent with the boundary conditions described in 
\cite{Iqbal:2009fd}. Though we will not perform the required careful 
analysis in this paper, it seems likely that the fermionic analogue 
of Appendix \ref{sbc} would find a one parameter set of inequivalent 
boundary actions that lead to the same boundary conditions. From 
the bulk viewpoint this ambiguity is likely related to the freedom 
associated with rotating a bulk spinor $\psi_1$ into $\Gamma_5 \psi_2$
($\Gamma_5$ is the bulk chirality matrix). We ignore this potential complication in the rest of this subsection, and focus on the simple canonical
case described in \cite{Iqbal:2009fd}.

Let the field theory operator proportional to $\psi^a$ be denoted by 
$\Psi^a$. Let us assume that we know from field theory that 
\begin{equation}\label{fnorm}
\langle \Psi^a(x) \bar{\Psi}^b(0) \rangle  = 
\delta^{ab} \frac{h_\psi 2N (\vec x\cdot\vec\sigma)}{(4\pi)^2 x^4}.
\end{equation}

We will now describe the boundary conditions dual to a field theory 
double trace deformation. Let the fermionic fields have the small $z$ 
expansion 
\begin{equation}\label{sze}
\begin{split}
\psi_1^a&= z^\frac{3}{2}\left( \zeta^a_{1+} + \zeta^a_{1-} \right)
+ {\cal O}(z^\frac{5}{2}), \\
\psi_2^a&= z^\frac{3}{2}\left( \zeta^a_{2+} + \zeta^a_{2-} \right)
+ {\cal O}(z^\frac{5}{2}). \\
\end{split}
\end{equation} 
Above the subscripts $+$ and $-$ denote the eigenvalue of the corresponding 
fermions under parity. 

Using the procedure of the previous subsection, the bulk dual of 
the field theory double trace deformation 
$$ \frac{\pi}{4k} \left[s_{ab} 
\left({\bar \Psi}^a + \Psi^a \right)\left({\bar \Psi}^b + \Psi^b \right)
- t_{ab}
\left({\bar \Psi}^a - \Psi^a \right)\left({\bar \Psi}^b - \Psi^b \right)
+u_{ab} \left({\bar \Psi}^a + \Psi^a \right)
~ i \left({\bar \Psi}^b - \Psi^b \right) \right]$$
is given by the modified boundary conditions 
\begin{equation} \label{mbc} \begin{split}
\frac{\zeta^a_{1+}}{g_a}&=\frac{N\pi\sqrt{h_\psi^a h_\psi^b}}{8k} 
\left( s_{ab} \frac{\zeta^b_{1-}}{g_b} + \frac{1}{2} u_{ab} 
\frac{\zeta^b_{2-}}{g_b} \right), \\
\frac{\zeta^a_{2+}}{g_a}&=\frac{N \pi \sqrt{h_\psi^a h_\psi^b}}{8k}
\left( t_{ab} \frac{\zeta^b_{2-}}{g_b} + \frac{1}{2} u_{ba} 
\frac{\zeta^b_{1-}}{g_b} \right).
\end{split}
\end{equation} 

\subsection{Gauging a global symmetry}\label{gglobal}

As originally introduced by Witten \cite{Witten:2003ya}, gauging a global symmetry with 
Chern-Simons term in the boundary CFT is equivalent to changing the boundary condition 
of the bulk gauge field corresponding to the boundary current of the global symmetry. 
We will review this relation in this subsection and in appendix B.

Let us start by considering a boundary CFT with $U(1)$ global symmetry. The current 
associated to this global symmetry is dual to a $U(1)$ gauge field $A_\m$ in the bulk. 
In the $A_z=0$ radial gauge, the action for the gauge field $A_\m$ is
\ie \label{ba}
{1\over 4g^2} \int {d^3\vec x dz\over z^4} F_{\mu\nu} F^{\mu\nu}
= \int d^3\vec x dz \,\left( {1\over 2g^2}\partial_z A_i \partial_z A_i+ {1\over 4g^2}F_{ij}F_{ij} \right).
\fe
Onshell the bulk action evaluates to 
\ie
\int d^3\vec x  \,\left( {1\over 2g^2}A_i \partial_z A_i \right).
\fe
where the integral is taken over a surface of constant $z$ for small 
$z$. 
 The equations of 
motion w.r.t. the boundary gauge field impose the 
{\it electric} boundary condition
\ie\label{bcgff}
{1\over g^2}\partial_z A_i\big|_{z=0}=0.
\fe
Near $z=0$, the most general solution to the gauge field equations of 
motion is 
$$A_i=A_i^1(x)+ z A_i^2(x).$$
The boundary condition \eqref{bcgff} forces $A_i^2$ to vanish but allows 
$A_i=A_i^1$, the value of the gauge field on the cut off surface, 
to fluctuate freely at the boundary $z=0$. The theory so obtained 
is the conceptual equivalent of the `alternate' quantized scalar theory 
described in Appendix \ref{sbc}.

If we add a boundary $U(1)$ Chern-Simons term to the bulk action
\footnote{This is the same as adding a term in the bulk action 
proportional to $\int F\wedge F$ as this term is the total derivative 
of the Chern Simons term}  (in Euclidean signature )
\ie\label{acs}
{ik\over 4\pi}\int d^3\vec x\,\epsilon_{ijk}A_i\partial_j A_k,
\fe
and allow arbitrary variation $\delta A_i$ at $z=0$, the equation 
of motion of the boundary field $A_i$ generates the modified 
 boundary condition
\ie
{1\over g^2}\partial_z A_i+{ik\over 2\pi}\epsilon_{ijk}\partial_jA_k\big|_{z=0}=0,
\fe
which is the electric-magnetic mixed boundary condition. By the AdS/CFT dictionary, this 
is also equivalent to adding the term (\ref{acs}) into the boundary theory, where $A_i$ 
is now interpreted as the three dimensional gauge field coupled to the $U(1)$ current.

This procedure can be straightforwardly generalized to $U(M)$. Adding the $U(M)$ 
Chern-Simons action on the boundary
\ie\label{naacs}
{ik\over 4\pi}\int d^3\vec x\epsilon_{ijk}\tr\left(A_i\partial_j A_k+{2\over 3}A_i A_j A_k\right).
\fe
modifies the electric boundary condition to
\ie \label{bcgt}
{1\over g^2}\partial_z A_i+{ik\over 2\pi}\epsilon_{ijk}\left(\partial_jA_k+A_j A_k\right)\big|_{z=0}=0.
\fe
Note that this mixed boundary condition is still gauge invariant. 

Of course $\partial_z A_i$ is determined in terms of $A_i$ by the 
equations of motion. As the equations of motion are linear, the 
relation between these quantities is linear - but nonlocal- and takes 
the form 
$$\partial_z A_i(q)= G_{ij}(q) A_j(q).$$
The function $G_{ij}(q)$ has a simple physical interpretation; 
it is the two point function of the current operator (with natural 
normalization) in the theory at $k=\infty$ (at this value of $k$
the boundary condition \eqref{bcgt} is simply the standard Dirichlet 
boundary condition). A simple computation yields 
\ie \label{gn}
\langle J_i(p) J_j(-q)\rangle ={1\over 2g^2}G_{ij}(q) \delta^3(p-q)=\
- {|p| \over 2 g^2} \left( \delta_{ij} - 
{p_i p_j\over p^2} \right) (2\pi)^3 \delta^3(p-q).
\fe
Note that here we have normalized the current coupled to the Chern-Simons gauge field 
according to the convention for nonabelian gauge group generators, 
${\rm Tr}(t^a t^b) = {1\over 2}\delta^{ab}$ for generators $t^a$, $t^b$ in the 
fundamental representation. This is also the normalization convention we use to 
define the Chern-Simons level $k$ (which differs by a factor of 2 from the natural 
convention for $U(1)$ gauge group).

Recall that \eqref{gn} yields the two point functions of the `ungauged' 
theory - i.e. the theory with $k=\infty$.  Our analysis of the dual 
boundary theory to this ungauged system, we find it convenient to work 
with currents normalized so that 
\ie \label{ftc}
\langle J_i(p) J_j(-q)\rangle =- {\widetilde N |p| \over 32} \left( \delta_{ij} - 
{p_i p_j\over p^2} \right) (2\pi)^3 \delta^3(p-q).
\fe
Our convention is such that in the free theory $\widetilde N$ counts the total number 
of complex scalars plus fermions (i.e. the two point function for the 
charge current for a free complex scalar is equal to that of the free 
complex fermion and is given by \eqref{ftc} upon setting 
${\widetilde N}=1$, see Appendix \ref{fft}). In order that \eqref{gn} and 
\eqref{ftc} match we must identify 
$$g^2=\frac{16}{\widetilde N},$$
so that the effective boundary conditions on gauge fields become 
\ie \label{finbc}
\frac{ \pi {\widetilde N}}{8 k} \partial_z A_i+ i \epsilon_{ijk}\partial_jA_k\big|_{z=0}=0.
\fe
In summary, gauging of the global symmetry is affected by the boundary 
conditions \eqref{finbc}. Note that the boundary conditions \eqref{finbc}
constrain only the boundary field strength $F_{ij}$. Holonomies around noncontractable cycles are 
unconstrained  and must be integrated over. In the finite temperature 
theory the integral over the Polyakov line of $U(M)$ enforces the $U(M)$, 
as we study in detail in section \ref{abjfpf}.


\subsection{Deconstruction of boundary conditions: general remarks}

\subsubsection{The bulk dual of the finite Chern Simons coupling}
\label{bdfcs}

With essential preliminaries taken care of we now turn to the main topic 
of this subsection, namely the deconstruction of the supersymmetric 
boundary conditions of the previous section.

The Vasiliev dual of free susy theories was described in subsection \ref{fdt}.
What is the Vasiliev dual to the free field theory deformed {\it only} by
turning on a finite Chern Simons t'Hooft coupling $\lambda =\frac{N}{k}$? The deformation we study is unaccompanied by any potential and Yukawa 
terms - in particular those needed to preserve supersymmetry - and so is 
not supersymmetric. Consequently the comparisons between susy Lagrangians
and boundary conditions, presented later in this section, does not directly
address the question raised here. As we will see, however, the answer to this 
question is partly constrained by symmetries, and receives indirect inputs 
from our analysis of susy theories below.

We first recall that it was conjectured in \cite{Giombi:2011kc} that the 
bulk dual to turning 
on $\lambda$ involves a modification of the {\it bulk} Vasiliev equations 
by turning on an appropriate parity violating phase, $\theta(X)$, as a 
function of $\lambda$. The results of the previous section clearly 
substantiate this conjecture \footnote{As those results are valid only 
for the linearized theory, they unfortunately cannot distinguish between 
a constant phase and a more complicated phase function; we return to 
this issue below.}. It is possible, however, that in addition to 
turning on the phase, a nonzero Chern Simons coupling also results in 
modified boundary conditions on bulk scalars and fermions. We now 
proceed to investigate this possibility.

A consideration of symmetries greatly constrains possible modifications 
of boundary conditions. Recall that the Vasiliev dual to free susy 
theories possesses a $U(2^{\frac{n}{2}-1})\times U(2^{\frac{n}{2}-1})$ 
global symmetry. In the dual boundary theory the 
$U(2^{\frac{n}{2}-1})\times U(2^{\frac{n}{2}-1})$ symmetry rotates the 
fundamental bosons and fermions respectively, and is preserved by 
turning on a nonzero Chern Simons coupling. A constant phase in 
Vasiliev's equations also preserves this symmetry. It follows that 
all accompanying boundary condition deformations must also preserve 
this symmetry. 

Parity even and odd bulk scalars respectively transform in the 
(adjoint + singlet, singlet) and  (singlet, adjoint+singlet) 
representations of the $U(2^{\frac{n}{2}-1})\times U(2^{\frac{n}{2}-1})$ 
symmetry. The only conformally invariant modifications of boundary condition 
that preserve this symmetry are those dual to the double trace coupling 
of the parity odd and parity even singlet scalars, and that dual to the 
triple trace deformation of three parity even singlet scalars. 

The conjectures of the previous section strongly constrain the double trace 
type deformation of boundary conditions induced by the Chern Simons 
coupling \footnote{Our analysis 
of boundary conditions in the previous section was insensitive to triple 
trace type boundary conditions, and so does not constrain the triple trace
type modification.}. Let us, for instance, compare Lagrangian 
and boundary conditions of the fixed line of ${\cal N}=1$ theories 
described in the previous subsection. The double trace scalar potential 
in these theories is listed in \eqref{sdtn1l} below and vanishes at 
$\omega=-1$. On the other hand the rotation $\gamma$ in the scalar boundary 
conditions for the dual Vasiliev system is listed in \eqref{N1n2bc}, and vanishes 
for the dual of $\omega=-1$. In other words the Vasiliev dual to the 
Chern-Simons theory with no scalar potential obeys boundary conditions 
such that all `parity even' scalars continue to have $\Delta=1$ boundary 
conditions, while all `parity odd' scalars continue to have $\Delta=2$ 
boundary conditions. While the argument presented above holds only for $n=2$,
the result continues to apply at $n=4$ and $n=6$ as well, 
as we will see in more detail in the detailed comparisons below.
\footnote{ For the case $n=4$ consider, for instance, the ${\cal N}=2$ 
theory with two 
fundamental chiral multiplets. The free theory has a $U(2) \times U(2)$ 
symmetry. The interacting theory preserves the diagonal $SU(2)$ subgroup   
of this symmetry (corresponding to rotations of the two chiral multiplets). 
The parity odd and even single trace operators in this theory each 
transform in the $1+3$ representations of this symmetry. The allowed 
double trace deformations of this interacting theory couple the parity 
even $3$ with the parity odd $3$ and the parity even scalar with the 
parity odd scalar. It so happens that these two terms 
appear with the same coefficient in both the field 
field theory potential \eqref{ntpot} and the 
corresponding Vasiliev boundary conditions (the fact that these terms appear 
with the same coefficient in \eqref{ntbc} is simply the fact that the 
singlet monomial $I$, appears on the same footing as the triplet 
monomials $\psi_2 \psi_3, \psi_3 \psi_4, \psi_4 \psi_2$ in the scalar 
boundary conditions). These facts together demonstrate that the 
Chern Simons term (which could have acted only on the singlet double 
trace term and so would have `split the degeneracy' between singlets 
and triplets) has no double trace type effect on scalar boundary conditions. 
}

We turn now to the fermions. Bulk fermions transform in the 
(fundamental, antifundamental) and (antifundamental, fundamental) of 
the free symmetry algebra. There is, of course, a natural double trace
type singlet boundary condition deformation with this field content 
(this deformation has the same effect on boundary conditions as a double 
trace field theory term $(\phi_a {\bar \psi^b})(\psi_b {\bar \phi}^a)$ 
where $a$ and $b$ are global symmetry indices and brackets denote the 
structure of gauge contractions). Perhaps surprisingly, we will 
now argue that 
merely turning on the Chern Simons term {\it does} induce such a 
boundary condition deformation. More precisely, it turns out that 
the bulk theory with 
trivial boundary conditions on fermions corresponds to a quantum 
field theory with fermion double trace potential equal to 
$$-\frac{6 \pi}{k} {\bar \Psi}\Psi$$
for every single trace Fermionic operator. 

We present a heuristic argument for this conclusion in Appendix 
\ref{fermdt} by comparing the Lagrangian and boundary conditions
of the line of ${\cal N}=1$ theories with a single chiral multiplet. 
However the most convincing argument for this conclusion is that 
it leads to consistent results between the Lagrangian and boundary 
conditions in {\it every} case we study in detail later in this section. 

In order to compensate for the shift described above, will find it useful, 
in our analysis below,  to compare Fermionic boundary conditions with a  shifted field theory 
Lagrangian: one in which we add by hand the double trace term 
$\frac{6 \pi}{k} {\bar \Psi}\Psi$ for every single trace fermionic field. 
Bulk fermionic fields have trivial boundary conditions only when the 
double trace deformations of the corresponding fermionic operators vanish
in the shifted field theory Lagrangian.

\subsubsection{Special Points in moduli space for scalars}

If we wish to specify the bulk dual for a 3d conformal field theory, 
it is insufficient to specify the the bulk action and the boundary 
conditions for bulk scalars (see Appendix \ref{sbc}). In order 
to specify the correlators of the dual theory we must, in addition, 
specify the precise nature of the boundary dynamics that gives 
rise the the resultant boundary conditions. Inequivalent boundary 
dynamics that lead to the same boundary conditions result in 
distinct correlation functions; in particular to different counterterms
in correlators.

Of the set of all boundary actions that lead to a particular boundary 
condition, one is particularly simple ($\theta_0=0$ in Appendix 
\ref{tsf}); this choice of boundary counterterms ensures that 
correlators between dimension one and dimension two operators vanish 
identically (including contact terms). Let us suppose that the dual of 
a particular quantum 
field theory is governed by this simple boundary dynamics. Then the 
dual of this theory deformed by a scalar double trace deformation 
cannot, in general, also be governed by the same simple boundary 
dynamics (see Appendix \ref{tsf}). 

In the moduli space of field theories obtained from one another by double 
trace deformations, it follows that there is a special point at which boundary 
scalar dynamics is governed by the simple $\theta_0=0$ rule. It certainly 
seems natural to conjecture that this special theory is governed by a 
Lagrangian with no double trace terms, i.e. the pure Chern Simons theory 
described in the previous subsection. As we will explain below, 
this assumption unfortunately appears to clash with an atleast equally 
natural assumption about the AdS/CFT implementation of the boundary Chern 
Simons gauging of a global symmetry, as we review below. 

\subsubsection{Identification of bulk and boundary Chern Simons terms}

As we have explained in section \ref{gglobal}, it is very natural to 
simply identify the boundary field theoretic Chern Simons 
term with a  Chern Simons
term for the boundary value of bulk gauge fields. If we make this assumption 
then it follows that  the boundary conditions for bulk vector uniquely specify 
its boundary dynamics and the comparison of gauge field structures between 
the bulk and the boundary establish a map between moduli spaces of field 
theories and the Vasiliev dual. As we have mentioned in the previous 
subsubsection, however, the results obtained in this manner clash with 
those obtained from the `natural' identification of the specially 
simple field theory as far as scalar double trace operators are concerned. 
As we explain, one way out of this conundrum is to abandon the `natural' 
assumption of the previous subsection. However we do not propose a d
definitive resolution to this clash in this paper, leaving 
this for future work.

In the rest of this section we present a detailed comparison between 
double trace deformations of the field theory Lagrangian and boundary 
conditions of the dual Vasiliev theory, for the various theories we 
study, starting with those theories that allow a nontrivial matching 
of gauge field terms. 

\subsection{${\cal N}=3$ fixed line with 1 hypermultiplet}

In this section we present a detailed comparison of 
the Lagrangian \ref{N31hypr} of a fixed line 
of one hypermultiplet ${\cal N}=3$ theories with boundary conditions
\eqref{N=3bcP} of its conjectured Vasiliev dual. 

\subsubsection{Boundary conditions for the vector}\label{thetantN}

As described in the section \ref{gglobal}, the Chern-Simons gauging of 
the boundary global current results in modifying the boundary conditions 
for the 
dual gauge field in the bulk. The modified boundary condition are given by 
\eqref{finbc} which can also be written as 
\begin{equation}\label{gfmbc}
\epsilon_{ijk}F_{jk}= \frac{i\pi\tilde{N}}{4k}F_{zi}.
\end{equation}
The form of boundary conditions for gauge field used in section 
\ref{pssbbc} 
\begin{equation}
B^{(1)}(\vec x,z|Y)\big|_{\cO(y^2,\bar y^2)} = z^2\left[e^{i\beta}(yFy)+ 
               \Gamma e^{-i\beta}(\bar y \overline F\bar y)\right] + {\cal O}(z^3) 
\end{equation}
are equivalent to 
\begin{equation}\label{gfbcss}
\epsilon_{ijk}F_{jk}=2i\tan(\beta-\theta_0)F_{zi}.
\end{equation}
Comparing \eqref{gfmbc} and \eqref{gfbcss} we get 
\begin{equation}\label{gfbcf}
\tan(\beta-\theta_0)= \frac{\pi\tilde{N}}{8k}.
\end{equation}
From \eqref{N=3bcP} we have
$$\beta= \theta_0 + (\tilde{\beta}- \theta_0) P_\Gamma, $$
where ${\tilde \beta}$ is the free parameter that parameterizes the 
fixed line of boundary conditions \eqref{N=3bcP}. In particular case 
of vectors proportional to $P_\Gamma\beta={\tilde \beta}$. 
Comparing \eqref{gfmbc}, \eqref{gfbcss} and \eqref{gfbcf} it follows that
\begin{equation}\label{formbeta}
\tan(\tilde{\beta}-\theta_0)= \frac{k_1}{k_2}\tan\theta_0,
\end{equation}
where 
\begin{equation}\label{tantheta}
\tan\theta_0 = \frac{\pi\tilde{N}}{8k_1}= \frac{\pi N h_A}{2k_1}.
\end{equation}
Here $h_A$ is the ratio of the two point function of current 
at the ungauged ${\cal N}=3$ point ($k_2= \infty$) to the two 
point function in the free theory. \eqref{formbeta} establishes 
a clear map between the parameter ${\tilde \beta}$ that labels 
boundary conditions in \eqref{N=3bcP} and the parameter $\frac{k_1}{k_2}$ 
that labels the fixed line of dual field theories.

\subsubsection{Scalar double trace deformation}
In this subsection we compare the scalar double trace operators 
in the field theory Lagrangian \eqref{N31hypr} with the boundary 
conditions for scalar fields \eqref{N=3bcP} in the Vasiliev dual. 
  
The scalar double trace deformation in the Lagrangian \eqref{N31hypr} 
is given by 
\begin{equation}\label{lagstl}\begin{split}
V_s &= \frac{2\pi}{k_1} \Phi^a_+\Phi^b_-\eta_{ab} 
        +\frac{2\pi}{k_2}\left( \Phi^0_+\Phi^0_- +\Phi^a_+\Phi^b_-\eta_{ab} \right), \\
    &= -\frac{2\pi}{k_1}\Phi^0_+\Phi^0_- 
       + \frac{2\pi}{k_1} \left( 1+ \frac{k_1}{k_2} \right) \Phi_+^i\Phi_-^i. \\
\end{split}
\end{equation} 
This potential interpolates between that of the ${\cal N}=3$ ungauged 
theory ($k_2=\infty$) and ${\cal N}=4$ theory ($k_2=-k_1$). 
The two point function of $\Phi^a_\pm$ are twice of those given in 
\eqref{dttpf} and thus matches with \eqref{norm}. 
The boundary conditions for scalar fields are described by the rotation 
angle
\begin{equation} \label{bcondit}
\gamma = \theta_0 P_1 + \tilde\beta P_{\psi_1\psi_4,\psi_2\psi_4,\psi_3\psi_4}. 
\end{equation}

The double trace term $\frac{2\pi}{k_1} ( 1+ \frac{k_1}{k_2}) 
\Phi_+^i\Phi_-^i$ couples two $SO(3)$ vectors. The rotation 
angle that multiples $P_{\psi_1\psi_4,\psi_2\psi_4,\psi_3\psi_4}$ in \eqref{bcsf}
is determined by the coefficient of this term. The precise 
relationship between these may be obtained as follows. Let us 
suppose that the formula \eqref{bcsf} applies starting from some 
as yet unknown point, ${\tilde \beta}={\tilde \beta}_0$, in the moduli 
space of theories. In other words we hypothesize that $\theta_0=0$ 
(in the language of Appendix \ref{tsf}) for the point in moduli 
space with ${\tilde \beta}= {\tilde \beta}_0$. Let us also suppose that 
$k_2=(k_2)_0$
corresponding field theory. It follows then from  \eqref{bcsf}, 
\eqref{bcondit} and \eqref{lagstl} that (see below for the numerical values of the proportionality constants)
$$ \tan ({\tilde \beta} - {\tilde \beta}_0) \propto 
\frac{1}{k_2}- \frac{1}{(k_2)_0}.$$

\noindent {\it Case:  ${\tilde \beta_0}=0$}:

Purely from the viewpoint of the scalars it is 
natural to conjecture that ${\tilde \beta}_0=0$ and $(k_2)_0=-k_1$. 
This 
conjecture is motivated by the following observations. The contact term 
in the two point function between $\Phi_+^i$ and $\Phi_-^i$ vanishes in the 
field theory dual to  bulk boundary conditions governed by the parameter 
${\tilde \beta}_0$.
At leading order in boundary perturbation theory (i.e. at order $1/k$) a 
naive computation yields a contact term proportional to the double trace 
coupling of  $\Phi_+^i$ and $\Phi_-^i$. Thus appears to imply that the 
special field theory have a vanishing double trace term; this occurs 
at the ${\cal N}=4$ point and so ${\tilde \beta}_0=0$. If we make this assumption it then follows that 
that \begin{equation}\label{N34rel1}
\tan \tilde\beta = \tan \theta_0 \left( 1+ \frac{k_1}{k_2} \right),
{\rm ~~~~with~~~} \tan\theta_0 = \frac{N\pi}{2k_1}\sqrt{h_+ h_-},
\end{equation}
where $h_+$ and $h_-$ is the ratio of two point function for $\Phi_+$ and 
$\Phi_-$ respectively in the interacting (${\cal N}=4$ point) to free 
theory. Unfortunately \eqref{N34rel1} conflicts with 
\eqref{formbeta}, so both relations cannot be simultaneously correct.

\bigskip

\noindent {\it Case: ${\tilde \beta}_0=\theta_0$}:

The conflict with \eqref{N34rel2} vanishes if we instead assume that 
\begin{equation}\label{identaa}
{\tilde \beta}_0=\theta_0.
\end{equation} 
This is dual to the `ungauged' ${\cal N}=3$ theory and so it follows that 
and $(k_2)_0= \infty$. Under this assumption it follows that  
\begin{equation}\label{N34rel2}
\tan ( \tilde\beta- \theta_0) = \tan\theta_0 \left( \frac{k_1}{k_2} \right), 
{\rm ~~~~with ~~~~} \tan\theta_0 = \frac{N\pi}{2k_1}\sqrt{h_+ h_-},
\end{equation}
where $h_+$ and $h_-$ is the ratio of two point function for $\Phi_+$ and 
$\Phi_-$ respectively in the interacting (`ungauged' ${\cal N}=3$ point) to free theory.
Note that \eqref{N34rel2} perfectly matches \eqref{tantheta} 
if $h_A=\sqrt{h_+ h_-}$. It is plausible that supersymmetry enforces
this relationship on field theory operators, but we will not attempt 
to independently verify this relationship in this paper. 

Perhaps the simplest resolution of the clash between  
\eqref{N34rel1} and \eqref{formbeta} is obtained by setting 
${\tilde \beta}_0=\theta_0$. Before accepting this suggestion we must 
understand why the contact term in the scalar- scalar two point function 
vanishes at the ${\cal N}=3$ rather than at the ${\cal N}=4$ point 
(where the double trace term in the Lagrangian vanishes). As discussions
relating to contact terms are famously full of pitfalls; we postpone the
detailed study of this question to later work.  

\bigskip

\noindent {\it Coefficient of the scalar double trace deformation}

The double trace term in \eqref{lagstl} that couples two $SO(3)$ scalars
is $\frac{2 \pi}{k_1}\Phi^0_+ \Phi^0_-$. Note that the coefficient 
of this term is independent of $k_2$, which matches with the fact that 
the coefficient of  $P_1$ in \eqref{bcondit} is independent of 
${\tilde \beta}$. 

If we assume that ${\tilde \beta}_0=\theta_0$ for this term as well 
we once again find the second of \eqref{N34rel1}, where $h_+$ and $h_-$
have the same meaning as in \eqref{N34rel1}, except that the two 
point function in question is that of the the scalar 
operator $\phi^0$. We conclude that $\phi^a$ and $\phi^0$ have 
equal values of $h_+h_-$.  If, instead, ${\tilde \beta}=0$ then a very 
similar equation holds; the only difference is that $h_+h_-$ would then 
compute ratios of the interacting and free two point functions at the 
${\cal N}=4$ point.

\subsubsection{Fermionic double trace deformation}

The fermionic double trace deformation for this fixed line is given by 
\begin{equation}\begin{split}
V_3 &= \frac{2\pi}{k_1} \left( \half \bar\Psi^a\Psi^b\delta^{ab} 
       -2\bar\Psi^0\Psi^0 -\bar\Psi^0\bar\Psi^0- \Psi^0\Psi^0 \right) 
       +\frac{2\pi}{k_2} \left( \bar\Psi^a\Psi^b\eta^{ab} +\half \bar\Psi^a\bar\Psi^b\eta_{ab} 
       +\half \Psi^a\Psi^b\eta^{ab} \right). \\
\end{split}
\end{equation}
Adding $\delta V_f= \frac{3\pi}{k}{\bar \psi}^a \psi^a$ in order to account the 
effect of finite Chern Simons level as described earlier, we obtain the 
shifted potential 
\begin{equation}
\begin{split}
V_3+ \delta V_f=& -\frac{\pi}{k_1} (\Psi^a - \bar\Psi^a)(\Psi^b - \bar\Psi^b)\delta^{ab}
     + \frac{\pi}{k_1}\left( 1 + \frac{k_1}{k_2} \right) 
     \left( \bar\Psi^a  + \Psi^a \right) \eta_{ab} \left( \bar\Psi^b + \Psi^b \right). \\
\end{split}
\label{N3quips}
\end{equation}
The two point function of $\langle \bar\Psi^a \Psi^b \rangle$ is twice of the 
that given in \eqref{dttpf} because $\Psi^a$ are constructed out of field doublets 
and thus matches with \eqref{fnorm}. 

The rest of the analysis closely mimics the study of scalar double 
trace deformations presented in the previous subsection. 
We associate(in the boundary conditions) the projector  $P_\psi^a$ with the 
real Lagrangian deformation $[i(\psi^a-\bar\psi^a)]^2$ and $P_{\Gamma \psi^a}$ 
with the other real Lagrangian deformation $(\psi^a +\bar \psi^a)^2$. 
As for the scalar double trace deformations, \eqref{mbc} yields results 
consistent with \eqref{formbeta} if and only if we assume that \eqref{mbc}
applies for deformations about the special point ${\tilde \beta}=\theta_0$.
Given this assumption \eqref{N=3bcP} 
and \eqref{mbc} matches with the identification \eqref{N34rel2} with 
$\sqrt{h_+h_-}= h_{\psi}$ and $h_{\psi}$ 
interpreted as the ratio of $\langle \bar\Psi^a\Psi^b \rangle$ at ${\cal N}=3$ point to the free theory.\footnote{If, on the other hand, \eqref{mbc} had applied for deformations
around ${\tilde \beta}=0$ we would instead have found agreement with 
\eqref{N34rel1} with $\sqrt{h_+h_-}= h_{\psi}$, where $h_{\psi}$ 
would have been interpreted as the ratio of 
$\langle \bar\Psi^a\Psi^b \rangle$ at ${\cal N}=4$. Of course these results 
contradict \eqref{formbeta}.}

\subsection{${\cal N}=3$ fixed line with 2 hypermultiplets}
In this section we compare the Lagrangian for the fixed line of two 
hypermultiplet theories presented in \eqref{N32hypr} with the boundary
conditions \eqref{N=3bcP2} of the conjectured Vasiliev duals. The 
field theories under study interpolate between the ungauged 
${\cal N}=3$ theory ($k_2=\infty$) and the ${\cal N}=6$ theory 
(at $k_2= -k_1$).

\subsubsection{Vector field boundary conditions}\label{thetantN2}
The comparison here is very similar to that performed in the 
previous subsection, and our presentation will be brief. Making the 
natural assumptions spelt out in the previous section, the gauge 
field boundary conditions listed in  \eqref{N=3bcP2} assert that 
$$ \beta= \theta_0 + (\tilde{\beta}-\theta_0)P_\Gamma. $$
Using \eqref{gfbcf} we find
\begin{equation}\label{stt}
\tan(\tilde{\beta}-\theta_0) = \frac{k_1}{k_2} \tan2\theta_0.
\end{equation}
with the identification 
$$ \tan(2\theta_0) = \frac{\pi\tilde{N}}{8k_1}= \frac{\pi N h_A}{k_1} $$
where $h_A$ is interpreted as the ratio of the two point function of the 
flavor current in the ungauged ${\cal N}=3$ theory to the free theory.

\subsubsection{Scalar double trace deformation}

The scalar double trace deformation for this case, in the notation defined in 
appendix \eqref{N32hypr}, is given by 
\begin{equation}\begin{split} \label{lssn6}
V_s &= \frac{\pi}{k_1}\Phi_+^{Ii}\Phi_-^{Jj}\eta^{IJ}\eta_{ij} 
      -\frac{2\pi}{k_2}\Phi_+^{I0}\Phi_-^{J0}\eta^{IJ} \\
    &= \frac{\pi}{k_1}\left( \Phi_+^{Ii}\Phi_-^{Jj}\eta^{IJ}\eta_{ij} 
      + 2\Phi_+^{I0}\Phi_-^{J0}\eta^{IJ} \right) 
      - \frac{2\pi}{k_1} \left( 1+\frac{k_1}{k_2} \right) \Phi_+^{I0}\Phi_-^{J0}\eta^{IJ}. \\
\end{split}
\end{equation} 
Due the fact that $\Phi^{Ii}_+$ and $\bar\Phi^{Ii}_-$ are made of two field 
doublets, there free two point function are four times of those given in \eqref{dttpf} 
and thus twice of those given in \eqref{norm}. The boundary 
conditions of the dual scalars listed in  \eqref{N32hypr} is governed by 
\begin{equation}\label{bcssn6}
\gamma = \theta_0P_{1,\psi_i\psi_a,\psi_a\psi_b}-\tilde\beta P_{\psi_i\psi_j}
, ~~~~P_{1,\psi_i\psi_j,\psi_i\psi_a,\psi_a\psi_b} \tilde f_{1,2} = \tilde f_{1,2} .
\end{equation}
As in the previous section the coefficient of the double trace deformations 
\eqref{lssn6} and the boundary conditions of scalars in \eqref{bcssn6}
are both respectively independent of $k_2$ and ${\tilde \beta}$ in 
every symmetry channel but one (i.e. (vector, scalar) under 
$SU(2) \times SU(2)$). Comparing coefficients in this special channel
we find that \eqref{bcssn6} and \eqref{N32hypr} agree with 
\eqref{gfbcf}if and only if we assume that \eqref{bcsf} applies for 
deformations of ${\tilde \beta}$ away from the special point 
${\tilde \beta}_0=\theta$ at which point $k_2=\infty$. 
\begin{equation}\label{N36rel2}
\tan( \tilde\beta- \theta_0 ) = \tan2\theta_0 \left( \frac{k_1}{k_2} \right)
{\rm ~~~~ with ~~~~} \tan 2\theta_0 = \frac{\pi N}{k_1}\sqrt{h_+h_-},
\end{equation} 
with $h_\pm$ interpreted as the ratio of two point function in ${\cal N}=3$ 
ungauged point to free theory.

On the other hand upon assuming  $\tilde \beta_0=0$ we find 
\begin{equation}\label{N36rel1}
\tan( \tilde\beta+ \theta_0 ) = \tan2\theta_0 \left( 1+ \frac{k_1}{k_2} \right) 
{\rm ~~~~with ~~~~} \tan 2\theta_0 = \frac{\pi N}{k_1}\sqrt{h_+h_-},
\end{equation}
with $h_\pm$ interpreted as the ratio of two point function in ${\cal N}=6$ point to free theory. This is in contradiction with \eqref{stt}.

We now turn to the comparison of the double trace terms and boundary 
conditions in all other channels (i.e. (scalar, scalar), (vector, vector)
and (scalar, vector) under $SO(3) \times SO(3)$. In each case if
 we assume that \eqref{bcsf} applies starting from the special point
${\tilde \beta}_0=\theta_0$, we find the second of \eqref{N36rel2} with 
with $h_\pm$ interpreted as the ratio of two point function in ${\cal N}=3$ 
ungauged point to free theory for the appropriate scalar. This suggests
that the product $h_+h_-$ is the same for scalars in all four symmetry 
channels; this product is also equal to $h_A^2$. It is possible that 
this equality is consequence of ${\cal N}=3$ supersymmetry of the 
field theory; we leave the verification of this suggestion to future work.

\subsubsection{Fermionic double trace deformation}
The fermionic double trace deformation for this case, in the notation defined in 
appendix \eqref{N32hypr}, after compensating by a for the chern simons shift
\footnote{The compensating factor in this case is $\delta V_f= \frac{3\pi}{2k_1} 
\bar\Psi^{Ii}\Psi^{Ii}$}, is given by 
\begin{equation}
\begin{split}
V_f+ \delta V_f &= \frac{\pi}{k_1} \bigg( \bar\Psi^{Ii}\Psi^{Jj}\delta^{IJ}\delta^{ij} 
     + \bar\Psi^{Ii}\Psi^{Jj}\eta^{IJ}\delta^{ij} + \left(\bar\Psi^{0i}\bar\Psi^{0j}\eta_{ij} 
     +\Psi^{0i}\Psi^{0j}\eta_{ij} \right) \bigg) \\
    & +\frac{\pi}{k_2} (\bar\Psi^{I0}+\Psi^{I0})(\bar\Psi^{J0}+\Psi^{J0})\eta_{IJ}. \\
    &= \frac{\pi}{k_1} \bigg( \bar\Psi^{Ii}\Psi^{Jj}\delta^{IJ}\delta^{ij} 
      + \bar\Psi^{Ii}\Psi^{Jj}\eta^{IJ}\delta^{ij} + \left(\bar\Psi^{0i}\bar\Psi^{0j}\eta_{ij} 
      + \Psi^{0i}\Psi^{0j}\eta_{ij} \right) \\
    & - (\bar\Psi^{I0}+\Psi^{I0})(\bar\Psi^{J0}+\Psi^{J0})\eta_{IJ} \bigg) 
      + \frac{\pi}{k_1}\left( 1+ \frac{k_1}{k_2} \right) (\bar\Psi^{I0}+\Psi^{I0})(\bar\Psi^{J0}+\Psi^{J0})\eta_{IJ} .\\
\end{split}
\end{equation}
The two point function  
$\langle \bar\Psi^{Ii}\Psi^{Jj} \rangle$ is twice of that given by \eqref{fnorm}. 

The bulk boundary conditions are generated by  
$$ \alpha = \theta_0( P_{\psi_i,\psi_a}+P_{\psi_i\psi_j\psi_a,\psi_i\psi_a\psi_b,\psi_4\psi_5\psi_6}
            -P_{\psi_a\Gamma})+\tilde\beta (P_{\psi_i\Gamma}- P_{\psi_1\psi_2\psi_3}) .$$
Consistency requires us to assume that \eqref{mbc} applies for deviations 
away from ${\tilde \beta}=0$ (i.e. from the ungauged ${\cal N}=3$ theory).
Applying \eqref{mbc} we recover \eqref{N36rel2} provided 
$h_\psi=\sqrt{h_+h_-}$ where $h_\psi$ is the ratio the two point function 
$\langle \bar\Psi^{Ii}\Psi^{Jj} \rangle$ 
at the ungauged ${\cal N}=3$ point to free theory.\footnote{If, instead, \eqref{mbc} had applied starting from ${\tilde \beta}=0$ we would have found consistency with \eqref{N36rel1} provided 
$h_\psi=\sqrt{h_+h_-}$ where 
$h_\psi$ interpreted as the ratio the two point function 
$\langle \bar\Psi^{Ii}\Psi^{Jj} \rangle$ at ${\cal N}=6$ point to 
free theory. This result contradicts the gauge field matching
and so cannot apparently cannot be correct. }

\subsection{Fixed Line of ${\cal N}=1$ theories}

We now turn to the comparison of the Lagrangian  \eqref{pN=1} of the 
large $N$ fixed line of ${\cal N}=1$ field theories with the boundary conditions 
\eqref{nobc} (a beta function is generated at finite $N$, the zeros of 
this beta function are the two ends of the line we study below). 
 We restrict attention 
to the case $M=1$. The field content of the theory is a single 
complex scalar $\phi$ together with a single complex fermion $\psi$.

\subsubsection{Scalar Double trace terms}

The (scalar)(scalar) double trace potential in \eqref{pN=1} is given by 
\begin{equation}\label{sdtn1l}
 \frac{2 \pi(1+\omega)}{k} {\bar \phi} \phi {\bar \psi} \psi.  
\end{equation}
$\omega=-1$ is the ${\cal N}=1$ theory with no superpotential while 
$\omega=1$ is the ${\cal N}=2$ theory. 
The two point functions of the constituent single trace operators, ${\bar \phi} \phi$ and 
$\bar{\psi} \psi$, are given, in the free theory, by \eqref{dttpf} (note that this corresponds to 
$h_+=h_-=\frac{1}{2}$ in \eqref{norm}).

The $n=2$ Vasiliev dual to this system is conjectured to have boundary 
conditions listed in \eqref{N1n2bc}. Specifically the boundary 
conditions require $B$ to  take the form 
\begin{equation}\label{bcl}
 B(x,z)= z f_1(x) \left( (1+\Gamma) \cos \gamma  + i (1-\Gamma) \sin \gamma \right) 
 + i f_2(x) z^2  \left( (1-\Gamma) \cos \gamma  + i (1+\Gamma) \sin \gamma \right) 
\end{equation}
where $f_1$ and $f_2$ are real constants and $\gamma$ ranges from zero (for the ${\cal N}=1$ 
theory with no superpotential) to $\gamma= \theta_0$ (for the ${\cal N}=2$ theory).
Notice that the shift change in phase between these two points is $\theta_0$, while the 
change in the coefficient of the corresponding double trace term in the Lagrangian 
\eqref{sdtn1l} is $\frac{4 \pi}{k}$. 

In order to establish a map between the Lagrangian parameter $\omega$ and 
the boundary condition parameter $\gamma$ we need to know the location 
of the special point, $\gamma_0$, in $\gamma$ parameter space
from which \eqref{bcsf} applies (this is the point with $\theta_0=0$ in 
the language of Appendix \ref{tsf}). Unlike the previous subsections, in 
this case we have no information from the gauge field boundary conditions, 
so the best we can do is to make a guess. We consider two cases. 

\bigskip

\noindent {\it Case ${\gamma_0}=\theta_0$}:

The results of the previous
subsection suggest that $\gamma_0=\theta_0$ so that the special 
point in the moduli space of Vasiliev theories is the ${\cal N}=2$ 
theory. If this is the case then 
$$\tan ( \theta_0-\gamma)= \tan \theta_0 \frac{1-\omega}{2}$$
where 
\begin{equation} 
 \tan \theta_0= \frac{\pi\lambda \sqrt{h_+ h_-}}{2}
\end{equation}
and $h_+$ gives the ratio of the interacting and free two point functions of ${\bar \phi} \phi$ 
for the ${\cal N}=2$ theory. 

\bigskip

\noindent 
{\it Case $\gamma_0=0$}:

Purely from the point of view of the scalar part of the Lagrangian, the 
most natural assumption is $\gamma_0=0$ in 
which case 
$$\tan \gamma= \tan \theta_0 \frac{1+\omega}{2}$$
where 
\begin{equation} 
 \tan \theta_0= \frac{\pi \lambda \sqrt{h_+ h_-}}{2}
\end{equation}
and $h_+$ gives the ratio of the interacting and free two point functions of ${\bar \phi} \phi$ 
for the ${\cal N}=1$ theory with no superpotential.

\subsubsection{Fermion double trace terms}
The (fermion)(fermion) double trace potential term after accounting for the shift 
described in 
\begin{equation}\label{fdtn1l}\begin{split}
V_f+\delta V_f &= V_f + \frac{6\pi}{k}\bar\psi\phi \bar\phi\psi \\
               &= \frac{\pi(\omega+1)}{k}(\bar\psi\phi+ \bar\phi\psi)^2 
               -\frac{2\pi}{k}(\bar\psi\phi- \bar\phi\psi)^2.
\end{split}
\end{equation}
Here $\omega=-1$ corresponds to the undeformed ${\cal N}=1$ theory and 
$\omega=1$ corresponds to the ${\cal N}=2$ theory. The two point function of 
the operator $\bar\psi\phi$ and $\bar\phi\psi$ are given in \eqref{dttpf}. 
Note that this corresponds to $h_\psi=\half$ in \eqref{norm}.
The boundary condition for fermions are given by \eqref{fbcc} with 
$$\alpha= \theta_0 P_{\psi_2} + \gamma P_{\psi_1}.$$ 
As explained in the previous section, the coefficient of the $P_{\psi_2}$ in 
the boundary conditions is associated with the coefficient of double trace 
deformation $(i(\bar\psi\phi-\bar\phi\psi))^2$ while the coefficient of 
$P_{\psi_1}$ is associated with the double trace deformation 
$(\bar\psi\phi+ \bar\phi\psi)^2$. Note that this matches with the fact 
that coefficient of the former are constant along the line while those 
of the later change along the fixed line. 

Using the analysis of section \eqref{fermbc} we can get a more quantitative 
match. As in the previous subsubsection it is natural to assume - and 
we conjecture - that 
If \eqref{mbc} applies starting from the ${\cal N}=2$ point, at which the 
first term in \eqref{fdtn1l} has coefficient $\frac{2\pi}{k}$. With 
this assumption
\begin{equation}
\tan (\theta_0- \gamma) = \tan \theta_0 \frac{1-\omega}{2}, 
{\rm ~~~~with ~~~~} \tan\theta_0 =\frac{\pi\lambda h_{\psi}}{2} ,
\end{equation}
where $h_{\psi}$ is the ratio of interacting to free two point function 
$\langle \bar\psi\phi~\bar\phi\psi \rangle$ in ${\cal N}=2$ theory.

If, on the other hand \eqref{mbc} were to apply starting from the pure
${\cal N}=1$ point we would find 
\begin{equation}
\tan \gamma = \tan \theta_0 \frac{1+\omega}{2}, 
{\rm ~~~~with~~~~} \tan\theta_0 = \frac{\pi\lambda h_{\psi}}{2}.
\end{equation}
where $h_{\psi}$ is the ratio of interacting and free two point function 
$\langle \bar\psi\phi~\bar\phi\psi \rangle$ in ${\cal N}=1$ theory with no 
superpotential. The results of the previous two subsections appear to 
disfavor this possibility over the one presented in the previous 
paragraph.

\subsection{${\cal N}=2$ theory with 2 chiral multiplets}

In the final subsection of this section we turn to the comparison of 
the Lagrangian \eqref{N2Mch} (with $M=2$) of the  ${\cal N}=2$ theory 
with 2 
fundamental chiral multiplets with the boundary conditions \eqref{ntbc}.
The theory we study admits no marginal superpotential deformations, and 
so appears as a fixed point rather than a fixed line at any given value 
of $k_1$. 

\subsubsection{Scalar double trace deformation}
The scalar double trace deformation in \eqref{N2Mch} is given by 
\begin{equation}
V_s = \frac{2\pi}{k}\Phi^a_+ \Phi^a_-,
\end{equation}
where $\Phi^a_+= \bar\phi^i \phi_j (\sigma^a)^j_{~i}$, 
$\Phi^a_-= \bar\psi^i \psi_j (\sigma^a)^j_{~i}$ and $a$ runs over 0,1,2,3. 
In Appendix \ref{fft} we have
computed the two point functions of the operators $\Phi^a_+$ and $\Phi_-^a$
in free field theory; the result is given by \eqref{dttpf} with an extra
factor of two to account for the fact that the operators $\Phi^a_\pm$ are 
constructed out of field doublets. In other words the two point functions 
of $\Phi^a_\pm$ exactly agree with those presented in \eqref{norm} with 
$h_+$ and $h_-$ interpreted as the ratio of the two point functions of 
$\Phi_\pm$ in the interacting theory and the free theory
\footnote{Here it is ambiguous what is the interacting theory i.e. 
what is the value of k in theory without the double trace deformations,
from where \eqref{bcsf} applies}. With this 
interpretation \eqref{bcsf} predicts the boundary conditions of the 
bulk scalars with $d_{a\alpha}=1$ (both for the singlet of $SU(2)$ as 
well as the triplet). Comparing these equations 
with the actual boundary conditions 
$$ \gamma = \theta_0, 
~~~~ P_{1,\psi_2\psi_3,\psi_2\psi_4,\psi_3\psi_4}\tilde f_{1,2}=\tilde f_{1,2}, $$
we conclude that 
$g_a= g_\alpha$ both for singlet scalars as well as for 
$SU(2)$ triplet scalars. 

In order to make a quantitative comparison between the Lagrangian and 
boundary conditions we need to make an assumption about which point 
in the moduli space of double trace deformations \eqref{bcsf} applies 
from. Given the results of the previous subsections it is natural 
to guess that \eqref{bcsf} applies for double trace deformations away 
from the ${\cal N}=2$ theory. Assuming that the theory with no double 
trace deformation has trivial scalar boundary conditions, we conclude that 
\begin{equation}\label{gam}
\tan\theta_0 =  \frac{\pi\lambda\sqrt{h_+ h_-}}{2} .
\end{equation}
where $h_{\pm}$ are the ratios of two point functions of the scalar 
operators in the ${\cal N}=2$ and free theories. 
This equation must hold separately for singlet as well as SU(2) vector 
sector. It seems very likely that $h_+=h_-=h_s$ for all scalars in which 
case
\begin{equation}\label{gaman}
\tan\theta_0 = \frac{ \pi  \lambda h_s}{2}. 
\end{equation}

\subsubsection{fermion double trace deformation}
The fermion double trace deformation in this case is given by 
\begin{equation}
V_f= \frac{\pi}{k} \bar\Psi^a\Psi^a,
\end{equation}
where $\Psi^a= \bar\phi^i \psi_j (\sigma^a)^j_{~i}$, 
$\bar\Psi^a= \bar\psi^i \phi_j (\sigma^a)^j_{~i}$ and $a$ runs over 0,1,2,3. 
In order to compare this double trace potential with boundary conditions, 
however, we must remove the effect of the Chern Simons term. In other 
words we should expect the fermion boundary conditions to match with 
an effective fermion double trace potential 
given by 
$$\delta S= \frac{4\pi}{k} {\bar \Psi^a} \Psi^a. $$
(it is easily verified that a shift by $- \frac{3 \pi}{k}$ in the 
coefficient of ${\bar \Psi^a} \Psi^a$ is equivalent to a shift of 
$- \frac{6 \pi}{k}$ 
in the coefficient of each fermion). 
The two point functions of these fields is given by (see Appendix \ref{fft})
$$ \vev{\Psi^a(x) {\bar \Psi^b(0)}}= 
\frac{ N \delta^{ab} h_\psi}{8 \pi^2} \frac{\vec x\cdot\vec \sigma}{x^4},$$
where $h_\psi$ is the ratio of the two point function in the interacting
and free theories. 

This matches onto the analysis leading up to \eqref{mbc} if we set 
$s=t=4$ and $u=0$. Here we assume that \eqref{mbc} applies for deformations
about the ${\cal N}=2$ point. 
 In this application of \eqref{mbc} all factors of $g_a$ 
relate to fields that are related by $SO(4)$ invariance, and so must be 
equal. Consequently factors of $g_a$ cancel from that equation. 
Comparing \eqref{mbc} with $s=t=4$ and $u=0$ with 
the actual fermion boundary conditions, in this case 
$$\alpha=\theta_0,$$ 
we recover the equation 
\begin{equation}\label{gamf}
\tan\theta_0 =  \frac{ \pi\lambda h_\psi}{2}. 
\end{equation}
We see that \eqref{gamf} is consistent with \eqref{gam} 
provided $h_\psi= \sqrt{h_+ h_-}$, with $h_\psi$ interpreted as the ratio
of the two point function in the ${\cal N}=2$ and free theories. 
It seems very likely to us that in fact $h_\psi=h_+=h_-=h_s$. \\

\section{The ABJ triality}

Having established the supersymmetric Vasiliev theories with various boundary 
conditions dual to Chern-Simons vector models, we will now use the relation between 
deformations of the boundary conditions and double trace deformations in the boundary 
conformal field theory to extract some nontrivial mapping of parameters. In the case 
of ${\cal N}=6$ theory, the triality between ABJ vector model, Vasiliev theory, 
and type IIA string theory suggests a bulk-bulk duality between Vasiliev theory and 
type IIA string field theory. We will see that the parity breaking phase $\theta_0$ 
of Vasiliev theory can be identified with the flux of flat Kalb-Ramond $B$-field in 
the string theory.


\subsection{From ${\cal N}=3$ to ${\cal N}=4$ Chern-Simons vector models}

Let us consider the ${\cal N}=3$ $U(N)_k$ Chern-Simons vector model with one hypermultiplets. Upon gauging the diagonal $U(1)$ flavor symmetry with another Chern-Simons gauge field at level $-k$, one obtains the ${\cal N}=4$ $U(N)_k\times U(1)_{-k}$ theory. In section \ref{thetantN}, by comparing the boundary conditions, we have found the relation
\begin{equation}
\tan\theta_0 = \frac{\pi\tilde{N}}{8k}= \frac{\pi \lambda h_A}{2}.
\end{equation}
By comparing the structure of three-point functions with the general results of \cite{Maldacena:2012sf}, we see that $\tan\theta_0$ is identified with $\widetilde\lambda$ of \cite{Maldacena:2012sf}. Therefore, by consideration of supersymmetry breaking by $AdS$ boundary conditions, we determine the relation between the parity breaking phase $\theta_0$ of Vasiliev theory and the Chern-Simons level of the dual ${\cal N}=3$ or ${\cal N}=4$ vector model to be
\ie\label{lnk}
\widetilde\lambda = {\pi\widetilde N\over 8k}.
\fe
Recall that $\widetilde N$ is defined as the coefficient of the two-point function of the $U(1)$ flavor 
current $J_i$ in the ${\cal N}=3$ Chern-Simons vector model, normalized so that $\widetilde N$ is $4$ 
for each {\it free} hypermultiplet. 
In notation similar to that of the previous section 
${\tilde N}=4 N h_A$ where $h_A$ is the ratio of the two point function of the 
flavour currents in the interacting and free theory. Consequently 
\eqref{lnk} may be rewritten as 
\ie\label{lnkk}
\widetilde\lambda = {\pi \lambda h_A\over 2}.
\fe

After gauging this current with $U(1)$ Chern-Simons gauge field $\widetilde A_\mu$ at level $-k$, passing to the ${\cal N}=4$ theory, the new $U(1)$ current which may be written as $J_{new}=-k*d\widetilde A$ has a different two-point function than $J_i$, as can be seen from section 3.1. The two-point function of  $J_{new}$ also contains a parity odd contact term, as was pointed out in \cite{Witten:2003ya}.

We would also like to determine the relation between $\theta_0$ and $\lambda=N/k$, which cannot be fixed directly by the consideration of supersymmetry breaking by boundary conditions. The two-loop result of \cite{Giombi:2011kc} on the parity odd contribution to the three-point functions also applies to correlators of singlet currents made out of fermion bilinears in supersymmetric Chern-Simons vector models, since the double trace and triple terms do not contribute to the parity odd terms in the three-point function at this order. From this we learn that $\theta_0 = {\pi\over 2}\lambda + {\cal O}(\lambda^3)$. Parity symmetry would be restored if we also send $\theta(X)\to -\theta(X)$ under parity, and in particular $\theta_0 \to -\theta_0$. Further, in the supersymmetric Vasiliev theory, $\theta_0$ should be regarded as a periodically valued parameter, with periodicity $\pi/2$. This is because the shift $\theta_0 \to \theta_0 + {\pi\over 2}$ can be removed by the field redefinition ${\cal A}\to \psi_1 {\cal 
A}\psi_1$, $B\to -i\psi_1 B\psi_1$, where $\psi_1$ is any one of the Grassmannian auxiliary variables. Note that the factor of $i$ in the transformation of the master field $B$ is required to preserve the reality condition. Essentially, $\theta_0 \to \theta_0 + {\pi\over 2}$ amounts to exchanging bosonic and fermionic fields in the bulk.

Giveon-Kutasov duality \cite{Giveon:2008zn} states that the ${\cal N}=2$ $U(N)_k$ Chern-Simons theory with $N_f$ fundamental and $N_f$ anti-fundamental chiral multiplets is equivalent to the IR fixed point of the ${\cal N}=2$ $U(N_f+k-N)_k$ theory with the same number of fundamental and anti-fundamental chiral multiplets, together with $N_f^2$ mesons in the adjoint of the $U(N_f)$ flavor group, and a cubic superpotential coupling the mesons to the fundamental and anti-fundamental superfields. Specializing to the case $N_f=1$ (or small compared to $N,k$), this duality relates the ``electric" theory: ${\cal N}=2$ $U(N)_k$ Chern-Simons vector model with $N_f$ pairs of $\Box,\overline\Box$ chiral multiplets at large $N$, to the ``magnetic" theory obtained by replacing $\lambda\to 1-\lambda$ and rescaling the value of $N$, together with turning on a set of double trace deformations and flowing to the critical point. In the holographic dual of this vector model, the double trace deformation in the definition of 
the magnetic theory simply amounts to changing the boundary condition on a set of bulk scalars and fermions. This indicates that the bulk theory with parity breaking phase $\theta_0(\lambda)$ should be equivalent to the theory with phase $\theta_0(1-\lambda)$, suggesting that the identification 
\ie\label{thlam}
\theta_0 = {\pi\over 2}\lambda
\fe
is in fact exact in the duality between Vasiliev theory and ${\cal N}=2$ Chern-Simons vector models of the Giveon-Kutasov type. By turning on a further superpotential deformation, this identification can be extended to the ${\cal N}=3$ theory as well. Together with (\ref{lnkk}), (\ref{thlam}) then implies that relation $\tan({\pi\over 2}\lambda) = {\pi\widetilde N\over 8k}
= \frac{\pi \lambda h_A}{2}$ in the ${\cal N}=3$ Chern-Simons vector model in the planar limit. Note that in the $k\to \infty$ limit where the theory becomes free, this relation becomes the simply $\widetilde N = 4N$, which follows from our normalization convention of the spin-1 flavor current.

A similar comparison between double trace deformations of scalar operators and the change of scalar boundary condition in the bulk Vasiliev theory lead to the same identification between $\theta_0$ and $\widetilde N$, $k$. Note that in the supersymmetric Chern-Simons vector model, $\widetilde N$ by our definition is the two-point function coefficient of a flavor current, which is related to the two-point function coefficient of gauge invariant scalar operators by supersymmetry. However, our $\widetilde N$ is a priori normalized {\it differently} from that of Maldacena and Zhiboedov \cite{Maldacena:2012sf}, where $\widetilde N$ was defined as the coefficient of two-point function of higher spin currents, normalized by the corresponding higher spin charges.\footnote{We thank Ofer Aharony for discussions on this point.}

A high nontrivial check would be to prove the relations (\ref{lnkk}) and (\ref{thlam}) directly in the field theory using the Schwinger-Dyson equations considered in \cite{Giombi:2011kc}. In the case of Chern-Simons-scalar vector model, this computation is performed in \cite{OferUpcoming}. It is found in \cite{OferUpcoming} that the relation $\theta_0=\pi\lambda/2$ holds, whereas the scalar two-point function is precisely proportional to $k\tan\theta_0$ up to a numerical factor that depends on the number of matter fields,\footnote{\cite{OferUpcoming} adopted the natural field theory normalization for the scalar operator, which would agree with our normalization for the flavor current, and differ from the normalization of \cite{Maldacena:2012sf} by a factor $\cos^2\theta_0$. } remarkably coinciding with our finding in the supersymmetric theory by consideration of boundary conditions and holography. We leave it to future work to establish these relations in the supersymmetric theory using purely large $N$ 
field 
theoretic technique.

\subsection{ABJ theory and a triality}

Now let us consider the ${\cal N}=3$ $U(N)_k$ Chern-Simons vector model with two hypermultiplets. Upon gauging the diagonal $U(1)$ flavor symmetry with another Chern-Simons gauge field at level $-k$, one obtains the ${\cal N}=6$ $U(N)_k\times U(1)_{-k}$ ABJ theory. By comparing the boundary conditions, in section \ref{thetantN2}, we have found the formula
\ie
\tan(2\theta_0) = {\pi\widetilde N\over 8k} = \pi\lambda h_A,
\fe
where $\widetilde N$ is the coefficient of the two-point function of the $U(1)$ flavor current in the ${\cal N}=6$ theory, and $h_A$, as usual, is the 
ratio of the flavor current two point function in the interacting and 
free theory. Note that the factor of 
2 in the argument of $\tan(2\theta_0)$ is precisely consistent with the fact that in the $k\to \infty$ limit, the $U(1)$ flavor current which is made out of twice as the ${\cal N}=2$ theory of one hypermultiplet considered in the previous subsection, so that $\widetilde N$ is enhanced by a factor of 2 (namely, $\widetilde N=8N$ in the free limit).

Now we can complete our dictionary of ``ABJ triality". We propose that the $U(N)_k\times U(M)_{-k}$ ABJ theory, in the limit of large $N,k$ and fixed $M$, is dual to the $n=6$ extended supersymmetric Vasiliev theory with $U(M)$ Chan-Paton factors, parity breaking phase $\theta_0$ that is identified with ${\pi\over 2}\lambda$, and the ${\cal N}=6$ boundary condition described in section 4.2.6. The {\it bulk} 't Hooft coupling can be identified as $\lambda_{bulk}\sim M/N$. In the strong coupling regime where $\lambda_{bulk}\sim {\cal O}(1)$, we expect a set of bound states of higher spin particles to turn into single closed string states in type IIA string theory in $AdS_4\times \mathbb{CP}^3$ with flat Kalb-Ramond $B_{NS}$-field flux
\ie
{1\over 2\pi\alpha'}\int_{\mathbb{CP}^1}B_{NS} = {N-M\over k} + {1\over 2}.
\fe
In the limit $N\gg M$, we have the identification
\ie
\theta_0 = {\pi\over 2}\lambda = {1\over 4\A'}\int_{\mathbb{CP}^1}B_{NS} - {\pi\over 4}.
\fe
Note that this is consistent with $B_{NS}\to -B_{NS}$ under parity transformation. This suggests that the RHS of Vasiliev's equation of motion involving the $B$-master field should be related to worldsheet instanton corrections in string theory (in the suitable small radius/tensionless limit).

\subsection{Vasiliev theory and open-closed string field theory}

A direct way to engineer ${\cal N}=3$ Chern-Simons vector model in string theory was proposed in \cite{Gaiotto:2009tk}. Starting with the $U(N)_k\times U(M)_{-k}$ ABJ theory, one adds $N_f$ fundamental ${\cal N}=3$ hypermultiplets of the $U(N)$. In the bulk type IIA string theory dual, this amounts to adding $N_f$ D6-branes wrapping $AdS_4\times \mathbb{RP}^3$, which preserve ${\cal N}=3$ supersymmetry. The vector model is then obtained by taking $M=0$. The string theory dual would be the ``minimal radius" $AdS_4\times \mathbb{CP}^3$, supported by the $N_f$ D6-branes and flat Kalb-Ramond $B$-field with
\ie
{1\over 2\pi\alpha'}\int_{\mathbb{CP}^1}B_{NS} = {N\over k} + {1\over 2}.
\fe
In this case, our proposed dual $n=4$ Vasiliev theory in $AdS_4$ with ${\cal N}=3$ boundary condition carries $U(N_f)$ Chan-Paton factors, as does the open string field theory on the D6-branes. This lead to the natural conjecture that the open-closed string field theory of the D6-branes in the ``minimal" $AdS_4\times \mathbb{CP}^3$ with flat $B$-field is the {\it same} as the $n=4$ Vasiliev theory, at the level of classical equations of motion. It would be fascinating to demonstrate this directly from type IIA string field theory in $AdS_4\times \mathbb{CP}^3$, say using the pure spinor formalism \cite{Berkovits:2007zk, Berkovits:2008ga, Mazzucato:2011jt}.

\section{The partition function of free ABJ theory on $S^2$ as a matrix integral}\label{abjfpf}

The ABJ theory is a  
supersymmetric Chern-Simons theory based on the gauge group 
$U(N) \times U(M)$, at level $k$ (for $U(N)$) and $-k$ (for $U(M)$) 
respectively. In addition to the gauge fields, this theory possesses
four chiral multiplets $A_1, A_2, B_1, B_2$ (in $d=3$ ${\cal N}=2$ language).
While $A_1$ and $A_2$ transform in the fundamental times antifundamental 
of $U(N) \times U(M)$, $B_1$ and $B_2$ transform in the antifundamental 
times fundamental of the same gauge group. The chiral fields all have
canonical kinetic terms, and interact with each other via a superpotential 
proportional to $ \epsilon^{ij} \epsilon^{mn} \Tr A_i B_m A_j B_n$. 
While ABJ Lagrangian classically enjoys invariance under the ${\cal N}=6$ 
superconfomal algebra (an algebra with 24 fermionic generators) 
for all values of parameters, it has been argued that, quantum mechanically, 
the theory exists as a superconformal theory only for $k \geq |N-M|$.

In this section we will study the partition function of the {\it free} ABJ theory on $S^3$. 
In other words we study the free theory and compute 
\begin{equation}\label{hj}
 Z=\Tr (x^{E}) 
\end{equation}
In more conventional notation $x=e^{-\beta}$ and $Z$ is the usual 
thermal partition function at $T=\frac{1}{\beta}$. 
Here we study the limit $ k \to \infty$. In this limit the 
ABJ theory is free and its partition function is given by the simple formula 
\cite{Sundborg:1999ue,Aharony:2003sx,Bhattacharya:2008bja}  
\begin{equation}\label{mint}
Z=\int DU DV \exp \left[ \sum_{n=1}^\infty \frac{\left( F_B(x^n) +(-1)^{n+1} 
F_F(x^n) \right)}{n} \left( \Tr U^n \Tr V^{-n} + \Tr V^n \Tr U^{-n} \right) \right]
\end{equation}
Here $U$ is an $N \times N$ unitary matrix, $V$ is an 
$M \times M$ unitary matrix. $F_B(x)$ and $F_F(x)$ are the  
bifundamental letter partition functions 
(equal to the antibifundamental letter partition function) 
over bosonic and fermionic fields respectively. The letter partition 
function receives contribution from all the basic bifundamental
(antibifundamental)fields and there derivatives after removing 
contribution from equation of motion and are given by  
\begin{equation}
\begin{split}
F_B(x) &= \Tr_{bosons} \left( x^{\Delta} \right), ~~~~~~~~~ ({\rm where ~\Delta ~is~the~ dilatation~operator})\\
     &= \frac{4x^{\half}(1-x^2)}{(1-x)^3},  \\
 F_F(x) &=\Tr_{fermions} \left( x^{\Delta} \right) = \frac{8x(1-x)}{(1-x)^3}, \\
F(x) &= \Tr_{all} \left( x^{\Delta} \right) =F_B(x)+F_F(x) = \frac{4 \sqrt{x}}{(1-\sqrt{x})^2}. \\
\end{split}
\end{equation}
In the rest of this section we will study the matrix integral \eqref{mint} 
as a function of $x$ in the large 
$M$ and $N$ limit. More precisely, we will focus on the limit 
$N  \to \infty$ and $M  \to \infty$ with $$A=\frac{N}{M}$$ 
held fixed. Note that we will always assume $A > 1$.

\subsection{Exact Solution of a truncated toy model}\label{tm}

The summation over $n$ in \eqref{ltpf} makes the matrix model in that equation 
quite complicated to study for $F(x)>1$. While this matrix model is in principle
`exactly solvable' using the work of \cite{Jurkiewicz:1982iz}, the 
implicit solution thus obtained can be turned into explicit formulae 
only in special limits (see below for more discussion). Instead of plunging 
into a discussion of this exact solution, in the rest of this section, 
we will analyze the model in various limits and approximations; these 
exercises will clearly reveal the qualitative nature of the solution to the 
matrix model \eqref{ltpf}.

In this section we analyze a toy model whose solution will qualitatively 
describe the full phase structure of \eqref{ltpf}.  
In quantitative terms we will explain below that  solution of toy model 
presented in this subsection agrees with the exact solution of the matrix 
model when $x=x_c$, and can be used as the starting point for developing a
perturbative expansion of this solution in a power series in $x-x_c$. 
In other words the toy model presented in this subsection qualitatively captures
the phase structure of the full matrix model; it also gives a quantitatively 
correct description of the first phase transition. 

\begin{figure}
\begin{tabular}{cc}
\includegraphics[width=70mm]{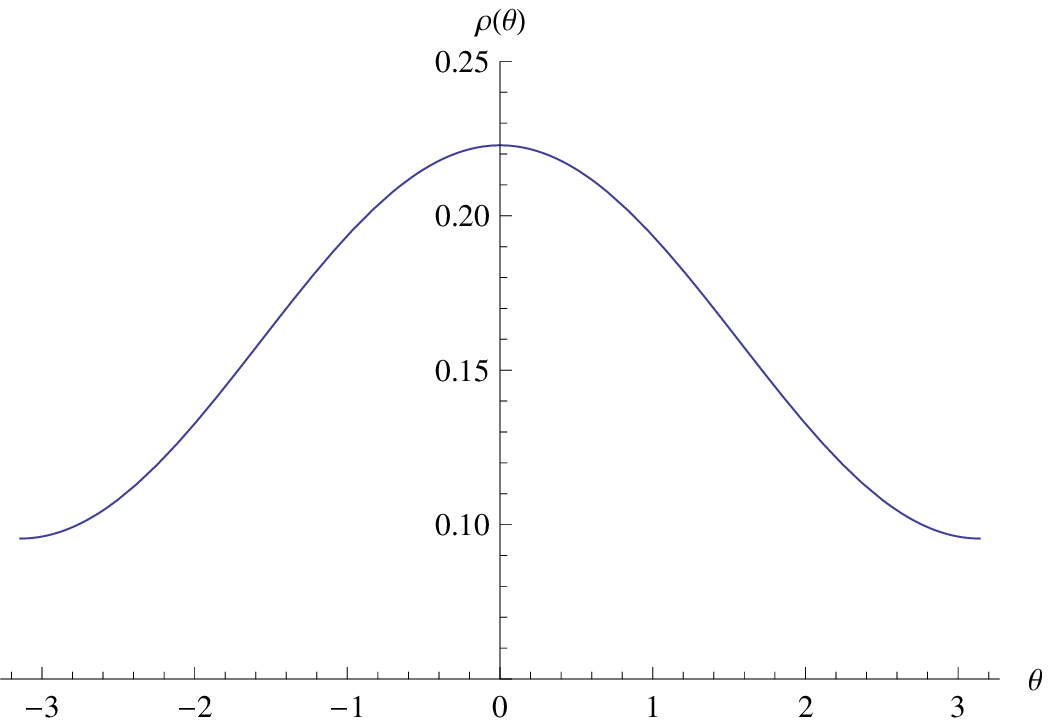} & \includegraphics[width=70mm]{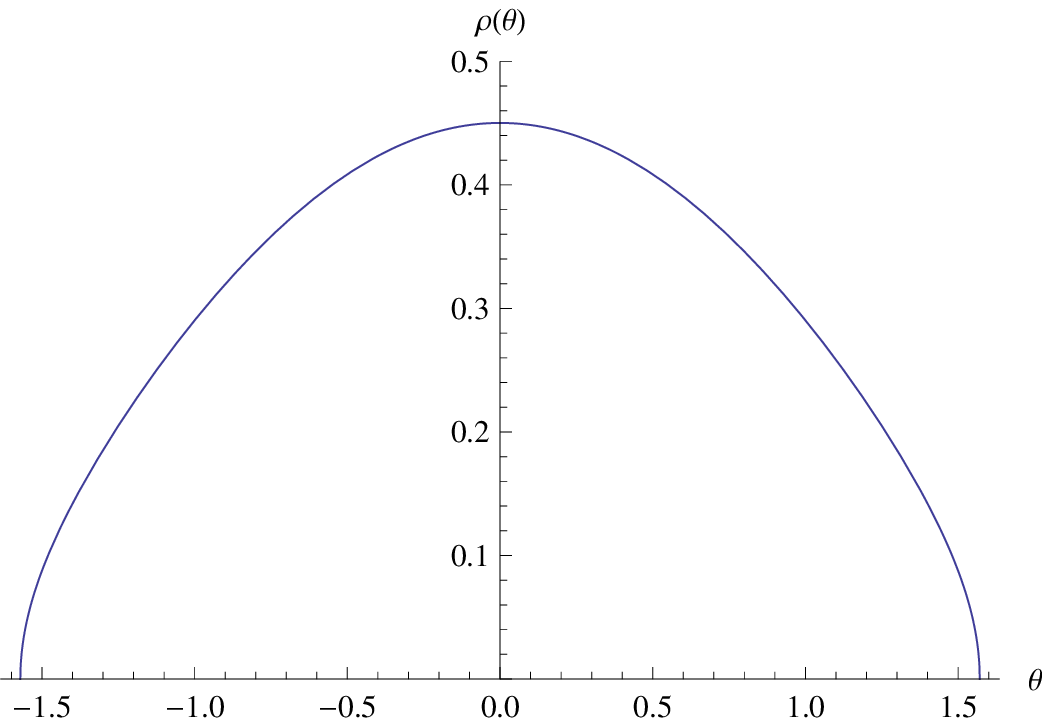}
\end{tabular}
\caption{Eigenvalue distribution for {\it wavy} $(\lambda=5)$ and {\it clumped} $(\lambda=1)$ 
phases of Gross-Witten-Wadia model. }
\label{fig:GWevplots}
\end{figure}

The toy model we study is the matrix model obtained from
\eqref{mint} by truncating to the $n=1$ part of its action,
\begin{equation}\label{ltpft}
 Z_t= \int DU DV \exp \left[ F(x)\left( \Tr U \Tr V^{-1} +\Tr V \Tr U^{-1} \right) \right].
\end{equation}
The general saddle point solution to \eqref{ltpft} obtained extremely easily. Let us assume that 
\eqref{ltpft} has a saddle point solution on which $ \Tr U= N \rho_1 $ and $\Tr V=M \chi_1$. The eigenvalue distribution
for $U$ is then the saddle point solution to the auxiliary matrix model 
\begin{equation}\label{auxu}
 \int DU \exp \left[ N \frac{F(x)}{A} \chi_1 \left( \Tr U + \Tr U^{-1} \right) \right].
\end{equation}
In a similar fashion the eigenvalue distribution of $V$ is given as the solution to the 
auxiliary matrix integral 
\begin{equation}\label{auxv}
 \int DV \exp \left[ M A F(x) \rho_1 \left( \Tr V+ \Tr V^{-1} \right) \right].
\end{equation}
The matrix integrals \eqref{auxu} and \eqref{auxv} are of the famous Gross-Witten-Wadia 
form \cite{Gross:1980he},\cite{Wadia:1980cp}. Here we briefly review the 
solution Gross-Witten-Wadia model. The relevant matrix integral is that of an $N\times N$ unitary 
matrix $W$, defined as 
\begin{equation}
 {\cal I} = \int DW \exp \left[ \frac{N}{\lambda} \Tr (W + W^{-1}) \right].
\end{equation}
where $\lambda$ is coupling constant. In the large N limit this model undergoes phase transition 
at $\lambda = 2$. For $\lambda > 2$ the eigenvalue density distribution is 
$$ \rho(\theta) = \frac{1}{2 \pi} \left( 1 + \frac{2}{\lambda} \cos \theta \right) .$$
We call this phase as the ``wavy" phase as the eigenvalue distribution sinusoidal and 
non-vanishing over the entire $\theta$ circle $-\pi < \theta \leq \pi$. 
For $\lambda < 2$ the eigenvalue distribution is given by 
$$  \rho(\theta) = \frac{2}{\pi \lambda} \cos \left( \frac{\theta}{2} \right) 
\left( \sin^2 \frac{\theta_c}{2} - \sin^2 \frac{\theta}{2} \right)^{\half} $$ 
where 
$$\sin^2 \frac{\theta_c}{2} = \frac{\lambda}{2}, ~~~~~  {\rm and}~~ -\theta_c < \theta < \theta_c. $$
We call this phase the ``clumped" phase as the eigenvalue distribution is non-vanishing only 
in subset of $\theta$ circle. Figure (\ref{fig:GWevplots}) shows the eigenvalue distribution 
for the two phases.

It follows from the results just presented that the eigenvalue distributions in the
models \eqref{auxu} and \eqref{auxv} are given by 
\begin{equation} \label{soltoyu}\begin{split}
\rho_{U}(\theta)&=\frac{1}{2 \pi} \left(1 + \frac{2 \chi_1 F(x)}{A} \cos(\theta)  \right),~~{\rm i.e}~\Tr U= N\frac{\chi_1 F(x)}{A}~~{\rm when}~\frac{2 \chi_1 F(x)}{A}<1,\\
\rho_{U}(\theta)&= \frac{2\chi_1 F(x)}{\pi A} \cos \frac{\theta}{2}  \sqrt{ \frac{A}{2 \chi_1 F(x)}-\sin^2 \frac{\theta}{2} },~~~ {\rm i.e}~ 
\Tr U=N \left(1-\frac{A}{4 \chi_1 F(x)} \right) ~~{\rm when} ~\frac{2 \chi_1 F(x)}{A}>1;\\
\end{split}
\end{equation}
\begin{equation} \label{soltoyv}\begin{split}
\rho_{V}(\theta)&=\frac{1}{2 \pi} \left( 1 + 2 \rho_1 A F(x) \cos(\theta)  \right),~~{\rm i.e.} ~\Tr V= M \rho_1 F(x) A ~~ {\rm when }~2 \rho_1 A  F(x) <1, \\
\rho_{V}(\theta)&= \frac{2 \rho_1 A F(x)}{\pi} \cos \frac{\theta}{2}  \sqrt{ \frac{1}{2 \rho_1 A  F(x)}-\sin^2 \frac{\theta}{2} },~~~{\rm i.e.}~
\Tr V=M \left(1-\frac{1}{4 A \rho_1 F(x)} \right)~~{\rm when} ~ 2 \rho_1 A F(x) >1.\\
\end{split}
\end{equation}
To complete the solution to the model we must impose the self consistency conditions 
$\Tr U=N\rho_1$ on \eqref{soltoyu} and $\Tr V= M\chi_1$ on \eqref{soltoyv}. 
Without loss of generality
let us assume that $N \geq M$ so that $A \geq 1$. As we have explained above, 
flat (constant) eigenvalue distributions for both $U$ and $V$ are always
solutions; this solution is stable for $F(x)<1$ and unstable for $F(x)>1$. 
 It is also easy to check that if the $U$ eigenvalue distribution 
is flat then the same must be true for the $V$ eigenvalue distribution, and 
vica versa. In addition we have four possibilities; each of the $U$ and $V$ 
matrix models may be in either the wavy or the clumped phases. 
We consider each in turn.

\subsubsection{$U$ wavy, $V$ wavy} 

In this case the self consistency equations are  
$$\rho_1 = \frac{\chi_1 F(x)}{A}, ~~~\chi_1 = \rho_1 F(x) A.$$
If $\rho_1$ and $\chi_1$ are nonzero, we have a solution only when $F(x)=1$; on this solution $\rho_1 =\frac{\chi_1}{A}$ 
where $\chi_1 \leq \frac{1}{2}$ (for self-consistency with the assumption that $V$ is wavy) but is otherwise arbitrary.

\subsubsection{$U$ wavy, $V$ clumped}

In this case the self consistency equations are 
$$\rho_1= \frac{\chi_1 F(x)}{A}, ~~~\chi_1= 1-\frac{1}{4 A \rho_1 F(x)}.$$
These equations admit real solutions only when $F(x)>1$. The solution 
is given by 
\begin{equation} \label{stum} \begin{split}
\chi_1&=\frac{1}{2 F(x)} \left(F(x) +\sqrt{F^2(x)-1} \right),\\
\rho_1&= \frac{1}{2 A } \left(F(x) +\sqrt{F^2(x)-1} \right).
\end{split}
\end{equation}
Clearly this solution exists only when $F(x)>1$. 
The assumption that $U$ is wavy is true only when 
\begin{equation}\label{ineq}
F(x) +\sqrt{F^2(x)-1} <A .
\end{equation}
In other words, the solution \eqref{stum} is selfconsistent when 
$$1 < F(x) < \frac{1}{2} \left(A + \frac{1}{A} \right). $$

\subsubsection{$U$ clumped, $V$ wavy}

It may be verified that there are no solutions of this nature when $A>1$.

\subsubsection{$U$ clumped, $V$ clumped} 

In this case the self-consistency equations are 
$$\rho_1=1-\frac{A}{4 \chi_1 F(x)} , ~~~v= 1-\frac{1}{4 A \rho_1 F(x)}. $$
The solution to these equations is given by 
\begin{equation} \label{stum1} \begin{split}
\rho_1&= \frac{\sqrt{\left(A^2-4 A F(x)-1\right)^2-16 A F(x)}-A^2+4 A F(x)+1}{8 A F(x)} , \\ 
\chi_1&= \frac{\sqrt{\left(A^2-4 A F(x)-1\right)^2-16 A F(x)}+A^2+4 A F(x)-1}{8 A F(x)}.
\end{split}
\end{equation}
This solution is consistent with the assumption that $U$ is wavy provided that 
$$ \rho_1 \geq \frac{1}{2}. $$
In other words this solution exists only when 
$$F(x) > \frac{1}{2} \left(A + \frac{1}{A} \right). $$ 

\subsubsection{Summary of Solution}

In summary, any given temperature (except for the special case $F(x)=1$) 
the toy model has a unique stable saddle point. This saddle point is 
listed in Table \ref{tab:toysol}. The model starts out in the flat-flat 
phase, transits to wavy-clumped via a first order phase transition at 
$F(x)=1$ and then transits to clumped-clumped via a third order phase 
transition at $F(x)=\frac{1}{2} \left(A + \frac{1}{A} \right)$. 
\begin{table}
\centering
\begin{tabular}{|c|c|c|}
\hline
   &   U   &   V   \\
\hline
$F(x) < 1$  &  flat ($\rho_1$=0)   &  flat ($\chi_1$=0)  \\
\hline
$1 < F(x) < \frac{1}{2}\left(A + \frac{1}{A} \right)$ & wavy & clumped \\
\hline
$F(x) > \frac{1}{2}\left(A + \frac{1}{A} \right)$ & clumped & clumped \\
\hline
\end{tabular}
\caption{Nature of eigenvalue distribution for U and V matrices in different temperature regimes}
\label{tab:toysol}
\end{table}

\subsubsection{Solution obtained by first integrating out $U$}

In the first two phases listed in Table \ref{tab:toysol} above, the 
eigenvalue distribution for the matrix $U$ is ungapped. In these phases 
the free energy is stationary with respect to a variation of the Fourier modes, 
$\rho_n$, of the eigenvalue distribution of $U$. These two phases may, 
therefore, simply be studied in an effective matrix model for the matrix 
$V$, obtained by integrating $\rho_n$ out classically, using their 
equations of motion. The part of the action in  \eqref{ltpft} that depends
on $\rho_n$ is given by
$$ N^2 \left[
 \frac{F(x)}{A} \left( \rho_1 \chi_{-1}+ \rho_{-1}\chi_1\right) 
 - \sum_{n=1}^\infty \frac{\rho_n \rho_{-n}}{n} \right].$$
On-shell we find $\rho_1= \frac{F(x)}{A} \chi_1$, 
$\rho_{-1}= \frac{F(x)}{A} \chi_{-1}$,  and $\rho_n=0$ ($|n| \geq 2$). 
Integrating out $\rho_n$ we obtain the following effective matrix model for
$V$ (note this is accurate only at leading order in $N$)
\begin{equation}\label{ltpftm}
 Z_t= \int DV \exp \left[ F(x)^2  \Tr V \Tr V^{-1}  \right]
\end{equation}
This model was solved in section 6.4 of \cite{Aharony:2003sx} ($m_1^2-1$ in that paper 
is our $F(x)^2$ and $b$ of that paper should be set to zero). The 
solution takes the following form. For $F(x)^2<1$ the $V$ eigenvalue 
distribution is flat, in agreement with Table \ref{tab:toysol}. 
At $F(x)^2=1$ the model undergoes a phase transition. The $V$ eigenvalue 
distribution is clumped in the high temperature phase. Using equations 
6.11 and 6.18 of \cite{Aharony:2003sx}, it is easily verified that the $V$ eigenvalue 
distribution agrees with that given in \eqref{soltoyv} and \eqref{stum}. 
In particular the value of $\chi_1$ on this solution is given by \eqref{stum}. 
From the fact that $\rho_1= \frac{F(x)}{A} \chi_1$, it follows also that 
$\rho_1$ on the solution takes the value presented in \eqref{stum}.
Consequently the assumption of this section, namely that the $U$ eigenvalue 
distribution is wavy rather than clumped, is self consistent only 
when the inequality \eqref{ineq} is true. When this inequality is violated, 
our system undergoes a further phase transition (in agreement with 
Table \ref{tab:toysol}). However this phase transition and the resultant 
high temperature phase are not accurately captured by the effective model 
\eqref{ltpftm}. 
 
\subsection{Effective description of the low and intermediate temperature 
phase of the full model}

We will argue self-consistently below that the  qualitative features of the 
phase diagram of the toy model are also true of the full matrix model \eqref{mint}. 
The full model also undergoes two transitions; the first from uniform-uniform to 
wavy-clumped and the second from wavy-clumped to clumped-clumped. Exactly as in 
the previous section, the low and intermediate temperature phases of the full model 
may be analyzed by integrating out the Fourier modes, $\rho_n$, of the holonomy 
$U$ using their equations of motion. Performing this integration (see the previous 
section for procedure) we obtain the effective matrix model 
\begin{equation}\label{minte}
Z=\int DV \exp \left[ \sum_{n=1}^\infty \frac{\left( F_B(x^n) +(-1)^{n+1} F_F(x^n) 
\right)^2}{n}  \, \Tr V^n \Tr V^{-n} \right].
\end{equation}
\eqref{minte} has a simple physical interpretation. Note that \eqref{minte} 
is the free partition function of a gauge theory based on the gauge group 
$U(M)$. Our effective theory has only adjoint matter, with effective bosonic
letter partition function $F_B(x)^2+F_F(x)^2$ and effective fermionic letter
partition function $2 F_B(x)F_F(x)$. But this is exactly the partition function
for mesonic fields of the sort $AB$. In other words \eqref{minte} describes 
a phase in which the gauge group $U(N)$ is completely confined, so that 
the effective letters $A$ and $B$ only appear in the combination $AB$; 
composite letters in the adjoint of $U(M)$. The entire effect of the 
integration over the $U(N)$ holonomy is to effect this complete confinement. 
The effective description of this phase is in terms of a single gauge 
group, $U(M)$, and adjoint letters $AB$. 

This matrix model \eqref{minte} 
has been studied in detail, in a perturbation expansion 
in $x-x_c$, in section 5.5 of \cite{Aharony:2003sx}. The qualitative behavior 
is similar to the toy model studied in the previous section. 

\subsubsection{$F(x)<1$}

In the case that $F(x)<1$ the saddle point is given by 
$$ \Tr V^n=0$$
for all $n \neq 0$. This also implies that 
$$\Tr U^n=0$$ for all $n \neq 0$. 
The free energy in this phase vanishes at leading order in $N$. At first subleading order, 
the partition function 
\eqref{mint} is obtained by computing the one loop determinant about this saddle point 
and is given by \cite{Aharony:2003sx}
\begin{equation}\label{ltpf}
 Z= \prod_{n=1}^{\infty} \frac{1}{1-(F_B(x^n) +(-1)^{n+1} F_F(x^n))^2}.
\end{equation}
Note that the result \eqref{ltpf} diverges when $F_B(x^n) +(-1)^{n+1} F_F(x^n)=1$ for any $n$. 
Now $F(x)$ is a monotonically 
increasing function of $x$ with $F(0)=0$ and $F(1)=\infty$. As $x^n<x$ for $x \in (0,1)$ 
it follows that $F(x) <1$ implies that $F(x^n)<1$ for all positive $n$. In other 
words, as $x$ is increased from zero (i.e. the temperature of the system is increased 
from zero) the partition function \eqref{ltpf} first diverges when $F(x)=1$, i.e. 
at $x=x_c=17-12 \sqrt{2}=0.0294\cdots$. As explained in \cite{Aharony:2003sx}, 
this divergence has a simple physical interpretation. The effective potential 
for the mode $\chi_1$ is proportional to $(1-F(x)^2) |\chi_1|^2$. This potential 
develops a zero at $F(x)=1$ and is tachyonic for $F(x)>1$.\footnote{We can see 
all this directly in the full matrix model \eqref{mint} involving both the 
$U$ and the $V$ variables. The potential for the modes $\Tr U$ and $\Tr V$ is given by \
$$ \Tr U \Tr U^{-1} + \Tr V \Tr V^{-1}- F(x) \Tr U \Tr V^{-1} -F(x) \Tr V \Tr U^{-1}.$$ 
Let $\Tr U = N \rho_1$ and $\Tr V = M \chi_1$ then this potential can be written as 
$$ V(\rho_1,\chi_1) = M^2 \left[ |A\rho_1 - F(x)\chi_1 |^2 + (1 - F(x)^2)
|\chi_1|^2 \right].$$
$\rho_1=\chi_1=0$ is a stable minimum of this potential for $F(x)<1$. At $F(x)=1$ 
the potential develops a flat direction that evolves into an unstable direction 
for $F(x)>1$. It follows that the trivial solution studied in this subsection is 
unstable for $F(x)>1$ providing an explanation for the divergence on $F(x) \rightarrow 1$. 
At $F(x)=1$ the system undergoes a phase transition to another phase. 
As we will see below this phase transition is of first order. }

\subsubsection{$F(x)>1$}

The $V$ eigenvalue distribution gets increasingly clumped as $x$ is increased.
Recall that the $U$ eigenvalue distribution is determined by the equations 
$$\rho_n = \frac{F_B(x^n) + (-1)^{n+1}F_F(x^n)}{A}\chi_n, $$
which gives the eigenvalue distribution for $U$ as 
$$ \rho_u(\theta) = \frac{1}{2\pi} \left( 1 + 2\sum_{n=1}^{\infty} 
\frac{F_B(x^n) + (-1)^{n+1}F_F(x^n)}{A}\chi_n \cos\theta \right). $$
As $x$ is increased this eigenvalue distribution eventually goes 
negative at $\theta=\pi$ for $x >x_{c_2}$ where $x_{c_2}$ is a second 
critical temperature that we will not in generality be able 
analytically compute in this paper. For $x>x_{c_2}$ the effective 
action \eqref{mint} is no longer accurate. As in the toy model 
of the previous section, of course, the physics of this second phase
transition is the clumping of the $U$ eigenvalue distribution.

\subsubsection{The second phase transition at large $A$}

When $A$ is very large, the second phase transition occurs at a 
high temperature of order $\sqrt{A}$ (as we will see below). 
This results in a key simplification; when the $U$ eigenvalue 
distribution undergoes the phase transition, the $V$ eigenvalue
distribution is well approximated by a $\delta$ function. In other 
words ${1\over M}{\Tr V^n}=1$ for all $n$. Consequently, at leading 
order in the $\frac{1}{A}$ expansion, the second phase transition 
is well described by the matrix model 
\begin{equation}\label{mintlrgA}
Z=\int DU \exp \left[ \frac{N}{A} \sum_{n=1}^{\infty} 
\frac{F_B(x^n) + (-1)^{n+1} F_F(x^n)}{n} 
\left( \Tr U^n  + \Tr U^{-n} \right) \right].
\end{equation}
When $x$ is of order unity, the saddle point to this matrix model is wavy with 
$$\frac{\Tr U^n}{N}= \frac{\Tr U^{-n}}{N}=\frac{F_B(x^n)+(-1)^{n+1} F_F(x^n)}{A},$$
so that 
\begin{equation} \label{lrgAldngU}
 \rho_{u}(\theta)= \frac{1}{2 \pi} \left[ 1+ 2 \sum_{n=1}^{\infty} 
\frac{ F_B(x^n) + (-1)^{n+1} F_F(x^n)}{A} \cos n \theta \right].
\end{equation}
The leading contribution to free energy computed using this saddle point distribution 
\begin{equation}\begin{split}
Z &= \exp \left[- \sum_{n=1}^{\infty} \frac{1}{n} 
\left( \Tr U^n \Tr U^{-n}-\frac{NF_B(x^n)+(-1)^{n+1} F_F(x^n)}{A} \left( \Tr U^n+\Tr U^{-n} \right) \right) \right] \\
  &= \exp \left[ -N^2 \sum_{n=1}^{\infty} \frac{1}{n} \left( \rho_n^2 -2 \rho_n 
\frac{F_B(x^n) +(-1)^{n+1} F_F(x^n) }{A}  \right) \right] \\
  &= \exp \left[ -N^2 \sum_{n=1}^{\infty} \frac{1}{n} \left[ \left( \rho_n -
 \frac{F_B(x^n) +(-1)^{n+1}F_F(x^n)}{A} \right)^2 - 
\left( \frac{F_B(x)^n +(-1)^{n+1} F_F(x^n)}{A} \right)^2 \right] \right]. \\
\end{split}
\end{equation}
The first term in the sum vanishes on the saddle point density distribution and we get 
\begin{equation} \label{mesons}
Z = \exp \left[ M^2 \sum_{n=1}^{\infty} \frac{1}{n}  \left( F_B(x)^n 
+(-1)^{n+1} F_F(x^n) \right)^2 \right]. \\
\end{equation}
\eqref{mesons} has a simple interpretation. Products of `letters' of the 
form $AB$ are singlets under the gauge group U(N), but transform in the 
adjoint of $U(M)$. The partition function over all bosonic operators 
is simply given by $M^2 (F_B^2(x)+F_F^2(x))$, while the partition function 
over all fermionic letters of the same form is given by 
$2 M^2 (F_B(x) F_F(x))$. 
\eqref{mesons} is precisely the Bose/Fermi exponentiation (multi-particling)
of these single meson partition functions. In other words  \eqref{mesons} 
is the partition function over a gas of non-interacting mesons of the form 
$AB$. In the limit described in this subsection it follows that the 
intermediate temperature phase may be thought of as a phase in which 
the gauge group $U(N)$ is completely confined while the gauge group 
$U(M)$ is completely deconfined. 

At high temperatures $T \gg 1$ the eigenvalue distribution \eqref{lrgAldngU}
attains its minimum at $\theta=\pi$. This minimum value decreases below 
zero when 
\begin{equation}\label{ptt}
\sum_{n=1}^\infty \left( (-1)^{n+1} F_B(x^n) + F_F(x^n) \right) >\frac{A}{2}.
\end{equation}
As $A$ is assumed large in this section, this condition can only be met 
in the limit that $x \to 1$, i.e. in the large $T$ limit. 
At leading order in the large temperature limit 
$x-1= -\frac{1}{T}$ (recall $x=e^{-1/T}$) and  
$$F_B(x^n)\approx \frac{8}{T^2 n^2}, ~~~F_F(x) \approx \frac{8}{T^2 n^2}, ~~~
F(x^n) \approx \frac{16 T^2}{n^2}, $$
and the eigenvalue distribution \eqref{lrgAldngU} reduces to 
$$\rho(\theta)= \frac{1}{2 \pi} \left( 1+ \sum_{n=1}^{\infty} \frac{32 T^2}
{A (2n+1)^2} \cos n \theta \right). $$
In this approximation the condition \eqref{ptt} for the 
eigenvalue distribution to go negative is given by
\begin{equation}\label{tt}
T^2 > T_c^2= \frac{A}{4\pi^2}.
\end{equation}
As it makes no sense for an eigenvalue distribution to be negative, it follows that the $U$ matrix undergoes 
the clumping transition at $T=T_c$. Note that $T_c$ is of order $\sqrt{A}$, and so is large, as promised at the 
beginning of this section. 

The condition \eqref{ptt} gives an expression for the phase transition 
temperature that may be power series expanded in $\frac{1}{\sqrt{A}}$ 
(\eqref{tt} is the leading term in that expansion). However \eqref{ptt}
was itself derived under the approximation that the $V$ eigenvalue distribution 
is a delta function. In reality (see below) the $V$ eigenvalue distribution
has a width of order $\frac{1}{T^2}$ which is $ \sim \frac{1}{A}$ near the 
phase transition temperature. It is possible to systematically account for 
the broadening of the $V$ eigenvalue distribution (and thereby develop a 
systematic procedure for computing the phase transition temperature to
arbitrary order in $\frac{1}{A}$). We demonstrate how this works in Appendix 
\ref{one} by computing the first correction to \eqref{ptt} 
resulting from the finite width of the eigenvalue distribution of the 
matrix $V$.

\subsubsection{Effect of interactions on first phase transition}

The most general form of effective action in large $N$ perturbation theory is 
\begin{equation}\begin{split}\label{zintfull}
Z = \int DU DV & \exp \bigg[ \sum_{n=1}^\infty \bigg( A_n^{(0)}(x) \left( \Tr U^n \Tr V^{-n} +  
   \Tr V^n \Tr U^{-n} \right) \bigg) \\ &+  
  \sum_{m,n} A^{(1)}_{m,n} \bigg( \Tr V^m \Tr V^n \Tr U^{-m-n} + \Tr U^m \Tr U^n \Tr V^{-m-n} \bigg)  \\ & 
+ \sum_{m,n,p} A^{(2)}_{m,n,p} \bigg( \Tr V^m \Tr V^n \Tr V^p \Tr U^{-m-n-p} + \Tr U^m \Tr U^n \Tr V^p \Tr V^{-m-n-p} \\ &
 + \Tr U^m \Tr U^n \Tr U^p \Tr V^{-m-n-p} \bigg) + \ldots \bigg].
\end{split}
\end{equation}
Moving to the Fourier basis and integrating out the $U$ modes we get an effective 
adjoint matrix model the $V$ matrix
\begin{equation}\begin{split}\label{veffadj}
Z_{eff} = \int DV \exp \bigg[ & \sum_{n=1}^n  B^{(0)}_n(x) 
\Tr V^n \Tr V^{-n}   +
\sum_{m,n} B^{(1)}_{m,n} \Tr V^m \Tr V^n \Tr V^{-m-n} \\ & + 
\sum_{m,n,p} B^{(2)}_{m,n,p} \Tr V^m \Tr V^n \Tr V^p \Tr V^{-m-n-p} + \ldots \bigg]
\end{split}
\end{equation}
where the $B$ coefficients can be determined in terms of the $A$ coefficients appearing in 
\eqref{zintfull}.
As explained in the \cite{Aharony:2003sx} the only interaction terms relevant for the phase 
transition in this adjoint matrix model are 
$$ \Tr V^2 (\Tr V^{-1})^2,~~\Tr V^{-2} (\Tr V)^2,~~(\Tr V \Tr V^{-1})^2 . $$ 
Now we will determine the coefficient of these term in the effective adjoint model in terms 
of the coefficients appearing in the original matrix model. The relevant part of the 
original action in Fourier modes is 
\begin{equation}\begin{split}\label{zeffV}
\frac{S}{N^2} =& - \left( \rho_1 \rho_{-1} +  \frac{1}{2} \rho_2 \rho_{-2} \right) - \frac{1}{A^2} \left( \chi_1 \chi_{-1} +
 \half \chi_2 \chi_{-2} \right) + \frac{m_1(x)}{A}(\rho_1\chi_{-1} + \rho_{-1}\chi_1) \\ & 
  + \frac{m_2(x)}{A}(\rho_2\chi_{-2}+\rho_{-2}\chi_2) + 
 \frac{a}{A} (\rho_1\rho_{-1}\chi_1\chi_{-1}) \\ & + \frac{b}{A}(\rho_1^2\chi_{-1}^2 + 
 \rho_{-1}^2\chi_1^2)  + \frac{c}{A}(\rho_2\chi_{-1}^2 + \rho_{-2}\chi_1^2).
\end{split} 
\end{equation}
Here $\rho_n$ and $\chi_n$ are the Fourier mode of eigenvalue distribution for $U$ 
and $V$ matrices respectively and $A=\frac{N}{M}$. Also $c \sim \lambda$ while 
$a,b \sim \lambda^2$ where $\lambda$ is the 't Hooft coupling. The coefficients 
$m_1(x)$ and $m_2(x)$ reduces to $F(x)$ and $F_B(x^2)+(-1)F_F(x^2)$ respectively 
in the free theory.The equation of motion for $\rho_1$ and $\rho_2$ are
\begin{equation}\begin{split}\label{rhoeom}
\rho_1 &= - \frac{m_1(x)}{A} \bigg[ \frac{\chi_1(-1+\frac{a}{A}\chi_1\chi_{-1})
 -\frac{2b}{A}\chi_1^2\chi_{-1}}
{(-1+\frac{a}{A}\chi_1\chi_{-1})^2 - (\frac{2b}{A}\chi_1\chi_{-1})^2} \bigg]~	, \\ 
\rho_2 &= \frac{m_2(x)}{A}\chi_2 - \frac{2c}{A}\chi_1^2 . \\
\end{split}
\end{equation}
Linearizing in $a$ and $b$ and substituting back the action we get the effective adjoint model 
(keeping only the terms relevant to phase transition)
\begin{equation}
\begin{split}
\frac{S_{eff}}{N^2} =& \bigg[ -\left(\frac{1-m_1(x)^2}{A^2}\right)\chi_1\chi_{-1} - \left(\frac{1-m_2(x)^2}{2 A^2}\right)\chi_2\chi_{-2} + 
        \frac{2c}{A^2}m_2(x)(\chi_2\chi_{-1}^2 + \chi_{-2}\chi_1^2) \\ & + 
        \frac{(a+2b)m_1(x)^2 + 4Ac^2}{A^3} \chi_1^2\chi_{-1}^2 \bigg].
\end{split}
\label{seffv}
\end{equation}
Now we can again integrate out $\chi_2$ to get 
\begin{equation}\begin{split}
\frac{S_{eff}}{N^2} = - \left(\frac{1-m_1(x)^2}{A^2}\right)\chi_1\chi_{-1} + 
    \left( \frac{(a+2b)m_1(x)^2}{A^3} + \frac{4c^2(1+m_2(x)^2)}{A^2(1-m_2(x)^2)} \right) \chi_1^2\chi_{-1}^2.
\end{split}
\end{equation}

The phase structure described by this effective action was 
described in \cite{Aharony:2003sx}. The first order transition of the 
free theory splits into two phase transitions ; the first of second order
(when $m_1(x)=1$) and the next of third order at a higher temperature -- 
if the coefficient of quartic term is negative. However the transition remains
a single transition of first order if it is positive; this transition occurs at 
$m_1(x)<1$ (see Appendix \ref{two}).  In Appendix \ref{two} below
we demonstrate all this in a somewhat more quantitative fashion by 
studying interaction in a truncated toy model.

\section{Conclusion}

In this paper, we proposed the higher spin gauge theories in $AdS_4$ described by 
supersymmetric extensions of Vasiliev's system and appropriate boundary conditions 
that are dual to a large class of supersymmetric Chern-Simons vector models. 
The parity violating phase $\theta_0$ in Vasiliev theory plays the key role 
in identifying the boundary conditions that preserve or break certain 
supersymmetries. In particular, our findings are consistent with the 
following conjecture: starting with the duality between parity invariant 
Vasiliev theory and the dual free supersymmetric $U(N)$ vector model at 
large $N$, turning on Chern-Simons coupling for the $U(N)$ corresponds 
to turning on the parity violating phase $\theta_0$ in the bulk, and at 
the same time induces a change of fermion boundary condition as described 
in section \ref{bdfcs}. We conjectured that the relation 
$\theta_0 = {\pi\over 2}\lambda$, where $\lambda=N/k$ is the 't Hooft 
coupling of the boundary Chern-Simons theory, suggested by two-loop 
perturbative calculation in the field 
theory and Giveon-Kutasov duality and ABJ self duality, is exact. 

Turning on various scalar potential and scalar-fermion coupling in the 
Chern-Simons vector model amounts to
double trace and triple trace deformations, which are dual to deformation 
of boundary conditions on spin 0 
and spin $1/2$ fields in the bulk theory. Gauging a flavor symmetry of the 
boundary theory with Chern-Simons 
amounts to changing the boundary condition on the bulk spin-1 gauge field 
from the magnetic boundary condition 
to a electric-magnetic mixed boundary condition. Consideration of supersymmetry 
breaking by boundary conditions 
allowed us to identify precise relations between $\theta_0$, the Chern-Simons 
level $k$, and two-point function 
coefficient $\widetilde N$ in ${\cal N}=3$ Chern-Simons vector models.

While substantial evidence for the dualities proposed in this paper is 
provided by the analysis of 
linear boundary conditions, we have not analyzed in detail the non-linear 
corrections to the boundary conditions, which
are responsible for the triple trace terms needed to preserve supersymmetry. 
Furthermore, we have not nailed down
the bulk theory completely, due to the possible non-constant terms in the 
function $\theta(X)=\theta_0 + \theta_2 X^2 + \theta_4 X^4+\cdots$ that controls 
bulk interactions and breaks parity. It seems that $\theta_2, \theta_4$ etc. 
cannot be removed merely by field redefinition, and presumably contribute to 
five and higher point functions at bulk tree level, and yet their presence 
would not affect the preservation of supersymmetry. This non-uniqueness at 
higher order in the bulk equation of motion is puzzling, as we know of no 
counterpart of it in the dual boundary CFT. Perhaps clues to resolving this 
puzzle can be found by explicit computation of say the contribution of $\theta_2$ 
to the five-point function. It is possible that a thorough analysis of the 
near boundary behavior of 
solutions to Vasiliev's equations (via a Graham Fefferman type analysis) 
could be useful in this regard.

We have also encountered another puzzle that applies to Vasiliev duals 
of all Chern Simons field theories, not necessarily supersymmetric. 
Our analysis of the bulk Vasiliev
description of the breaking of higher spin symmetry correctly 
reproduced those double trace terms in the divergence of higher spin 
currents that involve a scalar field on the RHS. However we were unable 
to reproduce the terms bilinear in two higher spin currents. The reason 
for this failure was very general; when acting on a state 
the higher spin symmetry generators never appear to violate the boundary 
conditions for any field except the scalar. It would be reassuring to 
resolve this discrepancy. 

The triality between ABJ theory, $n=6$ Vasiliev theory with $U(M)$ Chan-Paton 
factors, and type IIA string theory on $AdS_4\times \mathbb{CP}^3$ suggests 
a concrete way of embedding Vasiliev theory into string theory. In particular, 
the $U(M)$ Vasiliev theory is controlled by its {\it bulk} 't Hooft coupling 
$\lambda_{bulk} = g^2 M \sim M/N$. We see clear indication from the dual field 
theory that at strong $\lambda_{bulk}$, the nonabelian higher spin particles 
form color neutral bound states, that are single closed string excitations. 
Vice versa, in the small radius limit and with near critical amount of flat 
Kalb-Ramond $B$-field on $\mathbb{CP}^3$, the type IIA strings should break 
into multi-particle states of higher spin fields. This picture is further 
supported by the study of thermal partition function of ABJ theory in the free limit. 
The dual field theory mechanism for the disintegration of the string 
is very general, and so should apply more generally to the dual string 
theory description of any field theory with bifundamental matter, when 
the rank of one of the gauge groups is taken to be much smaller than 
the other \footnote{We thank K. Narayan for discussions on this point. }.

In this paper we have computed the thermal phase diagram of ABJ theory in 
the free limit. This phase diagram has three distinct phases; a low 
temperature string like phase, an intermediate temperature thermal Vasiliev 
like phase and a high temperature black hole like phase. It would be very 
interesting to extend these computations to the interacting theory. 
Order by order in $\frac{M}{N}$ such computations may be technically 
feasible nonperturbatively in $\lambda=\frac{N}{K}$ following the methods 
employed in \cite{Giombi:2011kc} and \cite{Jain:2012cs}. 

It has been argued that the vacuum of the ABJ model spontaneously breaks 
supersymmetry for $k<N-M$ \cite{Aharony:2008gk}. Requiring the existence of a 
supersymmetric vacuum, the maximum value of t'Hooft coupling 
in a theory with $M \neq N$ is  $\frac{N}{k_{min}}=\frac{1}{1-\frac{M}{N}}$.
As the radius of the dual AdS space in string units is proportional to a
positive power of the t'Hooft coupling, it follows that ABJ theories have 
a weakly curved string  description only in the limit $\frac{M}{N} \to 1$.
It is interesting that, in the free computations performed above, 
the new intermediate phase (a free gas of Vasiliev particles) 
continued to exist all the way upto $\frac{M}{N}=1$. If this continues 
to be the case in the strongly interacting theory, it may be possible to 
access this new phase at strong coupling via a string worldsheet computation. 
We find this a fascinating possibility.

More generally, the recasting of ABJ theory as a Vasiliev theory suggests 
that it would be interesting, purely within field theory, to study 
ABJ theory in a power expansion in $\frac{M}{N}$ but nonperturbatively 
in $\lambda$. At $\frac{M}{N}=0$ this would 
require a generalization of the results of \cite{Maldacena:2011jn} and \cite{Maldacena:2012sf}
to the supersymmetric theory. It may then be possible to systematically
correct this solution in a power series in $\frac{M}{N}$. This 
would be fascinating to explore.

Perhaps the most surprising recipe in this web of dualities is that the 
full classical equation of motion of the bulk higher spin gauge theory can 
be written down explicitly and exactly, thanks to Vasiliev's construction. 
One of the outstanding questions is how to derive Vasiliev's system directly 
from type IIA string field theory in $AdS_4\times \mathbb{CP}^3$, or to learn 
about the structure of the string field equations (in $AdS$) from Vasiliev's 
equations. As already mentioned, a promising approach is to consider the 
open-closed string field theory on D6-branes wrapped on 
$AdS_4\times \mathbb{RP}^3$, which should directly reduce to $n=4$ Vasiliev 
theory in the minimal radius limit.
It would also be interesting to investigate whether - and in what guise - 
the huge bulk gauge symmetry of Vasiliev's description survives in the 
bulk string sigma model description of the same system. We leave these 
questions to future investigation.

\bigskip

\acknowledgments

We would like to thank Xi Dong, Daniel Jafferis, Simone Giombi, Guy Gur-Ari, 
K. Narayan, Steve Shenker, Sandip Trivedi, Spenta Wadia, Ran Yacoby, and 
especially Ofer Aharony for useful discussions and comments. 
We would like to thank Ofer Aharony and Shuichi Yokoyama for help comments on 
preliminary draft of this paper. XY would like to thank Tata Institute of 
Fundamental Research, Aspen Center for 
Physics, Harish-Chandra Research Institute, Simons Center for Geometry and Physics, Stanford 
University, California Institute of Technology, Weizmann Institute, Kavli Institute for 
Theoretical Physics, and Lebedev Institute for hospitality during the course of this work. 
CC would like to thank National Taiwan
University and National Tsing Hua University for hospitality during
the course of this work. SM would like to thank Harish Chandra research institute 
and Cargese summer school 
for hospitality while this work was in progress.
The work of CC and XY are supported in part by the Fundamental Laws Initiative 
Fund at Harvard University, and by NSF Award PHY-0847457. The work of SM is 
supported in part by Swarnjayanti fellowship. SM and TS would like to acknowledge 
our debt to the people of India for their generous and steady support to research 
in the basic sciences.

\appendix
\appendixpage
\addappheadtotoc

\section{Details and explanations related to section 2}

\subsection{Star product conventions and identities}\label{star}

It follows from the definition of the star product that  
\begin{equation}\begin{split} \label{stareg}
y^\A * y^\B&= y^\A y^\B + \epsilon^{\A \B}; ~~~[y^\A, y^\B]_* = 2 
\epsilon^{\A\B}\\
z^\A * z^\B&= z^\A z^\B - \epsilon^{\A \B}; ~~~[z^\A, z^\B]_*
=-2 \epsilon^{\A\B} \\
y^\A * z^\B&= y^\A z^\B - \epsilon^{\A \B};
~~~z^\A * y^\B= z^\A y^\B + \epsilon^{\A \B}; ~~~[y^\A, z^\B]_* = 0\\ 
\end{split}
\end{equation}
Identical equations (with obvious modifications) apply to the bar variables.
Spinor indices are lowered using the $\epsilon$ tensor as follows
\begin{equation}
z_\A=z^\B\epsilon_{\B\A} , ~~~\epsilon_{12}=-\epsilon_{21}=\epsilon^{12}
=-\epsilon^{21}=1, ~~~\epsilon_{\A \gamma} \epsilon^{\gamma \B}= -\delta_\A^\B
\end{equation}
Note that for an arbitrary function $f$ we have
\begin{equation}
\begin{split}
z^\alpha * f = z^\A f + \epsilon^{\A\B}(\partial_{y^\B}f - \partial_{z^\B}f) \\
f*z^\A = z^\A f + \epsilon^{\A\B}(\partial_{y^\B}f + \partial_{z^\B}f) \\
\end{split}
\label{zf}
\end{equation}
Using \eqref{zf} 
we the following (anti)commutator
\begin{equation}
\begin{split}
[z^\A,f]_* &= -2\epsilon_{\A\B} \partial_{z^\B}f \\
\{ z^\A,f \}_* &= 2z^\A f + 2\epsilon^{\A\B}\partial_{y^{\B}}f
\end{split}
\label{zfca}
\end{equation}
It follows from \eqref{stareg} that 
\begin{equation}\label{starder}\begin{split}
[z_\alpha, f]_*=-2\frac{\partial f}{\partial z^\alpha}, ~~~
[y^\alpha, f]_*=2\epsilon^{\alpha \beta} \frac{\partial f}{\partial y^\beta}, ~~~
[y_\alpha, f]_*=2\frac{\partial f}{\partial y^\alpha}
\end{split}
\end{equation}
Similar expression(with obvious modifications) are true for (anti)commutators with 
$\bar{y}$ and $\bar{z}$.
Substituting $f=K$ into \eqref{zf} and 
using $\partial_{y^\A}K=-z_\A K$, one obtains
\begin{equation}
\{ z^\A,K \}_* = 0, ~~~{\rm i.e.} ~~~K *z^\A* K=-z^\A
\end{equation} 
In a similar manner we find 
$$\{ y^\A, K \}_* = 0, ~~~{\rm i.e.} ~~~K* y^\A* K=-y^\A$$
On the other hand $K$ clearly commutes with ${\bar y}_\da$ and ${\bar z}_\da$.
The second line of \eqref{kprop} follows immediately from these 
observations. 

The first line of \eqref{kprop} is also easily verified.

\subsection{Formulas relating to $\iota$ operation}\label{iotad}

We present a proof of \eqref{ipr} 
\begin{equation}\label{iotaprop} \begin{split}
\iota(f *g) &= \bigg( f(Y,Z) \exp\bigg[ \epsilon^{\A\B} \left( \overleftarrow\partial_{y^\A} 
  + \overleftarrow\partial_{z^\A} \right) \left( \overrightarrow\partial_{y^\B} 
  - \overrightarrow\partial_{z^\B} \right) \\
 & \hspace{45mm} + \epsilon^{\da\db} \left( \overleftarrow\partial_{\bar{y}^\da} 
  + \overleftarrow\partial_{\bar{z}^\da} \right) \left( \overrightarrow\partial_{\bar{y}^\db} 
  - \overrightarrow\partial_{\bar{z}^\db} \right)  \bigg] g(Y,Z) \bigg)_{(Y,Z) 
    \rightarrow ({\tilde Y}, {\tilde Z})} \\
&= f({\tilde Y},{\tilde Z}) \exp \bigg[ -\epsilon^{\A\B} 
  \bigg( \overleftarrow\partial_{y^\A} - \overleftarrow\partial_{z^\A} \bigg) 
  \bigg( \overrightarrow\partial_{y^\B} + \overrightarrow\partial_{z^\B} \bigg) \\
  & \hspace{45mm}  -\epsilon^{\da\db} \left( \overleftarrow\partial_{y^\da} - \overleftarrow
   \partial_{z^\da} \right) \left( \overrightarrow\partial_{y^\db} + 
   \overrightarrow\partial_{z^\db} \right)  \bigg] g({\tilde Y},{\tilde Z}) \\
&=\iota(g)* \iota(f)
\end{split}
\end{equation}
where $(Y, Z)=(y, {\bar y}, z, {\bar z}) $ and $({\tilde Y}, 
{\tilde Z})=(iy, i\bar y,-iz,-i\bar z,-idz,-id\bar z)$.

We now demonstrate that 
$$ \iota(C*D) = -\iota(D)*\iota(C)$$
if $C$ and $D$ are each oneforms. 
\begin{equation}\label{iotaform}\begin{split}
\iota(C*D) &= \iota \left(C_M*D_N dX^M dX^N )\right)
= \iota(D_N)*\iota(C_M) \iota(dX^M)\iota(dX^N) \\
&=-\iota(D_N)*\iota(C_M) \iota(dX^N)\iota(dX^M)
= -\iota(D)*\iota(C)
\end{split}
\end{equation}

\subsection{Different Projections on Vasiliev's Master Field}\label{projapn}

One natural projection one might impose on the Vasiliev master field 
is to restrict to real fields where reality is defined by  
\begin{equation} \label{nreal}
{\cal A}={\cal A}^*
\end{equation}
This projection preserves the reality of the 
field strength (i.e. ${\cal F}$ is real if ${\cal A}$ is). 
As we will see below, however, the projection \eqref{nreal} 
does not have a natural extension to the supersymmetric Vasiliev theory,
and is not the one we will adopt in this paper. 

The second `natural' projection on Vasiliev's master fields is given by 
\ie\label{mintrunc}
\iota(W)=-W,~~~\iota(S)= -S,~~~\iota(B)= K* B*K.
\fe
Note that the various components of ${\cal F}$ transform 
homogeneously under this projection
\begin{equation}\label{fsr} \begin{split}
& \iota \left(d_xW + W*W \right)  = -\left(d_xW + W*W \right),\\
& \iota \left(d_x \hat S+\{W, \hat S \}_*\right)=
-\left(d_x \hat S+\{W, \hat S \}_*\right),\\
& \iota \left(\hat S*\hat S \right)=
-\left(\hat S*\hat S \right),
\end{split}
\end{equation}
(the signs in \eqref{mintrunc} were chosen to ensure that all the 
quantities in \eqref{fsr} transform homogeneously).  
Note also that
\begin{equation} \label{bs}
\iota(B*K) =B*{K}, ~~~\iota(B*{\bar K})=B*{\bar K} .
\end{equation}
(we have used $K*K=1$).

As we have explained in the main text, in this paper we impose the 
projection \eqref{RCD} on all fields. \eqref{RCD} may be thought of 
as the product of the projections \eqref{nreal} and \eqref{mintrunc}. 
As we have mentioned in the main text ${\cal F}$ transforms homogeneously
under this truncation (see \eqref{et}); in components
\begin{equation}\label{fsra} \begin{split}
& \iota \left(d_xW + W*W \right)^* = -\left(d_xW + W*W \right),\\
& \iota \left(d_x \hat S+\{W, \hat S \}_*\right)^*=
-\left(d_x \hat S+\{W, \hat S \}_*\right),\\
& \iota \left(\hat S*\hat S \right)^*=
-\left(\hat S*\hat S \right). 
\end{split}
\end{equation}

\subsection{More about Vasiliev's equations}\label{veq}

Expanded in components the first equation in \eqref{veqns} reads
\ie\label{veqnsb}
&d_x W+W*W =0,\\
&d_x \hat S+\{W, \hat S \}_* =0,\\
&\hat S * \hat S  = f_*(B*K) dz^2 + \bar f_*(B*\bar K) d\bar z^2.
\fe
The second equation reads 
\ie\label{veqnsc}
&d_x B + W*B-B*\pi(W) = 0,\\
&{\hat S} * B-B*\pi({\hat S})=0.\\
\fe

We will now demonstrate that the second equation in \eqref{veqns} 
follows from the first (i.e. that \eqref{veqnsc} follows from 
\eqref{veqnsb}). Using \eqref{bianchi} and the first of \eqref{veqns}  
we conclude that 
\begin{equation} \label{biexp}
d_x \left(f_*(B*K) dz^2 + \bar f_*(B*\bar K) d\bar z^2 \right) 
+ {\hat A}*  \left(f_*(B*K) dz^2 + \bar f_*(B*\bar K) d\bar z^2 \right)=0.
\end{equation}

The components of \eqref{biexp} proportional to $dx dz^2$ yield, 
\begin{equation} \label{fb}
d_x B*K + [W, B*K]_* = 0
\end{equation}
Multiplying this equation by $K$ from the right and using 
$K*W*K=\pi(W)$ we find the first of \eqref{veqnsc}.
 
The components of \eqref{biexp} proportional to $dx d {\bar z}^2$ 
yield
\begin{equation} \label{sb}
d_x B*\bar K + [W, B*\bar K]_* = 0
\end{equation}
Multiplying this equation by ${\bar K}$ from the right and using 
${\bar K}*W* {\bar K}={\bar K}*W* {\bar K}=\pi(W)=$ (the second 
step uses the truncation condition \eqref{Rtrunc} on $W$)
we once again find the first of \eqref{veqnsc}. 

The term in \eqref{biexp} proportional to $dz^2 d\bar z$ and 
$dz d\bar z^2$ may be processed as follows. Let 
\begin{equation}\label{decomp}
{\hat S}={\hat S}_{z}+ {\hat S}_{{\bar z}}
\end{equation}
where ${\hat S}_{z}$ is proportional to $dz$ and ${\hat S}_{{\bar z}}$ 
is proportional 
to $d{\bar z}$. The part of \eqref{biexp} proportional to $dz^2 d\bar z$
yields
\begin{equation}\label{sah}
[S_{{\bar z}}, B*K]_* = 0
\end{equation}
Multiplying this equation with $K$ from the right 
and using $K*{\hat S}_{{\bar z}}*K=\pi({\hat S}_{{\bar z}}) $ we find 
\begin{equation}\label{saa}
{\hat S}_{{\bar z} } * B-B*\pi({\hat S}_{{\bar z} }) = 0
\end{equation}

Finally, the part of \eqref{biexp} proportional to $dz d\bar z^2$
yields
\begin{equation}\label{saht}
[S_{z  }, B*{\bar K}]_* = 0
\end{equation}
Multiplying this equation with ${\bar K}$ from the right 
and using 
$${\bar K} *{\hat S}_{z}*{\bar K}={\bar \pi}
({\hat S}_{z}) = \pi({\hat S}_{z})$$
(where we have used \eqref{proj}) we find   
\begin{equation}\label{sab}
{\hat S}_{z} * B-B*\pi({\hat S}_{z}) = 0
\end{equation}
Adding together \eqref{saa} and \eqref{sab} we find the second 
of \eqref{veqnsc}

The fact that $z$ and ${\bar z}$ each have only two components, mean 
that there are no terms in \eqref{biexp} proportional to $dz^3$ or 
$d {\bar z}^3$, so we have fully analyzed the content of \eqref{biexp}.

\subsection{Onshell (Anti) Commutation of components of 
Vasiliev's Master Field}
 
In this subsection we list some useful commutation and anticommutation 
relations between the adjoint fields $S_z$, $S_{\bar z}$, $B*K$ and 
$B*{\bar K}$. The relations we list can be derived almost immediately from 
Vasiliev's equations; we list them for ready reference
\begin{equation}\label{commut}\begin{split}
&[B*K, B*{\bar K}]_*=0\\
&\{S_{z}, S_{\bar z}\}_* = 0\\
&[S_{{\bar z}}, B*K]_* = 0\\
&[S_{z}, B*{\bar K}]_* = 0\\
&\{S_{{\bar z}}, B*K \}_*= 0\\
&\{S_{z}, B*{\bar K}\}_* = 0\\
\end{split}
\end{equation}
The derivation of these equations is straightforward. The first equation 
follows upon expanding the commutator and 
noting that $K*B*{\bar K}={\bar K}*B*K$ (this follows from 
\eqref{Rtrunc} together with the obvious fact that $K$ and ${\bar K}$ commute).
The second equation in \eqref{commut} 
follows upon inserting the decomposition \eqref{decomp}
into the third equation in \eqref{veqnsb}. The third and fourth equations 
in \eqref{commut}are simply \eqref{sah} and \eqref{saht} rewritten. 

The fifth equation in \eqref{commut} may be derived from the third equation 
as follows 
\begin{equation}\label{ed}\begin{split}
&S_{{\bar z}}* B *K=B *K*S_{{\bar z}}\\
&\Rightarrow S_{{\bar z}}* B =B *K*S_{{\bar z}}*K\\
&\Rightarrow S_{{\bar z}}* B = -B *{\bar K}*S_{{\bar z}}* {\bar K}\\
&\Rightarrow S_{{\bar z}}* B*{\bar K} = -B *{\bar K}*S_{{\bar z}} \\
\end{split}
\end{equation}
In the third line of \eqref{ed} we have used the truncation condition 
\eqref{Rtrunc}

The sixth equation in \eqref{commut} is derived in a manner very similar to the 
fifth equation. 

\subsection{Canonical form of $f(X)$ in Vasiliev's equations}\label{redef}

In this subsection we demonstrate that we can use the 
change of variables $X\to g(X)$ for some odd real function $g(X)$ 
together with multiplication by an invertible holomorphic even function 
to put any function $f(X)$ in the form \eqref{fxsol}, atleast provided 
that the function $f(X)$ admits a power series expansion about $X=0$ 
and that $f(0)\neq 0$.  

An arbitrary function $f(X)$ may be decomposed into its even and odd parts 
$$f(X)=f_e(X)+ f_o(X)$$
If $f_e(X)$ in invertible then the freedom of multiplication with an
even complex function may be used to put $f(X)$ in the form 
$$f(X)= 1+ {\tilde f}_o(X) $$ where ${\tilde f}_o(X)=\frac{f_o(X)}{f_e(X)}$.
Clearly ${\tilde f}_o(X)$ is an odd function that admits a power series
expansion. Atleast in the sense of a formal power series expansion of 
all functions, it is easy to convince oneself that any such function may be 
written in the form $g(X)e^{i \theta(X)}$ where $g(X)$ is an a real odd function 
and $\theta(X)$ is a real even function. We may now use the freedom of 
variable redefinitions to work with the variable $g(X)$ instead of $X$. 
This redefinition preserves the even nature of $\theta(X)$ and casts 
$f(X)$ in the form \eqref{fxsol}.

\subsection{Conventions for $SO(4)$ spinors}\label{spconv}

Euclidean $SO(4)$ $\Gamma$ matrices may be chosen as 
\begin{equation}\label{gd}
\Gamma_a=
\left( \begin{tabular} {cc}
$0$ & $\sigma_a$ \cr
${\bar \sigma}_a$ & $0$\cr
\end{tabular} \right)
\end{equation}
where $ a= 1 \ldots 4$ and 
\begin{equation}\label{sd}
\sigma_a=(\sigma_i, iI), ~~~{\bar \sigma}_a= -\sigma_2 \sigma_a^T 
\sigma_2=(\sigma_i, -i I)
\end{equation}
(where $i= 1 \ldots 3$ and $\sigma^i$ are the usual Pauli matrices).  
In the text below we will often refer to the fourth component of $\sigma^\mu$ 
as $\sigma^z$; in other words 
$$\sigma^z=iI$$
(we adopt this cumbersome notation to provide easy passage to 
different conventions).
The chirality matrix $\Gamma_5=\Gamma_1 \Gamma_2 \Gamma_3 \Gamma_4$ 
is given by 
\begin{equation}\label{gma}
\Gamma_5=\left( \begin{tabular} {cc}
$I$ & $0$ \cr
$0$ & $-I$\cr
\end{tabular} \right)
\end{equation}
$\Gamma$ matrices act on the spinors 
$$\left( \begin{tabular} {c}
$\chi_\alpha$ \cr
$\bar{\zeta}^{\dot \beta}$\cr
\end{tabular} \right)$$
whereas the row spinors that multiply $\Gamma$ from the left have the index 
structure 
$$\left( \begin{tabular} {cc}
$\chi^\alpha$ &
$\bar{\zeta}_{\dot \beta}$\cr
\end{tabular} \right)$$
As a consequence we assign the index structure 
$(\sigma_a)_{\alpha {\dot \beta}}$ and ${\bar \sigma}^{{\dot \alpha} \beta}.$
It is easy to check that
\begin{equation}\label{gd}
[\Gamma_a, \Gamma_b]= 2 \left( \begin{tabular} {cc}
$\sigma_{ab}$ & $0$ \cr
$0$ & ${\bar \sigma}_{ab}$\cr
\end{tabular} \right)
\end{equation}
where 
\begin{equation}
\begin{split}
\sigma_{ab} &= \half (\sigma_a\bar{\sigma}_b - \sigma_b\bar{\sigma}_a), ~~~
\bar{\sigma}_{ab} = \half (\bar{\sigma}_a\sigma_b - \bar{\sigma}_b\sigma_a) \\
\Rightarrow ~~ \sigma_{ij} &= i \epsilon_{ijk} \sigma^k, ~~~
{\bar \sigma}_{ij}= i \epsilon_{ijk} {\bar \sigma}^k,
~~~\sigma_{i4}=-i \sigma_i, ~~~{\bar \sigma}_{i4}= i \sigma_i
\end{split}
\end{equation}
Clearly the index structure above is $(\sigma_{ab})_\alpha ^{~\beta}$ and 
 $({\bar \sigma}_{ab})^\da _{~\db}$.
Spinor indices are raised and lowered according to the conventions
$$\psi_\alpha = \epsilon_{\alpha \beta} \psi^\beta, ~~~ 
\psi^\A = \psi_\B \epsilon^{\B \A}, ~~~\epsilon^{12}=\epsilon_{12}=1$$
The product of a chiral spinor $y^\alpha$ and an antichiral spinor 
$\bar{y}^{\dot \beta}$ is a vector. By convention
we define the associated vector as 
\begin{equation}\label{vfss}
V_\mu= y^\alpha (\sigma_\mu)_{\alpha {\dot \beta}} \bar{y}^{\dot \beta}
\end{equation}
The product of a chiral spinor $y$ with itself is a self dual antisymmetric 
2 tensor which we take to be 
\begin{equation}\label{ten}
V_{ab} =y^\alpha (\sigma_{ab})_\alpha ^{~\beta} y_\beta
\end{equation}
Similarly the product of an antichiral spinor with itself is an antiselfdual 
2 tensor which we take to be
 \begin{equation}\label{ten}
V_{ab} ={\bar y}_\da ({\bar \sigma}_{ab})^\da _{~\db} \bar{y}^\db
\end{equation}

\subsection{$AdS_4$ solution}\label{solverf}

In this appendix we will show that 
\begin{equation}\label{sol}
W_0 = (e_0)_{\alpha\dot{\beta}}y^{\alpha}\bar{y}^{\dot{\beta}} + (\omega_0)_{\alpha\beta}y^{\alpha}y^{\beta} + 
       (\omega_0)_{\dot{\alpha}\dot{\beta}}\bar{y}^{\dot{\alpha}}\bar{y}^{\dot{\beta}} 
\end{equation}
with the $AdS_4$ values for the vielbein and spin connection, satisfies the Vasiliev equation 
\begin{equation}\label{eqn}
d_x W_0 + W_0 * W_0 = 0.
\end{equation}
Substituting \eqref{sol} in \eqref{eqn} and collecting terms quadratic in $y$ and $\bar{y}$ we get 
\begin{equation}
\begin{split}
y^{\alpha} \bar{y}^{\dot{\alpha}} &:~~~~ d_x e_{\alpha\dot{\beta}} + 4 \omega_{\alpha}^{~\beta}\wedge e_{\beta\dot{\beta}} 
          - 4 e_{\alpha\dot{\gamma}} \wedge \omega^{\dot{\gamma}}_{~\dot{\beta}} = 0 \\
y^{\alpha} y^{\beta} &:~~~~ d_x \omega_{\alpha}^{~\beta} - 4 \omega_{\alpha}^{~\gamma}\wedge w_{\gamma}^{~\beta} 
                  - e_{\alpha\dot{\alpha}}\wedge e_{\beta\dot{\beta}}\epsilon^{\dot{\alpha}\dot{\beta}} = 0 \\ 
y^{\dot{\alpha}}y^{\dot{\beta}} &:~~~~ d_x \omega^{\dot{\alpha}}_{~\dot{\beta}} 
                 + 4 \omega^{\dot{\alpha}}_{~\dot{\gamma}}\wedge \omega^{\dot{\gamma}}_{~\dot{\beta}} 
               - e_{\alpha\dot{\alpha}}\wedge e_{\beta\dot{\beta}}\epsilon^{\alpha\beta} = 0 \\ 
\end{split}
\label{3eqn}
\end{equation} 
Let us consider the Vasiliev gauge transformations generated by 
$$ \epsilon(x|Y) = C_{1ab} ~(y\sigma_{ab}y) + C_{2ab} ~(\bar{y}\bar{\sigma}_{ab}\bar{y}) $$
Under these the vielbein and spin connection changes by
\begin{equation}
\begin{split}
\delta e_{\A\da} &= -4 C_{1ab} (\sigma_{ab})_\A^{~\B} e_{\B\da} - 4 C_{2ab} 
    ~e_{\A\db} (\bar{\sigma}_{ab})^\db_{~\da} \\
\delta \omega_\A^{~\B} &= d_x C_{1ab} (\sigma_{ab})_{\A}^{~\B} 
    - 8 C_{1ab} ~\omega_\A^{~\gamma} (\sigma_{ab})_{\gamma}^{~\B} \\
\delta \omega^\da_{~\db} &= d_x C_{2ab} (\bar{\sigma}_{ab})^\da_{~\db} 
    + 8 C_{2ab} ~\omega^\da_{~\dot{\gamma}} (\bar{\sigma}_{ab})^{\dot{\gamma}}_{~\db} \\
\end{split}
\label{ewgt}
\end{equation}
Notice that these are just the rotation of the vielbeins in the tangent space. 
The two homogeneous terms in $\delta e$ are just the rotation by under  
$SU(2)_L$ and $SU(2)_R$ of $SO(4)$ that acts on the tangent space. As expected under 
such rotation the spin connection transforms inhomogeneously. Substituting 
\eqref{ewgt} in \eqref{3eqn} it is easily verified that \eqref{3eqn} transforms 
homogeneously. 

In fact the first equation in \eqref{3eqn} is just the torsion free 
condition while the second and third equation expresses the selfdual and 
anti-selfdual part of curvature two form in term of vielbeins. 
Substituting the $AdS_4$ 
values of vielbeins and spin connection \eqref{ewsol} one can easily check 
that \eqref{3eqn} are satisfied.
 
Converting \eqref{3eqn} from bispinor notation to $SO(4)$ vector notation 
using the following conversion
\begin{equation}
\begin{split}
e_{\A\dot{\B}} = 2e_a (\sigma_a)_{\A\dot{\B}} ~, ~~
\omega_\A^{~\B} = \frac{1}{16} \omega_{ab} (\sigma_{ab})_\A^{~\B} ,~~
\omega^\da_{~\db} = -\frac{1}{16} \omega_{ab} (\sigma_{ab})^\da_{~\db} ,~~
\end{split}
\label{ewconv}
\end{equation}
we get 
\begin{equation}
\begin{split}
T_a & \equiv d_x e_a + \omega_{ab}\wedge e_b = 0 \\
R_{ab} & \equiv d_x \omega_{ab} + \omega_{ac}\wedge \omega_{cb} 
              + 6 e_a \wedge e_b = 0.\\
\end{split}
\label{TR}	
\end{equation}

\subsection{Exploration of various boundary conditions for scalars in the 
non abelian theory}\label{bdry}

The same theory in $AdS_4$ with $\Delta=2$ boundary condition on the 
$U(M)$-{\it singlet} bulk scalar is dual to the critical point of the $SU(N)$ 
vector model with $M$ flavors and the double trace deformation by
$(\bar\phi^{ia} \phi_{ia})^2$.
Alternatively, this critical point may be defined by introducing a Lagrangian 
multiplier $\A$ and adding the term
\ie
\A \bar\phi^{ia} \phi_{ia}
\fe
to the Lagrangian of the vector model.\footnote{The critical point can be 
conveniently defined using dimensional regularization.} As in the case of 
the $M=1$ critical vector model, higher spin symmetry is broken by $1/N$ 
effects. Note that the $SU(M)$ part of the spin-2 current is also broken 
by $1/N$ effects, i.e. there are no interacting colored massless gravitons, 
as expected. To see this explicitly from the boundary CFT, let us consider 
the spin-2 current
\ie
(J_{\mu\nu}^{(2)})^a{}_b = {1\over 2} \bar\phi^{ia} \overleftrightarrow\partial_\mu \overleftrightarrow\partial_\nu \phi_{ib} - 2 \partial_{(\mu} \bar\phi^{ia} \partial_{\nu)} \phi_{ib} + \delta_{\mu\nu} \partial^\rho \bar\phi^{ia} \partial_\rho \phi_{ib}.
\fe
Using the classical equation of motion
\ie
\Box \phi_i = \A \phi_i,
\fe
we have
\ie
\partial^\mu (J_{\mu\nu}^{(2)})^a{}_b =(\partial_\nu\A) \bar\phi^{ia} \phi_{ib} - \A \partial_\nu(\bar\phi^{ia}\phi_{ib}).
\fe
While the $SU(M)$-singlet part of $J_{\mu\nu}$, being the stress-energy tensor, 
is conserved ($\bar\phi^{ia}\phi_{ia}$ is set to zero by $\A$-equation of motion), 
the $SU(M)$ non-singlet part of $J_{\mu\nu}$ is not conserved, and acquires 
an anomalous dimension of order $1/N$ at the leading nontrivial order in the 
$1/N$ expansion. In the bulk, the colored gravitons become massive, and 
their longitudinal components are supplied by the bound state of the singlet 
scalar and a colored spin-1 field.

One could also consider the theory in $AdS_4$ with $\Delta=2$ boundary condition 
on {\it all} bulk scalars, that is, on both the singlet and adjoint of the $SU(M)$ 
bulk gauge group. The dual CFT is the critical point defined by turning on 
the double trace deformation $\bar\phi^{ia}\phi_{ib}\bar\phi^{jb}\phi_{ja}$ 
and flow to the IR, or by introducing the Lagrangian multiplier $\Lambda_a{}^b$, 
and the term
\ie
\Lambda_a{}^b \bar\phi^{ia}\phi_{ib}
\fe
in the CFT Lagrangian. Now the classical equations of motion
\ie
\Box\phi_{ia} = \Lambda_a{}^b \phi_{ib},~~~~\bar\phi^{ia}\phi_{ib}=0,
\fe
imply the divergence of the colored spin-2 currents
\ie
\partial^\mu (J_{\mu\nu}^{(2)})^a{}_b = \Lambda_b{}^c \bar\phi^{ia}\overleftrightarrow\partial_\nu \phi_{ic} - \Lambda_c{}^a \bar\phi^{ic} \overleftrightarrow \partial_\nu \phi_{ib}
= \Lambda_b{}^c (J^{(1)}_\nu)^a{}_c - \Lambda_c{}^a (J^{(1)}_\nu)^c{}_b.
\fe
Once again, the $SU(M)$ non-singlet spin-2 current is no longer conserved. 
In this case, the colored gravitons in the bulk are massive because their 
longitudinal component are supplied by the two-particle state of colored 
scalar and spin-1 fields.

\section{Supersymmetry transformations on bulk fields of spin $0$, ${1\over 2}$, and $1$}\label{susyappen}

We begin by rewriting the magnetic boundary condition on the spin-1 bulk fields in the supersymmetric Vasiliev theory. With the magnetic boundary condition, the $2^{n-1}$ vector gauge fields are dual to ungauged $U(2^{{n\over 2}-1})\times U(2^{{n\over 2}-1})$ ``R-symmetry" currents of boundary CFT that rotate the bosonic and fermionic flavors separately. Supersymmetrizing Chern-Simons coupling will generally break this flavor symmetry to a subgroup. We will see this as the violation of magnetic boundary condition by the supersymmetry variation of the bulk spin-1 fields. If we do not gauge the flavor symmetries of the Chern-Simons vector model, then all bulk vector fields should be assigned the magnetic boundary condition. We will see later that in this case only up to ${\cal N}=3$ supersymmetry can be preserved, whereas by relaxing the magnetic boundary condition on some of the bulk vector fields, it will be possible to preserve ${\cal N}=4$ or $6$ supersymmetry.

In terms of Vasiliev's master field $B$ which contains the field strength, the general electric-magnetic boundary condition may be expressed as 
\ie\label{FbarFone}
B\big|_{\cO(y^2,\bar y^2)}\to z^2\left[e^{i\B}(yFy)+e^{-i\B}(\bar y \overline F\bar y)\Gamma\right],~~~~z\to 0,
\fe
where $F\equiv F_{\mu\nu}\sigma^{\mu\nu}$ and its complex conjugate $\overline{F}$ are functions of $\psi_i$, and are constrained by the linear relation 
\ie\label{FbarF}
F=-\sigma^z \overline F\sigma^z.
\fe
With this choice of boundary condition, the boundary to bulk propagator for the spin-1 components of the $B$ master field is given by the standard one,
\ie
B^{(1)}& = {z^{2}\over (\vec x^2+z^2)^{3}} e^{-y\Sigma\bar y} \left[ e^{i\B} (\lambda  {\bf x}\sigma^zy)^{2} + e^{-i\B} ( \lambda\sigma^z{\bf x} \sigma^z\bar y)^{2}\Gamma \right]
\\
& \equiv  \widetilde B^{(1)} \left[ e^{i\B} (\lambda {\bf x} \sigma^z y)^{2} + e^{-i\B} ( \lambda\sigma^z{\bf x} \sigma^z\bar y)^{2}\Gamma \right].
\fe
It indeed obeys (\ref{FbarF}), with $F$ and $\overline F$ given by
\ie
&F_\A{}^\B=-(\lambda  {\vec x\cdot\vec\sigma}\sigma^z)_\A(\lambda {\vec x\cdot\vec\sigma}\sigma^z)^\B,
\\
& \overline F_{\dot\A}{}^{\dot\B}=-( \lambda\sigma^z{\vec x\cdot\vec\sigma}\sigma^z)_{\dot\A}( \lambda\sigma^z{\vec x\cdot\vec\sigma} \sigma^z)^{\dot\B}=-( \lambda{\vec x\cdot\vec\sigma})_{\dot\A}( \lambda{\vec x\cdot\vec\sigma} )^{\dot\B},
\fe
and
\ie
(\sigma^z \overline F\sigma^z)_\A{}^\B=-( \lambda{\vec x\cdot\vec\sigma})_{\dot\A}( \lambda{\vec x\cdot\vec\sigma})^{\dot\B}(\sigma^z)_\A{}^{\dot\A} (\sigma^z)^{\B}{}_{\dot\B}=( \lambda{\vec x\cdot\vec\sigma}\sigma^z)_{\A}( \lambda{\vec x\cdot\vec\sigma}\sigma^z )^{\B}=-F_\A{}^\B.
\fe
In the next four subsections, we give the explicit formulae for the supersymmetry variation $\delta_\epsilon$ (i.e. spin $3/2$ gauge transformation of Vasiliev's system) of bulk fields of spin $0, 1/2, 1$, sourced by boundary currents of spin $0, 1/2, 1$.

\subsection{$\delta_\epsilon$: spin 1 $\to$ spin ${1\over 2}$ }

Let us start with the $B$ master field sourced by a spin-1 boundary current at $\vec x=0$, i.e. the spin-1 boundary to bulk propagator $B^{(1)}(x|Y)$, and consider its variation under supersymmetry, which is generated by $\epsilon(x|Y)$ of degree one in $Y=(y,\bar y)$:
\ie
\delta_\epsilon B^{(1)}(x|Y) =&-\epsilon * e^{i\B}(\lambda {\bf x}  \sigma^z y)^2\widetilde B^{(1)}+ e^{i\B}(\lambda  {\bf x}  \sigma^zy)^2\widetilde B^{(1)}*\pi(\epsilon)
\\
& -\epsilon *  e^{-i\B}(\lambda\sigma^z {\bf x}\sigma^z\bar y)^2\Gamma\widetilde B^{(1)}+ e^{-i\B}(\lambda \sigma^z{\bf x}\sigma^z\bar y)^2\Gamma\widetilde B^{(1)}*\pi(\epsilon).
\fe
Carrying out the $*$ products explicitly, we find
\ie
&-\epsilon * (\lambda  {\bf x} \sigma^zy)^2\widetilde B^{(1)}+ (\lambda  {\bf x} \sigma^zy)^2\widetilde B^{(1)}*\pi(\epsilon)
\\
&=-(\Lambda y+\overline\Lambda \bar y)*  (\lambda  {\bf x} \sigma^zy)^2  \widetilde B^{(1)}+ (\lambda {\bf x}\sigma^z y)^2 \widetilde B^{(1)}*(-\Lambda y+\overline \Lambda \bar y)
\\
&=-\{y_\A, (  {\bf x} \sigma^zy)_\B (  {\bf x} \sigma^zy)_\C \widetilde B^{(1)}\}_*\{\Lambda^\A,\lambda^\B\lambda^\C\}-[y_\A, (  {\bf x} \sigma^zy)_\B(  {\bf x} \sigma^zy)_\C  \widetilde B^{(1)}]_*[\Lambda^\A,\lambda^\B\lambda^\C]
\\
&~~~~-[\bar y_{\dot\A}, ( {\bf x}\sigma^z y)_\B( {\bf x}\sigma^z y)_\C  \widetilde B^{(1)}]_*\{\overline\Lambda^{\dot\A},\lambda^\B\lambda^\C\}-\{\bar y_{\dot\A}, (  {\bf x}\sigma^z y)_\B  ( {\bf x}\sigma^z y)_\C\widetilde B^{(1)}\}_*[\overline\Lambda^{\dot\A},\lambda^\B\lambda^\C]
\\
&=-2\{\Lambda y,\lambda^\B\lambda^\C\} (  {\bf x}\sigma^z y)_\B  ( {\bf x}\sigma^z y)_\C\widetilde B^{(1)}-2[\Lambda\partial_y,\lambda^\B\lambda^\C] ({\bf x}\sigma^z y)_\B ({\bf x}\sigma^z y)_\C \widetilde B^{(1)}
\\
&~~~~-2\{\overline\Lambda\partial_{\bar y},\lambda^\B\lambda^\C\} ({\bf x}\sigma^z y)_\B  ( {\bf x}\sigma^z y)_\C\widetilde B^{(1)}-2[\overline\Lambda\bar y,\lambda^\B\lambda^\C] ({\bf x}\sigma^z y)_\B({\bf x}\sigma^z y)_\C  \widetilde B^{(1)}
\\
&=2\{\overline\Lambda\Sigma y-\Lambda y, (\lambda  {\bf x}\sigma^z y)^2 \} \widetilde B^{(1)}+2[\Lambda\Sigma\bar y-\overline\Lambda\bar y,(\lambda  {\bf x}\sigma^z y)^2]   \widetilde B^{(1)}-4[ ( {\bf x} \sigma^z\Lambda)_\B,\lambda^\B(\lambda {\bf x} \sigma^zy)]  \widetilde B^{(1)},
\fe
and
\ie
&-\epsilon *  (\lambda \sigma^z{\bf x}\sigma^z\bar y)^2\Gamma\widetilde B^{(1)}+ (\lambda\sigma^z {\bf x}\sigma^z\bar y)^2\Gamma\widetilde B^{(1)}*\pi(\epsilon)
\\
&=-2\{\Lambda y,\lambda^\B\lambda^\C\Gamma\} ( \sigma^z{\bf x}\sigma^z\bar y)_\B  (\sigma^z {\bf x}\sigma^z\bar y)_\C\widetilde B^{(1)}-2[\Lambda\partial_y,\lambda^\B\lambda^\C\Gamma] ( \sigma^z{\bf x}\sigma^z\bar y)_\B  (\sigma^z {\bf x}\sigma^z\bar y)_\C\widetilde B^{(1)}
\\
&~~~~-2\{\overline\Lambda\partial_{\bar y},\lambda^\B\lambda^\C\Gamma\} (\sigma^z {\bf x}\sigma^z\bar y)_\B (\sigma^z {\bf x}\sigma^z\bar y)_\C \widetilde B^{(1)}-2[\overline\Lambda\bar y,\lambda^\B\lambda^\C\Gamma] ( \sigma^z{\bf x}\sigma^z\bar y)_\B  ( \sigma^z{\bf x}\sigma^z\bar y)_\C\widetilde B^{(1)}
\\
&=2\{\overline\Lambda\Sigma y-\Lambda y, (\lambda\sigma^z {\bf x}\sigma^z\bar y)^2 \Gamma\} \widetilde B^{(1)}+2[\Lambda\Sigma\bar y-\overline\Lambda\bar y,(\lambda\sigma^z {\bf x}\sigma^z\bar y)^2\Gamma]   \widetilde B^{(1)}-4\{ (\sigma^z {\bf x}\sigma^z\overline \Lambda)_\B,\lambda^\B(\lambda\sigma^z {\bf x}\sigma^z\bar y)\Gamma\}  \widetilde B^{(1)}.
\fe
Note that the commutators and anti-commutators in above formula are due to the $\psi_i$-dependence only, and do not involve $*$ product.
$\delta_\epsilon B^{(1)}$ contains supersymmetry variation of fields of spin $1/2$ and $3/2$. We will focus on the variation spin $1/2$ fields, since they can be subject to a family of different boundary conditions, corresponding to turning on fermionic double trace deformations (i.e. $({\rm fermion~singlet})^2$) in the boundary CFT. So we restrict to terms linear in $(y,\bar y)$,
\ie\label{VspV}
&\delta B^{(1)}\big|_{\cO(y,\bar y)}=-4[ ( {\bf x} \sigma^z\Lambda)_\B,\lambda^\B(\lambda {\bf x} \sigma^zy)]  \widetilde B^{(1)}-4\{ (\sigma^z {\bf x}\sigma^z\overline \Lambda)_\B,\lambda^\B(\lambda\sigma^z {\bf x}\sigma^z\bar y)\Gamma\}  \widetilde B^{(1)}
\\
&\to-4e^{i\B}{z^{{3\over 2}}\over (\vec x^2+z^2)^3}[ ( {\vec x\cdot\vec\sigma}\sigma^z \Lambda_+)_\B,\lambda^\B(\lambda{\vec x\cdot\vec\sigma}\sigma^zy)]  +4e^{-i\B}{z^{{3\over 2}}\over (\vec x^2+z^2)^3}[ ( {\vec x\cdot\vec\sigma} \sigma^z\Lambda_+)_{\B},\lambda^{\B}(\lambda {\vec x\cdot\vec\sigma}\bar y)]\Gamma  
\\
\fe
where in the second line we kept the leading terms, of order $z^{3\over 2}$, in the $z\to 0$ limit.

\subsection{$\delta_\epsilon$: spin ${1\over 2}$ $\to$ spin $1$ }

The general conformally invariant boundary condition on spin $1/2$ fermions, in terms of Vasiliev's $B$ master field, takes the form
\ie\label{fbcc}
B\big|_{\cO(y,\bar y)}\to z^{3\over 2}\left[e^{i\A}(\chi y) - \Gamma e^{-i\A}(\bar \chi \bar y)\right],
\fe
Here $\chi$ and its complex conjugate $\bar\chi$ are chiral and anti-chiral spinors that are odd functions of the Grassmannian variables $\psi_i$. They are further constrained by the linear relation
\ie\label{chre}
\chi= \sigma^z\bar\chi.
\fe
$\A$ is generally a linear operator that acts on the vector space spanned by odd monomials in the $\psi_i$'s, i.e. it assigns phase angles to fermions in the bulk $R$-symmetry multiplet.
A choice of the spin-$1/2$ fermion boundary condition is equivalent to a choice of the ``phase angle" operator $\A$. 


The fermion boundary to bulk propagator that satisfies the above boundary condition is:
\ie
B^{({1\over 2})}& = {z^{3\over 2}\over (\vec x^2+z^2)^{2}} e^{-y\Sigma\bar y} \left[ e^{i\A} (\lambda {\bf x} \sigma^z y) -\Gamma e^{-i\A}(\lambda\sigma^z{\bf x}  \sigma^z\bar y)\right]
\\
&\equiv \left[  e^{i\A}(\lambda {\bf x}  \sigma^zy) - \Gamma e^{-i\A}(\lambda\sigma^z{\bf x} \sigma^z \bar y)\right] \widetilde B^{({1\over 2})}.
\fe
Here the linear operator $\A$ is understood to act on $\lambda$ only, the latter being an odd function of $\psi_i$'s.

Next, we make super transformation on the fermion boundary to bulk propagator. The supersymmetry transformation reads
\ie
\delta B^{({1\over 2})}=&-e^{i\A}\epsilon * (\lambda  {\bf x}\sigma^z y)\widetilde B^{({1\over 2})}+ e^{i\A}(\lambda {\bf x}\sigma^z y)\widetilde B^{({1\over 2})}*\pi(\epsilon)
\\
& -e^{-i\A}\epsilon *  (\lambda\sigma^z {\bf x}\sigma^z\bar y)\Gamma\widetilde B^{({1\over 2})}+e^{-i\A} (\lambda\sigma^z {\bf x}\sigma^z\bar y)\Gamma\widetilde B^{({1\over 2})}*\pi(\epsilon),
\fe
where $\epsilon = \Lambda y + \bar\Lambda\bar y$, $\Lambda$ is an odd supersymmetry parameter $\eta$ multiplied by an odd function of the $\psi_i$'s. $\eta$ in particular anti-commutes with all $\psi_i$'s, and therefore anti-commutes with $\lambda$ which involves an odd number of $\psi_i$'s.

Carrying out the $*$ algebra, we have
\ie
&-\epsilon * (\lambda {\bf x}\sigma^z y)\widetilde B^{({1\over 2})}+ (\lambda  {\bf x}\sigma^z y)\widetilde B^{({1\over 2})}*\pi(\epsilon)
\\
&=2\{\overline\Lambda\Sigma y-\Lambda y, (\lambda {\bf x}\sigma^z y) \} \widetilde B^{({1\over 2})}+2[\Lambda\Sigma\bar y-\overline\Lambda\bar y,(\lambda{\bf x}\sigma^z y)]   \widetilde B^{({1\over 2})}-2[ ( {\bf x} \sigma^z\Lambda)_\B,\lambda^\B]  \widetilde B^{({1\over 2})},
\fe
and
\ie
&-\epsilon *  (\lambda\sigma^z {\bf x}\sigma^z\bar y)\Gamma\widetilde B^{({1\over 2})}+ (\lambda\sigma^z {\bf x}\sigma^z\bar y)\Gamma\widetilde B^{({1\over 2})}*\pi(\epsilon)
\\
&=2\{\overline\Lambda\Sigma y-\Lambda y, (\lambda\sigma^z {\bf x}\sigma^z\bar y) \Gamma\} \widetilde B^{({1\over 2})}+2[\Lambda\Sigma\bar y-\overline\Lambda\bar y,(\lambda\sigma^z {\bf x}\sigma^z\bar y)\Gamma]   \widetilde B^{({1\over 2})}-2\{ ( \sigma^z{\bf x}\sigma^z\overline \Lambda)_\B,\lambda^\B\Gamma\}  \widetilde B^{({1\over 2})}.
\fe

The supersymmetry variation of the spin-1 field strengths are extracted from $\cO(y^2,\bar y^2)$ terms in $\delta B^{({1\over 2})}$, namely
\ie\label{FspF}
&\delta_\epsilon B^{({1\over 2})}(x|Y)\big|_{\cO(y^2,\bar y^2)}=2 \{\overline\Lambda\Sigma y-\Lambda y, e^{i\A}(\lambda  {\bf x}\sigma^z y) \} \widetilde B^{({1\over 2})} - 2[\Lambda\Sigma\bar y-\overline\Lambda\bar y, \Gamma e^{-i\A}(\lambda\sigma^z {\bf x}\sigma^z\bar y)]   \widetilde B^{({1\over 2})}
\\
&\to-4 {z^{2}\over (\vec x^2+z^2)^3}\{\Lambda_0 {\vec x\cdot\vec\sigma}\sigma^zy, e^{i\A} (\lambda \sigma^z  {\vec x\cdot\vec\sigma} y) \} - 4 {z^{2}\over( \vec x^2+z^2)^3}[\Lambda_0 {\vec x\cdot\vec\sigma}\bar y,\Gamma e^{-i\A}(\lambda{\vec x\cdot\vec\sigma}\bar y)]  .
\fe
In the second line, we have taken the small $z$ limit and kept the leading terms, of order $z^2$.


\subsection{$\delta_\epsilon$: spin ${1\over 2}$ $\to$ spin $0$ }

The supersymmetry variation of the scalar field due to a spin-${1\over 2}$ fermionic boundary source is extracted from $\delta_\epsilon B^{({1\over 2})}$ of the previous subsection, restricted to $y=\bar y=0$:
\ie\label{fermiontoboson}
&\delta_\epsilon B^{({1\over 2})}\big|_{y,\bar y=0}(\vec x,z)=-2[ (  {\bf x} \sigma^z\Lambda)_\B, e^{i\A}\lambda^\B]  \widetilde B^{({1\over 2})}-2 \Gamma [ (\sigma^z {\bf x}\sigma^z\overline \Lambda)_\B,e^{-i\A}\lambda^\B]  \widetilde B^{({1\over 2})}
\\
&~~~~+2z^{-{1\over 2}} \Gamma [ (\sigma^z {\bf x} \Lambda_+)_\B,e^{-i\A}\lambda^\B]  \widetilde B^{({1\over 2})}-2 z^{1\over 2} \Gamma [ (\sigma^z {\bf x} \Lambda_-)_\B,e^{-i\A}\lambda^\B]  \widetilde B^{({1\over 2})}
\\
&=
2(e^{i\A}+ \Gamma e^{-i\A}){z\over (\vec x^2+z^2)^2}[ ( \sigma^z {\vec x\cdot\vec\sigma} \Lambda_+)_\B,\lambda^\B] -2(e^{i\A}-\Gamma e^{-i\A}){z^2\over (\vec x^2+z^2)^2}[ (  \Lambda_+)_\B,\lambda^\B]  
\\
&~~~~-2(e^{i\A}-\Gamma e^{-i\A}){z^2\over (\vec x^2+z^2)^2}[ (  {\vec x\cdot\vec\sigma} \sigma^z\Lambda_-)_\B,\lambda^\B]  + {\cal O}(z^3).
\fe
In the last two lines, $\A$ as a linear operator is understood to act on $\lambda$ only (and not on $\Lambda_\pm$).

\subsection{$\delta_\epsilon$: spin $0$ $\to$ spin ${1\over 2}$ }

The general conformally invariant linear boundary condition on the bulk scalars $B^{(0)}(\vec x,z) = B(\vec x,z|y=\bar y=0)$ may be expressed as
\ie\label{scbc}
B^{(0)}(\vec x,z) = (e^{i\C}+ \Gamma e^{-i\C})\tilde f_1 z+(e^{i\C}-\Gamma e^{-i\C})\tilde f_2 z^2 + {\cal O}(z^3)
\fe
in the limit $z\to 0$. Here $\tilde f_1,\tilde f_2$ are real and even function in $\psi_i$, and are subject to a set of linear relations that eliminate half of their degrees of freedom. The phase $\C$ is generally a linear operator acting on the space spanned by even monomials in the $\psi_i$'s (analogously to $\A$ in the fermion boundary condition). We will determine our choice of $\C$ and the linear constraints on $\tilde f_{1,2}$ later. 

The boundary-to-bulk propagator for the scalar components of the $B$ master field, subject to the above boundary condition, is now written as
\ie
B^{(0)} &=f_1(\psi)\widetilde B^{(0)}_{\Delta=1} +f_2(\psi) \widetilde B^{(0)}_{\Delta=2},
\fe
where
\ie
f_1(\psi)=(e^{i\C}+ \Gamma e^{-i\C})\tilde f_1(\psi),~~~~f_2(\psi)=(e^{i\C}-\Gamma e^{-i\C}))\tilde f_2(\psi).
\fe
A straightforward calculation gives the supersymmetry variation of the spin-${1\over 2}$ fermion due to a scalar boundary source at $\vec x=0$,
\ie\label{VgbV}
&\delta_\epsilon \widetilde B^{(0)}(\vec x,z)\big|_{\cO(y,\bar y)}\to-4{z^{3\over 2}\over (\vec x^2+z^2)^2}\{\Lambda_0\sigma^z{\vec x\cdot \vec\sigma}y,f_1\}-4{z^{3\over 2}\over (\vec x^2+z^2)^2} [\Lambda_0{\vec x\cdot \vec\sigma}\bar y,f_1]
\\
&~~~~~~+2{z^{3\over 2}\over (\vec x^2+z^2)^2}[\Lambda_+\sigma^z \bar y,f_2]+2{z^{3\over 2}\over (\vec x^2+z^2)^2}\{\Lambda_+ y,f_2\}
\\
&=-4{z^{3\over 2}\over (\vec x^2+z^2)^2}\left(e^{i\C}\{\Lambda_0\sigma^z{\vec x\cdot \vec\sigma}y,\tilde f_1\}-\Gamma e^{-i\C}[\Lambda_0\sigma^z{\vec x\cdot \vec\sigma}y,\tilde f_1]+e^{i\C} [\Lambda_0{\vec x\cdot \vec\sigma}\bar y,\tilde f_1]-\Gamma e^{-i\C} \{\Lambda_0{\vec x\cdot \vec\sigma}\bar y,\tilde f_1\}\right)
\\
&~~~~~~+2{z^{3\over 2}\over (\vec x^2+z^2)^2}\left(e^{i\C}[\Lambda_+\sigma^z \bar y,\tilde f_2]+ \Gamma e^{-i\C}\{\Lambda_+\sigma^z \bar y,\tilde f_2\}+e^{i\C}\{\Lambda_+ y,\tilde f_2\}+ \Gamma e^{-i\C}[\Lambda_+ y,\tilde f_2]\right).
\fe
We have taken the small $z$ limit, and kept terms of order $z^{3\over 2}$.
Again, in the last two lines $\C$ as a linear operator should be understood as acting on $\tilde f_{1,2}(\psi)$ only and not on $\Lambda$.

\section{The bulk dual of double trace deformations and 
Chern Simons Gauging}\label{more}

\subsection{Alternate and Regular boundary conditions for 
scalars in $AdS_{d+1}$}\label{sbc}

In this section we review the AdS/CFT implementation alternate and 
regular boundary conditions for scalars, in the presence of multitrace
deformations. The material reviewed here is well known 
(see e.g. \cite{Witten:2001ua,Gubser:2002vv,Mueck:2002gm,Sever:2002eg,
Sever:2002fk,Witten:2003ya} - we most closely follow the 
approach of the paper \cite{Mueck:2002gm}); our brief review focuses 
on aspects we will have occasion to use in the main text of our paper. 

\subsubsection{Multi-trace deformations in large $N$ field theories}

In this brief subsection we will address the following question: 
how is the generating function of correlators of a large $N$ field theory 
modified by the addition of a multi-trace deformation to the action of 
the theory?

Consider any large $N$ field theory whose single trace operators are 
denoted by $O_i$. Let $W(J)$ denote the generating function of correlators\footnote{More precisely this equation should have read 
\begin{equation}\label{gfc}
\langle e^{\int d^d x J_i(x) O^i(x)} \rangle =e^{-W[J_i(x)]}.
\end{equation}
However for ease of readability, in all the formal discussions of this 
section we will use compact notation in which we suppress the position 
dependence of operators and fields, and do not explicitly indicate 
integration.}
\begin{equation}\label{gfc}
\langle e^{J_i O_i} \rangle =e^{-W[J_i]}.
\end{equation}
Note that $W[J_i]$ is of order $N^2$ in a matrix type
large $N$ theory, while it is of order $N$ in a vector type large $N$ 
theory. For formal purposed below we will find it useful to Legendre  
transform $W$ to define an effective action for the operators $O_i$
\begin{equation}\label{eeo}
I[O^i]=W[J_i]+O^iJ_i .
\end{equation}
$I[O^i]$ is a function only of $O^i$ (and not of $J_i$) in the following 
sense. The RHS of \eqref{eeo} is viewed as an action for the dynamical 
variable $J_i$. The equation of motion for $J_i$ follows from varying 
this action and is 
\begin{equation}
\label{eeon}
\frac{\partial W}{\partial J_i}= -O^i.
\end{equation}
The RHS of \eqref{eeo} is evaluated with the onshell value of $J_i$. 

$I[O^i]$ plays the role of the effective action for the trace 
operators $O^i$. In the large $N$ limit the dynamics of the operators 
$O^i$ is generated by the classical dynamics of the action $I(O^i)$. 
 
Of course $W[J^i]$ may equally be thought of as the Legendre transform 
of $I[O^i]$
\begin{equation}\label{eeonn}
W[J_i]=I[O^i] -O^iJ_i ,
\end{equation}
where $O^i$ is the function of $J^i$ obtained by solving the equation 
of motion 
\begin{equation}
\label{eeon}
\frac{\partial I}{\partial O^i}=J_i.
\end{equation}  

Now let us suppose that the action $S$  of the original large $N$ field 
theory is deformed by the addition of a multitrace term 
$S \rightarrow S+ P(O^i)$ where $P(O^i)$ is an arbitrary function of 
$O^i$. The effective action for this deformed 
theory is simply given by ${\tilde I}(O^i)$
\begin{equation}\label{eea}
{\tilde I}(O^i)= I(O^i)+ P(O^i).
\end{equation}
The generating function of correlators of the deformed theory is 
once again given by the Legendre transform 
\eqref{eeonn} with $I[O^i]$ replaced by ${\tilde I}[O^i]$.

\subsubsection{Bulk dual to multi trace deformations in regular and 
alternate quantization}

Consider a real scalar field propagating in $AdS_{d+1}$ according to the 
action 
\begin{equation} \label{actsct1} 
S=\frac{1}{2} \int d^{d+1}x \sqrt{g}\left(  \partial_\mu \phi 
\partial^\mu \phi + m^2 \phi^2  \right).
\end{equation}
It is well known that these scalars admit two distinct conformally 
invariant boundary conditions - sometimes referred to as 
alternate and standard quantization - in the mass range 
$- \left( \frac{d^2}{4}-1 \right) > m^2 > 
-\frac{d^2}{4}$. In this subsection we will review 
the very well known rules for the computation of correlation 
functions for scalars with alternate and standard boundary conditions.  

The action \eqref{actsct1} is ambiguous as it generically receives 
divergent contributions from the boundary, as we now explain. 
We use coordinates so that the metric of AdS space is given by 
\eqref{metric}. Near $z=0$ the general solution to the 
equation motion from \eqref{actsct1} takes the form 
\begin{equation}\label{falloff}
\phi=\frac{\phi_1 z^{\frac{d}{2}- \zeta}}{2 \zeta}  + 
\phi_2  z^{\frac{d}{2}+\zeta},
\end{equation}
where $\zeta$ is the positive root of the equation 
$\zeta^2= m^2+\frac{d^2}{4}$.  Let us cut of the action 
\eqref{actsct1} at a small value, $z_c$ of the 
coordinate $z$. Onshell \eqref{actsct1} 
evaluates to 
\begin{equation}\label{osa}
S =-\frac{1}{2} \int d^dx \frac{1}{z_c^{d-1}} \,\phi\, \partial_z \phi,
\end{equation}
where the integral is evaluated over the boundary surface $z=z^c$. 
It is easily verified that the action $S$ has a divergence proportional 
to $z_c^{2 zeta}$ when evaluated on the generic solution \eqref{falloff}. 
To cure this divergence we supplement \eqref{actsct1}  with a 
diffeomorphically invariant boundary action for the $d$ dimensional 
boundary field $\phi(z_c, x)$  
\begin{equation}\label{covcount}
\delta S= \frac{1}{2} 
\int d^{d}x \sqrt{g}\left( \frac{d}{2}-\zeta \right) \phi^2
\end{equation}
where, once again, the integral is taken over the boundary surface $z=z_c$ 
and $g$ is the induced metric on this boundary. It is easily verified that
\begin{equation}\label{osa}
S+ \delta S= -\frac{1}{2} \int d^d x  \phi_1(x) \phi_2(x).
\end{equation}
Regularity in the interior of $AdS$ relates $\phi_2$ to
$\phi_1$. The relationship is clearly linear and so takes the 
form 
\begin{equation}\label{pts}
\phi_2(x)= \int d^dx G(x-y) \phi_1(y).
\end{equation}
In the rest of this subsection we use abbreviated notation so that 
\eqref{osa} is written as $S=-\frac{1}{2} \phi_1 \phi_2$ and 
\eqref{pts} is written as $\phi_2=G\phi_1$. It follows that the onshell 
action is given by 
\begin{equation}\label{osa}
S=-\frac{1}{2} \phi_1 G \phi_1.
\end{equation}

In the case of alternate quantization the boundary action \eqref{osa}, 
thought of as a functional of the dynamical field $\phi_1= 
\lim_{z_c \to 0} \frac{\phi}{z_c^{\frac{d}{2}-\zeta}}$,  is identified with 
the single trace effective action $I[O]$ defined in \eqref{eeo}. 
The generator of correlators of this theory is obtained by coupling 
$\phi_1=\frac{\phi}{z_c^{\frac{d}{2}-\zeta}}$ to a source $J$:
\begin{equation}\label{osan}
S=-\frac{1}{2} \phi_1 G \phi_1 - J \phi_1.
\end{equation}
The resulting equation of motion for $\phi_1$ yields 
\begin{equation}\label{eomj}
G \phi_1=-J.
\end{equation}
Integrating out $\phi_1$ we find the action 
$$S=J G^{-1} J.$$
It follows that the two point function of the dual operator is 
$-G^{-1}$. 
It also follows from \eqref{eomj} that 
$$ \phi_2= -J.$$
in particular $\phi_2$ vanishes wherever $J$ vanishes. 
Consequently, alternate quantization is associated with the boundary 
condition $\phi_2=0$. 

The multi trace deformation $P(O)$ of the dual theory is implemented, in 
alternate quantization,  by adding the term  
$P(\phi_1)$to the boundary effective action 
\eqref{osa}, in perfect imitation of \eqref{eea}. Correlation functions of 
the deformed theory are obtained by the Legendre transform of this 
augmented boundary action. The resultant equation of motion is 
$G \phi_1 +J- P'(\phi_1)=0$ yields the bulk boundary conditions  
$$ \phi_2 +J - P'(\phi_1)=0.$$

In the case of regular quantization we supplement the action 
\eqref{osa} with an additional degree of freedom ${\tilde \phi}_2$
so that the full boundary action takes the form 
\begin{equation}\label{osam}
S=-\frac{1}{2} \phi_1 G \phi_1 + {\tilde \phi}_2 \phi_1 .
\end{equation}
The dynamical field $\phi_1$ is then integrated out using its equation 
of motion 
\begin{equation}\label{dfp} 
G \phi_1={\tilde \phi}_2.
\end{equation}
On shell, therefore ${\tilde \phi}_2= \phi_2$. The resultant action
\begin{equation}\label{osam}
S=\frac{1}{2} {\tilde \phi}_2 G^{-1} {\tilde \phi}_2 
\end{equation}
as a function of ${\tilde \phi_2}$ is identified with $I(O)$ 
in \eqref{eeo}. The generator of correlators of the theory is obtained 
by coupling ${\tilde \phi_2}$ to a source $J$ 
$$S=\frac{1}{2} {\tilde \phi}_2 G^{-1} {\tilde \phi}_2  - J {\tilde \phi}_2\, ,
$$
and then integrating this 
field  out according to its equations of motion. This allows us, in 
particular, to identify the two point function of the dual theory with 
$G$. Note also that the resultant equation of motion, 
$G^{-1} {\tilde \phi}_2=J$ implies
$$\phi_1=J,$$ 
so that $\phi_1$ vanishes wherever $J$ vanishes. 
In other words standard quantization is associated with the boundary 
condition $\phi_1=0$. The multitrace deformation $P(O)$ of the dual 
theory is implemented, in standard quantization, by adding 
$P({\tilde \phi_2})$ to the action \eqref{osam}. The resultant boundary
condition is 
$$\phi_1 -J +P'(\phi_2)=0.$$

\subsubsection{Marginal multitrace deformation with two scalar field in 
opposite quantization}\label{tsf}

Consider two scalar fields in $AdS_4$, $\phi$ and $\chi$, with $\phi$ 
quantized with alternate quantization and $\chi$ with regular quantization. 
In the compact notation defined in earlier subsection, the generating 
function 
of correlation function of the dual field theory deformed by double trace 
operator $\tan\theta_0 O_1 O_2$ is 
\begin{equation}\label{2fact}
S= -\half G \phi_1^2 - \half G \chi_1^2 + \chi_1\tilde{\chi}_2- J_1 \phi_1 
     - J_2 \tilde{\chi}_2 + \tan\theta_0\tilde{\chi}_2\phi_1.
\end{equation} 
The action is linear in ${\tilde \chi}_2$; the equation of motion for 
this field immediately yields 
\begin{equation}\label{fbc}
J_2= \frac{1}{\cos\theta_0}( \sin\theta_0\phi_1 + \cos\theta_0\chi_1 ).
\end{equation}
Using \eqref{fbc} to eliminate $\phi_1$ in favor of $\chi_1$, $S$
simplifies to a function of $\phi_1$. The resultant equation of 
motion yields 
\begin{equation}\label{p12bc}
J_1= -\frac{1}{\cos\theta_0}G(\cos\theta_0 \phi_1 - \sin\theta_0 \chi_1 ).
\end{equation}
Using $G \phi_1=\phi_2$ and $G\chi_1=\chi_2$, \eqref{p12bc} may be 
rewritten as 
\begin{equation}\label{p12bcn}
J_1= -\frac{1}{\cos\theta_0} (\cos\theta_0 \phi_2 - \sin\theta_0 \chi_2 ).
\end{equation}

Upon setting $J_1=J_2=0$, \eqref{fbc} and \eqref{p12bcn} express the 
boundary conditions of the trace deformed model. These boundary conditions
may, most succinctly be expressed as follows. Let us define new 'rotated' 
bulk fields 
$$ \phi'=\cos\theta_0 \phi - \sin\theta_0 \chi, ~~~
\chi'=\sin\theta_0\phi + \cos\theta_0\chi. $$
Note that the rotated fields have same bulk action as the original fields.
The boundary conditions \eqref{fbc} and \eqref{p12bcn} reduce to 
$$ \phi'_2=0, ~~~~\chi'_1=0.$$

In summary dual to the the double trace deformed field theory has 
the same action as well as boundary conditions for $\phi'$ and $\chi'$ 
as the dual to the undeformed theory had for $\phi$ and $\chi$. 
Despite this fact, the double trace deformed theory is {\it not} 
field redefinition equivalent to the original theory. This can be 
seen in many ways. Most simply, the full 
action \eqref{2fact} does not have a simple rotational invariance, and 
does not take a simple form when reexpressed in terms of $\phi'$ 
and $\chi'$. This lack of equivalence also shows itself up in the
generator of two point functions of the operators dual to $\phi'$
and $\chi'$.  This generating function is obtained by plugging \eqref{fbc}
and \eqref{p12bc} into \eqref{2fact}; we find  
\begin{equation}\label{corgen}
-S= -\cos^2\theta_0 \frac{J_1^2}{2 G} + \cos^2\theta_0 \frac{J_2^2 
G}{2} + \sin\theta_0 \cos\theta_0 J_1 J_2.
\end{equation}
The fact that $\theta_0$ does not disappear from \eqref{corgen} 
demonstrates the lack of equivalence of the trace deformed model 
from the trace undeformed  model ($\theta_0=0$). Note in particular 
that the double trace deformed theory has a contact cross two point function
$$ \langle O_\phi(x) O_\chi(y) \rangle = \sin\theta_0 \cos \theta_0 
\delta(x-y),$$
which is absent in the trace undeformed theory. On the other hand 
the direct correlators $\langle O_\phi(x) O_\phi(y) \rangle$ and 
$\langle O_\chi(x) O_\chi(y) \rangle$ have the same spacetime structure
in the deformed and undeformed theories, but have different 
normalizations.

\subsection{Gauging a $U(1)$ symmetry}\label{ggs}

Let us begin with a three dimensional CFT with a $U(1)$ global symmetry, generated by the current $J_i$, where $i$ is the three-dimensional vector index. This theory will be referred to as CFT${}_\infty$, as opposed to the theory obtained by gauging the $U(1)$ with Chern-Simons gauge field at level $k$, which we refer to as CFT${}_k$.
Suppose CFT${}_\infty$ is dual to a weakly coupled gravity theory in $AdS_4$. The global $U(1)$ current $J_i$ of the boundary CFT is dual to a gauge field $A_\mu$ in the bulk. The two-derivative part of the bulk action for the gauge field is
\ie
{1\over 4} \int {d^3\vec x dz\over z^4} F_{\mu\nu} F^{\mu\nu}
= \int d^3\vec x dz \,\left( {1\over 2}F_{zi}F_{zi} + {1\over 4}F_{ij}F_{ij} \right).
\fe
Working in the radial gauge $A_z=0$, we have
\ie
F_{zi} = \partial_z A_i, ~~~~F_{ij} = \partial_i A_j-\partial_j A_i.
\fe
Consider the linearized, i.e. free, equation of motion
\ie
(\partial_z^2 + \partial_j^2) A_i - \partial_i \partial_j A_j = 0,
\fe
together with the constraint
\ie\label{aconstraint}
\partial_z \partial_i A_i=0.
\fe
Near the boundary, a solution to the equation of motion has two possible asymptotic behaviors, $A_i\sim z+{\cal O}(z^2)$, or $A_i\sim 1+{\cal O}(z^2)$. Equivalently, they can be expressed in gauge invariant form as the magnetic boundary condition
\ie
\left.F_{ij}\right|_{z=0}= 0,
\fe
and the electric boundary condition
\ie
\left.F_{zi} \right|_{z=0}= 0,
\fe
respectively.
With the magnetic boundary condition, $A_\mu$ is dual to a $U(1)$ global current in the boundary CFT, i.e. CFT${}_\infty$. The family of CFT${}_k$, on the other hand, is dual to the same bulk theory with the mixed boundary condition (still conformally invariant)
\ie\label{mixbc}
\left. \left( {1\over 2}\epsilon_{ijk} F_{jk} + {i\A\over k} F_{zi} \right)\right|_{z=0}= 0.
\fe
Here $\A$ is a constant. It will be determined in terms of the two-point function of the current $J_i$.

Let us now solve the bulk Green's function with the mixed boundary condition. The bulk linearized equation of motion with a point source at $z=z_0$, after a Fourier transformation in the boundary coordinates $\vec x$, is
\ie\label{seceqn}
(\partial_z^2-p^2)A_i + p_i p_j A_j = \delta(z-z_0) \xi_i.
\fe
Due to the constraint (\ref{aconstraint}), the source $\xi_i$ is restricted by $p_i\xi_i=0$. The boundary condition is
\ie\label{Abc}
\left.\left(\epsilon_{ijk}p_j A_k + {\A\over k} \partial_z A_i \right)\right|_{z=0}=0.
\fe
Without loss of generality, let us consider the case $\vec p=(0,0,p)$, and assume $p=p_3>0$. The Green equation is now written as
\ie
&\partial_z^2 A_3 = 0,
\\
&(\partial_z^2-p^2) A_i = \delta(z-z_0) \xi_i ,~~~i=1,2,
\fe
and the boundary condition as
\ie
\partial_z A_3|_{z=0}=0,~~~ \left.\left( p \epsilon_{ij}A_j - {\A\over k} \partial_z A_i\right)\right|_{z=0}=0,~~~i=1,2.
\fe
The $z$-independent part of $A_3$ can be gauged away. We may then take the solution
\ie
& A_3 =0,
\\
& A_i = \theta(z-z_0) \left[g_i(p)+h_i(p) \right] e^{-p (z-z_0)} + \theta(z_0-z)\left[ g_i(p) e^{-p (z-z_0)} + h_i(p) e^{p(z-z_0)} \right],
\fe
where $g_i(p)$ and $h_i(p)$ obey
\ie
& -p(g_i+h_i) -(-pg_i+ph_i) = \xi_i.
\\
&\epsilon_{ij}(g_j e^{pz_0} + h_j e^{-pz_0}) + {\A\over k} (g_i e^{pz_0} - h_i e^{-pz_0}) = 0.
\fe
The solutions are
\ie
g_i = {e^{-2p z_0}\over 2(1+{\A^2\over k^2})p}\left[ (1-{\A^2\over k^2})\xi_i + 2{\A\over k}\epsilon_{ij}\xi_j \right],~~~~ h_i = -{\xi_i\over 2p}.
\fe
The nontrivial components of Green's function are thus given by
\ie
& G_{33} =0,
\\
& G_{ij} = {1\over 2p}\left[ e^{-p(z+z_0)} {(1-{\A^2\over k^2})\delta_{ij} + 2{\A\over k}\epsilon_{ij}\over 1+{\A^2\over k^2}}  \right] - {\delta_{ij}\over 2p} \bigg[ \theta(z-z_0) e^{-p(z-z_0)} + \theta(z_0-z) e^{p(z-z_0)} \bigg].
\fe
In particular, we find the change of the bulk Green's function due to the changing of the boundary condition,
\ie
G^{(k)}_{ij} - G^{(\infty)}_{ij} \equiv \Delta_{ij}(p,z,z_0)= {\A \over k p}  { \epsilon_{ij} -{\A\over k}\delta_{ij} \over 1+{\A^2\over k^2}} e^{-p(z+z_0)}.
\fe
The boundary to bulk propagator for $k=\infty$ can be obtained by taking $z_0\to 0$ limit on $z_0^{-1}G^{(\infty)}$, giving
\ie
& K_{33} = 0,
\\
& K_{ij} = - e^{-pz} \delta_{ij} .
\fe
We observe that $\Delta_{ij}$ factorizes into the product of two boundary to bulk propagators, $K(p,z)$ and $K(p,z_0)$, multiplied by
\ie\label{pth}
M_{ij}(p)={\A \over k p}  { \epsilon_{ij} -{\A\over k}\delta_{ij} \over 1+{\A^2\over k^2}} .
\fe
This is reminiscent of the change of scalar propagator due to boundary conditions \cite{Hartman:2006dy, Giombi:2011ya}.
So far we worked in the special case $p=p_3$. Restoring rotational invariance, (\ref{pth}) is replaced by
\ie
M_{ij}(p) &= {\A \over k |p|}  { \epsilon_{ijk}{p^k\over |p|} -{\A\over k}(\delta_{ij} - {p_ip_j\over p^2})\over 1+{\A^2\over k^2}} 
\\
&= {\A/k\over 1+{\A^2/ k^2}}\epsilon_{ijk} {p^k\over p^2} - {\A^2/k^2\over 1+{\A^2/k^2}} {\delta_{ij}-{p_ip_j\over p^2}\over |p|}.
\fe
In the boundary CFT, the change of boundary condition amounts to coupling the $U(1)$ current $J^i$ to a boundary gauge field $A_i$ at Chern-Simons level $k$. $M_{ij}(p)$ is proportional to the two-point function of $A_i$ in the Lorentz gauge $\partial_j A^j=0$. Namely,
\ie
\langle A_i(p) A_j(-q) \rangle = {32\over \widetilde N} M_{ij}(p) (2\pi)^3\delta^3(p-q) ,
\fe
where $\widetilde N$ is the overall normalization factor in the two-point function of the current $J_i$,
\ie
\langle J_i(p) J_j(-q)\rangle =- {\widetilde N |p|\over 32} \left( \delta_{ij} - {p_i p_j\over p^2} \right) (2\pi)^3 \delta^3(p-q).
\fe
Our convention is such that in the free theory $\widetilde N$ counts the total number of complex scalars and fermions. Note that here we are normalizing the current coupled to the Chern-Simons gauge field according to the convention for nonabelian gauge group generators, ${\rm Tr}(t^a t^b) = {1\over 2}\delta^{ab}$ for generators $t^a$, $t^b$ in the fundamental representation. This is also the normalization convention we use to define the Chern-Simons level $k$ (which differs by a factor of 2 from the natural convention for $U(1)$ gauge group).

To see this, note that the inverse of the matrix $M_{ij}$ in (\ref{pth}), restricted to directions transverse to $\vec p=p_3 \hat e_3$, is
\ie
(M_\perp^{-1})_{ij} = {kp\over \A} \epsilon_{ij} + \delta_{ij} p.
\fe
After restoring rotational invariance, this is
\ie
(M_\perp^{-1})_{ij} = {k\over \A} \epsilon_{ijk}p^k + \left(\delta_{ij} - {p_i p_j\over p^2}\right) |p|
\fe
which for $\A={\pi\over 8} \widetilde N$ precisely matches $32\widetilde N^{-1}$ times the kinetic term of the Chern-Simons gauge field plus the contribution to the self energy of $A_i$ from $\langle J_i(p) J_j(-p)\rangle_{CFT_\infty}$.

\section{Supersymmetric Chern-Simons vector models at large $N$}
\label{lags}

In this appendix, we review the Lagrangian of Chern-Simons vector models with various numbers of supersymmetries and/or superpotentials. The scalar potentials and scalar-fermion coupling resulting from the coupling to auxiliary fields in the Chern-Simons gauge multiplet and superpotentials can be expressed in terms of bosonic or fermionic singlets under the $U(N)$ Chern-Simons gauge group as double trace or triple trace terms. These can be matched with the change of boundary conditions in the holographically dual Vasiliev theories in $AdS_4$, described in section 4.

\subsection{$\cN=2$ theory with $M$ $\Box$ chiral multiplets}\label{N2Mch}

The action of the $\cN=2$ pure Chern-Simons theory in Lorentzian signature is
\ie
S^{\cN=2}_{CS}={k\over 4\pi}\int \text{Tr}(A\wedge dA+{2\over 3}A^3-\bar\chi\chi+2D\sigma),
\fe
where $\chi,\bar\chi$ and $D,\sigma$ are fermionic and bosonic auxiliary fields. The $M$ chiral multiplets in the fundamental representation couple to the gauge multiplet through the action
\ie
S_{m}&=\int\sum^{M}_{i=1}\left[D_\m\bar \phi^{i}D^\m \phi_{i}+\bar\psi^{i}({\slash\!\!\!\! D}+\sigma)\psi_{i}+\bar \phi^i(\sigma^2-D)\phi_i+\bar\psi^i\bar\chi\phi_i+\bar\phi^i\chi\psi_i-\bar F F\right].
\fe
We will focus on the matter coupling
\ie
{k\over 4\pi}\text{Tr}(-\bar\chi\chi+2D\sigma)+\int\sum^{M}_{i=1}\left[\bar\psi^{i}\sigma\psi_{i}+\bar \phi^i(\sigma^2-D)\phi_i+\bar\psi^i\bar\chi\phi_i+\bar\phi^i\chi\psi_i-\bar F F\right].
\fe
Integrating out the auxiliary fields, we obtain the scalar potential and scalar-fermion coupling,
\ie
V&={4\pi^2\over k^2}\bar \phi^i\phi_j\bar\phi^j\phi_k\bar\phi^k\phi_i+{4\pi\over k}\bar\phi^j\phi_i\bar\psi^i\psi_j+{2\pi\over k}\bar\psi^i\phi_j\bar\phi^j\psi_i.
\fe
For the purpose of comparing with vector models of other numbers of supersymmetries, it is useful to consider the $M=2$ case. Let us define bosonic and fermionic gauge invariant bilinears in the matter fields,
\ie
&\Phi_{+}^a=\bar\phi^i\phi_j (\sigma^a)^j{}_i,~~~~\Phi_{-}^a=\bar\psi^i\psi_j(\sigma^a)^j{}_i,~~~~\Psi^i{}_j=\bar\phi^i\psi_j,
\fe
where $\sigma^a=({\bf 1},\sigma^1,\sigma^2,\sigma^3)$. The non-supersymmetric theory with two flavors and without matter self-interaction $V$ would have had $SU(2)_b\times SU(2)_f$ flavor symmetry acting on the bosons and fermions separately. With respect to this symmetry, $\Phi_{+}^a$, $\Phi_{-}^a$ and $\Psi^i{}_j$ are in the representation $({\bf 1}\oplus{\bf 3},{\bf 1})$, $({\bf 1},{\bf 1}\oplus{\bf 3})$ and $({\bf 2},{\bf 2})$ respectively. Expressed in terms of the bosonic and fermionic singlets, $V$ can be written as
\ie \label{ntpot}
V=&{\pi^2\over 2k^2}\Phi^a_+\Phi^b_+\Phi^c_+\text{Tr}\left(\sigma^a\sigma^b\sigma^c\right)+{2\pi \over k}\Phi^a_+\Phi^a_-+{2\pi \over k}\bar\Psi^i{}_j\Psi^j{}_i.
\fe
Note that the $({\rm fermion~ singlet})^2$ terms is invariant under $SU(2)_b\times SU(2)_f$, whereas the $({\rm bosonic~singlet})^2$ term and the scalar potential explicitly break $SU(2)_b\times SU(2)_f$ to the diagonal flavor $SU(2)$.

Indeed, the boundary conditions of the conjectured holographic dual described in section 4.3.1 are such that the fermionic boundary condition (characterized by $\C$) is invariant under the $SO(4)\sim SU(2)_b\times SU(2)_f$ that rotates the four Grassmannian variables of supersymmetric Vasiliev theory, while the scalar boundary condition only preserve an $SU(2)\sim SO(3)$ subgroup.

\subsection{$\cN=1$ theory with $M$ $\Box$ chiral multiplets}

The $\cN=2$ theory in the previous section admits a one-parameter family of exactly marginal deformations that preserves $\cN=1$ supersymmetry. The matter coupling of this $\cN=1$ theory is given by
\ie\label{pN=1}
V=&{4\pi^2\omega^2\over k^2}\bar \phi^i\phi_j\bar\phi^j\phi_k\bar\phi^k\phi_i+{2\pi(1+\omega)\over k}\bar\phi^j\phi_i\bar\psi^i\psi_j+{2\pi\omega\over k}\bar\psi^i\phi_j\bar\phi^j\psi_i
\\
&+{\pi(\omega-1)\over k}(\bar\psi^i\phi_j\bar\psi^j\phi_i+\bar\phi^i\psi_j\bar\phi^j\psi_i),
\fe
where $\omega$ is a real deformation parameter. The ${\cal N}=2$ theory is given by $\omega=1$.

\subsection{The ${\cal N}=2$ theory with $M$ $\Box$ chiral multiplets and $M$ $\overline\Box$ chiral multiplets}

Now we turn to the ${\cal N}=2$ Chern-Simons vector model with an equal number $M$ of fundamental and anti-fundamental chiral multiplets. This model differs from the ${\cal N}=2$ theory with $2M$ fundamental chiral multiplets through the scalar-fermion coupling and scalar potential only.
The part of the Lagrangian that couples matter fields to the auxiliary fields in the gauge multiplet is given by
\ie
&{k\over 4\pi}\text{Tr}(-\bar\chi\chi+2D\sigma)+\sum^{M}_{i=1}\left[\bar\psi^{i}\sigma\psi_{i}+\bar \phi^i(\sigma^2-D)\phi_i+\bar\psi^i\bar\chi\phi_i+\bar\phi^i\chi\psi_i-\bar F F\right]
\\
&+\sum^{M}_{i=1}\left[-\tilde\psi^{i}\sigma\bar{\tilde\psi}_{i}+\tilde \phi^i(\sigma^2+D)\bar{\tilde\phi}_i-\tilde\psi^i\chi\bar{\tilde\phi}_i-\tilde\phi^i\bar\chi\bar{\tilde\psi}_i-\tilde F \bar{\tilde F}\right].
\fe
Integrating out the auxiliary fields, we obtain
\ie\label{d-term}
V_d=&{4\pi^2\over k^2}(\bar \phi^k\phi_i\bar\phi^i\phi_j\bar\phi^j\phi_k-\bar \phi^k\bar{\tilde\phi}_i\tilde\phi^i\bar{\tilde\phi}_j\tilde\phi^j\phi_k-\bar \phi^k\bar{\tilde\phi}_i\tilde\phi^i\phi_j\bar\phi^j\phi_k+\tilde \phi^k\bar{\tilde\phi}_i\tilde\phi^i\bar{\tilde\phi}_j\tilde\phi^j\bar{\tilde\phi}_k)
\\
&+{4\pi\over k}(\bar\phi^j\phi_i\bar\psi^i\psi_j-\bar\phi^j\bar{\tilde\psi}_i\tilde\phi^i\psi_j-\tilde\psi^j\phi_i\bar\psi^i\bar{\tilde\phi}_j+\tilde\psi^i\bar{\tilde\psi}_j\tilde\phi^j\bar{\tilde\phi}_i)
\\
&+{2\pi\over k}(\bar\psi^i\phi_j\bar\phi^j\psi_i-\bar\psi^i\bar{\tilde\phi}_j\tilde\phi^j\psi_i-\tilde\psi^i\phi_j\bar\phi^j\bar{\tilde\psi}_i+\tilde\psi^i\bar{\tilde\phi}_j\tilde\phi^j\bar{\tilde\psi}_i).
\fe

\subsection{The ${\cal N}=3$ theory with $M$ hypermultiplets}

The ${\cal N}=3$ Chern-Simons vector model with $M$ hypermultiplets can be obtained from the ${\cal N}=2$ theory described in the previous subsection by adding the superpotential \cite{Kao:1993gs, Gaiotto:2007qi}
\ie
W=-{k\over 8\pi}\tr\varphi^2+\tilde\Phi^i\varphi\Phi_i
\fe
where $\varphi$ is an auxiliary ${\cal N}=2$ chiral superfield. Integrating out $\varphi$, we obtain a quartic superpotential
\ie
W={2\pi\over k}(\tilde\Phi^i \Phi_j)(\tilde\Phi^j\Phi_i).
\fe
After integrating over the superspace, we obtain
\ie
\int d^2\theta~W+c.c.={2\pi\over k}\left[2\tilde\phi^i\phi_j(\tilde\phi^j F_i+\tilde F^j\phi_i+\tilde\psi^j\psi_i )+(\tilde\psi^i\phi_j+\tilde\phi^i\psi_j)(\tilde\psi^j\phi_i+\tilde\phi^j\psi_i)+c.c.\right].
\fe
Integrating out the auxiliary fields $F,\tilde F$, the $W$-term potential is
\ie
V_{w}=&{2\pi\over k}\left[2(\tilde\phi^i\phi_j)(\tilde\psi^j\psi_i)+(\tilde\psi^i\phi_j+\tilde\phi^i\psi_j)(\tilde\psi^j\phi_i+\tilde\phi^j\psi_i)+c.c\right]
\\
&+{16\pi^2\over k^2}(\bar\phi^j\bar{\tilde\phi}_i)(\tilde\phi^i\phi_k)(\tilde\phi^k\bar{\tilde\phi}_j)+{16\pi^2\over k^2}(\bar\phi^j\bar{\tilde\phi}_i)(\bar\phi^i\phi_k)(\tilde\phi^k\phi_j).
\fe
The total potential is given by the $D$-term plus $W$-term potentials:
\ie
V=V_d+V_w.
\fe
To make the $SO(3)$ R-symmetry manifest, we rewrite the potential in terms of the $SO(3)$ doublets:
\ie\label{hyper}
(\phi^{A}_i)=\begin{pmatrix}\phi_i\\ \bar{\tilde\phi}_i\end{pmatrix},~~~(\psi_{A,i})=\begin{pmatrix}\psi_i\\ \bar{\tilde\psi}_i\end{pmatrix}.
\fe
The $D$-term and $W$-term potentials are
\ie
V_d=&{4\pi^2\over k^2}\left[ (\bar\phi_1\phi^1)(\bar\phi_1\phi^1)(\bar\phi_1\phi^1)- (\bar\phi_1\phi^2)(\bar\phi_2\phi^2)(\bar\phi_2\phi^1)- (\bar\phi_1\phi^2)(\bar\phi_2\phi^1)(\bar\phi_1\phi^1)+ (\bar\phi_2\phi^2)(\bar\phi_2\phi^2)(\bar\phi_2\phi^2)\right]
\\
&+{4\pi\over k}\left[(\bar\phi_1\phi^1)(\bar\psi^1\psi_1)-(\bar\phi_1\psi_2)(\bar\phi_2\psi_1)-(\bar\psi^2\phi^1)(\bar\psi^1\phi^2)+(\bar\psi^2\psi_2)(\bar\phi_2\phi^2)\right]
\\
&+{2\pi\over k}\left[(\bar\psi^1\phi^1)(\bar\phi_1\psi_1)-(\bar\psi^1\phi^2)(\bar\phi_2\psi_1)-(\bar\psi^2\phi^1)(\bar\phi_1\psi_2)+(\bar\psi^2\phi^2)(\bar\phi_2\psi_2)\right],
\fe
and
\ie
V_{w}=&{2\pi\over k}\left[2(\bar\phi_2\phi^1)(\bar\psi^2\psi_1)+(\bar\psi^2\phi^1+\bar\phi_2\psi_1)(\bar\psi^2\phi^1+\bar\phi_2\psi_1)+c.c\right]
\\
&+{16\pi^2\over k^2}(\bar\phi_1\phi^2)(\bar\phi_2\phi^1)(\bar\phi_2\phi^2)+{16\pi^2\over k^2}(\bar\phi_1\phi^2)(\bar\phi_1\phi^1)(\bar\phi_2\phi^1).
\fe
We have also suppressed the flavor indices. The total potential can be written in a $SO(3)$ R-symmetry manifest way:
\ie
V=V_1+V_2+V_3,
\fe
where $V_1$ contains the double trace operator of the form $({\rm bosonic~singlet})^2$,
\ie
V_1
={4\pi\over k}(\bar\phi_A\phi^B)(\bar\psi^{A}\psi_B),
\fe
$V_2$ is the scalar potential in the form of a triple trace term,
\ie
V_2
&={16\pi^2\over 3k^2} (\bar\phi_A\phi^B)(\bar\phi_B\phi^C)(\bar\phi_C\phi^A)-{4\pi^2\over 3k^2} (\bar\phi_B\phi^C)(\bar\phi_A\phi^B)(\bar\phi_C\phi^A),
\fe\
$V_3$ is the double trace term of the form $({\rm fermionic~singlet})^2$,
\ie
V_3
=&-{2\pi\over k}(\bar\psi^{A}\phi_B)(\bar\phi^{B}\psi_A)+{4\pi\over k}(\bar\psi^{A}\phi_A)(\bar\phi^{B}\psi_B)+{2\pi\over k}(\bar\psi^{A}\phi_A)(\bar\psi^{B}\phi_B)+{2\pi\over k}(\bar\phi^{A}\psi_A)(\bar\phi^{B}\psi_B),
\fe
where $\phi_A,\psi^A$ are defined as
\ie
\phi_A=\phi^B\epsilon_{BA},~~~\psi^A=\epsilon^{AB}\psi_B,
\fe
and $\epsilon^{AB},\epsilon_{AB}$ are antisymmetric tensors with $\epsilon_{12}=\epsilon^{12}=1$. 

For reference in main text we will record the double trace part of the potential 
in $SO(3)$ vector notation. Let us define
\begin{equation}
\begin{split}
\Phi_+^a= \bar\phi_A \phi^B (\sigma^a)^{~A}_{B} ~~~&\Leftrightarrow ~~~ 
           \bar\phi_A \phi^B= \half \Phi_+^a (\bar\sigma^a)^{~B}_{A} \\
\Phi_-^a= \bar\psi^A \psi_B (\sigma^a)^{~B}_{A} ~~~&\Leftrightarrow ~~~ 
           \bar\psi^A \psi_B= \half \Phi_-^a (\bar\sigma^a)^{~A}_{B} \\
\Psi^a= \bar \phi_A \psi_B (\epsilon\sigma^a)^{AB} ~~~&\Leftrightarrow ~~~ 
           \bar \phi_A \psi_B= -\half \Psi^a (\sigma^a\epsilon)_{AB} \\
\bar \Psi^a= -\bar \psi^A\phi^B (\sigma^a\epsilon)_{AB} ~~~&\Leftrightarrow ~~~ 
           \bar\psi^A\phi^B= \half \bar\Psi^a (\epsilon\bar\sigma^a)^{AB} \\
\end{split}
\label{N3quidef}
\end{equation}
where 
$$ (\sigma^a)_A^{~B}= (\sigma^i, iI)_A^{~B}, ~~~
(\bar\sigma^a)_A^{~B}= (\epsilon(\sigma^a)^T\epsilon)_A^{~B}= (\sigma^a, -iI)_A^{~B},~~~ 
\epsilon^{12}=\epsilon_{12}=1 .
$$
Here $\sigma^i$ are Pauli sigma matrices. The a,b indices runs over 1,2,3,0. A,B runs over 1,2. 
$\Psi^a$ and $\bar\Psi^a$ transform under the as vectors of $SO(4)$ which under $SO(3)$ transform 
as singlet(a=0) and a vector(a=1,2,3) while $\phi^A, \psi_A$ transform as doublets of $SU(2)$.
\begin{equation}
\begin{split}
V_1=& \frac{2\pi}{k}\Phi^a_+ \Phi_-^b \eta_{ab},\\
V_3=& \frac{2\pi}{k}\left( \half \bar\Psi^a\Psi^b\delta^{ab}- 
2\bar\Psi^0\Psi^0- \bar\Psi^0\bar\Psi^0- \Psi^0\Psi^0 \right). \\
\end{split}
\label{N3pot}
\end{equation}

\subsection{A family of $\cN=2$ theories with a $\Box$ chiral multiplet and a $\overline\Box$ chiral multiplet}
If we deformed the superpotential in the above subsection as
\ie
W={2\pi\omega\over k}(\tilde\Phi^i \Phi_j)(\tilde\Phi^j\Phi_i),
\fe
the $\cN=3$ supersymmetry is broken to $\cN=2$. In this case, the potential is
\ie
V=V_1+V_2+V_3,
\fe
where $V_1$ contains the double trace operator of the form $({\rm bosonic~singlet})^2$,
\ie
V_1={4\pi\over k}\left[(\bar\phi_1\phi^1)(\bar\psi^1\psi_1)+(\bar\phi_2\phi^2)(\bar\psi^2\psi_2)+\omega(\bar\phi_2\phi^1)(\bar\psi^2\psi_1)+\omega(\bar\phi_1\phi^2)(\bar\psi^1\psi_2)\right],
\fe
$V_2$ is the scalar potential in the form of a triple trace term,
\ie
V_2=&{4\pi^2\over k^2}\left[ (\bar\phi_1\phi^1)(\bar\phi_1\phi^1)(\bar\phi_1\phi^1)- (\bar\phi_2\phi^1)(\bar\phi_1\phi^2)(\bar\phi_2\phi^2)- (\bar\phi_1\phi^2)(\bar\phi_2\phi^1)(\bar\phi_1\phi^1)+ (\bar\phi_2\phi^2)(\bar\phi_2\phi^2)(\bar\phi_2\phi^2)\right]
\\
&~~~~+{16\pi^2\omega\over k^2}(\bar\phi_1\phi^2)(\bar\phi_2\phi^1)(\bar\phi_2\phi^2)+{16\pi^2\omega\over k^2}(\bar\phi_1\phi^2)(\bar\phi_1\phi^1)(\bar\phi_2\phi^1),
\fe\
$V_3$ is the double trace term of the form $({\rm fermionic~singlet})^2$,
\ie
V_3=&{2\pi\over k}\left[(\bar\psi^1\phi^1)(\bar\phi_1\psi_1)-(\bar\psi^1\phi^2)(\bar\phi_2\psi_1)-(\bar\psi^2\phi^1)(\bar\phi_1\psi_2)+(\bar\psi^2\phi^2)(\bar\phi_2\psi_2)\right]
\\
&+{4\pi\over k}\left[-(\bar\phi_1\psi_2)(\bar\phi_2\psi_1)-(\bar\psi^2\phi^1)(\bar\psi^1\phi^2)\right]+{2\pi\omega\over k}\left[(\bar\psi^2\phi^1)(\bar\psi^2\phi^1)+2(\bar\phi_2\psi_1)(\bar\psi^2\phi^1)+(\bar\phi_2\psi_1)(\bar\phi_2\psi_1)\right.
\\
&\left.+(\bar\phi_1\psi_2)(\bar\phi_1\psi_2)+2(\bar\psi^1\phi^2)(\bar\phi_1\psi_2)+(\bar\psi^1\phi^2)(\bar\psi^1\phi^2)\right].
\fe

\subsection{The ${\cal N}=4$ theory with one hypermultiplet}

As shown by \cite{Gaiotto:2008sd}, ${\cal N}=3$ $U(N)_k$ Chern-Simons vector model 
with $M$ hypermultiplets can be deformed to an ${\cal N}=4$ quiver type Chern-Simons 
matter theory by gauging (a subgroup of) the flavor group with another ${\cal N}=3$ 
Chern-Simons gauge multiplet, at the opposite level $-k$. Here we will focus on the 
case where the entire $U(M)$ is gauged, so that the resulting ${\cal N}=4$ theory 
has $U(N)_k\times U(M)_{-k}$ Chern-Simons gauge group and a single bifundamental 
hypermultiplet. This ${\cal N}=4$ theory will still be referred to as a vector model, 
as we will be thinking of the 't Hooft limit of taking $N,k$ large and $M$ kept 
finite. As we have seen, turning on the finite Chern-Simons level for the flavor 
group $U(M)$ amounts to simply changing the boundary condition on the $U(M)$ vector 
gauge fields in the bulk Vasiliev theory.

The part of the Lagrangian that couples matter fields to the auxiliary fields 
in the gauge multiplet is given by
\ie
&{k\over 4\pi}\text{Tr}(-\bar\chi\chi+2D\sigma)-{k\over 4\pi}\text{Tr}(-\bar{\hat\chi}\hat\chi+2\hat D\hat\sigma)
\\
&+\left[\bar\psi\sigma\psi+\bar \phi(\sigma^2-D)\phi+\bar\psi\bar\chi\phi+\bar\phi\chi\psi-
\hat\sigma\bar\psi \psi+\left(\hat\sigma^2+\hat D\right)\bar \phi\phi-\bar\psi\phi\bar{\hat\chi}-
\hat\chi\bar\phi\psi-2 \hat\sigma\bar\phi\sigma\phi-\bar F F\right]
\\
&+\left[-\tilde\psi\sigma\bar{\tilde\psi}+\tilde \phi(\sigma^2+D)\bar{\tilde\phi}-
\tilde\psi\chi\bar{\tilde\phi}-\tilde\phi\bar\chi\bar{\tilde\psi}+
\hat\sigma\tilde\psi \bar{\tilde\psi}+\left(\hat\sigma^2-\hat D\right)\tilde \phi\bar{\tilde\phi}+
\bar{\hat\chi}\tilde\phi\bar{\tilde\psi}+\tilde\psi\bar{\tilde\phi}\hat\chi-
2\hat\sigma\tilde\phi\sigma\bar{\tilde\phi}-\tilde F \bar{\tilde F}\right],
\fe
where we suppressed the both $SU(N)$ and $SU(M)$ indices. 
Integrating out the auxiliary fields, we obtain the potential:
\ie\label{pN=4}
V=&{2\pi\over k}\bar\phi_A\phi^A\bar\psi^B\psi_B+
{4\pi^2\over 3k^2}(\bar\phi_A\phi^B\bar\phi_B\phi^C\bar\phi_C\phi^A+
\bar\phi_A\phi^A\bar\phi_B\phi^B\bar\phi_C\phi^C-2\bar\phi_B\phi^C\bar\phi_A\phi^B\bar\phi_C\phi^A)
\\
&+{2\pi\over k}\left(-\bar\psi^A\phi^B\bar\phi_B\psi_A+\bar\phi^A\psi^B\bar\phi_A\psi_B+
\bar\psi^A\phi^B\bar\psi_A\phi_B\right).
\fe
The complex scalar $\phi^A$ and the fermion $\psi_A$ transform as $(2,1)$ and 
$(1,2)$ under the $SO(4)=SU(2)\times SU(2)$ R-symmetry. The potential (\ref{pN=4}) 
is manifestly invariant under the R-symmetry. 

For reference to main text we now record the double trace part of this potential in $SO(4)$ 
vector notation. Using the definitions \eqref{N3quidef}, the (scalar singlet)$^2$ part($V_1$) 
and (fermion singlet)$^2$ part($V_3$) are given by
\begin{equation}
\begin{split}
V_1=& -\frac{2\pi}{k}\Phi^0_+ \Phi^0_- ,\\
V_2=& -\frac{\pi}{k} \left( \bar\Psi^a\Psi^a + \bar\Psi^a\bar\Psi^a + \Psi^a\Psi^a \right).
\end{split}
\label{N4dtpot}
\end{equation}

\subsection{$\cN=3$ $U(N_{k_1})\times U(M)_{k_2}$ theories with one hypermultiplet}\label{N31hypr}
The $\cN=4$ theory in the previous section sits in a discrete one parameter 
family of $\cN=3$ $U(N)_{k_1}\times U(M)_{k_2}$ theories with one hypermultiplet.
The potential can be written in an $SO(3)$ R-symmetry manifest way:
\ie
V=V_1+V_2+V_3,
\fe
where $V_1$ contains the double trace operator of the form $(\text{bosonic singlet})^2$,
\ie\label{lsp}
V_1
=&{4\pi\over k_1}\bar\phi_A\phi^B\bar\psi^A\psi_B+{2\pi\over k_2}\left[ 
    \bar\phi_A\phi^A\bar\psi_B\psi^B+2\bar\phi_A\phi^B\bar\psi^A\psi_B\right],
\fe
$V_2$ is the scalar potential in the form of triple trace term. $V_3$ is the double 
trace term of the form $(\text{fermionic singlet})^2$,
\ie\label{N3qp}
V_3=&{2\pi\over k_1}\left[-\bar\psi^{A}\phi_B\bar\phi^{B}\psi_A+
2\bar\psi^{A}\phi_A\bar\phi^{B}\psi_B+\bar\psi^{A}\phi_A\bar\psi^{B}\phi_B+
\bar\phi^{A}\psi_A\bar\phi^{B}\psi_B\right]
\\
&+{2\pi\over k_2}\left[2\bar\psi^A\phi^B\bar\phi_A\psi_B+
  \bar\psi^A\phi_B\bar\psi^B\phi_A+ 
  \bar\phi_A\psi^B\bar\phi_B\psi^A\right].
\fe
In the notation defined in \eqref{N3quidef} $V_1$ and $V_3$ becomes
\begin{equation}\label{N4vcpot}
\begin{split}
V_1=& \frac{2\pi}{k_1} \Phi^a_+\Phi^b_-\eta_{ab} 
        +\frac{2\pi}{k_2}\left( \Phi^0_+\Phi^0_- +\Phi^a_+\Phi^b_-\eta_{ab} \right), \\
V_3=& \frac{2\pi}{k_1} \left( \half \bar\Psi^a\Psi^b\delta^{ab} 
       -2\bar\Psi^0\Psi^0 -\bar\Psi^0\bar\Psi^0- \Psi^0\Psi^0 \right) 
       +\frac{2\pi}{k_2} \left( \bar\Psi^a\Psi^b\eta^{ab} +\half \bar\Psi^a\bar\Psi^b\eta_{ab} 
       +\half \Psi^a\Psi^b\eta^{ab} \right). \\
\end{split}
\end{equation}

\subsection{The $\cN=6$ theory}
The above $\cN=4$ theory can be generalized to a quiver $\cN=3$ theory with $\tilde n$ hypermultiplets by starting with the $\cN=3$ $U(N)_k$ Chern-Simons vector model with $\tilde nM$ hypermultiplets and only gauging the $U(M)$ subgroup, of the $U(\tilde nM)$ flavor group, at level $-k$ with another $\cN=3$ Chern-Simons gauge multiplet. The resulting theory has $SU(\tilde n)$ flavor symmetry. For generic value of $\tilde n$, the theory has $\cN=3$ sypersymmetry, but for $\tilde n=1,2$, the theory exhibits $\cN=4,6$ sypersymmetry, respectively. Let us focus on the $\tilde n=2$ case. The part of the Lagrangian that couples matter fields to the auxiliary fields in the gauge multiplet is given by
\ie
&{k\over 4\pi}\text{Tr}(-\bar\chi\chi+2D\sigma)-{k\over 4\pi}\text{Tr}(-\bar{\hat\chi}\hat\chi+2\hat D\hat\sigma)
\\
&+\left[\bar\psi_a\sigma\psi^a+\bar \phi_a(\sigma^2-D)\phi^a+\bar\psi_a\bar\chi\phi^a+\bar\phi_a\chi\psi^a-\hat\sigma\bar\psi_a \psi^a\right.
\\
&\left.~~~+\left(\hat\sigma^2+\hat D\right)\bar \phi_a\phi^a-\bar\psi_a\phi^a\bar{\hat\chi}-\hat\chi\bar\phi_a\psi^a-2 \hat\sigma\bar\phi_a\sigma\phi^a-\bar F_a F^a\right]
\\
&+\left[-\tilde\psi_{\dot a}\sigma\bar{\tilde\psi}^{\dot a}+\tilde \phi_{\dot a}(\sigma^2+D)\bar{\tilde\phi}^{\dot a}-\tilde\psi_{\dot a}\chi\bar{\tilde\phi}^{\dot a}-\tilde\phi_{\dot a}\bar\chi\bar{\tilde\psi}^{\dot a}+\hat\sigma\tilde\psi_{\dot a} \bar{\tilde\psi}^{\dot a}\right.
\\
&\left.~~~+\left(\hat\sigma^2-\hat D\right)\tilde \phi_{\dot a}\bar{\tilde\phi}^{\dot a}+\bar{\hat\chi}\tilde\phi_{\dot a}\bar{\tilde\psi}^{\dot a}+\tilde\psi_{\dot a}\bar{\tilde\phi}^{\dot a}\hat\chi-2 \hat\sigma\tilde\phi_{\dot a}\sigma\bar{\tilde\phi}^{\dot a}-\tilde F_{\dot a} \bar{\tilde F}^{\dot a}\right],
\fe
where $a,\dot a=1,2$ are the $SU(2)\times SU(2)$ indices. 
There is also an superpotential
\ie
W=-{2\pi\over k}\Tr(\tilde\Phi^{\dot a} \Phi^b \tilde\Phi_{\dot a}\Phi_b).
\fe
After integrating over the superspace, we obtain
\ie
\int d^2\theta~W+c.c.=&-{2\pi\over k}\left[2\tilde\phi^{\dot a}\phi^b(\tilde\phi_{\dot a} F_b+\tilde F_{\dot a}\phi_b+\tilde\psi_{\dot a}\psi_b )+(\tilde\psi^{\dot a}\phi^b+\tilde\phi^{\dot a}\psi^b)(\tilde\psi_{\dot a}\phi_b+\tilde\phi_{\dot a}\psi_b)+c.c.\right].
\fe
After integrating out all the auxiliary fields, the resulting potential can be written in a $SO(6)$ R-symmetry manifest way:
\ie
V=V_1+V_2+V_3,
\fe
where $V_1$ contains the double trace operator of the form $(\text{bosonic singlet})^2$,
\ie
V_1
=&-{2\pi\over k}(\bar\phi_{1a}\phi^{1a}\bar\psi^{2\dot b}\psi_{2\dot b}+\bar\phi_{1a}\phi^{1a}\bar\psi^{1b}\psi_{1b}+\bar\phi_{2\dot a}\phi^{2\dot a}\bar\psi^{2\dot b}\psi_{2\dot b}+\bar\phi_{2\dot a}\phi^{2\dot a}\bar\psi^{1b}\psi_{1b})
\\
&+{4\pi\over k}(\bar\phi_{2\dot a}\phi^{1b}\bar\psi^{2\dot a}\psi_{1b}+\bar\phi_{1b}\phi^{2\dot a}\bar\psi^{1b}\psi_{2\dot a}+\bar\phi_{1a}\phi^{1b}\bar\psi^{1a}\psi_{1b}+\bar\phi_{2\dot a}\phi^{2\dot b}\bar\psi^{2\dot a}\psi_{2\dot b})
\\
=&-{2\pi\over k}\bar\phi_{A}\phi^{A}\bar\psi^{B}\psi_{B}+{4\pi\over k}\bar\phi_{A}\phi^{B}\bar\psi^{A}\psi_{B}
\fe
where we have rewrite the potential in terms of the $SO(3)$ doublets (\ref{hyper}), and $A,B=(11,12,21,22)$ are the $SO(6)$ spinor indices. $V_2$ is the scalar potential in the form of triple trace term. $V_3$ is the double trace term of the form $(\text{fermionic singlet})^2$,
\ie\label{fpN=6}
V_3
=&{2\pi\over k}\left(\bar\psi^{A}\phi^{B}\bar\phi_{B}\psi_{A}-2\bar\psi^{A}\phi^{B}\bar\phi_{A}\psi_{B}\right)+{2\pi\over k}(\epsilon_{ABCD}\bar\psi^{A}\phi^{B}\bar\psi^{C}\phi^{D}+\epsilon^{ ABCD}\bar\phi_{A}\psi_{B}\bar\phi_{C}\psi_{D})
\fe
where $\epsilon_{11,12,21,22}=\epsilon^{11,12,21,22}=1$.

\subsection{$\cN=3$ $U(N)_{k_1}\times U(M)_{k_2}$ theories with two hypermultiplets}\label{N32hypr}
The $\cN=6$ theory in the previous section sits in a discrete one parameter family of $\cN=3$ $U(N)_{k_1}\times U(M)_{k_2}$ theories with two hypermultiplets. The superpotential of these theories are
\ie
W={2\pi\over k_1}\Tr(\tilde\Phi^{ a} \Phi_b \tilde\Phi^{ b}\Phi_a)+{2\pi\over k_2}\Tr(\tilde\Phi^{ a} \Phi_a \tilde\Phi^{ b}\Phi_b),
\fe
where $a,b=1,2$ are the $SU(2)$ flavor indices. The potential can be written in an $SO(3)$ R-symmetry and $SU(2)$ flavor symmetry manifest way:
\ie
V=V_1+V_2+V_3,
\fe
where $V_1$ contains the double trace operator of the form $(\text{bosonic singlet})^2$,
\ie
V_1=& {4\pi\over k_1}\bar\phi_{Aa}\phi^{Bb}\bar\psi^A_b\psi^a_B+{2\pi\over k_2}(\bar\phi_{Aa}\phi^{Aa}\bar\psi_{B b}\psi^{B b}+2\bar\phi_{Aa}\phi^{Ba}\bar\psi^A_b\psi^b_B)
\fe
where we have rewrite the potential in terms of the $SO(3)$ doublets (\ref{hyper}), and $A,B=1,2$ are the $SO(3)_R$ spinor indices. $V_2$ is the scalar potential in the form of triple trace term. $V_3$ is the double trace term of the form $(\text{fermionic singlet})^2$,
\ie\label{fpN=6}
V_3
=&{2\pi\over k_1}(\bar\psi^{Aa}\phi^{Bb}\bar\phi_{Bb}\psi_{Aa}-2\bar\psi^{A a}\phi^{B b}\bar\phi_{A b}\psi_{B a})+{2\pi\over k_1}\epsilon_{AB}\epsilon_{CD}\bar\psi_a^A\phi^{Bb}\bar\psi^{C}_b\phi^{D a}+{2\pi\over k_1}\epsilon^{AB}\epsilon^{CD}\bar\phi_{Aa}\psi_{B}^b\bar\phi_{Cb}\psi_D^{ a}
\\
&+{4\pi\over k_2}\bar\psi^{A}_a\phi^{Ba}\bar\phi_{Ab}\psi_{B}^b+{2\pi\over k_2}\epsilon_{AD}\epsilon_{CB}\bar\psi_a^A\phi^{Ba}\bar\psi^{C}_b\phi^{D b}+{2\pi\over k_2}\epsilon^{AD}\epsilon^{CB}\bar\phi^{ a}_A\psi_{aB}\bar\phi^b_{C }\psi_{Db}.
\fe

Now we record the double trace parts of the potential in vector notation of 
$SO(3)_R \times SU(2)_{flavor}$ symmetry. Let us define
\begin{equation}
\begin{split}
\Phi_+^{Ii} = \bar\phi_{Aa}\phi^{Bb}(\sigma^I)^A_{~B}(\sigma^i)^a_{~b} ~~~&\Leftrightarrow ~~~ 
  \bar\phi_{Aa}\phi^{Bb} = \frac{1}{4}\Phi_+^{Ii}(\sigma^I)^B_{~A}(\sigma^i)^b_{~a} \\
\Phi_-^{Ii} = \bar\psi^{A}_a\psi_B^{b}(\sigma^I)^B_{~A}(\sigma^i)^a_{~b} ~~~&\Leftrightarrow ~~~ 
  \bar\psi^{A}_a\psi_B^{b} = \frac{1}{4}\Phi_-^{Ii}(\sigma^I)^A_{~B}(\sigma^i)^b_{~a} \\
\Psi^{Ii} = \bar\phi_{Aa}\psi_B^{b}(\sigma^I\epsilon)^{AB}(\sigma^i)^a_{~b} ~~~&\Leftrightarrow ~~~ 
  \bar\phi_{Aa}\psi_B^{b} = -\frac{1}{4}\Psi^{Ii}(\epsilon\sigma^I)_{AB}(\bar\sigma^i)^b_{~a} \\
\bar\Psi^{Ii} = -\bar\psi^{A}_a\phi^{Bb}(\epsilon\bar\sigma^I)_{AB}(\bar\sigma^i)^a_{~b} ~~~&\Leftrightarrow ~~~ 
  \bar\psi^{A}_a\phi^{Bb} = -\frac{1}{4}\bar\Psi^{Ii}(\bar\sigma^I\epsilon)_{AB}(\sigma^i)^b_{~a} \\
\end{split}
\label{N36def}
\end{equation}
Here both set of indices I,J as well i,j run over 1,2,3,0. I,J are the vector indices of $SO(3)_R$
while i,j are vector indices of $SU(2)_{flavor}$. The 0 component corresponds to the singlet while 1,2,3 
represents the vector part. In this notation the double trace potential part of the becomes 
\begin{equation}
\begin{split}
V_1=& \frac{\pi}{k_1}\Phi_+^{Ii}\Phi_-^{Jj}\eta^{IJ}\eta_{ij} -\frac{2\pi}{k_2}\Phi_+^{I0}\Phi_-^{J0}\eta^{IJ}, \\
V_3=& \frac{2\pi}{k_1} \left( -\frac{1}{4}\bar\Psi^{Ii}\Psi^{Jj}\delta^{IJ}\delta^{ij} 
     +\half \bar\Psi^{Ii}\Psi^{Jj}\eta^{IJ}\delta^{ij} +\half \left(\bar\Psi^{0i}\bar\Psi^{0j}\eta_{ij} 
     +\Psi^{0i}\Psi^{0j}\eta_{ij} \right) \right) \\
    & +\frac{2\pi}{k_2} \left( \bar\Psi^{I0}\Psi^{J0}\eta^{IJ} +\half \bar\Psi^{I0}\bar\Psi^{J0}\eta^{IJ}
     +\half \Psi^{I0}\Psi^{J0}\eta^{IJ} \right). \\
\end{split}
\label{N36pot}
\end{equation}
The double potentials for ${\cal N}=6$ theory is obtained from \eqref{N36pot} on setting $k_2=-k_1=-k$. 

\section{Argument for a Fermionic double trace shift}\label{fermdt}

In this Appendix compare the boundary conditions and Lagrangian for the 
fixed line of ${\cal N}=1$ theories to argue for the effective shift of 
fermionic boundary conditions induced by the Chern Simons term. 

Let us use the notation ${\bar \phi} \psi =\Psi$ and 
$\bar \psi \phi= {\bar \Psi}$ for field theory single trace operators. 
We know that a double trace deformation proportional to 
$(\Psi +{\bar \Psi})^2$ is dual to fermion boundary condition \eqref{nobc} 
with $\alpha  \propto P_{\psi_1}$. On the other hand the double trace 
deformation $(i\Psi -i{\bar \Psi})^2$ is dual to the fermion boundary 
condition with $\alpha  \propto P_{\psi_2}$. Now in the zero potential 
theory  ($w=-1$) the relevant terms in \eqref{pN=1} are
$$-\frac{2 \pi}{k} \left( \Psi \Psi + {\bar \Psi} {\bar \Psi}
+ {\Psi} {\bar \Psi} \right),$$
 while $\alpha= \theta_0 P_{\psi_2}$.
At the ${\cal N}=2$ point, on the other hand, the fermion double trace term is
$$+\frac{2 \pi}{k} {\Psi} {\bar \Psi} $$
while $\alpha= \theta_0 (P_{\psi_1} + P_{\psi_2})$.
Subtracting these two data points we conclude that the double trace 
deformation by 
$$\frac{2 \pi}{k} \left( \Psi + {\bar \Psi} \right)^2$$
is dual to a boundary condition deformation with 
$\alpha= \theta_0 P_{\psi_1}$.
By symmetry it must also be that the double trace deformation by 
$$-\frac{2 \pi}{k} \left( \Psi - {\bar \Psi} \right)^2$$
is dual to a boundary condition deformation with 
$\alpha= \theta_0 P_{\psi_2}$.
Adding these together, it follows that a double trace deformation by 
$$\frac{8 \pi}{k} {\bar \Psi}\Psi$$
is dual to  the boundary condition deformation with 
$\alpha= \theta_0 (P_{\psi_1} + P_{\psi_2})$. 
But the ${\cal N}=2$ theory with this boundary condition has a double 
trace potential equal only to 
$$\frac{2 \pi}{k} {\bar \Psi}\Psi.$$
For consistency, it must be that the Chern Simons interaction itself induces
a change in fermion boundary conditions equal to that one would have 
obtained from a double trace deformation 
\begin{equation}\label{csshft}
-\frac{6 \pi}{k} {\bar \Psi}\Psi.
\end{equation}

\section{Two-point functions in free field theory}\label{fft}

Consider the action for free $SU(N)$ theory of a boson and a fermion in 
the fundamental representation, in flat 3 dimensional euclidean space
\begin{equation}\label{action}
S = \int d^3x \left( \partial_{\mu}\bar{\phi} \partial_{\mu}\phi 
    + \bar{\psi} \sigma^{\mu}\partial_{\mu}\psi \right)
\end{equation}
where the $SU(N)$ in indices are suppressed and will continue to be in what 
follows.
The Green's functions for the scalar and fermions are given by 
\begin{equation}\begin{split}
G_s(x) &= \langle \bar{\phi}(x)\phi(0) \rangle = \frac{1}{4\pi|x|} \\
G_f(x) &= \langle \bar{\psi}(x)\psi(0) \rangle = \frac{x.\sigma}{4\pi|x|^3} \\
\end{split}
\label{gsgf}
\end{equation}

Let us define the 'Single Trace' operators 
\begin{equation} \label{opd}
\begin{split}
\Phi_+ = \bar{\phi}\phi, ~~~ \Phi_- = \bar{\psi}\psi, ~~~ 
\Psi = \bar{\phi}\psi, ~~~ \bar{\Psi} = \bar{\psi}\phi, ~~~
J_B^\mu=i {\bar \phi} \partial^\mu \phi -\partial^\mu {\bar \phi} \phi, ~~~
J_F^\mu=i {\bar \psi} \sigma^\mu \psi.
\end{split}
\end{equation}
In the free theory 
\begin{equation}
\begin{split}
\langle \Phi_+(x) \Phi_+(0) \rangle &= \frac{N}{(4\pi)^2 x^2}, \\
\langle \Phi_-(x) \Phi_-(0) \rangle &= \frac{2N}{(4\pi)^2 x^4}, \\
\langle \Psi(x) \bar{\Psi}(0) \rangle  &= \frac{N (x.\sigma)}{(4\pi)^2 x^4}\\
J_B^\mu(x) J_B(0)^\nu&= \frac{N}{8 \pi^2} 
\frac{\delta^{\mu\nu}-\frac{2 x^\mu x^\nu}{x^2} }{x^4}\\
J_F^\mu(x) J_F(0)^\nu&= \frac{N}{8 \pi^2} 
\frac{\delta^{\mu\nu}-\frac{2 x^\mu x^\nu}{x^2} }{x^4}\\
\end{split}
\label{dttpf}
\end{equation}

\section{Corrections at large $A$}\label{one}

The expression for $T_c$ presented in \eqref{ptt} receives corrections in a 
power series expansion in $\frac{1}{A}$. In this Appendix we compute the 
first correction to the expression for the second phase transition temperature
presented in \eqref{ptt} at small $\frac{1}{A}$. 

\eqref{ptt} receives corrections once we take into account 
the fact that the $V$ eigenvalue distribution is not quite a delta function 
in the neighborhood of the phase transition. 
To compute the leading correction to eigenvalue distribution of V-matrices 
we substitute 
$$\frac{1}{N}\Tr U^n = \frac{1}{N}\Tr U^{-n} = \frac{F(x^n)}{A} $$ 
for odd $n$, and 
$$\frac{1}{N}\Tr U^n = \frac{1}{N}\Tr U^{-n} = \frac{F_B(x^n) -F_F(x^n)}{A} $$ 
for even $n$
(see \eqref{lrgAldngU}).  It follows that the effective matrix integral for 
V-matrices is given by 
\begin{equation}\label{lrgAldngZV}
Z=\int DV \exp \left[ \frac{N}{A} \sum_{n=1}^n 
\frac{\left( F_B(x^n) +(-1)^n F_F(x^n) \right)^2  }{n} 
\left( \Tr V^n  + \Tr V^{-n} \right) \right].
\end{equation}
The saddle point equation for this model is 
\begin{equation}\label{}
2 \sum_{n=1}^{\infty}\left( F_B(x^n) +(-1)^n F_F(x^n) \right)^2 \sin(n\alpha) ={\cal P}v\int d\beta \cot \left( \frac{\alpha-\beta}{2} \right).
\end{equation}
To leading order in $\frac{1}{A}$ the V-eigenvalues are clumped into a delta function around zero. To first subleading 
order we expect that the eigenvalues will spread but only in a small region around zero and vanishes outside. Since all the 
eigenvalues are small the above saddle point equation reduces to hermitian wigner model
\begin{equation}\label{}
\left( \sum_{n=1}^{\infty}n \left( F_B(x^n) +(-1)^n F_F(x^n) \right)^2 \right) \alpha ={\cal P}v\int d\beta \frac{\rho(\beta)}{\alpha - \beta} .
\end{equation}
The solution to the above Wigner model is 
\begin{equation}\label{}
\rho_v(\alpha) = \frac{2}{a^2} \sqrt{a^2 - \alpha^2}~, ~~~ a^2 = \frac{2}{\sum_{n=1}^{\infty} n \left( F_B(x^n) +(-1)^n F_F(x^n) \right)^2}.
\end{equation}
Using this one can compute 
\begin{equation}
\frac{1}{M}\Tr V^n = \frac{1}{M}\Tr V^{-n} = \frac{2}{an}J_1(an),
\end{equation}
where $J_1(x)$ is Bessel function.
Substituting these into \eqref{mint} and using saddle point approximation one gets the corrected eigenvalue distribution 
for U-matrices to be 
\begin{equation}\label{}
\rho_{u}(\theta)= \frac{1}{2 \pi} \left( 1 + \sum_{n=1}^{\infty} 
\frac{4  \left( F_B(x^n) +(-1)^n F_F(x^n) \right) 
J_1(an)}{an} \cos n \theta \right).
\end{equation}
At leading order at high temperatures we substitute 
$x \rightarrow 1-\frac{1}{T}$. The leading correction to the 
$U$ eigenvalue distribution (from the finite width of the $V$ eigenvalue 
distribution) is given by  
\begin{equation}\label{}
\delta \rho_u(\theta) \rightarrow \frac{1}{2\pi A} \left[ 64 T^2\sum_{n=1}^{\infty} \frac{1}{(2n-1)^2} \left( \frac{J_{1}(an)}{an}  -\frac{1}{2} \right) 
\cos n\theta \right],~~~
a \rightarrow \frac{1}{\sqrt{112\zeta(3)}T^2}.
\end{equation}
where $\zeta$ is the Riemann zeta function with $\zeta(3)$ = 1.202. 

The shift in the eigenvalue distribution evaluated at $\pi$ is given, to 
leading order in large $T$, by 
$$-\frac{1}{2\pi A} 32 T^2 a \int_0^\infty \frac{dx}{x^2} 
\left( \frac{J_1(x)}{x}-\frac{1}{2} \right) . $$
This results is a shift of the phase transition temperature (about the 
result \eqref{ptt} at leading order at large $A$)   
$$\delta T_c^2=-\frac{8}{\sqrt{112 \zeta(3)}} \int_0^\infty \frac{dx}{x^2} 
\left( \frac{J_1(x)}{x}-\frac{1}{2} \right).$$
Thus the finite width of the $V$ eigenvalue distribution gives rise to 
a fractional correction of order $\frac{1}{A}$  second 
phase transition temperature.

\section{Truncated toy matrix model including interaction effects}\label{two}

In this Appendix we study the toy model 
\begin{equation}\begin{split}\label{ztoyint}
Z = \int DU DV \exp \bigg[ -F(x) & \left( \Tr U \Tr V^{-1} +  \Tr V \Tr U^{-1} \right) - 
   a \Tr U \Tr U^{-1} \Tr V \Tr V^{-1} \\ &- b \left( (\Tr U)^2 (\Tr V^{-1})^2 + (\Tr U^{-1})^2 (\Tr V)^2 \right) \bigg]
\end{split}
\end{equation}
in a neighborhood of $F(x)=1$, with $a$ and $b$ taken to be small.

The saddle point equations for the for U and V eigenvalues are
\begin{equation}\begin{split}\label{sdltoy}
{\cal P}v \int d\lambda_1 \rho_u(\lambda_1) \cot \left( \frac{\lambda-\lambda_1}{2} \right) + 
\frac{2\chi_1}{A} \bigg( F(x)+(a+2b)\rho_1\chi_1 \bigg) \sin \lambda = 0 , \\
{\cal P}v \int d\alpha_1 \rho_v(\alpha_1) \cot \left( \frac{\alpha-\alpha_1}{2} \right) + 
\frac{2\rho_1}{A} \bigg( F(x)+(a+2b)\rho_1\chi_1 \bigg) \sin \alpha = 0   .      
\end{split}
\end{equation}
These equations are of the Gross-Witten-Wadia form with the 
Gross-Witten-Wadia coupling dependent on the 
$\rho_1$ and $\chi_1$ themselves. We now search 
for self consistent solutions to these equations.

\subsection{$U$ flat, $V$ flat}
we see that the $\rho_n$ = $\chi_n$ = 0 for all n, is always a solutions.

\subsubsection{$U$ flat, $V$ wavy or vice-versa}
Substituting either $\rho_1$ or $\chi_1$ to zero we see that the other one is necessarily zero. Thus 
flat-wavy or wavy-flat is not a solution.

\subsection{$U$ wavy, $V$ wavy:}
In this case we will have
\begin{equation}\begin{split}\label{spe}
(\lambda^{(u)}_{GW})^{-1} = \rho_1 = \frac{\chi_1}{A} \left( F(x) + (a+2b) 
\rho_{-1} \chi_1 \right) , \\
(\lambda^{(v)}_{GW})^{-1} = \chi_1 = A \rho_1 \left( F(x) + (a+2b) 
\rho_1 \chi_{-1} \right).
\end{split}
\end{equation}
These equations may be used solve for $\chi_1$ and 
$ \rho_{-1}$; without loss of generality we may choose 
$\chi_1$ and $\rho_1$ each to be real so that $\chi_1=\chi_{-1}$ and 
$\rho_1=\rho_{-1}$. We find 
\begin{equation}
\frac{\chi_1}{A} =  \rho_1 = \sqrt{ \frac{1-F(x)}{A(a+2b)}}.
\end{equation}
When $a+2b$ is positive this solution only makes sense for $F(x)\leq 1$. 
On the other hand when $a+2b$ is negative, the solution only makes 
sense for $F(x) \geq 1$.

Consistency of the solution(positivity of eigenvalue density distribution) further requires 
\begin{equation}
\chi_1 \leq \frac{1}{2},~~~~~~  \rho_1 \leq \frac{1}{2} .
\end{equation}
As $\chi_1 \geq \rho_1$ the first of these two conditions is stronger. 
When $a+2 b$ is positive this condition amounts to the requirement that 
$$F(x) \geq 
1-\frac{a+2 b}{4 A}.$$
When $a+2b$ is negative this condition amounts to the requirement that 
$$F(x) \leq 1 -\frac{a+2 b}{4  A}.$$

In summary, when $a+2b$ is positive the wavy-wavy solution exists for 
$$1-\frac{a+2 b}{4 A} \leq F(x) \leq 1.$$
At the lower end of this range the $V$ eigenvalue distribution is on the 
border of being gapped, while at the upper end of this range the $U$ 
and $V$ eigenvalue distributions are both flat.

When $a+2b$ is negative, on the other hand,  the wavy-wavy solution exists for 
$$1\leq F(x) \leq 1-\frac{a+2 b}{4 A} .$$
At the lower end of this range the $V$ eigenvalue distribution becomes flat, 
while at the upper end of this range the $V$ eigenvalue distribution
is at the edge of being gapped.

\subsection{$U$ wavy, $V$ clumped}
In this case we have 
\begin{equation}
 \begin{split}
 \rho_1 &= (\lambda^{(u)}_{GW})^{-1} = \frac{\chi_1}{A} \left[ F(x) + (a+2b) \rho_1 \chi_1 \right], \\
 \chi_1 &= 1 - \frac{\lambda^{(v)}_{GW}}{4} = 1 - \frac{1}{4 A \rho_1 [ F(x) + (a+2b) \rho_1 \chi_1 ]} .
 \end{split}
\end{equation}
We may solve for $\rho_1$ in terms of $\chi_1$ and then obtain an equation for
$\chi_1$ as follows 
\begin{equation}
 \begin{split}
  \rho_1 &= \frac{-\chi_1 F(x)}{-A + (a+2b)\chi_1^2} , \\
  \chi_1 &= 1 + \frac{(-A + (a+2b)\chi_1^2)^2}{4 A^2 \chi_1 F(x)^2}.
 \end{split}
\end{equation}
Again consistency of solution requires 
\begin{equation}
 \begin{split}
  \rho_1 \leq \half ~~~{\rm and }~~~ \chi_1 \geq \half.
 \end{split}
\end{equation}
The second condition is satisfied if and only if 
\begin{equation}
 F(x) \geq  1 -\frac{a + 2b}{4 A} .
\end{equation}
When this inequality is saturated, the $V$ eigenvalue distribution 
is on the border between wavy and clumped. The first condition 
is saturated at a higher temperature when the $U$ eigenvalue 
distribution first begins to clump (i.e. near the second phase transition of 
the free model). As the quartic interaction are not particularly important 
for this transition, we do not study this transition in detail. 

It is possible to verify that solutions of the $U$ clumped $V$ wavy form 
do not exist. As mentioned above, clumped-clumped solutions do exist, but 
the quartic interaction terms do not play an important role in determining 
their properties, and we do not consider them further here.

\subsection{Summary}

The quartic interaction terms of this subsection qualitatively modify the 
nature of the first phase transition of the free theory.

When $a+2b$ is positive the flat-flat configuration is the only solution 
to the saddle point equations when $F(x) < 1-\frac{a+2 b}{4 A}$. At this 
temperature two new solutions are nucleated. The first is a wavy wavy 
solution is a local maximum and so is unstable throughout the range of 
its existence. The second is a wavy-clumped solution and is locally solution. 
At $F(x)=$ the free energy of the wavy-clumped solution decrease below that 
of the flat-flat solution and the system undergoes a first order phase 
transition. At the higher temperature $F(x)=1$ the wavy-wavy solution merges 
with the flat flat solution and ceases to exist thereafter. At higher 
temperatures the flat-flat solution is unstable and the wavy-clumped solution 
is the unique stable saddle point. At still higher temperatures this saddle
point undergoes a 3rd order phase transition to the clumped-clumped saddle.

When $a+2b$ is negative the flat-flat configuration is the only solution 
to the saddle point equations when $F(x)<1$. At this 
temperature the flat-flat saddle goes unstable, but a wavy-wavy 
solution is nucleated,  and is stable at higher temperatures. 
The system undergoes a second order phase transition (from the flat-flat 
saddle to the wavy-wavy saddle) at $F(x)=1$. At $F(x) = 1-\frac{a+2 b}{4 A}$ (
recall this is a higher temperature than $F(x)=1$ because $a+2 b$ is negative)
the wavy-wavy saddle turns into a wavy-clumped saddle through a third 
order phase transition. At still higher temperatures 
the wavy-clumped saddle point undergoes a 3rd order phase transition to the 
clumped-clumped saddle.

\bibliographystyle{JHEP}

\bibliography{abj}

\end{document}